%% file: Caching-and-coded-multicasting-with-random-demands_v12.tex
\documentclass[11pt,draftclsnofoot,peerreviewca,letterpaper,onecolumn]{IEEEtran}

\usepackage{cite}
\usepackage{graphicx,color,epsfig,rotating,psfrag}
\usepackage{amsfonts,amsmath,amssymb,bbm}
\usepackage{algorithm,algorithmic}
\usepackage{subfigure}
\usepackage{mdwtab}
\usepackage{placeins}
\usepackage[latin1]{inputenc}
\usepackage{mathrsfs}
\usepackage{empheq}
\usepackage{soul}

\input{macros}

\newtheorem{defn}{Definition}%[section]
\newtheorem{example}{Example}%[section]
\newtheorem{theorem}{Theorem}%[section]
\newtheorem{lemma}{Lemma}%[section]
\newtheorem{corollary}{Corollary}%[section]
\begin{document}

\title{Order-Optimal Rate of Caching and Coded Multicasting with Random Demands}

\author{Mingyue Ji,~\IEEEmembership{Student Member,~IEEE}, 
Antonia M. Tulino,~\IEEEmembership{Fellow,~IEEE},\\
Jaime Llorca,~\IEEEmembership{Member,~IEEE}, 
Giuseppe Caire,~\IEEEmembership{Fellow,~IEEE}, 
\thanks{Mingyue Ji and Giuseppe Caire are with the Department of Electrical Engineering,
University of Southern California, Los Angeles, CA 90089, USA. Antonia M. Tulino and Jaime Llorca are with Alcatel Lucent, Bell labs, Holmdel, NJ, USA. (e-mail: \{mingyuej, caire\}@usc.edu, \{a.tulino, jaime.llorca\}@alcatel-lucent.com).}
\thanks{A short version of this work was presented at ISWCS 2014, Barcelona, August 26-29, 2014.}
\thanks{The work of Mingyue Ji and Giuseppe Caire was partially supported by the VAWN project (funded by Intel, Cisco and Verizon) and by the NSF Grant CCF 1161801.}
}

%% Create the title:
\maketitle

\newpage

\begin{abstract}
We consider the canonical {\em shared link network} formed by a source node, hosting a library of $m$ information 
messages (files), connected via a noiseless common link to $n$ destination nodes (users), 
each with a cache of size $M$ files. Users request files at random and independently, according to a given a-priori demand 
distribution $\qv$. A coding scheme for this network consists of a caching placement (i.e., a mapping of the library files into the user caches) 
and delivery scheme (i.e., a mapping for the library files and user demands into a common multicast codeword) such that, after 
the codeword transmission, all users can retrieve their requested file. The rate of the scheme is defined as the {\em average} codeword length normalized with respect 
to the length of  one file, where expectation is taken over the random user demands. 
For the same shared link network, in the case of deterministic demands, the optimal min-max rate 
has been characterized within a uniform bound, independent of the network parameters. 
In particular,  fractional caching (i.e., storing file segments) and using linear network coding has been shown to provide a 
min-max rate reduction proportional to $1/M$ with respect to standard schemes such as 
unicasting or ``naive'' uncoded multicasting.  
The case of random demands was previously considered by applying the same order-optimal min-max  
scheme separately within groups of files requested with similar probability. 
However, no order-optimal guarantee was provided for random demands under the average rate performance criterion.  
In this paper, we consider the random demand setting and provide general achievability and converse results.
In particular, we consider a family of schemes that combine random fractional caching according to a probability distribution $\pv$
that depends on the demand distribution $\qv$, with a linear coded delivery scheme based on 
chromatic number index coding. For the special but relevant case where $\qv $ is a Zipf distribution with parameter $\alpha$, 
we provide a comprehensive characterization of the order-optimal rate for all regimes of the system parameters $n,m,M$, $\alpha$. 
We complement our scaling law analysis with numerical results that confirm the superiority of our schemes with respect to previously 
proposed schemes for the same setting. 
\end{abstract}


\newpage

%%%%%%%%%%%%%%%%%%%%%%%%%%%%%%%%%%%%%%%%%%%%%%%%%%%%%%%%
\section{Introduction} \label{sec:intro}

Content distribution services such as video on demand (VoD), catch-up TV, 
and internet video streaming are premier drivers of the exponential traffic growth experienced in today's wireless 
networks \cite{cisco13}. A key feature of this type of services is the time-shifted nature of user requests for the same 
content, or {\em asynchronous content reuse} \cite{ji2013throughput}: while there exists a relatively small number of popular files that account 
for most of the traffic, users access them at arbitrary times, such that \emph{naive multicasting}\footnote{Naive multicasting refers to the transmission 
of a common, not-network-coded stream of data packets, simultaneously received and decoded by multiple users.}
as implemented in Media Broadcasting Single Frequency Networks (MBSFN) \cite{mediaFLO}, is not useful. 
In fact, because of the large asynchronism of the user demands, present technology (e.g., DASH, Dynamic Adaptive Streaming over HTTP \cite{muller2012evaluation}) 
employs \emph{conventional unicasting}, i.e.,  each user request is treated as an independent information message, thus missing the opportunity of exploiting the redundancy
of the user demands. 
 
Due to the increasing cost and scarcity of wireless bandwidth, an emerging and promising approach for improving over both naive multicasting and conventional unicasting 
consists of using storage resources to {\em cache} popular content directly at the wireless edge, e.g., at small-cell base stations or end 
user devices.\footnote{Note that the storage capacity has become exceedingly cheap: for example, a 2 TByte hard disk, enough to store 1000 movies, costs less than \$ 100.} 

Caching has been widely studied in several wireline contexts, primarily for web proxy caching systems and content distribution networks (CDNs) \cite{Baev01,Baev08,borst2010distributed,Krishnan00,Koropolu02,Cao97,wang1999survey,breslau1999web}.
In these works, a range of interrelated problems, such as accurate prediction of demand, intelligent content placement, and efficient online replacement, is considered. 
The {\em data placement problem} was introduced in \cite{Baev01}, where the objective is to find the placement of data objects in an arbitrary 
network with capacity constrained caches, such that the total access cost is minimized. It was shown that this problem is a generalization of the metric uncapacitated facility location problem and hence is NP-Hard \cite{Baev08}.
%NP-hard by reduction to an uncapacitated facility location problem \cite{Baev04}. 
Tractable approaches in terms of LP relaxation \cite{Baev08,borst2010distributed}  or greedy algorithms \cite{Krishnan00,Koropolu02} have been proposed, 
by exploiting special assumptions such as network symmetry and hierarchical structures.  
On the other hand, an extensive line of work has addressed the content replacement problem, where the objective is to adaptively refresh the cache(s) 
content while a certain user data request process evolves in time \cite{wang1999survey, Cao97,breslau1999web}. 
The most common cache replacement/eviction algorithms are {\em least frequently used} (LFU) and {\em least recently used} (LRU), 
by which the least frequently/recently used content object is evicted upon arrival of a new object to a network cache. A combination of placement and replacement 
algorithms is also possible and in fact used in today's CDNs, which operate by optimizing the placement of content objects over long time periods for which 
content popularity can be estimated, and using local replacement algorithms to handle short time-scale demand variations.

In a more recent set of works (a non-exhaustive list of which includes \cite{golrezaei2011femtocaching, golrezaei2012femtocaching,shanmugam2013femtocaching,llorcatulino132,llorcatulino14, maddah2012fundamental, maddah2013decentralized,niesen2013coded,ji2013throughput, ji2013fundamental, ji2013wireless, molisch2014caching,gitzenis2013asymptotic}), 
an information theoretic view of caching has provided insights into the fundamental limiting performance of caching networks of practical relevance. 
In this framework, the underlying assumption is that there exists a fixed library of $m$ possible information messages (files) and 
a given network topology that includes nodes that host a subset of messages (sources), request a subset of messages (users), and/or have constrained cache capacity (helpers/caches).  
The {\em caching phase} is part of the code set-up, and consists of filling up the caches with (coding) functions of the messages whose entropy is constrained to be not larger than  the corresponding cache capacity.  
After this set-up phase, the network is ``used'' for an arbitrary long time, referred to as the {\em delivery phase}.  At each request round, a subset of the nodes (users) request 
subsets of the files in the library and the network must coordinate transmissions such that these requests are satisfied, i.e., at the end of each round all destinations must 
decode the requested set of files.  The performance metric here is the number of time slots necessary to  satisfy all the demands. 
In the case of symmetric links, the number of time slots can be normalized by the number of times lots necessary to send a single file across a point to point link. 
Therefore, the performance metric is {\em rate} defined as in the index coding setting \cite{birkkol98,bar2011index,el2010index,lubetzky2009nonlinear,blasiak2010index, chaudhry2011complementary,jafar2013topological,haviv2012linear,arbabjolfaei2013capacity,6620404}, 
i.e., the number of equivalent file transmissions.

%%%%%%%%%%%%%%%%%%%%%%%%%%%%%%%%%%%%%%%%%%%%%%%%%%%%%
\subsection{Related work}

%%%%%%%%%%%%%%%%%%%%%%%%%%%%%%%%%%%%%%%%%%%%%%%%%%% 
Focusing on the subset of current works directly relevant to this paper, in \cite{golrezaei2011femtocaching} (see also the successively published papers \cite{golrezaei2012femtocaching,shanmugam2013femtocaching}) a bipartite network formed by helper nodes (with caches), user nodes (without caches), and capacitated noiseless links, was studied in the case of random i.i.d. requests according to some known demand distribution. This is a special case of the data placement problem with trivial routing \cite{Baev04}. The problem in \cite{golrezaei2011femtocaching,golrezaei2012femtocaching,shanmugam2013femtocaching} consists of minimizing the 
average rate necessary to satisfy all users, where averaging is over the random requests. 

In \cite{llorcatulino132,llorcatulino14}, the data placement problem is generalized to the (coded) content distribution problem (CDP), 
where information can, not only be stored and routed, but also coded, over the network. 
The authors showed an equivalence between the CDP and the network coding problem over a so-called caching-demand augmented graph, 
which proved the polynomial solvability of the CDP under uniform demands (each user requests the same subset of files), and the hardness of the CDP under arbitrary 
demands. The authors further showed via simulations on simple networks, the potential of network coding to enable cache cooperation 
gains between caches sharing a multicast link from a content source. While this work suggested the benefit of cooperative caching via both the direct exchange of information between 
neighbor nodes as well as via coded multicast transmissions from a common source, the analytical characterization of the optimal caching performance in arbitrary  networks remains a hard and open problem. To this end, significant progress has been made by considering specific network models that capture scenarios 
of practical relevance, especially in wireless networks.

In \cite{ji2013throughput}, the authors considered a Device-to-Device (D2D)  network with caching nodes that are at the same time helpers and users, 
communicating with each other under the interference avoidance ``protocol model'' of \cite{gupta2000capacity}. 
In this setting, under i.i.d. random requests following a Zipf distribution \cite{breslau1999web, cha2007tube} (with Zipf parameter $\alpha<1$), 
it was shown that decentralized random caching and {\em uncoded}\footnote{We refer to ``uncoded'' the schemes that send 
packets of individual files,  in contrast to ``coded" schemes that send mixtures of packets from different files (inter-session network coding).} 
unicast delivery achieves order-optimal average per-user throughput,\footnote{The per-user throughput expressed in bits per time slot is inversely proportional to the rate expressed 
in number of equivalent file transmissions, through a system constant that is irrelevant as far as scaling laws are concerned. Hence, in this context, maximizing throughput or minimizing rate 
are equivalent goals.} shown to scale as $\Theta\left(\frac{M}{m}\right)$,\footnote{We will use the following standard {\em order} notation: given two functions $f$ and $g$, we say that: 1)  $f(n) = O\left(g(n)\right)$ if there exists a constant $c$ and integer $N$ such that  $f(n)\leq cg(n)$ for $n>N$; 2) $f(n)=o\left(g(n)\right)$ if $\lim_{n \rightarrow \infty}\frac{f(n)}{g(n)} = 0$;  3) $f(n) = \Omega\left(g(n)\right)$ if $g(n) = O\left(f(n)\right)$; 4) $f(n) = \omega\left(g(n)\right)$ if $g(n) = o\left(f(n)\right)$;  5) $f(n) = \Theta\left(g(n)\right)$ if $f(n) = O\left(g(n)\right)$ and $g(n) = O\left(f(n)\right)$.} 
when both the number of users $n$ and the library size $m$ grow large and satisfy $nM \geq m$ (i.e., the aggregate cache across the network can contain 
the whole file library). 

Concurrently, another line of work in \cite{maddah2012fundamental, maddah2013decentralized}
considered a different network topology, here referred to as the {\em shared link network}. This is formed by a single source node (a server or base station) 
with all $m$ files, connected  via a shared noiseless link to $n$ user nodes, each with cache of size $M$ files. 
In \cite{maddah2012fundamental, maddah2013decentralized}, the authors addressed the min-max rate problem, i.e., minimizing (over the coding scheme) the worst-case 
rate (over the user demands).  Both deterministic and random caching schemes, with corresponding {\em coded}  multicast delivery schemes,  
were shown to provide approximately optimal min-max rate, i.e., within a multiplicative constant, independent of $n,m,M$, 
from an information theoretic lower bound.  Interestingly, when translating the results of \cite{maddah2012fundamental, maddah2013decentralized} in terms of 
per-user throughput, for the case $nM \geq m$ this scales also as $\Theta\left(\frac{M}{m}\right)$. 

The ensemble of these results show the remarkable fact that, both for the D2D and for the shared link network topologies, caching in the user devices 
can turn {\em memory into bandwidth}:  Moore's law (scaling of silicon integration) reflects directly in terms of a per-user throughput gain, in the sense that 
doubling the user device storage capacity $M$ yields seamlessly a two-fold increase in the per-user throughput. 
The D2D approach of  \cite{ji2013throughput} exploits the spatial reuse of D2D local communication, since caching allows each user to access the desired 
content within a short range. Instead, the approach of \cite{maddah2012fundamental,maddah2013decentralized} exploits the multiplexing 
gain of  global communication, creating network-coded symbols that are simultaneously useful to a large number of users. 
In an effort of combining these two gains, \cite{ji2013fundamental} considered the same D2D wireless network of \cite{ji2013throughput},   
with a caching and coded delivery scheme inspired by \cite{maddah2012fundamental,maddah2013decentralized}.
Somehow counterintuitively, it was shown that spatial reuse and the coded multicasting gains are not cumulative. 
An informal explanation of this fact follows by observing that  D2D spatial reuse and coded multicasting have contrasting goals. 
On one hand, spatial reuse is maximized by keeping communication as ``local'' 
as possible,  such that the same time slot can be reused with high density in space. On the other hand, coded multicasting produces 
codewords that are useful to many users, so that it is advantageous to have (coded) transmissions as ``global'' as possible.

While several variants and extensions of these basic setups have been recently considered \cite{pedarsani2013online, niesen2013coded, sengupta2013fundamental, ji2014multiple, karamchandani2014hierarchical, hachem2014multi, hachem2014coded, niesen2014coded, bacstuug2014living, bacstuug2014cache, molisch2014caching, altieri2014fundamental}, 
in this work we focus on the combination of the random requests aspect (as in 
\cite{golrezaei2011femtocaching,ji2013throughput,llorcatulino132,llorcatulino14}) and the single source 
shared link network (as in \cite{maddah2012fundamental, maddah2013decentralized}). 
This problem has been treated in \cite{niesen2013coded}, which considered a strategy based on partitioning the file library into subsets of approximately 
uniform request probability, and applying to each subset the strategy for the min-max approach of \cite{maddah2013decentralized}. 
This is motivated by observing that the average rate with random uniform demands is related, within a constant factor, to the min-max rate under arbitrary demands.
Then, by partitioning the set of files and allocating the cache memory across such subsets, the problem is decomposed into 
subproblems, each of which can be separately addressed by reusing the arbitrary demand strategy. 
Due to the difficulty of finding the optimal file partitioning and corresponding cache memory allocation, 
\cite{niesen2013coded} restricts its analysis to a scheme in which for any  two files in the same partition, the file popularities differ by at most a factor of two.

%{\RED ADD A FEW WORDS ON THE RECENT PAPER APPEARED AT ITA ????}

%%%%%%%%%%%%%%%%%%%%%%%%%%%%%%%%%%%%%%%%%%%%%%%%%%%%%%
\subsection{Contributions}

While in \cite{niesen2013coded} this approach is studied for a general demand distribution, our scaling order-optimality results 
apply to the specific case of a Zipf demand distribution. This is a very relevant case in practice since 
the popularity of Internet content has been shown, experimentally,  to follow a Zipf power 
law \cite{breslau1999web, cha2007tube} (or its variations \cite{hefeeda2008traffic}) defined as follows: a file $f = 1, \ldots, m$ is requested with probability 
\be
\label{eq: Zipf}
q_f = \frac{f^{-\alpha}}{\sum_{i=1}^m i^{-\alpha}}, \, \forall f = \{1, \cdots, m\},
\ee
where $\alpha \geq 0$ is the Zipf parameter.  
%Focusing on a specific demand distribution allows us to provide sharper results than those in \cite{niesen2013coded}.
In this context, our objective is to characterize the scaling laws of the optimal {\em average} rate and provide 
simple and explicit order-optimal schemes. Specifically, the contributions of this work are as follows:
\begin{enumerate}
\item We recognize that the sub-optimality of the scheme analyzed in \cite{niesen2013coded} is due to the fact that 
files are partitioned according to their local popularity without considering the effects of the remaining system 
parameters ($n,m,M$) on the "aggregate user demand distribution". In particular, the probability with which each user requests files can be very different from the probability with which 
each file is requested by the aggregate users. The other limitation of  \cite{niesen2013coded} is that the
scheme for coded delivery (see details in Section \ref{sec: Achievable Delivery Scheme})
is applied separately for each file group, resulting in missed coding opportunities between different groups. 
We propose a different way to optimize the random caching placement, according to a caching distribution that depends on all system parameters, 
and not just the ``local'' demand distribution $\qv$. Also, we consider ``chromatic number'' index coding delivery scheme applied to all requested packets. 
We refer to this scheme as {\em RAndom Popularity-based} (RAP) caching, with {\em Chromatic-number Index Coding} (CIC). 
\item For the proposed RAP-CIC, we provide a new upper bound on the achievable rate by bounding the average chromatic number of the induced random conflict graph.
By numerically optimizing this bound, we demonstrate the efficiency of our method and the gains over the method of \cite{niesen2013coded} by simulation. 
However, a direct analysis of the proposed scheme appears to be elusive. 
\item For the sake of analytical tractability, we further focus on a simpler caching placement scheme where the caching distribution is a step function
(some files are cached with uniform probability, and others are not cached at all) and a polynomial-time approximation of CIC, referred to as greedy constrained coloring (GCC).
We refer to this scheme as Random Least-Frequently-Used (RLFU) caching, given its analogy with 
the standard LFU caching policy,\footnote{LFU discards the least frequently requested file upon the arrival of a new file to a full cache of size $M$ files. 
In the long run, this is equivalent to caching the $M$ most popular files.}  
with GCC delivery, or RLFU-GCC. 
\item We provide an information theoretic lower bound on the (average) rate achieved by any caching scheme, and show the order-optimality of
the proposed achievability schemes for the special case of a Zipf demand distribution.  To the best of our knowledge, these are the first order-optimal results under this 
network model for nontrivial popularity distributions. In addition, our technique for proving the converse is not restricted to the Zipf distribution, 
such that it can be used to verify average rate order-optimality in other cases.   
\item Our analysis identifies the regions in which conventional schemes (such as LFU with naive multicasting) can still preserve order-optimality, 
as well as exposes the wide range of opportunities for performance improvements via RAP or RLFU, combined with CIC or GCC. 
We show that, as in the D2D setting of \cite{ji2013throughput}, when the Zipf parameter is $0 \leq \alpha < 1$, the average rate with random demands
and the min-max rate with arbitrary demands are order-equivalent.  On the other hand, when $\alpha >1$, the 
average rate can exhibit order gains with the respect to the min-max rate. 
\end{enumerate}

%Though naive multicasting is allowed for the sake of analytical simplicity, in line with \cite{maddah2012fundamental, maddah2013decentralized,niesen2013coded}, it is straightforward to see the  our results hold for asynchronous content reuse in all the regimes of parameters studied in this paper.   
We remark that while we consider simultaneous requests, the argument made in \cite{maddah2012fundamental, maddah2013decentralized} to handle streaming 
sessions formed by multiple successive requests starting at different times holds here as well.

Finally, it is interesting to note that, while RLFU-GCC becomes a special case of the general scheme described in \cite{niesen2013coded},
the optimization carried out in this paper and the corresponding performance analysis are new and non-trivial extensions. 
As pointed out in \cite{niesen2013coded}, one would think that an approach based on uniformly caching only the $\widetilde m \leq m$ most popular files 
has the disadvantage that ``{\em ... the difference in the popularities among the $\widetilde{m}$ cached files is ignored. Since these files can have widely different 
popularities, it is wasteful to dedicate the same fraction of memory to each one of them.  As a result, this approach does not perform well in general}.''  
In contrast, we show that for the Zipf case,  the approach is order-optimal provided that the cache threshold $\widetilde m$ is carefully optimized as a function of the system 
parameters.  Also, in \cite{niesen2013coded} it was also pointed out that ``{\em Another option is to dedicate a different amount of memory 
to each file in the placement phase. For example the amount of allocated memory could be proportional to the popularity of a file. 
While this option takes the different file popularities into account, it breaks the symmetry of the content placement. 
As a result, the delivery phase becomes intractable and the rate cannot be quantified ...}''
In contrast, using the proposed RAP caching optimization and CIC delivery across all requested packets, 
it is possible to find schemes that significantly outperform previous heuristics and, again for the Zipf case, 
are provably order-optimal  for all regimes of the system parameters $n,m,M,\alpha$.

The paper is organized as follows. In Section \ref{section: network model}, we present the network model and the problem formulation. 
The random caching and coded multicasting scheme is introduced in Section \ref{sec: Achievable Delivery Scheme} 
and a general converse result for the achievable average rate is given in Section \ref{sec: The General Lower Bound of the Achievable Rate}. 
In Section \ref{order opt}, we prove and discuss the order-optimality of the proposed 
scheme for the Zipf request distribution.  Further results, simulations and conclusive remarks  
are presented  in  Section \ref{sec: Discussion} and \ref{sec: conclusions}. 

%%%%%%%%%%%%%%%%%%%%%%%%%%%%%%%%%%%%%%%%%%%%%%%%%%%%%%
\section{Network Model and Problem Formulation}
\label{section: network model}

Consider a {\em shared link network} \cite{maddah2012fundamental, maddah2013decentralized,niesen2013coded} with 
file library $\Fc=\{1,\cdots, m\}$, where each file (i.e., message) has entropy equal to $F$ bits, 
and user set $\Uc=\{1,\cdots, n\}$,  where each user has a cache (storage memory) of capacity $MF$ bits. 
Without loss of generality, the files are represented by 
binary vectors $W_f \in \FF_2^F$. The system setup is as follows: 
\begin{enumerate}
\item At the beginning of time, a realization $\{W_f : f \in \Fc\}$ of the library is revealed to the encoder. 
\item The encoder computes the cached content, by using a set of $|\Uc |$ functions 
$\{Z_u: \FF_2^{mF} \rightarrow \FF_2^{MF} : u \in \Uc\}$,  such that $Z_u(\{W_f : f \in \Fc\})$ denotes the codeword stored in 
the cache of user $u$.  The operation of computing $\{Z_u : u \in \Uc\}$ and filling the caches does not cost any rate, i.e., it is done once for all at the 
network setup, referred to as the {\em caching phase}. 
\item After the caching phase, the network is repeatedly used. At each use of the network, 
a realization of the random request vector $\fsf  = (\fsf_1, \ldots, \fsf_n) \in \Fc^n$  is generated. We assume that $\fsf$ has i.i.d. components
distributed according to a probability mass function $\qv = (q_1, \ldots, q_m)$, referred to as the
{\em demand distribution}. This is known a priori and, without loss of generality up to index reodering, 
has non-increasing components $q_1 \geq \cdots \geq q_m$. 
\item We let $\fv = (f_1, \ldots, f_n)$ denote the realization of the random request vector $\fsf$. This is revealed to the encoder, 
which computes a {\em multicast codeword} as a function of the library files and the request vector 
In this work we consider a fixed-to-variable almost-lossless framework. 
Hence, the multicast encoder is defined by a fixed-to-variable encoding function 
$X : \FF_2^{mF} \times \Fc^n \rightarrow \FF_2^*$ (where $\FF_2^*$ denotes the set of finite length binary sequences),  
such that $X(\{W_f : f \in \Fc\},\fv)$ is the transmitted codeword.  We denote by $L(\{W_f : f \in \Fc\},\fv)$
the length function (in binary symbols) associated to the encoding function $X$. 
\item  Each user receives $X(\{W_f : f \in \Fc\},\fv)$ through the noiseless shared link, and decodes its requested file $W_{f_u}$ as 
$\widehat{W}_{f_u} = \lambda_u(X,Z_u,\fv)$, where  $\lambda_u : \FF_2^* \times \FF_2^{MF} \times \Fc^n \rightarrow \FF_2^{F}$
denotes the decoding function of user $u$. 

\item The concatenation of 1) demand vector generation, 2) multicast encoding and transmission over the shared link and 3) decoding, 
is referred to as the {\em delivery phase}. 
\end{enumerate}
Consistently with the existing information theoretic literature on caching networks (see Section \ref{sec:intro}), we refer to a content distribution scheme, formed by both caching and delivery phases, directly as a caching scheme, and measure the system performance in terms of the rate during the delivery phase only. %Overall, a caching scheme is formed by caching and delivery phases. Consistently with the existing information theoretic literature on caching networks (see Section \ref{sec:intro}),  we consider the system performance restricted to the delivery phase only.
In particular, we define the rate of the scheme as
\begin{equation} \label{average-rate}
R^{(F)} = \sup_{\{W_f : f \in \Fc\}} \; \frac{\EE[ L(\{W_f : f \in \Fc\},\fsf)]}{F},
\end{equation} 
where the expectation is with respect to the random request vector.\footnote{Throughout this paper, we directly use ``rate" to refer to the average rate defined by \eqref{average-rate} and explicitly use ``average (expected) rate" if needed for clarity.}

This definition of rate has the following operational meaning.  
Assume that the download of a single file through the shared link takes 
one ``unit of time''. Then, (\ref{average-rate}) denotes the 
worst-case (over the library) average (over the demands) download time for the whole network, 
when the users place i.i.d. random requests according to the demand  distribution $\qv$.  
The underlying assumption is that the content library (i.e., the realization of the files) changes very slowly in time, 
such that it is generated or refreshed at a time scale much slower than the time scale at which the users download the files.
Hence, it is meaningful to focus only on the rate of the delivery phase, and disregard the cost of filling the caches (i.e., the cost of the caching phase), 
which is included in the code construction. Users make requests, and the network satisfies them by sending 
a variable length transmission until every user can successfully decode. After all users have decoded, a new round of requests is made.
This forms a renewal process where the recurrent event is the event that all users have decoded their files. 
In the spirit of fixed-to-variable source coding, $R^{(F)}$ is the coding rate (normalized coding length) expressed in file ``units of time''.
Also, by the renewal theorem, it follows that $1/R^{(F)}$ yields (up to some fixed proportionality factor)  
the channel throughput in terms of per-user decoded bits per unit time. 
Finally, since the content library changes very slowly, averaging also over the realization of the files has little 
operational meaning. Instead, we take the worst-case over the file library realization. 

Consider a sequence of caching schemes defined by cache encoding functions $\{Z_u\}$,  
multicast coding function $X$, and decoding functions $\{\lambda_u\}$, for increasing file size $F = 1,2,3,\ldots$. 
For each $F$, the worst-case (over the file library) probability of error of the corresponding caching scheme is defined as
\begin{equation} \label{perr}
P_e^{(F)} (\{Z_u\},X, \{\lambda_u\}) = 
\sup_{\{W_f : f \in \Fc\}} \;  \PP \left ( \bigcup_{u \in \Uc} \Big \{ \lambda_u(X,Z_u,\fsf) \neq W_{\fsf_u}  \Big \} \right ). 
\end{equation}
A sequence of caching schemes is called {\em admissible} if  $\lim_{F \rightarrow \infty} P_e^{(F)} (\{Z_u\},X, \{\lambda_u\}) = 0$. 
{\em Achievability} for our system is defined as follows:

\begin{defn} \label{def:achievable-rate}
A rate $R(n,m,M,\qv)$ is {\em achievable} for the shared link network with $n$ users, library size $m$, cache capacity $M$, 
and demand distribution $\qv$, if there exists a sequence of admissible caching schemes with rate $R^{(F)}$ such that 
\[ \limsup_{F \rightarrow \infty} R^{(F)} \leq  R(n,m,M,\qv). \]
\hfill $\lozenge$ 
\end{defn}

We let $R^*(n,m,M,\qv)$ denote the infimum (over all caching schemes) of the achievable rates. 
The notion of ``order-optimality'' for our system is defined as follows: 

%\begin{defn}
%\label{def: order-optimal}
%A sequence of caching schemes for the shared link network with $n$ users, library size $m$, cache capacity $M$, and demand distribution $\qv$,
%is order-optimal if its rate $R(n,m,M,\qv)$ satisfies  
%\begin{eqnarray}
%\limsup_{n \rightarrow \infty} \frac{R(n,m,M,\qv)}{R^*(n,m,M,\qv)} \leq \nu,
%\end{eqnarray}
%for some constant $1 \leq \nu < \infty$, independent of $m,n,M$.
%\hfill $\lozenge$
%\end{defn}
{ 
\begin{defn}
\label{def: order-optimal}
Let $n,M$ be functions of $m$, such that $\lim_{m \rightarrow \infty} n(m) = \infty$. 
A sequence of caching schemes for the shared link network with $n$ users, 
library size $m$, cache capacity $M$, and demand distribution $\qv$,
is order-optimal if its rate $R(n,m,M,\qv)$ satisfies  
\begin{eqnarray}
\limsup_{m \rightarrow \infty} \frac{R(n,m,M,\qv)}{R^*(n,m,M,\qv)} \leq \nu,
\end{eqnarray}
for some constant $1 \leq \nu < \infty$, independent of $m,n,M$.
\hfill $\lozenge$
\end{defn}

Notice that in the definition of order-optimality we let first $F \rightarrow \infty$ (required by the definition of achievable rate) and then
we let $m \rightarrow \infty$. 
In this second limit, we let $n$ and $M$ be functions of $m$, indicating that the notion of ``order", throughout this paper, is with respect to 
the library size $m$. Depending on how $n$ and/or $M$ vary with respect to $m$, we can identify different system operating regimes.\footnote{The case of constant $m$ while $n \rightarrow \infty$ is also considered and treated  separately in Section \ref{sec: Discussion}.} 
}

%%%%%%%%%%%%%%%%%%%%%%%%%%%%%%%%%%%%%%%%%%%%%%%%%%%%%%%%%%%%%%%%%%
\section{Random Fractional Caching and Linear Index Coding Delivery}
\label{sec: Achievable Delivery Scheme}

In this section we focus on a particular class of admissible schemes where the caching functions $\{Z_u\}$ are random and independent 
across the users \cite{ji2013throughput,maddah2013decentralized} and the multicast encoder is based on linear index coding \cite{birkkol98,blasiak2010index}. 
With random coding functions, two flavors of results are possible: 1) by considering the average rate with respect to the random coding ensemble, one can prove the existence of deterministic sequences of caching schemes 
achieving rate not worse than average;  2) by considering the concentration of the rate conditioned on the 
random caching functions, we obtain a stronger result: namely, in the limit of large file size $F$, the (random) rate is
smaller than a given threshold with high  probability. This implies achievability of such rate threshold by the random 
scheme itself (not only in terms of a non-constructive existence argument based on random coding).  
Here, we prove achievability in this second (stronger) sense. 

%%%%%%%%%%%%%%%%%%%%%%%%%%%%%%%%%%%%%%%%%%%%%%%%%%%%%%%
\subsection{Random Fractional Caching Placement}

The caching placement phase works as follows: 
\begin{enumerate}
\item For some given integer $B$, each file $W_f$ is divided into {\em packets} of equal size $F/B$ bits, denoted by 
$\{W_{f,b} : b = 1,\ldots, B\}$.\footnote{Since we eventually take the limit for $F \rightarrow \infty$, for simplicity we 
neglect the fact that $B$ may not divide $F$.}
\item Each user randomly selects and stores in its cache a collection of $p_f M B$ distinct packets from each file $f \in \Fc$, where $\pv = (p_1, \ldots, p_m)$ is a 
vector with components $0 \leq p_f \leq 1/M$, such that $\sum_{f=1}^m p_f = 1$, referred to as the {\em caching distribution}.\footnote{Note that $p_f$ represents the fraction of the memory $M$ allocated to file $f$. Hence, we let $p_f$ be a function of $n,m,M,\qv$, but not a function of $B$.}
\end{enumerate}
It follows that, for each user $u$, 
\begin{equation}
Z_u = \left ( W_{1,b^u_{1,1}}, \ldots, W_{1,b^u_{1,p_1MB}}, W_{2,b^u_{2,1}}, \ldots, W_{2,b^u_{2,p_2MB}}, \ldots, \ldots, 
W_{m,b^u_{m,1}}, \ldots, W_{m,b^u_{m,p_mMB}} \right )
\end{equation}
where $b^u_{f,i}$ is the index of the $i$-th packet of file $f$ cached by user $u$, and where
the tuples of indices $(b^u_{f,1}, \ldots, b^u_{f, p_fMB})$ are chosen independently across the users $u \in \Uc$ and the files $f \in \Fc$, with uniform probability over all
${B \choose p_fMB}$ distinct subsets of size $p_fMB$ of the set of packets of size $B$.  
The collection of the cached packet indices (over all users and all files) is a random vector, denoted in the following by $\Csf$. 
For later use, a given cache configuration, i.e., a realization of $\Csf$, will be denoted by $\Cm$. 
Also, we shall denote by $\Cm_{u,f}$ the vector of indices of the packets of file $f$ cached by user $u$. Finally, for the sake of notation simplicity, we shall not distinguish between ``vectors''
(ordered lists of elements) and the corresponding ``sets'' (unordered lists of elements), such that we write $b \notin \Cm_{u,f}$ (resp., $b \in \Cm_{u,f}$) to indicate
that the $b$-th packet of file $f$ is not present (resp., present) in the cache of user $u$. 
Observe that a random fractional caching scheme is completely characterized by the caching distribution $\pv$, where $p_f$ denotes the fraction of file $f$ cached by each user. 
In Section \ref{sec: Proposed Scheme}, we shall describe how to 
design the caching distribution as a function of the system parameters.

%%%%%%%%%%%%%%%%%%%%%%%%%%%%%%%%%%%%%%%%%%%%%%%%%%%
\subsection{Linear Index Coding Delivery}

Finding a delivery scheme for the caching problem in the shared link network 
is equivalent to finding an index code with side information given by the cache configuration $\Cm$. 
It is clear that under the caching functions defined before, each user $u$ requesting file $f_u$ needs to obtain all the packets 
$W_{f_u, b}$ with $b \notin \Cm_{u,f_u}$. It follows that a demand vector $\fv$, given the cache configuration $\Cm$, can be translated into a 
packet-level demand vector $\Qm$, containing the packets needed by each user. Symmetrically with the notation introduced for the cache configuration, 
we denote by $\Qsf$ the corresponding random vector, and by $\Qm_{u,f}$ the packet-level demand restricted to user $u$ and file $f$. 
In particular, if user $u$ requests file $f_u$, then $\Qm_{u,f}$ is empty for all $f \neq f_u$ and it contains the complement set of $\Cm_{u,f_u}$ 
for $f = f_u$. 

Following the vast literature on index coding (see for example \cite{jafar2013topological,birkkol98,blasiak2010index}), we define the side-information graph 
$\Sc_{\Cm, \Qm}$ corresponding to the index coding problem defined by $(\Cm, \Qm)$ as follows: 
\begin{itemize}
\item Vertices of $\Sc_{\Cm, \Qm}$: For each packet in $\Qm$ (i.e., requested by some user), form all possible distinct 
labels of the form $v = \{ \mbox{packet identity}, \mbox{user requesting}, \mbox{users caching} \}$, where ``packet identity'' is the pair $(f,b)$ of file index and packet index, 
``user requesting'' is the index of some user $u$ such that $b \in \Qm_{u,f}$, and ``users caching'' is the set of {\em all} users $u'$ such that
$b \in \Cm_{u',f}$.  Then, a vertex in $\Sc_{\Cm, \Qm}$ is associated to each of such distinct labels. For simplicity of notation, we do not distinguish between label and vertex, 
and refer to ``label $v$'' or ``vertex $v$'' interchangeably, depending on the context. 
Notice that while ``packet identity'' and ``users caching'' are fixed by 
the packet identity and by the cache realization, the second label component (``user requesting'') can take multiple values, 
since several users may request the same packet.
\item Edges of  $\Sc_{\Cm, \Qm}$:
For each vertex $v$, $\rho(v)$, $\mu(v)$ and $\eta(v)$ denote the three fields in its label (namely, ``packet identity'', ``user requesting'', and ``users caching'').
Any two vertices $v_1$ and $v_2$ are connected by an edge if at least one of the following two conditions are satisfied: 
1) $\rho(v_1) = \rho(v_2)$, or 2) $\mu(v_1) \in \eta(v_2)$ and $\mu(v_2) \in \eta(v_1)$. 
\end{itemize}
The complement graph $\Hc_{\Cm, \Qm}$ of the side information graph $\Sc_{\Cm, \Qm}$ is known as the {\em conflict graph}. In particular, 
$\Hc_{\Cm, \Qm}$ has the same vertices of $\Sc_{\Cm, \Qm}$ and any two vertices $v_1$ and $v_2$ in $\Hc_{\Cm, \Qm}$ are connected by an edge 
if both the following conditions are satisfied:  1) $\rho(v_1) \neq \rho(v_2)$, and 2) $\mu(v_1) \notin \eta(v_2)$ or $\mu(v_2) \notin \eta(v_1)$. 

\begin{figure}[ht]
\centerline{\includegraphics[width=10cm]{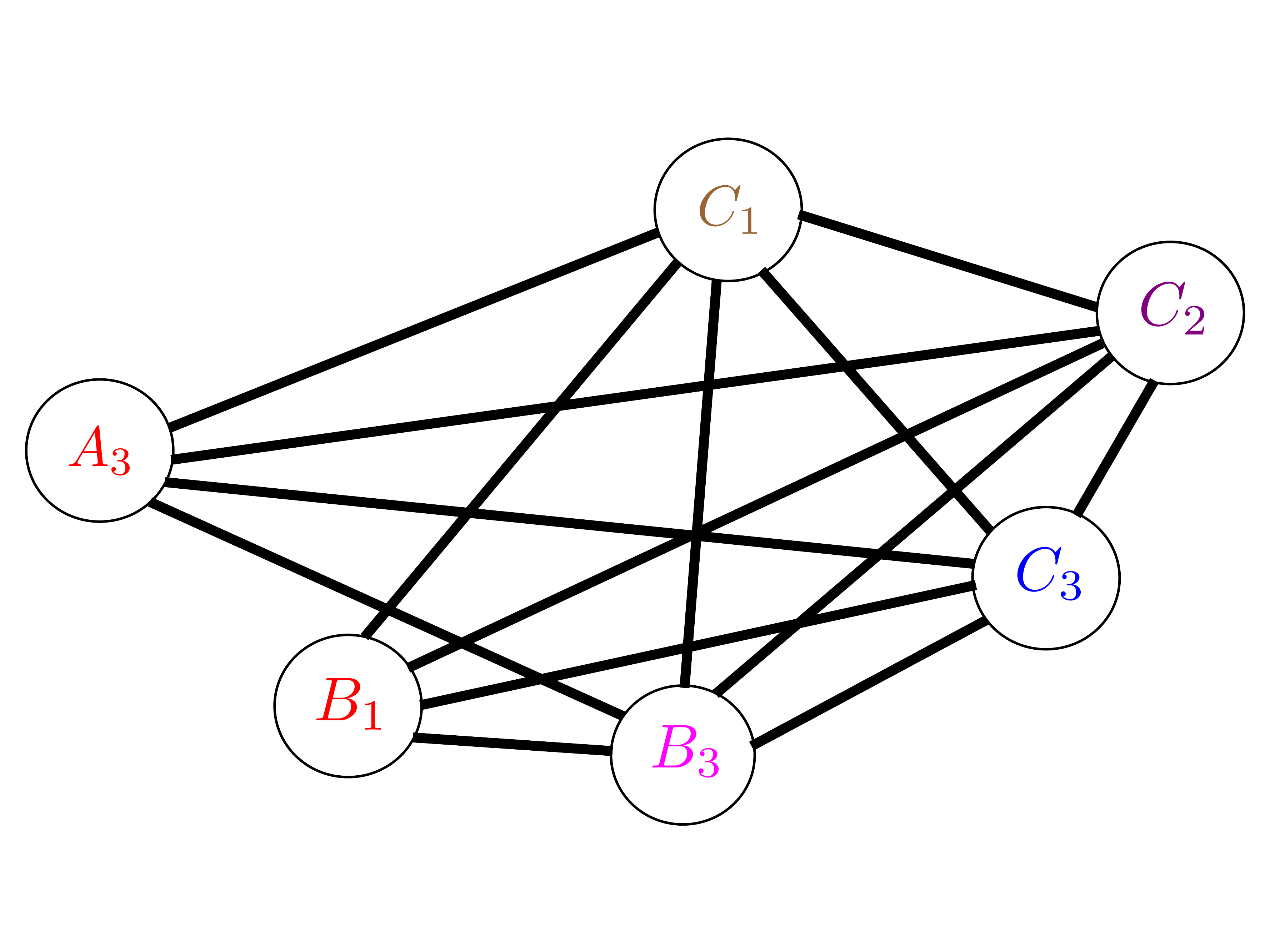}}
\caption{An illustration of the conflict graph, where $n=3$, $\mathcal U = \{1,2,3\}$, $m=3$, $\mathcal F = \{\mathrm A, \mathrm B, \mathrm C\}$ and $M=1$. Each file is 
partitioned into $3$ packets. 
The caching realization $\Cm$ and the packet-level demand vectors are given in Example \ref{example: algorithm}. 
The color for each vertex in this graph represents the vertex coloring scheme  obtained by Algorithm \ref{algorithm: coloring 1}. 
In this case, this vertex coloring is the minimum vertex coloring, and therefore it achieves the graph chromatic number.}
\label{fig: conflict_ex_coloring}
\end{figure}

\begin{example}
\label{example: algorithm}
We consider a network with $n=3$ users denoted as $\mathcal U = \{1,2,3\}$ and $m=3$ files denoted as $\mathcal F = \{\mathrm{A}, \mathrm{B},\mathrm{C}\}$. We assume $M=1$ and 
partition each file into $B = 3$ packets. 
For example, $\mathrm{A} = \{\mathrm{A}_1, \mathrm{A}_2, \mathrm{A}_3\}$. Let $\pv = \{\frac{2}{3}, \frac{1}{3}, 0\}$, which means that two packets of $\mathrm{A}$, one packet of $\mathrm{B}$ and none of $\mathrm{C}$ will be stored in each user's cache. We assume a caching realization $\Cm$ is given by: 
$\Cm_{1,\mathrm{A}} = \{\mathrm{A}_1,\mathrm{A}_2\}$, $\Cm_{1,\mathrm{B}} = \{\mathrm{B}_1\}$, $\Cm_{1,\mathrm{C}} = \emptyset$; 
$\Cm_{2,\mathrm{A}} = \{\mathrm{A}_1,\mathrm{A}_3\}$, $\Cm_{2, \mathrm{B}} = \{\mathrm{B}_2\}$, $\Cm_{2,\mathrm{C}} = \emptyset$; 
$\Cm_{3,\mathrm{A}} = \{\mathrm{A}_1,\mathrm{A}_2\}$, $\Cm_{3, \mathrm{B}} = \{\mathrm{B}_3\}$, $\Cm_{3,\mathrm{C}} = \emptyset$.  
Suppose that user $1$ request $\mathrm{A}$, 
user $2$ request $\mathrm{B}$ and user $3$ request $\mathrm{C}$ ($\fv = \{\mathrm{A}, \mathrm{B},\mathrm{C}\}$), 
such that $\Qm = \{\mathrm{A}_3, \mathrm{B}_1,\mathrm{B}_3,\mathrm{C}_1,\mathrm{C}_2,\mathrm{C}_3\}$. 
The corresponding conflict graph $\Hc_{\Cm, \Qm}$ is shown in Fig. \ref{fig: conflict_ex_coloring}. 
\hfill $\lozenge$
\end{example}

A well-known general index coding scheme consists of coloring the vertices of the conflict graph $\Hc_{\Cm, \Qm}$ and transmitting the concatenation of 
the packets obtained by EXOR-ing the packets corresponding to vertices with same color. For any vertex coloring of the conflict graph, 
vertices with the same color form (by definition) an independent set. Hence, the corresponding packets can be EXOR-ed together and sent over the shared link 
at the cost of the transmission of a single packet. 
%{\BLUE [It would be good to use the example to illustrate the index coding advantage when random fractional caching i used, as opposed to local caching schemes such as LFU that do not create coded multicast opportunities.]}

Letting  $\chi(\mathcal H_{\Cm, \Qm})$ denote the chromatic number of $\Hc_{\Cm, \Qm}$, 
the corresponding normalized code length is 
\begin{equation}  \label{CIC-rate 1}
\frac{L(\{W_f : f \in \Fc\},\fv)}{F} = \frac{\chi(\mathcal H_{\Cm, \Qm})}{B}, 
\end{equation}
since each coded packet corresponds to $F/B$ coded binary symbols, and we have a total of $\chi(\mathcal H_{\Cm, \Qm})$ coded packets (i.e., colors). 
For ease of reference, we denote this coding scheme as {\em Chromatic-number Index Coding} (CIC). In passing, we observe that, by design, 
the CIC scheme allows coding over the full set of requested packets $\Qm$, unlike the scheme proposed in 
\cite{niesen2013coded}, where coding is allowed within packets of specific file groups. 
Notice also that, by construction, CIC allows all users to decode their requested packets. Therefore, any sequence of CIC schemes yields probability of error identically zero, for all file lengths $F$, and not just vanishing probability of error in the limit. 
Hence, while in our problem definition we have considered a fixed-to-variable length ``almost-lossless'' coding
framework, this class of algebraic coding schemes are fixed-to-variable length {\em exactly} lossless. 

The graph coloring problem is NP-complete and hard to approximate in general \cite{zuckerman2006linear}. 
However, it is clear from the above presentation that {\em any} coloring scheme (possibly using a larger number of colors), yields a lossless caching scheme with possibly
larger coding length. In particular, exploiting  the special structure of the conflict graph originated by the caching problem, we present in the following 
an algorithm referred to as Greedy Constrained Coloring (GCC), which has polynomial-time complexity of $n,B$ and achieves asymptotically smaller or equal
rate with respect to the exponentially complex greedy coloring algorithm proposed in \cite{maddah2013decentralized} for the case of arbitrary demands. 
The proposed GCC is the composition of two sub-schemes, referred to as 
GCC$_1$ and GCC$_2$, given in Algorithms \ref{algorithm: coloring 1} and \ref{algorithm: coloring 2}, respectively. 
Eventually, GCC chooses the coloring with the smallest number of colors between the outputs of  
GCC$_1$ and GCC$_2$ (i.e., the shortest codeword). 

%%%%%%%%%%%%%%%%%%%%%%%%%%%%%%%%%%%%%%%%%%%%%%%%%%%%%%
\begin{algorithm}[ht]
\caption{GCC$_1$}
\label{algorithm: coloring 1}
\begin{algorithmic}[1]
\STATE Initialize $\Vc = $Vertex-set$({\mathcal H}_{\Cm, \Qm})$.
\WHILE{$\Vc \neq \emptyset$}
\STATE Pick any $v \in \Vc$, and let $\Ic = \{v\}$.
\FORALL{$v' \in \Vc/\Ic$}
\IF {\{There is no edge between $v'$ and $\Ic$ \} $\cap$ \{ $\{\mu(v'), \eta(v')\} = \{\mu(v), \eta(v)\}$ \}}
\STATE $\Ic = \Ic \cup \{v'\}$.
\ENDIF
\ENDFOR
\STATE Color all the vertices of the resulting set $\Ic$ by an unused color.
\STATE Let $\Vc \leftarrow \Vc \setminus \Ic$.
\ENDWHILE
\end{algorithmic}
\end{algorithm}
%%%%%%%%%%%%%%%%%%%%%%%%%%%%%%%%%%%%%%%%%%%%%%%%%%%%%%%%%%

Notice that $\{\mu(v), \eta(v)\}$ denotes the (unordered) set of the users either requesting or caching the packet corresponding to vertex $v$.
Notice also that each set $\Ic$ produced by Algorithm \ref{algorithm: coloring 1} is an independent set containing vertices with the 
{\em same} set of users either requesting or caching the corresponding packets. In fact, starting from a ``root'' node $v$ among those not
yet selected by the algorithm, the corresponding set $\Ic$ is formed by all the independent vertices $v'$ such that 
$\{\mu(v'), \eta(v')\} = \{\mu(v), \eta(v)\}$.

%%%%%%%%%%%%%%%%%%%%%%%%%%%%%%%%%%%%%%%%%%%%%%%%%%%%%%
\begin{algorithm}[ht]
\caption{GCC$_2$}
\label{algorithm: coloring 2}
\begin{algorithmic}[1]
\STATE Initialize $\Vc = $Vertex-set$({\mathcal H}_{\Cm, \Qm})$.
\WHILE{$\Vc \neq \emptyset$}
\STATE Pick any $v \in \Vc$, and let $\Ic = \{v\}$.
\FORALL{$v' \in \Vc/\Ic$}
\IF {$\rho(v') = \rho(v)$}
\STATE $\Ic = \Ic \cup \{v'\}$.
\ENDIF
\ENDFOR
\STATE Color all the vertices of the resulting set $\Ic$ by an unused color.
\STATE Let $\Vc \leftarrow \Vc \setminus \Ic$.
\ENDWHILE
\end{algorithmic}
\end{algorithm}
%%%%%%%%%%%%%%%%%%%%%%%%%%%%%%%%%%%%%%%%%%%%%%%%%%%%%%%%%%

It is also worthwhile to notice 
that GCC$_2$ is nothing else than ``naive multicasting'', that we have included here, for the sake of completeness, in a form symmetric to
that of GCC$_1$. In fact, it produces a set $\Ic$ (and a color) for each requested packet, 
which is then transmitted (uncoded) and simultaneously received by all requesting users. 

We can see that both the outer ``while-loop'' starting at line 2 and the inner ``for-loop'' starting at line 4 of Algorithm \ref{algorithm: coloring 1} iterate at most $nB$ times, respectively.  The operation in line 5 of Algorithm \ref{algorithm: coloring 1} costs at most complexity $n$. 
Therefore, the complexity of Algorithm \ref{algorithm: coloring 1} is 
$O(n^3B^2)$ (polynomial in $n$ and $B$). Also, it is easy to see that this complexity dominates that 
of Algorithm \ref{algorithm: coloring 2}. Therefore, the overall complexity of GCC is $O(n^3B^2)$.

%%%%%%%%%%%%%%%%%%%%%%%%%%%%%%%%%%%%%%%%%%%%%%%%%%%%%%%%%%%
\subsection{Achievable Rate}
\label{sec: Achievable Expected Rate}

As anticipated before, we shall consider the concentration of the (random) rate of the scheme described above, 
where the randomness follows from the fact that the caching functions, and therefore the conflict graph, are random. 
It is clear from the delivery phase construction that the output length of CIC or GCC does not depend on  
$\{W_f : f \in \Fc\}$  but only on $\Cm$ and $\Qm$ (see (\ref{CIC-rate 1})), since
the conflict graph is determined by the realization of the caches and of the demands. 
Therefore, without loss of generality, we can treat the files as fixed  arbitrary binary vectors and disregard the sup 
over $\{W_f : f \in \Fc\}$ in  the rate definition (see (\ref{average-rate})). 
Given $n,m,M$, the demand distribution $\qv$ and the caching distribution $\pv$, we define
\begin{equation} \label{CIC-rate} 
R^{\rm CIC}(n,m,M,\qv,\pv) \eqdef \frac{\EE[\chi(\Hc_{\Csf, \Qsf}) | \Csf]}{B}
\end{equation}
to be the conditional rate achieved by CIC. Similarly, we let $R^{\rm GCC}(n,m,M,\qv,\pv)$ 
denote the conditional rate achieved by GCC, defined by (\ref{CIC-rate}) after replacing the chromatic number with the 
number of colors produced by GCC. The same definition applies to $R^{{\rm GCC}_1}(n,m,M,\qv,\pv)$ and $R^{{\rm GCC}_2}(n,m,M,\qv,\pv)$.
%For notation simplicity, we omit here the dependence on $F$ (file length in bits) and on $B$ (file length in packets). 
In the following, it is intended that we consider the limit of the CIC and GCC schemes for $F, B \rightarrow \infty$ with 
fixed packet size $F/B \rightarrow $ constant. 
The performance of the proposed caching schemes is given by the following result. 

\begin{theorem}
\label{thm:up}   
For the shared link network with $n$ users, library size $m$, cache capacity $M$,  and demand distribution $\qv$, fix a caching distribution $\pv$. 
Then, for all $\epsilon > 0$, 
%\begin{equation}  
%\label{eq:2}
%\lim_{F \rightarrow \infty} \PP \left (| R^{\rm GCC}(n,m,M,\qv,\pv) - \min\{ \psi(\qv, \pv), \bar{m} \}|  \leq  \epsilon \right )  = 1,
%\end{equation}
\begin{equation}  
\label{eq:2}
\lim_{F \rightarrow \infty} \PP \left ( R^{\rm GCC}(n,m,M,\qv,\pv) \leq \min\{ \psi(\qv, \pv), \bar{m} \} + \epsilon \right )  = 1,
\end{equation}
where 
\be
\label{eq: m bar}
\bar m \eqdef \sum_{f = 1}^m \left(1 - \left(1 - q_f \right)^{n} \right),
\ee
and where
\begin{equation}
\label{eq: psi}
\psi(\qv,\pv) \eqdef \sum_{\ell=1}^n {n \choose \ell}  \sum_{f=1}^m \rho_{f,\ell} (p_f M)^{\ell-1} (1 - p_f M)^{n-\ell+1},  
\end{equation}
with 
\begin{equation} \label{rhorho} 
\rho_{f,\ell} \eqdef \PP\left (f = \arg\! \max_{j \in \Dc} \,\,\, (p_jM)^{\ell-1}(1-p_jM)^{n-\ell+1} \right ), 
\end{equation}
where $\Dc$ is a random set of $\ell$ elements selected in an i.i.d. manner from $\Fc$ (with replacement). 
\end{theorem}

\begin{IEEEproof} See Appendix \ref{sec: Proof of Theorem up}. \end{IEEEproof}

{\em Remarks:} 
\begin{enumerate}

\item Since by construction $R^{\rm CIC}(n,m,M,\qv,\pv) \leq R^{\rm GCC}(n,m,M,\qv,\pv)$ for any realization of $\Cm, \Qm$, then
also $R^{\rm GCC}(n,m,M,\qv,\pv)$ stochastically dominates $R^{\rm CIC}(n,m,M,\qv,\pv)$. Therefore, Theorem \ref{thm:up} immediately implies 
$\lim_{F \rightarrow \infty} \PP \left ( R^{\rm CIC}(n,m,M,\qv,\pv) \leq \min\{ \psi(\qv, \pv), \bar{m} \} + \epsilon \right )  = 1$. 

\item As mentioned earlier, both $R^{\rm CIC}(n,m,M,\qv,\pv)$ and $R^{\rm GCC}(n,m,M,\qv,\pv)$
are functions of the random caching placement $\Csf$. Hence, not only 
$\EE[R^{\rm GCC}(n,m,M,\qv,\pv) ] \leq \min\{\psi(\qv,\pv),\bar m\}$, but also the (conditional) rate 
$R^{\rm GCC}(n,m,M,\qv,\pv)$ concentrates its probability mass all to the left of the bound %concentrates its probability mass in  
$\min\{\psi(\qv,\pv),\bar m\}$, in the limit of  
$F, B \rightarrow \infty$ and $F/B \rightarrow$ constant. This means that in the large file limit, choosing a configuration of the caches that ``misbehaves'', 
i.e., that yields a rate larger than the bound of 
%that yields a rate larger than rate given in 
Theorem \ref{thm:up}, is an event of vanishing probability. 
%{\BLUE Hence, for ease of presentation, in the remainder of the body of this paper, we directly use $R^{\rm (\cdot)}(n,m,M,\qv,\pv)$ to denote the rate achieved by scheme $(\cdot)$ in the limit of $F, B \rightarrow \infty$.}
%In particular, we can also show that $\lim_{F \rightarrow \infty} \PP \left (| R^{\rm GCC_1}(n,m,M,\qv,\pv) - \psi(\qv, \pv)|  \leq  \epsilon \right )  = 1$ (See Appendix \ref{sec: Proof of Theorem up}), which means that $R^{\rm GCC_1}(n,m,M,\qv,\pv)$ concentrates its probability mass in $\psi(\qv,\pv)$, as $F, B \rightarrow \infty$ and $F/B \rightarrow $constant. Later, we can see that by designing $\pv$ carefully, $\psi(\qv, \pv) \leq \bar m$ in most regimes of parameter, which means that we can characterize almost the precise value of $R^{\rm GCC}(n,m,M,\qv,\pv)$ in most parameter regimes of our interest. In the following, for the ease of presentation, we just refer the bound given by Theorem \ref{thm:up} as the rate of $R^{\rm GCC}(n,m,M,\qv,\pv)$. }
\label{p2}

\item The achievable rate in Theorem \ref{thm:up} is given by the minimum between two terms. 
The first term, $\psi(\qv,\pv)$, follows from the analysis of GCC$_1$
given in Appendix \ref{sec: Proof of Theorem up}. In particular, we show that \\ %$R^{\rm GCC_1}(n,m,M,\qv,\pv) = \psi(\qv, \pv)$
$\lim_{F \rightarrow \infty} \PP \left (|R^{\rm GCC_1}(n,m,M,\qv,\pv) - \psi(\qv, \pv) | \leq  \epsilon \right )  = 1$, which means that $R^{\rm GCC_1}(n,m,M,\qv,\pv)$ concentrates its probability mass at 
$\psi(\qv,\pv)$, as $F, B \rightarrow \infty$ and $F/B \rightarrow $ constant. %Based on point \ref{p2} above, we directly write $R^{\rm GCC_1} = \psi(\qv, \pv)$.
%{\BLUE In particular, in Appendix \ref{sec: Proof of Theorem up}, we show that $R^{\rm GCC_1}(n,m,M,\qv,\pv) = \psi(\qv,\pv)$. }
The second term, $\bar{m}$, %follows from the analysis of GCC$_2$. In fact, $\bar m$
is simply the average number of distinct requested files, which is a natural upper bound of $R^{\rm GCC_2}$, the average number (normalized by $B$) of distinct requested (uncached) packets. %Hence $R^{\rm GCC_2} \leq \bar m$.}
%{\BLUE Appendix \ref{sec: Proof of Theorem up} is dedicated to the analysis of GCC$_1$. 
%As far as GCC$_2$ is concerned, its analysis is completely trivial. The number of colors produced by Algorithm \ref{algorithm: coloring 2} is equal to the number of distinct packets requested. This is given by $\sum_{f=1}^m \sum_{b=1}^B 1\{ (f,b) \in \Qsf \}$. Averaging with respect to $\Qsf$, since users make i.i.d. requests to files according to the demand distribution $\qv$, {\BLUE we have $\EE \left [ \sum_{f=1}^m \sum_{b=1}^B 1\{ (f,b) \in \Qsf \} \right ] \leq B \sum_{f=1}^m (1 - (1 - q_f)^n)$. Finally, dividing by $B$, we obtain the rate {\BLUE $\bar{m}= \sum_{f=1}^m (1 - (1 - q_f)^n)$, corresponding to the average number of distinct requested files}. 
As will be shown later, after careful design of the caching distribution $\pv$, the only case in which $\bar m < \psi(\qv, \pv)$ %and hence $R^{\rm GCC_2} < R^{\rm GCC_1}$ 
is in regimes of very small $M$, in which caching is shown to provide no order gains with respect to non-caching approaches such as conventional unicasting or naive multicasting of all requested files. Morover, in this regime, $\bar m$ becomes a tight upper bound of $R^{\rm GCC_2}$.
%\st{Later, we can see that by designing $\pv$ carefully, $\psi(\qv, \pv) \leq \bar m$ in most regimes of parameters, which means that we can characterize almost the precise value of $R^{\rm GCC}(n,m,M,\qv,\pv)$ in most cases. In the following,} 
Accordingly, we disregard $\epsilon$ and the fact that (\ref{eq:2}) involves a limit for $F \rightarrow \infty$ and identify
the rate achieved by GCC directly as 
%\begin{equation}
%\label{eq: abuse}
$R^{\rm GCC}(n,m,M,\qv,\pv) = \min\{ \psi(\qv, \pv), \bar{m} \}$. 
%\end{equation} 

\item The events underlying the probabilities $\rho_{f, \ell}$, defined in the statement of Theorem  \ref{thm:up}, can be illustrated as follows.
Let $\Dc$ be a random vector obtained by selecting in an i.i.d. fashion $\ell$ elements from $\Fc$ with probability $\qv$. Notice that $\Dc$ may contain repeated entries. 
By construction, $\PP(\Dc = (f_1, \ldots, f_\ell) ) = \prod_{i=1}^\ell q_{f_i}$. Then, 
$\rho_{f,\ell}$ is the probability that the element in $\Dc$ which maximizes the 
quantity  $(p_j M)^{\ell-1}(1-p_j M)^{n-\ell+1}$ is $f$. 

\item For the sake of the numerical evaluation of 
$\psi(\qv,\pv)$, it is worthwhile  to note that the probabilities $\rho_{f, \ell}$ can be easily computed as follows. 
Let $J, J_1, \ldots, J_\ell$ denote $\ell+1$ i.i.d. random variables distributed over $\Fc$ with same pmf $\qv$, and define (for simplicity of notation)
$g_\ell(j) \eqdef (p_j M)^{\ell-1} (1-p_j M)^{n-\ell+1}$. 
Since $g_\ell(J_1), \cdots, g_\ell(J_\ell)$ are i.i.d., the CDF of $Y_\ell \eqdef \max\{g_\ell(J_1), \cdots, g_\ell(J_\ell)\}$ is given by 
\be
\PP\left(Y_\ell \leq y \right) = \left (\PP\left( g_\ell(J) \leq y \right) \right)^\ell =  \left ( \sum_{j \in \Fc : g_\ell(j) \leq y} q_j \right )^\ell. 
\ee
Hence, it follows that 
\be
\label{rho}
\rho_{f,\ell} = \PP( Y_\ell = g_\ell(f) ) = \left(\sum_{j \in \Fc : g_\ell(j) \leq g_\ell(f)} q_j \right)^\ell - \left( \sum_{j \in \Fc : g_\ell(j) < g_\ell(f)}  q_j\right)^\ell,
\ee
\end{enumerate}
which can be easily computed by sorting the values $\{g_\ell(j) : j \in \Fc\}$. 

%%%%%%%%%%%%%%%%%%%%%%%%%%%%%%%%%%%%%%%%%%%%%%%%%%%%%%%%%%%%%%%%%%
%%%%%%%%%%%%%%%%%%%%%%%%%%%%%%%%%%%%%%%%%%%%%%%%%%%%%%%%%%%%%%%%%%
\subsection{Random Caching Optimization}
\label{sec: Proposed Scheme}

Driven by Theorem \ref{thm:up}, we propose to use as caching distribution the one that minimizes the rate $R^{\rm GCC}(n,m,M,\qv,\pv)$, i.e., 
\begin{align}
\label{eq:pstar}
& \pv^{*} = \underset{\pv:p_f\leq1/M,\sum_{f} p_f = 1}{ \arg\! \min} \min\{\psi(\qv,\pv), \bar m\},
\end{align}
where $\psi(\pv,\qv)$ is given by \eqref{eq: psi} with $\rho_{f,\ell}$ in \eqref{rho} and $\bar m$ is given by (\ref{eq: m bar}). 
In the following, we refer to random caching  according to the distribution $\pv^{*}$ as {\em RAndom Popularity-based} (RAP) caching placement. 
Consequently, the caching schemes with RAP placement and CIC or GCC delivery will be referred to as RAP-CIC and RAP-GCC, respectively. 

The distribution $\pv^*$ resulting from $\eqref{eq:pstar}$ may not have an analytically tractable expression 
in general. This makes a direct analysis of the performance of RAP-CIC and RAP-GCC difficult, if not impossible. 
To this end, in the following we also consider a simplified caching  placement according to the {\em truncated uniform
distribution} $\widetilde \pv$ defined by:
\begin{align}
\label{ptilde}
& \widetilde p_f = \frac{1}{\widetilde m}, \quad f \leq \widetilde m \notag\\
& \widetilde p_f = 0, \quad f \geq \widetilde m + 1
\end{align}
where the cut-off index $\widetilde m \geq M$ is a function of the system parameters. 

The form of  $\widetilde \pv$ in (\ref{ptilde}) is intuitive: each user caches the same fraction of (randomly selected) packets from each 
of the most $\widetilde m$ popular  files and does not cache any packet from the remaining $m - \widetilde{m}$ least popular files. 
If $\widetilde m = M$, this caching placement coincides with the least frequently used (LFU) caching policy \cite{lee2001lrfu}. 
For this reason, we refer to this caching placement as {\em Random LFU} (RLFU), and the corresponding 
caching schemes as RLFU-CIC and RLFU-GCC.   For later analysis purposes, we shall use a simplified upper bound  on the rate of RLFU-GCC given by the following corollary of Theorem \ref{thm:up}:

\begin{lemma}
\label{lemma: achievable}
For any $\epsilon > 0$, the rate achieved by RLFU-GCC satisfies 
\begin{equation}  
\label{eq: general achievable}
\lim_{F \rightarrow \infty} \PP \left ( R^{\rm GCC}(n,m,M,\qv,\widetilde{\pv}) \leq \min\left \{ \widetilde{\psi}(\qv,\widetilde{m}), \bar m\right \} + \epsilon \right )  = 1,
\end{equation}
where
\begin{equation}
\widetilde{\psi}(\qv,\widetilde{m}) =  \left( \frac{{\widetilde m}} {M}-1\right) \left ( 1 -\left(1-\frac{M}{\widetilde m} \right)^{ n \, G_{\widetilde m}} \right) 
+  n (1 - G_{\widetilde m}),
\end{equation}
with $G_{\widetilde m} \eqdef \sum_{f=1}^{\widetilde m} q_f$,  and where $\bar m$ is defined in (\ref{eq: m bar}). 
\end{lemma}

\begin{IEEEproof} See Appendix \ref{sec: Proof of Lemma achievable}. \end{IEEEproof}

For convenience, we disregard $\epsilon$ and the fact that (\ref{eq: general achievable}) involves a limit for $F \rightarrow \infty$ and refer directly 
to the achievable rate upper bound as
\be
\label{eq: ub RLFU-GCC}
R^{\rm ub}(n,m,M,\qv, \widetilde m) \eqdef \min\left \{ \widetilde{\psi}(\qv,\widetilde{m}), \bar m\right \}, 
\ee
where it is understood that the upper bound holds with high probability, as $F \rightarrow \infty$. 

While RLFU-GCC is generally inferior to RAP-GCC, we shall show in Section~\ref{order opt} that RLFU-GCC is sufficient to 
achieve order-optimal rate when $\qv$ is a Zipf distribution. In order to further shed light on the relative merits of the various approaches, 
in  Section \ref{sec: Discussion} we shall compare them in terms of actual rates (not just scaling laws),  obtained by simulation. 

%%%%%%%%%%%%%%%%%%%%%%%%%%%%%%%%%%%%%%%%%%%%%%%%% 
\section{Rate Lower Bound}
\label{sec: The General Lower Bound of the Achievable Rate}

In order to analyze the order-optimality of RLFU-GCC, we shall compare $R^{\rm ub}(n,m,M,\qv, \widetilde m)$ 
with a rate lower bound on the optimal  achievable rate $R^*(n,m,M,\qv)$. This is given by:

\begin{theorem}
\label{theorem: general lower bound}
Th rate $R(n,m,M,\qv)$ of any admissible scheme for the shared link network with $n$ users, library size $m$, cache capacity $M$,  
and demand distribution $\qv$ must satisfy
\begin{align}
\label{eq: general lower bound}
R(n,m,M,\qv) &\geq R^{\rm lb}(n,m,M,\qv) \notag\\
&\eqdef  %\max_{\ell, r, z} \;\; z P_1(\ell,r) P_2(\ell,r,z)  \left ( 1 - \frac{M}{\left\lfloor \frac{\ell}{{z}} \right\rfloor} \right ), 
\max_{\ell,r,\widetilde z}\left\{ P_1(\ell,r) P_2(\ell,r,\widetilde z)  \max_{z \in \{1, \cdots, \left\lceil \min\{\widetilde{z}, r\}\right\rceil\}}  z(1-M/{\lfloor \ell/z \rfloor})1\{\widetilde z,r \geq 1\},  \right. \notag\\
&  \left. \phantom{\max_{z \in \{1, \cdots, \left\lceil \min\{\widetilde{z}, r\}\right\rceil\}} } P_1(\ell,r)  P_2(\ell,1,\widetilde z) (1-M/{\ell})1\{\widetilde z \in (0,1)\}\right\}. 
\end{align}
where $\ell \in \{1, \ldots, m\}$, 
$r \in \RR_+$ with $r \leq n \ell q_\ell$ and 
$\widetilde z \in \RR_+$ with $\widetilde z \leq \min\left \{r, \ell \left(1 - \left(1 - \frac{1}{\ell}\right)^{r} \right) \right \}$, and where
\be
\label{eq: P1} 
P_1(\ell,r) \eqdef 1 - \exp\left(- \frac{\left(n \ell q_{\ell} - r \right)^2}{2n \ell q_{\ell}}\right),
\ee 
and
\be
\label{eq: P2} 
P_2(\ell,r,\widetilde z) \eqdef 1- \exp\left(- \frac{\left(\ell \left(1 - \left(1 - \frac{1}{\ell}\right)^{r} \right) - \widetilde z \right)^2}{2 \ell \left(1 - \left(1 - \frac{1}{\ell}\right)^{r} \right)}\right).
\ee
\end{theorem}

\begin{IEEEproof}
See Appendix \ref{sec: proof of theorem: general lower bound}.
\end{IEEEproof}

%%%%%%%%%%%%%%%%%%%%%%%%%%%%%%%%%%%%%%%%%%%%%%%%%%%%%%%%%%  HERE HERE HERE
%%%%%%%%%%%%%%%%%%%%%%%%%%%%%%%%%%%%%%%%%%%%%%%%%%%%%%%%%%
\section{Order-optimality}
\label{order opt}

The focus of this section is to prove the order-optimality of the RLFU-GCC scheme introduced in Section \ref{sec: Proposed Scheme}, 
when $\qv$ is a Zipf distribution (see (\ref{eq: Zipf})). 
%Using Lemma \ref{lemma: achievable} and Theorem \ref{theorem: general lower bound}, we shall consider the ratio
%$R^{\rm ub}(n,m,M,\qv, \widetilde m)/R^{\rm lb}(n,m,M,\qv)$ for $n \rightarrow \infty$, where the scaling of the other system parameters
%(namely, $m$ and $M$) as functions of $n$ is specified on a case by case basis, and where $\widetilde{m}$ is suitably chosen as a function of
%$n,m,M$ and of the Zipf parameter $\alpha$. When $m,M$ are not explicitly given in terms of $n$, it means that the corresponding scaling law holds for
%$m,M$ equal to any arbitrary functions of $n$, including $m,M$ constant, as a particular case. 
{
Using Lemma \ref{lemma: achievable} and Theorem \ref{theorem: general lower bound}, we shall consider the ratio
$R^{\rm ub}(n,m,M,\qv, \widetilde m)/R^{\rm lb}(n,m,M,\qv)$ for $n,m \rightarrow \infty$, where 
$m \rightarrow \infty$ and $n,M$ are functions  of $m$ as in Definition  \ref{def: order-optimal}.\footnote{When $M$ is not explicitly given in terms of $m$, it means that the corresponding scaling law holds for $M$ equal to any arbitrary functions of $m$, including $M$ constant, as a particular case.}} 
According to Definition \ref{def: order-optimal}, RLFU-GCC is order-optimal if the ratio
$R^{\rm ub}(n,m,M,\qv, \widetilde m)/R^{\rm lb}(n,m,M,\qv)$  is uniformly bounded for all sufficiently large $m$. 
Obviously, order-optimality of RLFU-GCC implies order-optimality of all ``better'' schemes, employing the optimized RAP distribution 
and/or CIC coded delivery.
%\footnote{The relationship between the rates of RAP and RLFU by using CIC, 
%$R^{\rm CIC}(n,m,M,\qv, \pv^*)$ and $R^{\rm CIC}(n,m,M,\qv,\widetilde \pv)$ is not known. However, it follows from Theorem \ref{thm:up} that
%$R^{\rm GCC}(n,m,M,\qv,\pv^*) \leq R^{\rm GCC}(n,m,M,\qv,\widetilde \pv)$ (with high probability) and, obviously, 
%$R^{\rm CIC}(n,m,M,\qv,\pv^*) \leq R^{\rm GCC}(n,m,M,\qv,\pv^*)$ and 
%$R^{\rm CIC}(n,m,M,\qv,\widetilde \pv) \leq R^{\rm GCC}(n,m,M,\qv, \widetilde \pv)$.} 

We shall also compare the (order optimal) rate achieved by RLFU-GCC with the rate achieved by other possibly suboptimal schemes,
such as conventional LFU caching with naive multicasting,\footnote{Recall that with conventional LFU every user caches the $M$ most popular files and hence there are no coded multicast opportunities. In fact, it is straight forward to show that if we fix the placement scheme to  (conventional) LFU in the shared link network, the best delivery scheme is naive multicasting.} 
and the scheme designed for arbitrary demands, achieving the order-optimal min-max rate \cite{maddah2012fundamental, maddah2013decentralized}. 
We shall say that a scheme A has an {\em order gain}  with respect to another scheme B if the rate achieved by A is $o(\cdot)$ of the rate achieved by B.  
We shall say that a  scheme A  has a  constant gain with respect to another scheme B if the rate of A is $\Theta(\cdot)$ of the rate of B, and their ratio converges
to some  $\kappa < 1$ as $m \rightarrow \infty$.  In addition, 
we shall say that some scheme A exhibits a {\em multiplicative caching gain} if its rate is inversely proportional to an increasing function of 
$M$. Specifically, we say that the multiplicative caching gain is sub-linear, linear, or super-linear if such function is sub-linear, linear, or super-linear in $M$, 
respectively.

We notice that the behavior of the Zipf distribution is fundamentally different in the two regions of the Zipf parameter 
$0 \leq \alpha < 1$ and $\alpha >1$.\footnote{%In this paper, we do not consider the case when the Zipf parameter $\alpha=1$.
{The regime $\alpha = 1$ requires not more difficult but somehow different analysis because of the bounding of the Zipf distribution (see Lemma 1 in \cite{ji2013throughput}). For the sake of brevity, given the fact that the analysis is already quite heavy, also motivated by the fact that most experimental data on content demands show $\alpha \neq 1$ \cite{breslau1999web}, in this paper, we omit this case.}}
In fact, for $\alpha < 1$, as $m \rightarrow \infty$, the probability mass is ``all in the tail'', i.e., the probability $\sum_{f=1}^{\widetilde{m}} q_f$ 
of the most probable $\widetilde{m}$ files vanishes, for any finite $\widetilde{m}$. 
In contrast, for $\alpha > 1$, the probability mass is ``all in the head'', i.e., 
for sufficiently large (finite) $\widetilde{m}$, the set of most probable $\widetilde{m}$ files contain almost all the probability mass, 
irrespectively of how large the library size $m$ is. In the following, we consider the two cases separately.

%%%%%%%%%%%%%%%%%%%%%%%%%%%%%%%%%%%%%%%%%%%%%%%%%%%%%%%%%%%%%%%%%
\subsection{Case $0 \leq \alpha < 1$}
\label{sec: gamma < 1}

In this case, we have:
\begin{theorem}
\label{theorem: gamma < 1}
For the shared link network with $n$ users, library size $m$, cache capacity $M$,  and random requests following a Zipf distribution $\qv$
with parameter $0 \leq \alpha<1$, RLFU-GCC with $\widetilde{m} = m$ yields order-optimal rate. 
The corresponding (order-optimal) achievable rate upper bound is given by $R^{\rm ub}(n,m,M,\qv, m) = \min \left\{ \left( \frac{m} {M}-1\right) \left ( 1 -\left(1-\frac{M}{m} \right)^{ n } \right), \bar m \right\}$.
\end{theorem}

\begin{IEEEproof} See Appendix \ref{sec: Proof of Theorem gamma < 1}. \end{IEEEproof}

RLFU with $\widetilde{m} = m$ corresponds to
caching packets at random, independently across users, with uniform distribution across all files in the library. 
Not surprisingly, the order-optimal rate given by Theorem \ref{theorem: gamma < 1} is order-equivalent to the 
min-max rate under deterministic demands \cite{maddah2013decentralized}.
We will refer to RLFU with $\widetilde{m} = m$ also as {\em uniform placement (UP)} and to UP-GCC as the scheme with UP as caching placement 
and GCC as delivery scheme.
Intuitively, this result is due to the heavy tail property of the Zipf distribution with $0 \leq \alpha < 1$ such that, in this case, the random demands are approximately uniform 
over the whole library and, from Lemma \ref{lemma: uniform} in Appendix \ref{sec: proof of theorem: general lower bound}, 
we know that the average rate under uniform random demands is order-equivalent 
to the min-max rate under arbitrary demands.

Nevertheless, making use of the knowledge of the Zipf parameter may yield fairly large constant rate gains, especially for 
$\alpha$ close to 1. In particular, we can optimize the parameter $\widetilde{m}$ as follows. 
Define $H(\alpha,x,y) \eqdef \sum_{i=x}^y i^{-\alpha}$ and consider the bounds on
the tail of the Zipf distribution given by the following lemma, proved in \cite{ji2013throughput}:

%%%%%%%%%%%%%%%%%%%%%%%%%%%
\begin{lemma}
\label{lemma: H}
If $\alpha \neq 1$, then
\begin{align}
\frac{1}{1-\alpha}(y+1)^{1-\alpha} - \frac{1}{1-\alpha}x^{1-\alpha} \leq H(\alpha,x,y) \leq 
\frac{1}{1-\alpha}y^{1-\alpha} - \frac{1}{1-\alpha}x^{1-\alpha} + \frac{1}{x^{\alpha}}.
\end{align}
\hfill$\square$
\end{lemma}
%%%%%%%%%%%%%%%%%%%%%%%%%%%%%%%%%%%%%%%%%%%%%%%
Notice that for the Zipf distribution with parameter $\alpha$, the term $G_{\widetilde{m}}$ in Lemma \ref{lemma: achievable}
is written explicitly as $G_{\widetilde{m}} = \frac{H(\alpha,1,\widetilde{m})}{H(\alpha,1,m)}$.
Then, using Lemma \ref{lemma: H} in Lemma \ref{lemma: achievable}, we can write
%{\BLUE [PLEASE CHECK THIS SINCE IN YOUR ORIGINAL VERSION THERE WERE PROBLEMS IN THESE STEPS .... THERE WERE SUMMATIONS
%WITH RESPECT TO $f$ THAT SHOULD  HAVE NOT BEEN THERE ...]}
\begin{eqnarray}
\label{eq: upper alpha<1}
R^{\rm ub}(n,m,M,\qv,\widetilde{m}) &\leq& \widetilde{\psi}(\qv, \widetilde{m})
\notag\\
& = & \left(\frac{{\widetilde m}}{M} - 1\right)\left(1 - \left(1 - \frac{M}{\widetilde m}\right)^{n \frac{H(\alpha,1,\widetilde m)}{H(\alpha,1,m)}}\right) + 
n \left(1-\frac{H(\alpha,1,\widetilde m)}{H(\alpha,1,m)} \right)  \notag\\
%&\leq& \left(\frac{{\widetilde m}}{M} - 1\right)\left(1 - \left(1 - \frac{M}{\widetilde m}\right)^{n\frac{\frac{1}{1-\alpha}(\widetilde m)^{1-\alpha} - \frac{1}{1-\alpha} + 1}{\frac{1}{1-\alpha}(m+1)^{1-\alpha} - \frac{1}{1-\alpha}}}\right) + n \left(1- \frac{\frac{1}{1-\alpha}(\widetilde m+1)^{1-\alpha}-\frac{1}{1-\alpha}}{\frac{1}{1-\alpha}m^{1-\alpha}-\frac{1}{1-\alpha}+1} \right)  \notag\\
&\buildrel (a) \over \leq& \left(\frac{\widetilde m}{M} -1 + \left(1-\left(\frac{\widetilde m}{m}\right)^{1-\alpha}\right)n\right),
\end{eqnarray}
where (a) follows from the fact that %is because that
\be
\left(\frac{{\widetilde m}}{M} - 1\right)\left(1 - \left(1 - \frac{M}{\widetilde m}\right)^{ n\frac{H(\alpha,1,\widetilde m)}{H(\alpha,1,m)}}\right) \leq \left(\frac{{\widetilde m}}{M} - 1\right),
\ee
and
\begin{eqnarray}
\frac{H(1,\alpha, \widetilde m)}{H(1,\alpha, m)} &\geq& \left(1- \frac{\frac{1}{1-\alpha}(\widetilde m+1)^{1-\alpha}-\frac{1}{1-\alpha}}{\frac{1}{1-\alpha}m^{1-\alpha}-\frac{1}{1-\alpha}+1} \right) \notag\\
&\buildrel m,\widetilde m \rightarrow \infty \over \longrightarrow&  \left(1-\left(\frac{\widetilde m}{m}\right)^{1-\alpha}\right)n. 
\end{eqnarray}

\begin{figure}[ht]
\centerline{\includegraphics[width=12cm]{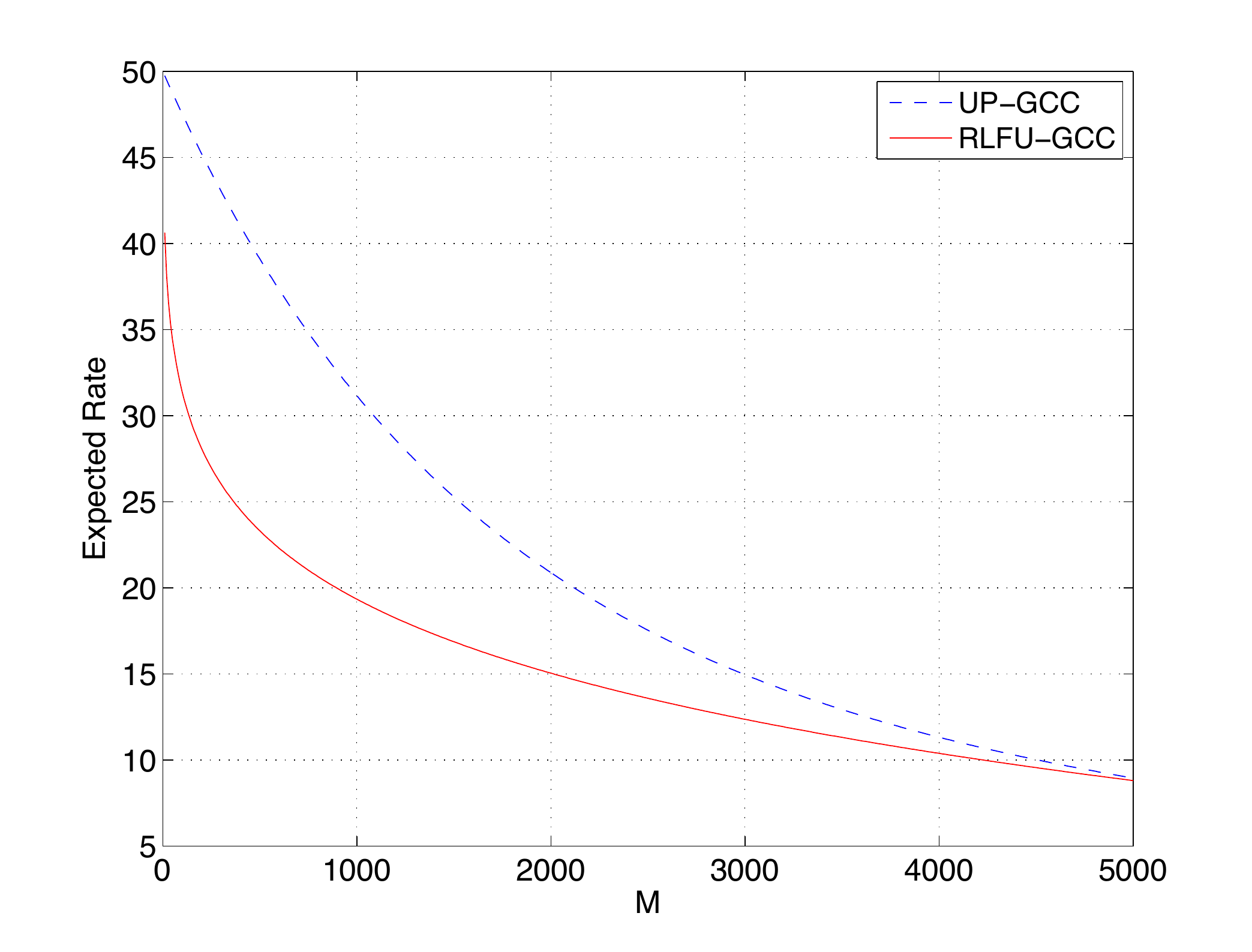}}
\caption{Rate versus cache size for UP-GCC and RLFU-GCC with  
optimized $\widetilde{m}$ (see (\ref{optimized-tile-m})), for $n=50$, $m=50000$, and Zipf parameter $\alpha=0.9$.}
\label{fig: gap RLFU and Uniform}
\end{figure}

Minimizing the upper bound given by (\ref{eq: upper alpha<1}) with respect to $\widetilde{m}$, subject to
$M \leq \widetilde m \leq m$, and treating $\widetilde{m}$ as a continuous variable, we obtain
\be  \label{optimized-tile-m}
\widetilde m= \min\left\{\max\left\{\left(\frac{n(1-\alpha)M}{m}\right)^{\frac{1}{\alpha}} m, M\right\}, m\right\}.
\ee
Fig.~\ref{fig: gap RLFU and Uniform}, shows the significant gains that can be achieved by using RLFU-GCC with optimized $\widetilde m$ (as given by \eqref{optimized-tile-m}) 
compared to UP-GCC  for a network with $m=50000$,  $n=50$ and $\alpha = 0.9$. 
For example, given a target rate of 20, UP-GCC  requires 
a cache capacity $M \approx 2000$, whereas RLFU-GCC with
optimized $\widetilde{m}$ requires only $M \approx 800$.

%%%%%%%%%%%%%%%%%%%%%%%%%%%%%%%%%%%%%%%%%%%%%%
\subsection{Case $\alpha > 1$}

This case is more intricate and we need to consider different sub-cases depending on how the number of users scales with the library size: 
namely,  we distinguish the cases of $n = \omega\left(m^{\alpha}\right)$, $n = \Theta\left(m^{\alpha}\right)$, 
and $n = o\left(m^{\alpha}\right)$. 

%%%%%%%%%%%%%%%%%%%%%%%%%%%%%%%%%%%%%%%%%%%%%%%%%%%%%%
\subsubsection{Regime of ``very large'' number of users: $n = \omega\left(m^{\alpha}\right)$}

\begin{theorem}
\label{theorem: gamma > 1 achievable 1}
For the shared link network with library size $m$, cache capacity $M$, 
random requests following a Zipf distribution $\qv$ with parameter $\alpha>1$, 
if $m \rightarrow \infty$ and the number of users scales as $n = \omega\left(m^{\alpha}\right)$,
UP-GCC (i.e., $\widetilde{m} = m$) achieves order-optimal rate.
\end{theorem}
%{\BLUE [Intuition based on multicast or aggregate distribution??]}
\begin{IEEEproof}
Theorem \ref{theorem: gamma > 1 achievable 1} can be proved by following the steps of 
the proof of Theorem \ref{theorem: gamma < 1} in Appendix \ref{sec: Proof of Theorem gamma < 1}. 
This is omitted for brevity. 
\end{IEEEproof}

%%%%%%%%%%%%%%%%%%%%%%%%%%%%%%%%%%%%%%%%%%%%%%%%%%%%%%%%%%%%%%
%%%%%%%%%%%%%%%%%%%%%%%%%%%%%%%%%%%%%%%%%%%%%%%%%%%%%%%%%%%%%%
\subsubsection{Regime of ``large" number of users $n = \Theta\left(m^{\alpha}\right)$}

\begin{theorem}
\label{theorem: gamma > 1 achievable 2}
For the shared link network with library size $m$, cache capacity $M$, 
random requests following a Zipf distribution $\qv$ with parameter $\alpha>1$, 
if $m \rightarrow \infty$ and the number of users scales as $n = \Theta\left(m^{\alpha}\right)$,
RLFU-GCC achieves order-optimal rate with the values of $\widetilde{m}$ given in Table \ref{table: table_2}, for different sub-cases of the 
system parameters. 
The corresponding (order-optimal) achievable rate upper bound $R^{\rm ub}(n,m,M,\qv,\widetilde{m})$ is also provided in Table \ref{table: table_2}.
\end{theorem}

\begin{IEEEproof} See Appendix \ref{Proof of Theorem gamma > 1 achievable 2}. \end{IEEEproof}

\begin{table}[ht]
\centerline{\includegraphics[width=12cm]{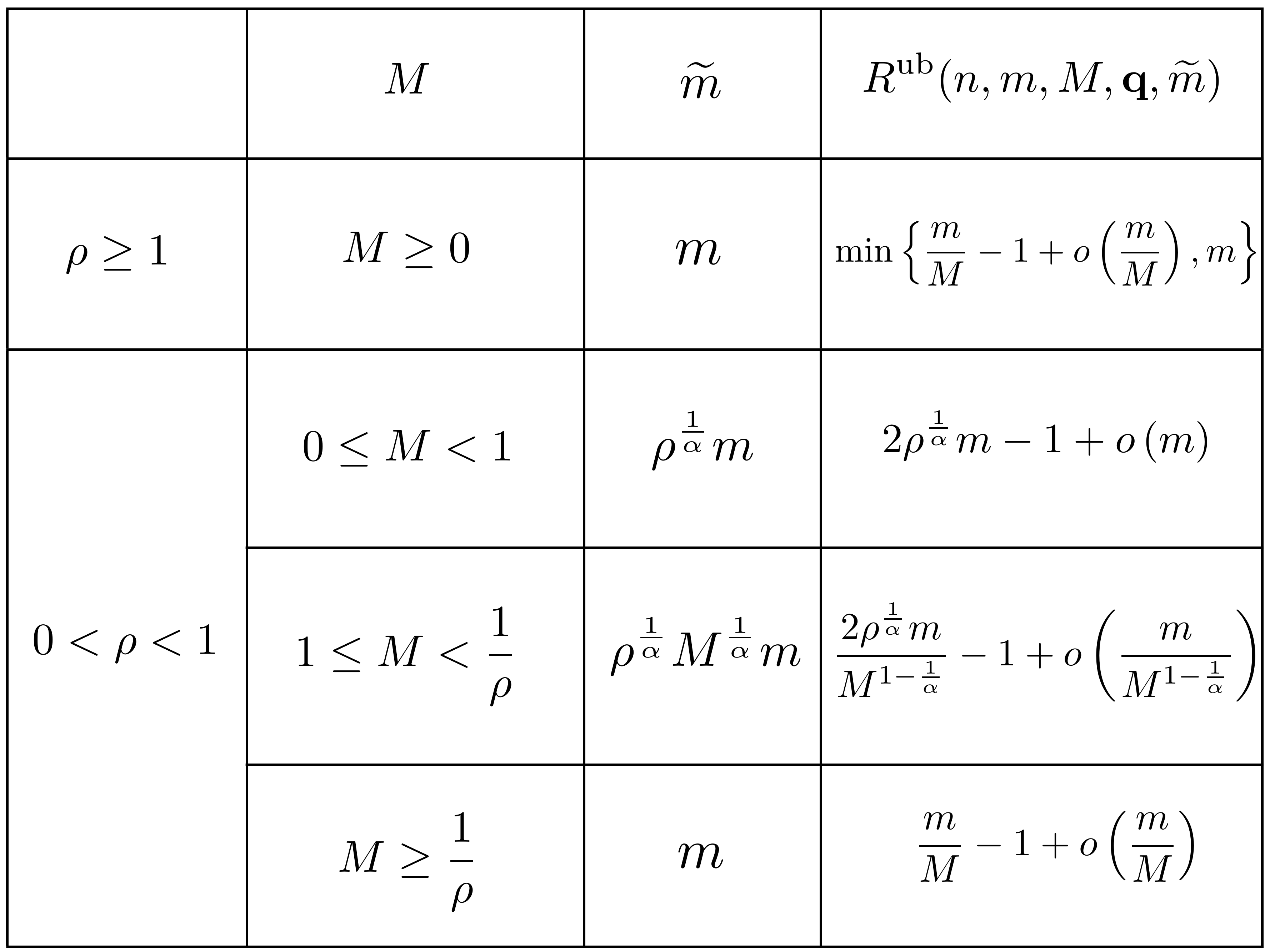}}
\caption{Order-optimal choice of $\widetilde{m}$ and the corresponding achievable rate upper bound $R^{\rm ub}(n,m,M,\qv, \widetilde m)$ for RLFU-GCC with $\alpha > 1$ and $n = \Theta\left(m^{\alpha}\right)$.}
\label{table: table_2}
\end{table}

%{\BLUE The ``boundary" case $n = \Theta\left(m^{\alpha}\right)$ has a similar, somehow simpler, behavior to both previous cases. }
Without loss of generality, we let the leading term of $n$ in terms of $m$ to be $\rho m^\alpha$ for some $\rho > 0$, i.e., 
$n = \rho m^\alpha + o(m^\alpha)$ for $m \rightarrow \infty$. Hence, we distinguish the following case:

\begin{itemize}
\item For $\rho > 1$, the network behaves similarly to the case of $n = \omega(m^{\alpha})$,
and UP-GCC achieves order-optimal rate, 
which scales as $\Theta\left(\frac{m}{M}\right)$ for $M \geq 1$. 

\item For $0 < \rho < 1$, we distinguish three regimes of $M$, namely, 
$0 \leq M < 1$, $1 \leq M < \frac{1}{\rho}$, and $M \geq \frac{1}{\rho}$. 
The corresponding order-optimal value of $\widetilde m$ varies from $\rho^{\frac{1}{\alpha}} m$, via $\rho^{\frac{1}{\alpha}}M^{\frac{1}{\alpha}}m$ 
to $m$. This corresponds to the order-optimal caching placement varying from RLFU to UP. 
Correspondingly, the scaling law of the rate varies from $\Theta\left(\rho^{\frac{1}{\alpha}} m\right)$, via $\Theta\left(\frac{\rho^{\frac{1}{\alpha}} m}{M^{1-\frac{1}{\alpha}}}\right)$ to $\Theta\left(\frac{m}{M}\right)$, where the multiplicative caching gain of RLFU-GCC (with order optimal $\widetilde m$) %over conventional unicasting and naive multicasting increases and the dependency of the throughput on $M$ 
varies from sub-linear to linear.
\end{itemize}

%%%%%%%%%%%%%%%%%%%%%%%%%%%%%%%%%%%%%%%%%%%%%%%%%%%%%%%%%%%%%%%%%
%%%%%%%%%%%%%%%%%%%%%%%%%%%%%%%%%%%%%%%%%%%%%%%%%%%
\subsubsection{Regime of ``small-to-moderate'' number of users $n = o\left(m^{\alpha}\right)$}

\begin{theorem}
\label{theorem: gamma > 1 achievable 3}
For the shared link network with library size $m$, cache capacity $M$, 
random requests following a Zipf distribution $\qv$ with parameter $\alpha>1$, 
if $m \rightarrow \infty$ and the number of users scales as $n = o\left(m^{\alpha}\right)$,
RLFU-GCC achieves order-optimal rate with the values of $\widetilde{m}$ given in Tables \ref{table: table_1_1} and
\ref{table: table_1_2}, for different sub-cases of the system parameters. 
The corresponding (order-optimal) achievable rate upper bound $R^{\rm ub}(n,m,M,\qv,\widetilde{m})$ is also 
provided in the Tables \ref{table: table_1_1} and
\ref{table: table_1_2}.
\end{theorem}
\begin{IEEEproof} See Appendices \ref{Proof of Theorem gamma > 1 3} and \ref{sec: proof of table II}. \end{IEEEproof}

%{\BLUE [NOTICE: IN THE TABLES YOU HAVE TO MAKE THE NOTATION CONSISTENT AND 
%IN PARTICULAR CHANGE $R^{\rm ub}(n,m,M, \widetilde{\pv}, \qv)$ INTO $R^{\rm ub}(n,m,M, \qv, \widetilde{m})$]}

\begin{table}[ht]
\centerline{\includegraphics[width=14cm]{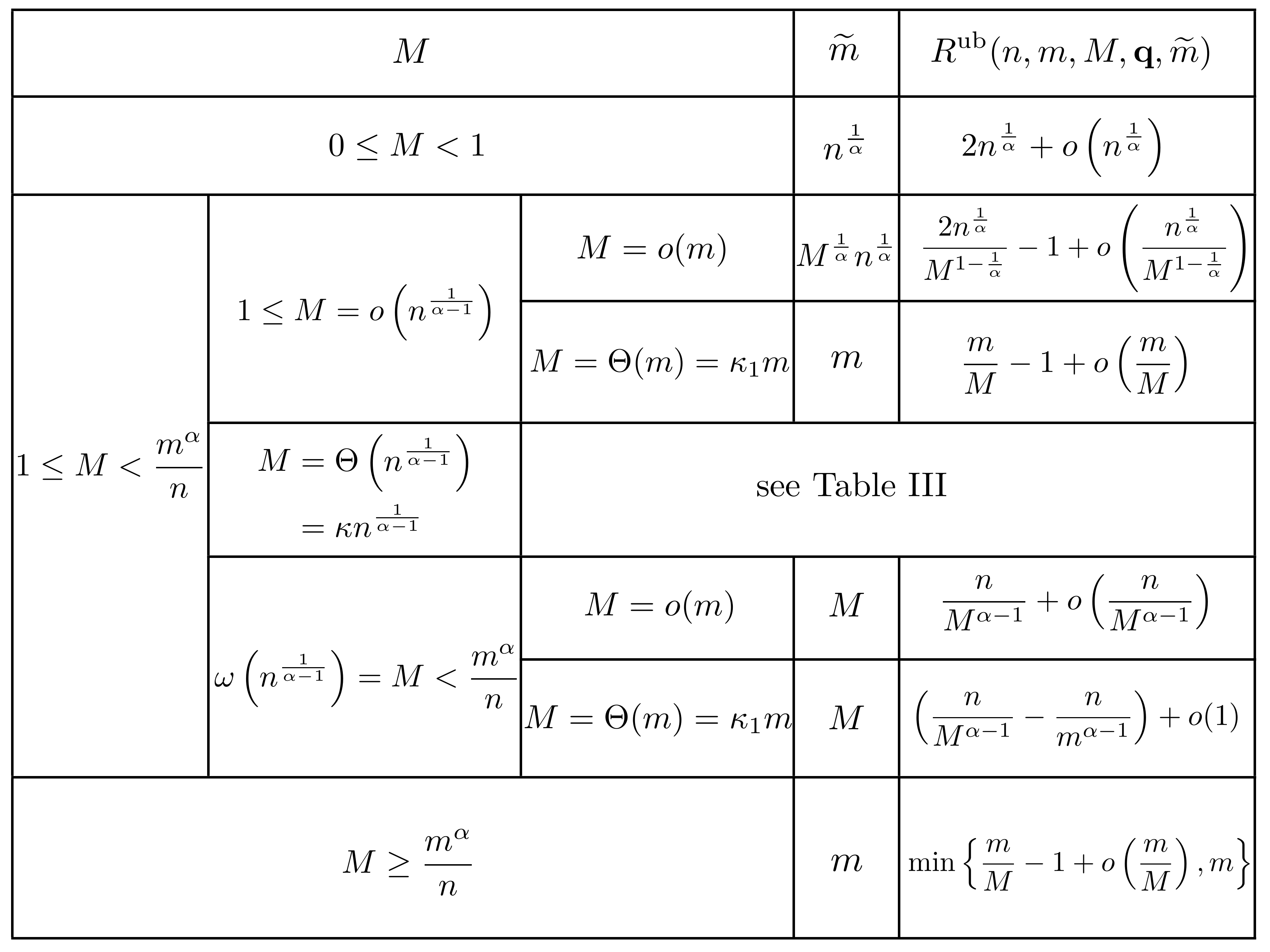}}
\caption{Order-optimal choice of $\widetilde m$ and the corresponding achievable rate upper bound $R^{\rm ub}(n,m,M, \qv, \widetilde{m})$ 
for RLFU-GCC with $\alpha > 1$ and $n = o\left(m^{\alpha}\right)$. Here, $0<\kappa_1<1$ indicates a fixed positive constant.}
\label{table: table_1_1}
\end{table}

\begin{table}[ht]
\centerline{\includegraphics[width=14cm]{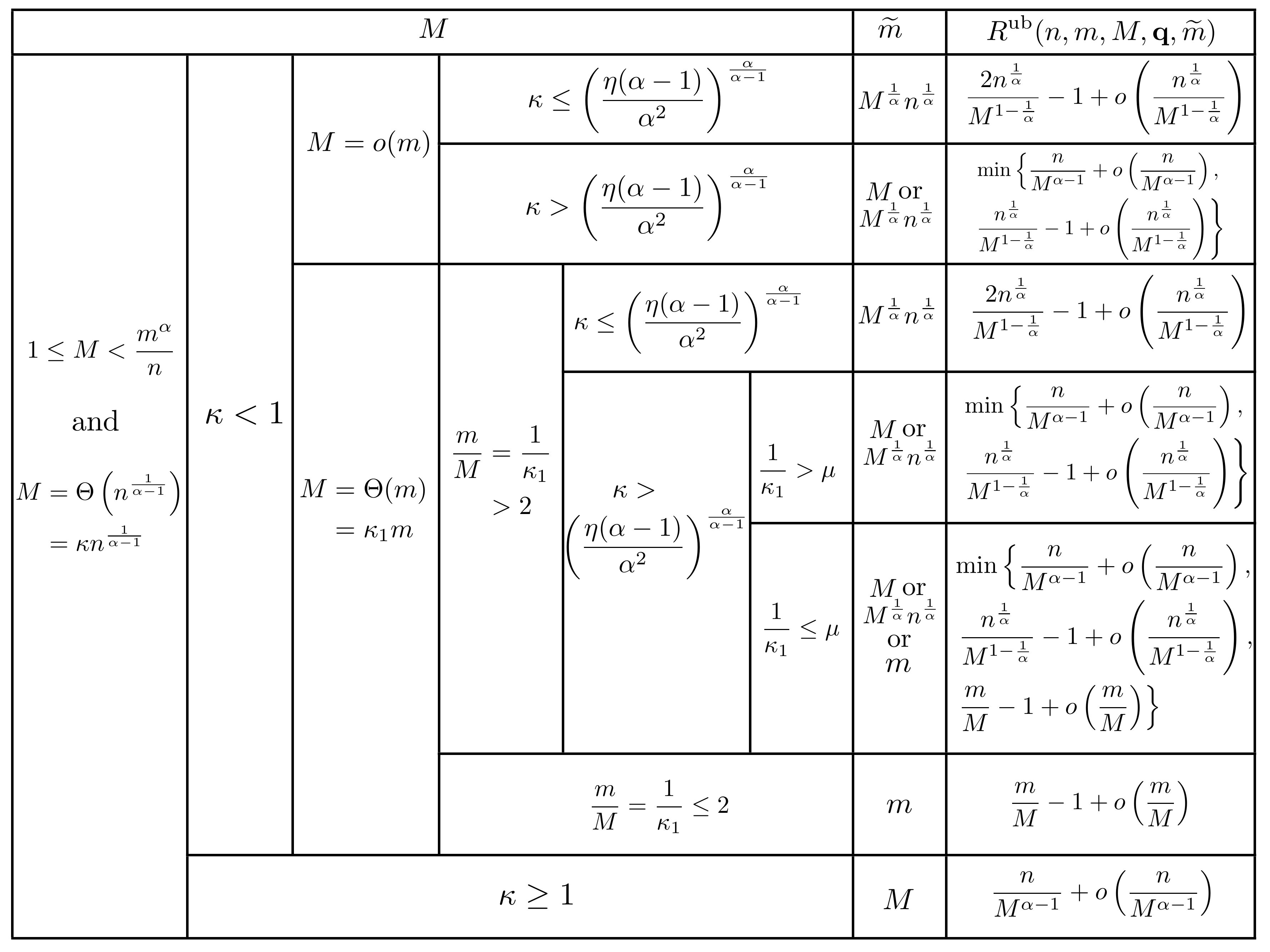}}
\caption{Sub-case of $1 \leq M < \frac{m^\alpha}{n}$ and $M = \Theta\left(n^{\frac{1}{\alpha-1}}\right)$, for the same regime of Table 
\ref{table: table_1_1}. Here, $\kappa, \kappa_1, \eta$ indicate fixed positive constants, and $\mu$ is an arbitrary positive constant larger than 2.}
\label{table: table_1_2}
\end{table}

Since Theorem \ref{theorem: gamma > 1 achievable 3} contains several regimes, it is useful to discuss separately 
some noteworthy behaviors. 
We start by consider the case of $n=o(m^{\alpha-1})$, for which 
there are two relevant regimes of $M$  (see Table \ref{table: table_1_1}), namely,    
$0 \leq M < 1$ and $1 \leq M < \frac{m^\alpha}{n}$. In particular:  

\begin{itemize}
\item 
If $0 \leq M < 1$, the achievable rate upper bound is $2n^{1/\alpha}$. 
{
This rate scaling can also be achieved by using naive multicasting for
all the requested files from the file set $\{1, \cdots, n^{\frac{1}{\alpha}}\}$ and conventional unicasting for the requested files from the remaining set $\{n^{\frac{1}{\alpha}}+1, \cdots, m\}$. 
It is not difficult to show (details are omitted) that the average number of 
distinct files requested from the set $\{n^{\frac{1}{\alpha}}+1, \cdots, m\}$ is $n^{\frac{1}{\alpha}} + o( n^{\frac{1}{\alpha}})$. Hence, 
both the naive multicasting and the conventional unicasting of the requested files from the respective sets require rate equal to $n^{1/\alpha}$ in the leading order, 
such that the concatenation of the two delivery schemes requires rate $2n^{1/\alpha}$. 
In order to achieve this (order-optimal) rate scaling, caching is not needed at all.} 
We conclude that, in this regime of ``small storage capacity'' ($M<1$), 
caching does not achieve any significant gain over the simple non-caching strategy described above, based on combining 
naive multicasting for the most popular files and conventional unicasting of the remaining less popular files. 

\item
For the case $1 \leq M < \frac{m^\alpha}{n}$, we notice that the assumption  $n = o(m^{\alpha-1})$ yields  
$\frac{m^\alpha}{n} = \omega( m)$. Hence,  the constraint $M < \frac{m^\alpha}{n}$ is dominated by the obvious
condition $M \leq m$, which always holds by definition.\footnote{If $M \geq m$, then each user can cache the whole library and the rate is trivially zero.} 
Considering that $n = o(m^{\alpha-1})$ implies that $m = \omega(n^{\frac{1}{\alpha-1}})$, 
we distinguish the following three sub-cases:  $1\leq M = o\left(n^{\frac{1}{\alpha-1}}\right)$, 
$M = \Theta\left(n^{\frac{1}{\alpha-1}}\right)$, and $\omega\left(n^{\frac{1}{\alpha-1}}\right)=M \leq m$.
We now discuss in more details the two regimes $1\leq M = o\left(n^{\frac{1}{\alpha-1}}\right)$ and $\omega\left(n^{\frac{1}{\alpha-1}}\right)=M\leq m$ 
shown in Table \ref{table: table_1_1}, while we refer the reader to Table \ref{table: table_1_2} for the case $M=\Theta\left(n^{\frac{1}{\alpha-1}}\right)$. 
In the case $1\leq M = o\left(n^{\frac{1}{\alpha-1}}\right)$ or, equivalently, $\omega(M^{\alpha-1}) = n 
< m^{\alpha-1}$, if $M = o(m)$, then the order-optimal RLFU parameter is $\widetilde m = M^{\frac{1}{\alpha}}n^{\frac{1}{\alpha}}$.
In this case, the rate is $\Theta\left(\frac{n^{\frac{1}{\alpha}}}{M^{1-\frac{1}{\alpha}}}\right)$, which exhibits order gain 
with respect to the rate obtained with UP, given by $\Theta\left(\min\left\{\frac{m}{M},m,n\right\}\right)$.
This also shows that the order-optimal average rate in this regime yields an order gain with respect to the min-max order-optimal rate \cite{maddah2012fundamental, maddah2013decentralized}.

We interpret this order gain as the benefit due to caching according to popularity. 
Intuitively, when $\alpha > 1$ and the number of users is not very large 
($n=o(m^{\alpha-1})$), only a limited number of files are  requested with non-vanishing probability. 
Meanwhile, $1\leq M = o\left(n^{\frac{1}{\alpha-1}}\right)$ and $n = o\left(m^{\alpha-1}\right)$ imply that
the cache capacity is $M = o(m)$, i.e., only a sublinear number of files can be cached. 
Hence, it is critically important to be able to focus on the files that deserve to be cached. 
We conclude that, in this case, caching according to the knowledge of the demand distribution makes a significant 
difference (in fact, a difference in the rate scaling order) with respect to UP.

In the other regime, $\omega\left(n^{\frac{1}{\alpha-1}}\right)=M\leq m$,  the cache size $M$ can be large. 
In this case, LFU (obtained by letting $\widetilde{m} = M$) combined with the naive multicasting of the uncached requested files 
achieves order-optimal rate, which scales as $\Theta\left(\frac{n}{M^{\alpha-1}}\right)$ (see Table \ref{table: table_1_1}). 
Again, this rate exhibits an order gain with respect to the min-max order-optimal rate.

Intuitively, this is due to the fact that, in this case, users request relatively few files, most of which are the popular ones. 
Since the storage capacity is large, then LFU caching covers most of the requests and the source node 
only needs to serve the unpopular requests, which account for a vanishing rate 
($\Theta\left(\frac{n}{M^{\alpha-1}}\right) = o(1)$).  
In addition, we observe that the multiplicative caching gain becomes super-linear for $\alpha>2$. % of the per-user throughput with the cache size $M$ does not hold in general (it only holds for $\alpha=2$). 

\end{itemize}

Then, we examine the case of $o(m^\alpha) = n = \omega\left(m^{\alpha-1}\right)$,\footnote{We do not discuss the case of $n = \Theta\left(m^{\alpha-1}\right)$ 
for the sake of brevity and ease of presentation. The corresponding result can be found in Table \ref{table: table_1_1} and \ref{table: table_1_2}.} 
where the number of users is relatively large. The relevant regimes of $M$ (see Table \ref{table: table_1_1}) in this case are 
$0 \leq M < 1$, $1 \leq M < \frac{m^\alpha}{n}$, and $\frac{m^\alpha}{n} \leq M < m$.
Keeping the order of $n$ in $m$ fixed and increasing the 
order of $M$ in $m$, the order-optimal $\widetilde m$ varies from $n^{\frac{1}{\alpha}}$ 
via $M^{\frac{1}{\alpha}}n^{\frac{1}{\alpha}}$ to $m$, 
indicating that the caching placement converges to UP
instead of LFU, in contrast to the case of $n = o\left(m^{\alpha-1}\right)$ considered before. 
This shows that in this regime, with the exception of the ``small storage capacity" regime $M < 1$, 
LFU with naive multicasting fails to achieve order-optimality.  In addition, as the order of $M$ in $m$ increases, 
the scaling law of the rate varies from  $\Theta\left(n^{\frac{1}{\alpha}}\right)$, via $\Theta\left(\frac{n^{\frac{1}{\alpha}}}{M^{1-\frac{1}{\alpha}}}\right)$ 
to $\Theta\left(\frac{m}{M}\right)$. This indicates that the multiplicative caching gain % of the throughput on cache memory $M$ 
goes from sub-linear to linear.

\subsection{Remark}

We finally remark that in this paper, for the sake of presentation clarity, we let $n$ be a function of $m$ such that $n \rightarrow \infty$ as $m \rightarrow \infty$. 
However, when $m$ is a constant independent of $n$, and $n \rightarrow \infty$, the ratio between $R^{\rm ub}(n,m,M,\qv,\widetilde{m})$ and $R^{\rm lb}(n,m,M,\qv)$ is also upper bounded by a constant, which is shown by the following corollary.
\begin{corollary}
\label{corollary: m constant}
For the shared link network with $n$ users, a library of constant size $m$, cache capacity $M$,  
and random requests following a Zipf distribution $\qv$ with parameter $\alpha \geq 0$,  UP-GCC achieves order-optimal  with the following gap guarantee:
\begin{eqnarray}
\limsup_{n \rightarrow \infty} \frac{R^{\rm ub}(n,m,M,\qv,m)}{R^{\rm lb}(n,m,M,\qv)} \leq \frac{12}{1-\epsilon},
\end{eqnarray}
for some arbitrarily small $\epsilon > 0$, independent of $m,n,M$.
\end{corollary}
\begin{IEEEproof} 
See Appendix \ref{sec: proof of corollary m constant}.
\end{IEEEproof}

%%%%%%%%%%%%%%%%%%%%%%%%%%%%%%%%%%%%%%%%%%%%%%%%%%%%%%%%%%%%%%
\section{Discussions and Simulation Results}
\label{sec: Discussion}

In Section \ref{order opt}, we have seen that, under a Zipf demand distribution, 
RLFU-GCC with $\widetilde m$ given in Tables  \ref{table: table_2}, \ref{table: table_1_1}, and \ref{table: table_1_2}  achieves order-optimal rate and so do RLFU-CIC, RAP-GCC and RAP-CIC. In all these schemes, once the cache configuration is given, 
the delivery phase reduces to an index coding problem. Despite the fact that for general index coding no graph coloring scheme is known to be sufficient to guarantee order-optimality \cite{haviv2012linear}, for the specific problem at hand we have the pleasing result that CIC and even the simpler GCC are sufficient for order optimality. % is achieved. 

\begin{figure}[ht]
\centerline{\includegraphics[width=12cm]{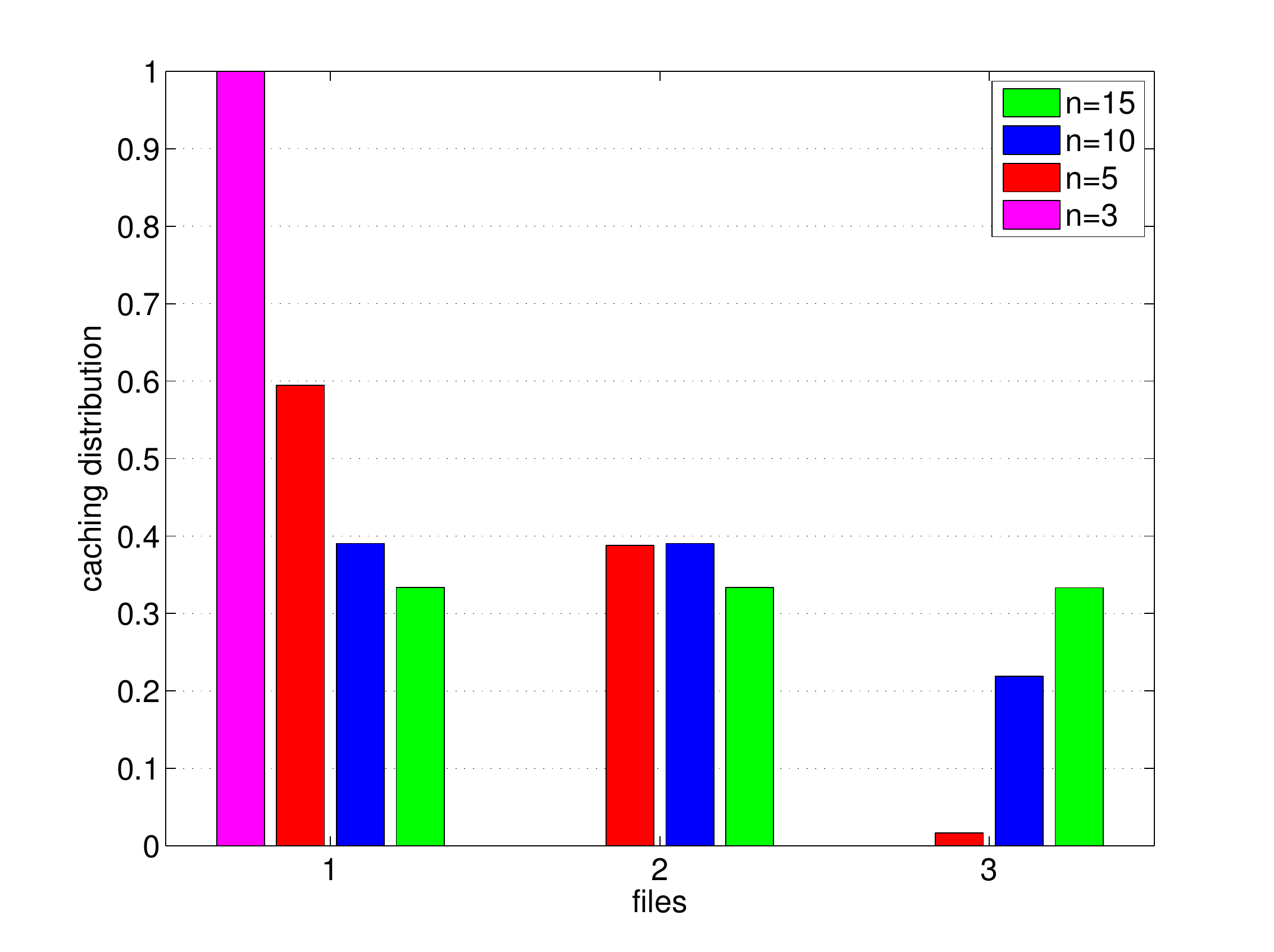}}
\caption{The optimal caching distribution $\pv^*$ for a network, where $m=3$, $M=1$ and $n=3,5,10,15$ and the demand distribution is $\qv = [0.7,0.21, 0.09]$.}
\label{fig: p_opt_versus_files}
\end{figure}

{While this result is proved by considering the RLFU-GCC scheme, for the sake of analytical simplicity, 
one would like to directly  use RAP-GCC or, better, RAP-CIC, to achieve some further gain in terms of actual rate, beyond the scaling law. 
While the minimization of  $\min\{ \psi(\qv, \pv), \bar{m} \}$ given in (\ref{eq:2}) with respect to $\pv$ is a non-convex problem without
a appealing structure, it is possible to use brute-force search or branch and bound methods \cite{land1960automatic} 
to search for good choices of the caching distribution $\pv$. 
Fig.~\ref{fig: p_opt_versus_files} shows $\pv^*$ obtained by numerical minimization of the bound
$\min\{ \psi(\qv, \pv), \bar{m} \}$ for a toy case with $m=3$, $M=1$, $n=3,5,10,15$ and demand distribution $\qv = [0.7,0.21, 0.09]$.  
We observe how the caching distribution $\pv^*$, which does not necessarily coincide with $\qv$,} adjusts according to the system 
parameters to balance the local caching and coded multicasting gains. In particular, 
$\pv^*$ goes from caching the most popular files (as in LFU) for $n=3$ to UP for $n=15$. 
Recall from Theorems \ref{theorem: gamma > 1 achievable 1}-\ref{theorem: gamma > 1 achievable 3}  that the optimized $\tilde \pv$ follows this same 
trend, going from LFU ($\widetilde m=M$) to  UP ($\widetilde m=m$) as $n$ increases, while constrained to be a step function.
This, perhaps surprising, behavior arises from the fact that 
even if the ``local" demand distribution $\qv$ is fixed, when the number of users increases, the ``aggregate" demand distribution, i.e., the probability that a file gets requested 
at least by one user, flattens. This effectively uniformizes the ``multicast weight" of each file, requiring caching distributions that 
flatten accordingly.

\begin{figure}[ht]
\centerline{\includegraphics[width=12cm]{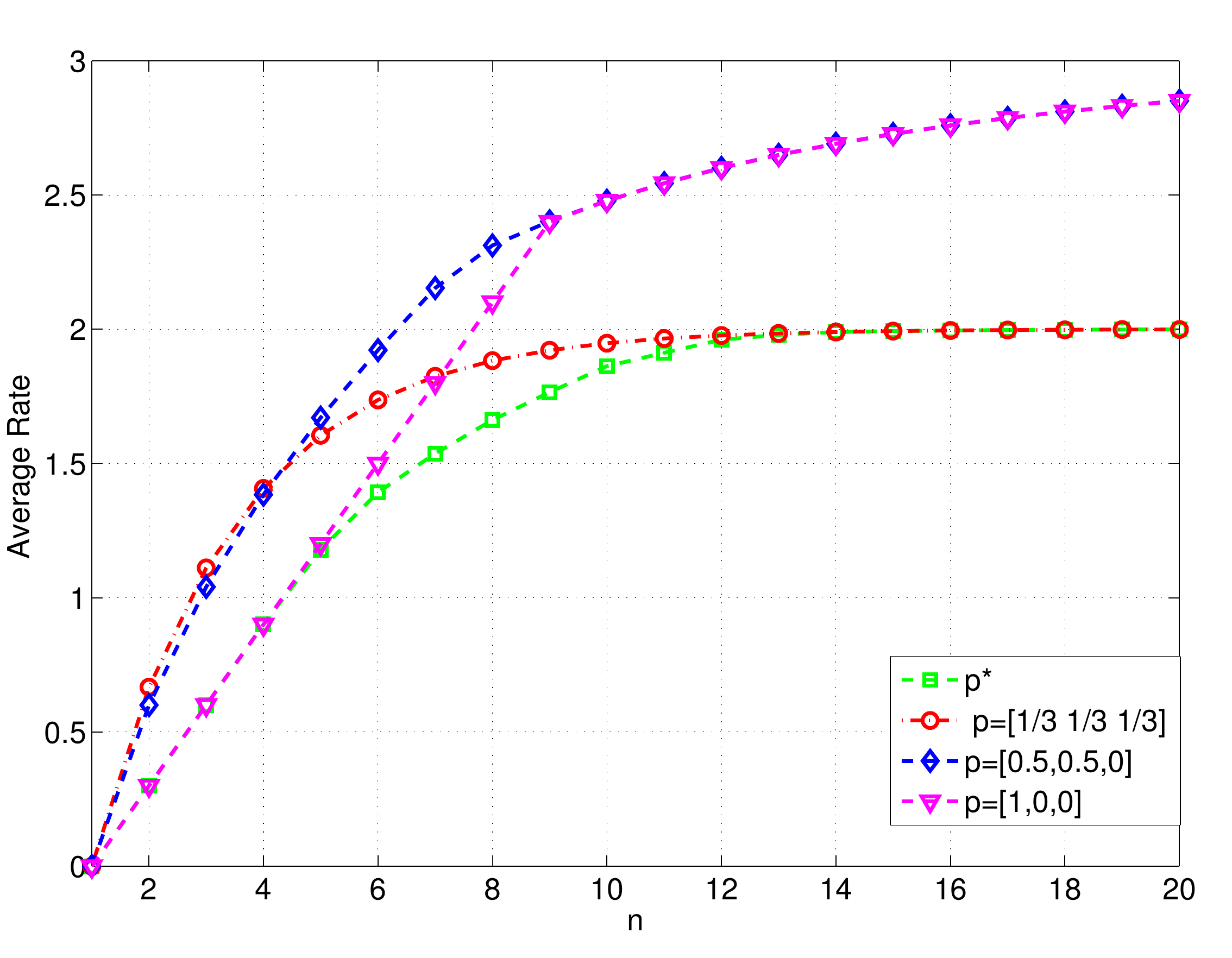}}
\caption{$R^{\rm GCC}(n,m,M,\qv,\pv)$ for different caching distributions $\pv$ 
 and for a network with  $m=3$, $M=1$ and $n=3,5,10,15$ and  demand distribution $\qv = [0.7,0.21, 0.09]$. Note that $\pv = [1/3,1/3,1/3]$, $\pv = [0.5,0.5,0]$, and $\pv = [1,0,0]$ correspond to RLFU with $\widetilde m=3$, $\widetilde m=2$, and $\widetilde m=1$, respectively.}
\label{fig: Rate_optimized}
\end{figure}

The corresponding achievable rate, $R^{\rm GCC}(n,m,M,\qv,\pv^*)$ with $\pv^*$ given by \eqref{eq:pstar}, is shown in Fig.~\ref{fig: Rate_optimized}, confirming the performance improvement provided by RAP-GCC. 
For comparison, Fig.~\ref{fig: Rate_optimized} also shows the rates achieved by RLFU-GCC, $R^{\rm GCC}(n,m,M,\qv, \tilde\pv)$ with $\tilde m = 1,2,$ and $3$.

As discussed in Theorems \ref{theorem: gamma > 1 achievable 1}-\ref{theorem: gamma > 1 achievable 3}, 
the fact that the caching distribution adjusts to changes in all system parameters, and not just the demand distribution $\qv$, is a key aspect of our order-optimal 
schemes and one of the reasons for which previously proposed schemes have failed to provide order-optimal guarantees. 

In addition, unlike uncoded delivery schemes that transmit each non-cached packet separately, or the scheme suggested in \cite{niesen2013coded}, where files are grouped into subsets and coding is performed within each subset, another key aspect of the order-optimal schemes presented in this paper is the fact that coding is allowed within the entire set of requested packets. 
When treating different subsets of files separately,  missed coding opportunities can significantly degrade efficiency of coded multicasting. 

For example, in the setting of Fig. \ref{fig: Rate_optimized} with $M=1.5$ and $n=20$,
by following the recipe given in \cite{niesen2013coded},~\footnote{The achievable rate for the scheme proposed in \cite{niesen2013coded} is computed based on a grouping of the files,  an optimization of the memory assigned to each group, and a separate coded transmission scheme for each group, as described in \cite{niesen2013coded}.} each of the $m=3$ files becomes a separate group, delivered independently of each other, %get assigned a different memory and have to be delivered separately, which, after optimization by brute force searching, 
yielding an expected rate of $1.5$, which can also be achieved by conventional LFU with naive multicasting. 
On the other hand, for this same setting, RLFU-GCC uses a uniform caching distribution and GCC over all requested 
packets, yielding a rate of $0.5$. 

\begin{figure}[ht]
\centering
\subfigure[]{
\centering \includegraphics[width=7.5cm]{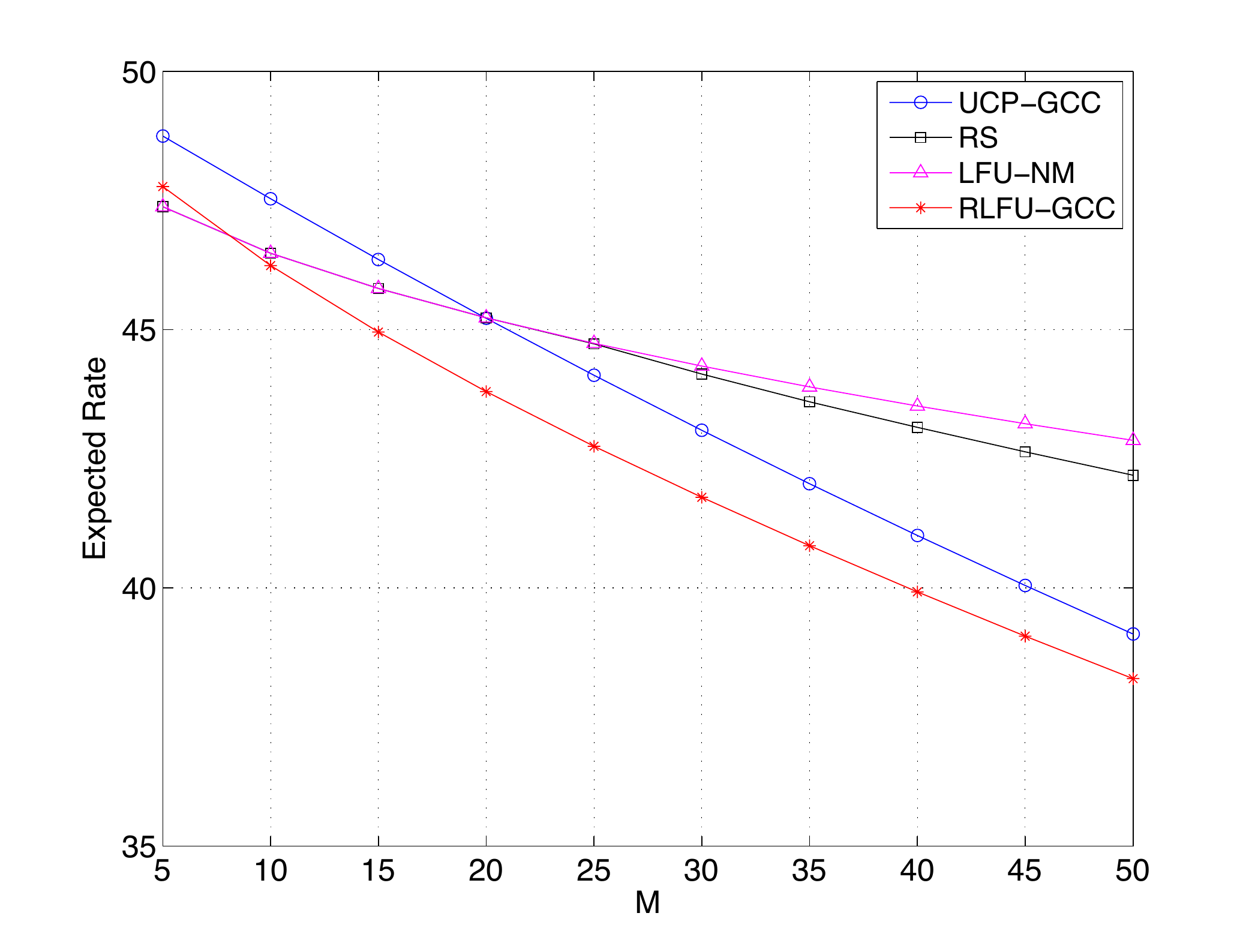}
%%\label{fig: }
}
\subfigure[]{
\centering \includegraphics[width=7.5cm]{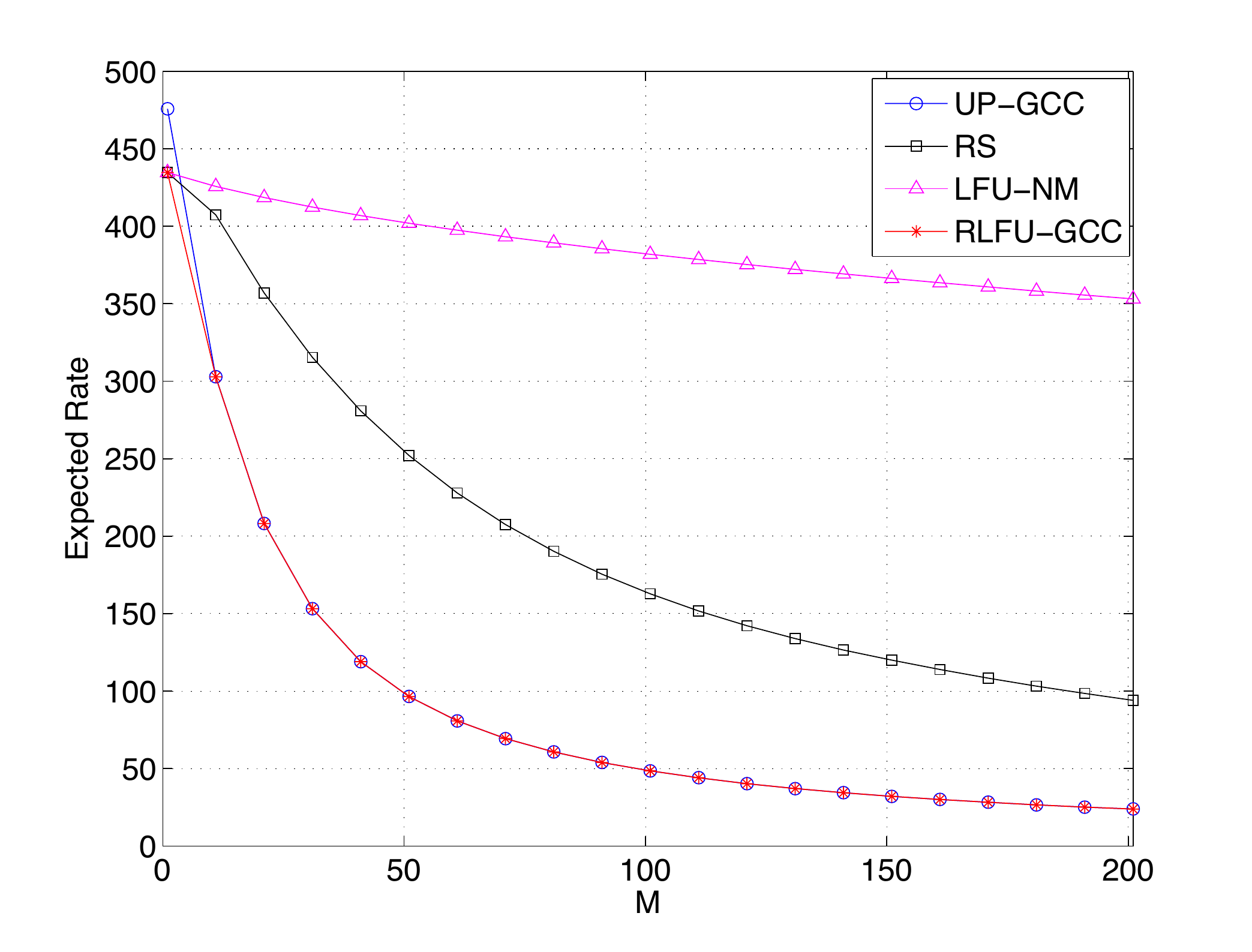}
%%\label{fig: }
}
\subfigure[]{
\centering \includegraphics[width=7.5cm]{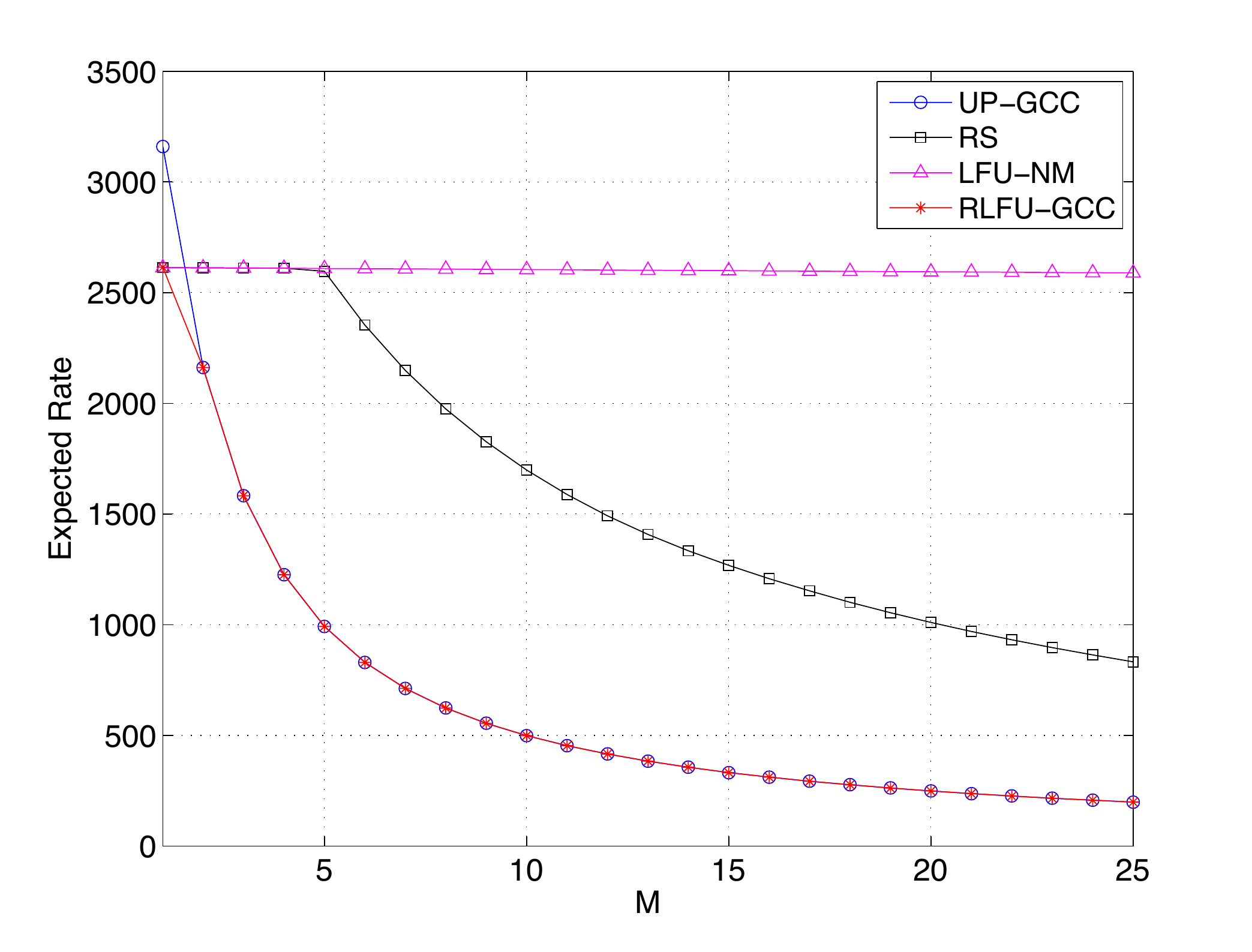}
%%\label{fig: }
}
\subfigure[]{
\centering \includegraphics[width=7.5cm]{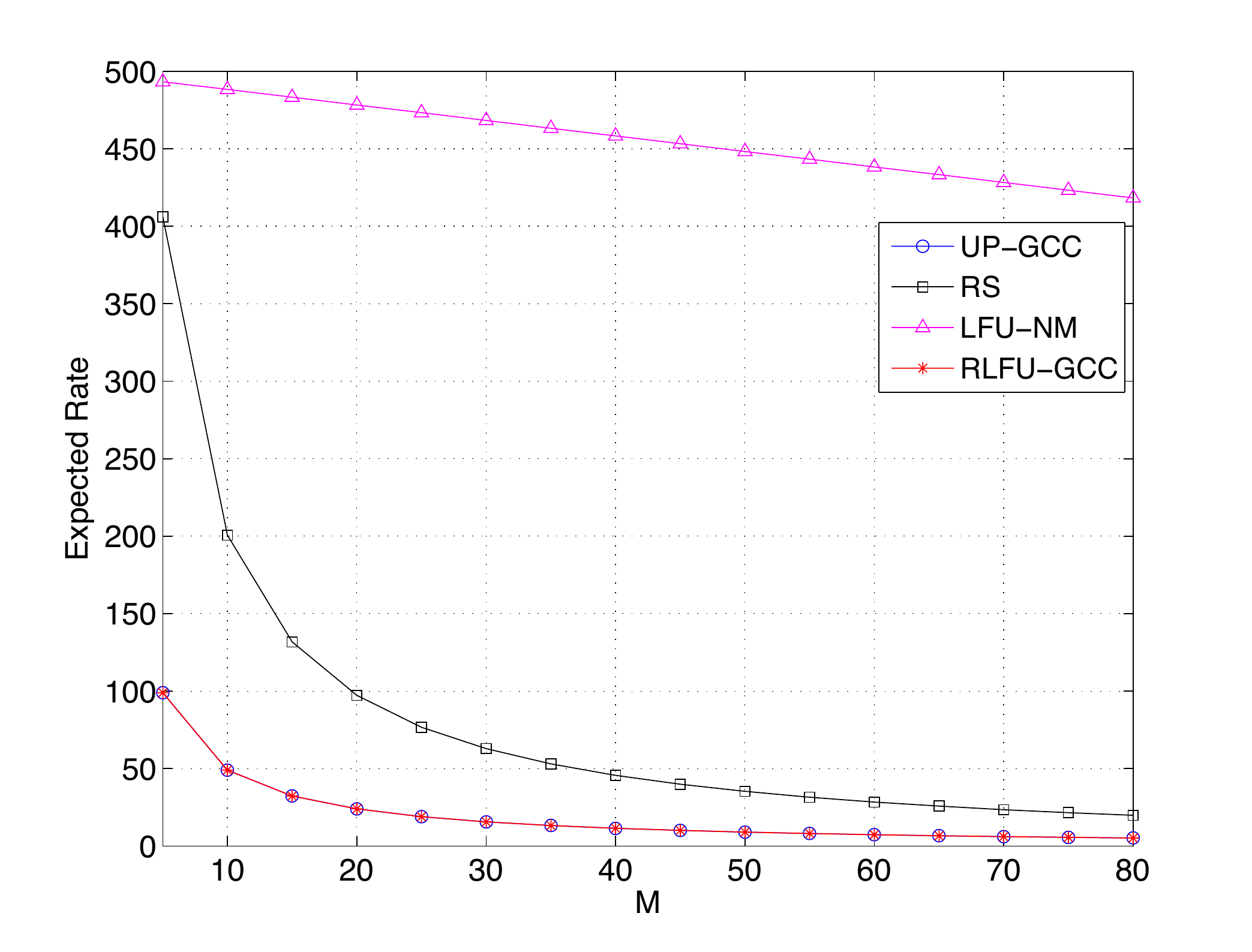}
%%\label{fig: }
}
\caption{ Simulation results for $\alpha=0.6$. a)~$m=5000,n=50$. b)~$m=5000,n=500$. c)~$m=5000,n=5000$. d)~$m=500,n=5000$. RLFU in this figure corresponds to the RLFU with optimized $\widetilde{m}$ given by (\ref{eqmopt}). %RS denotes the  reference scheme presented in \cite{niesen2013coded}. %Except for the reference scheme, the caching placements are denoted by the legends and the delivery scheme is GCC for all the simulations.
}
\label{fig: result 1}
\end{figure}

\begin{figure}[ht]
\centering
\subfigure[]{
\centering \includegraphics[width=7.5cm]{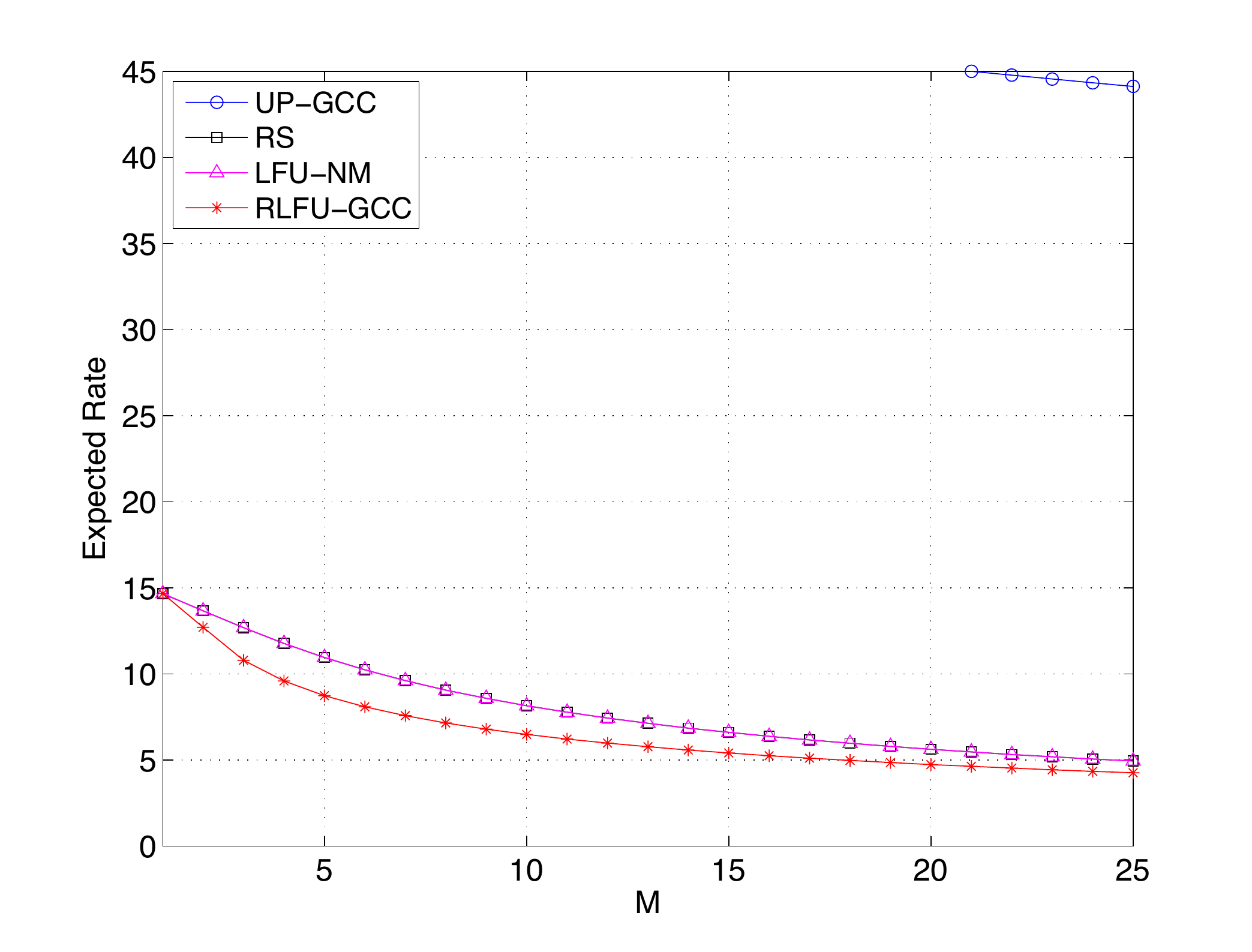}
%\label{fig: }
}
\subfigure[]{
\centering \includegraphics[width=7.5cm]{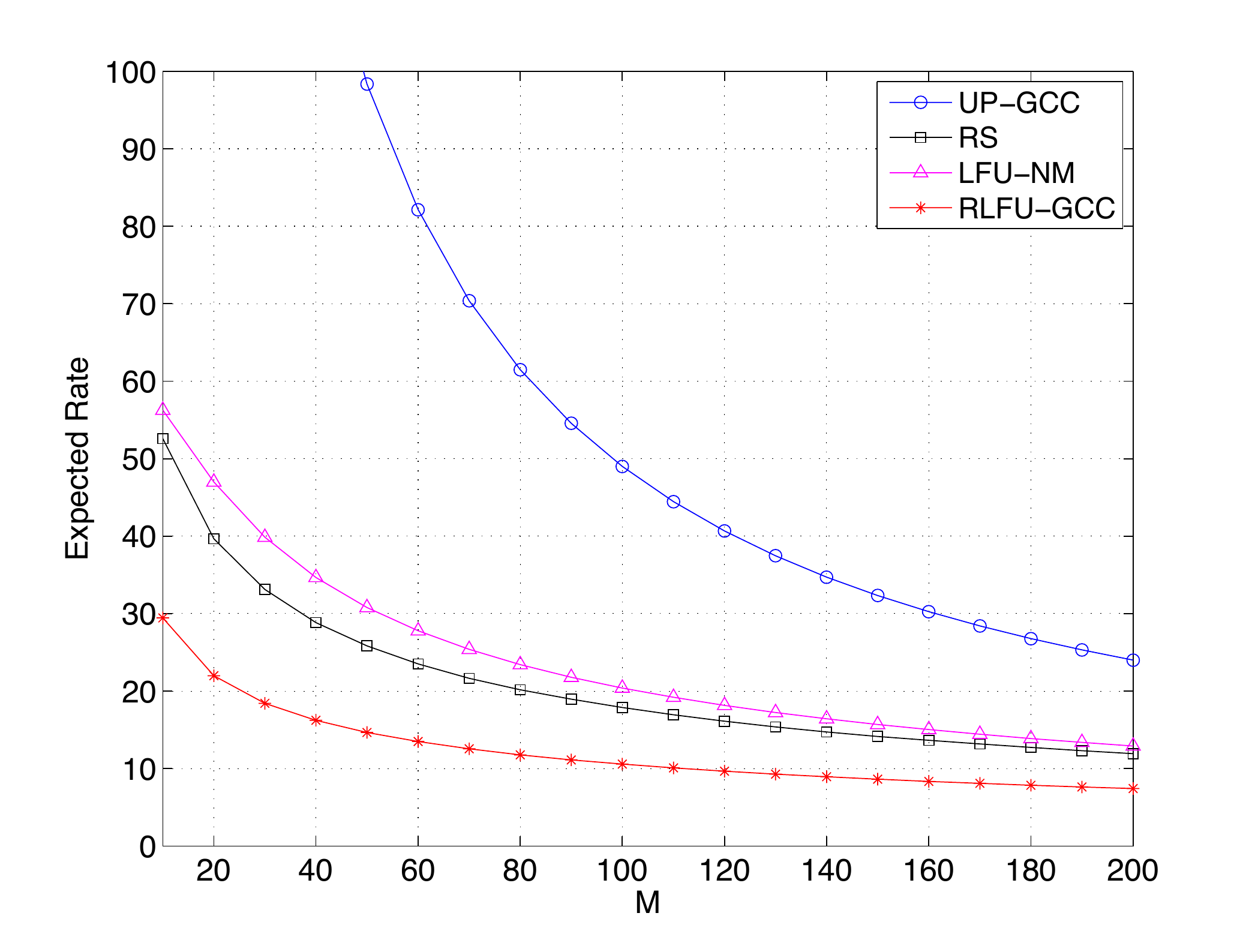}
%\label{fig: }
}
\subfigure[]{
\centering \includegraphics[width=7.5cm]{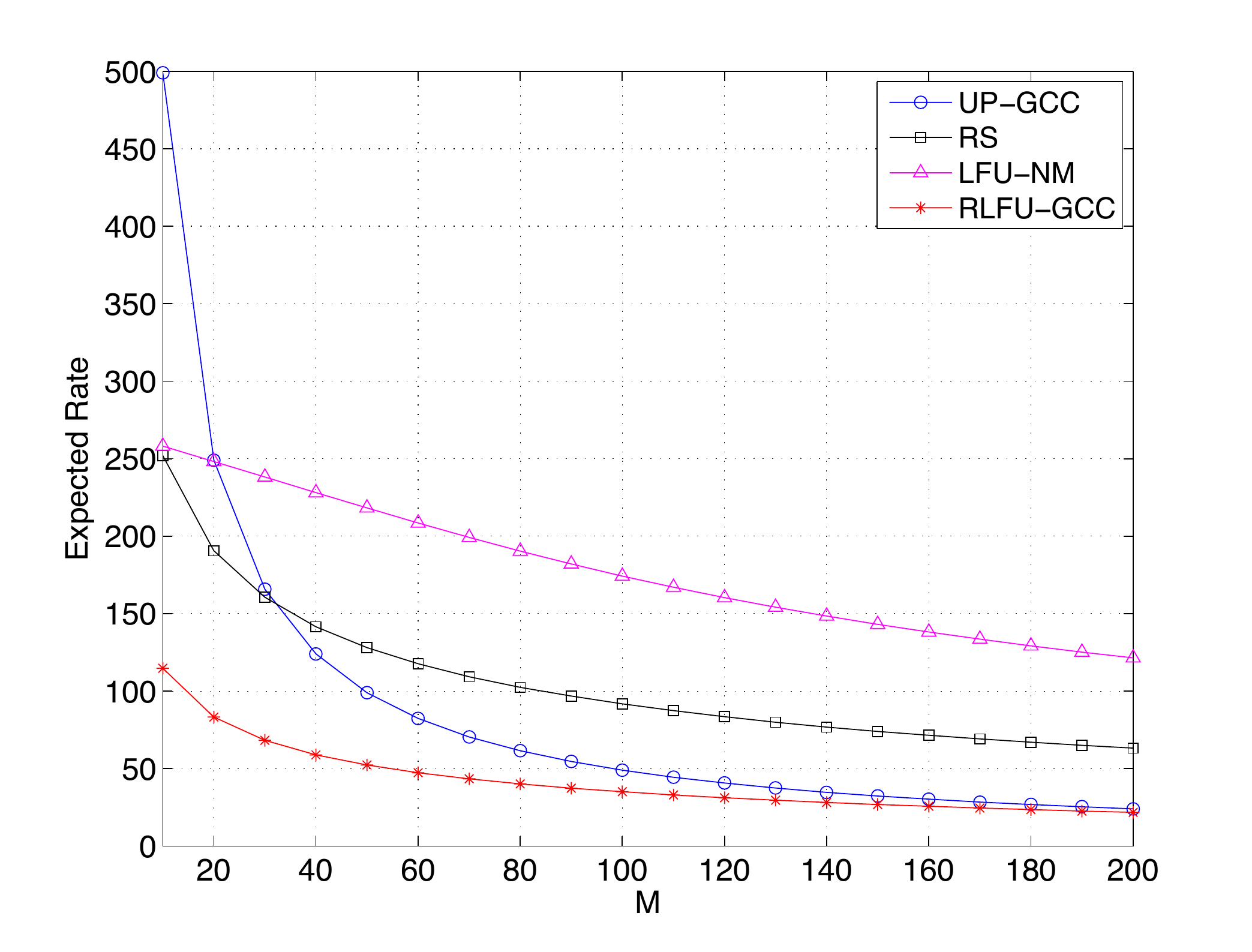}
\label{fig: alpha1p6_m5000_n5000_v5}
}
\subfigure[]{
\centering \includegraphics[width=7.5cm]{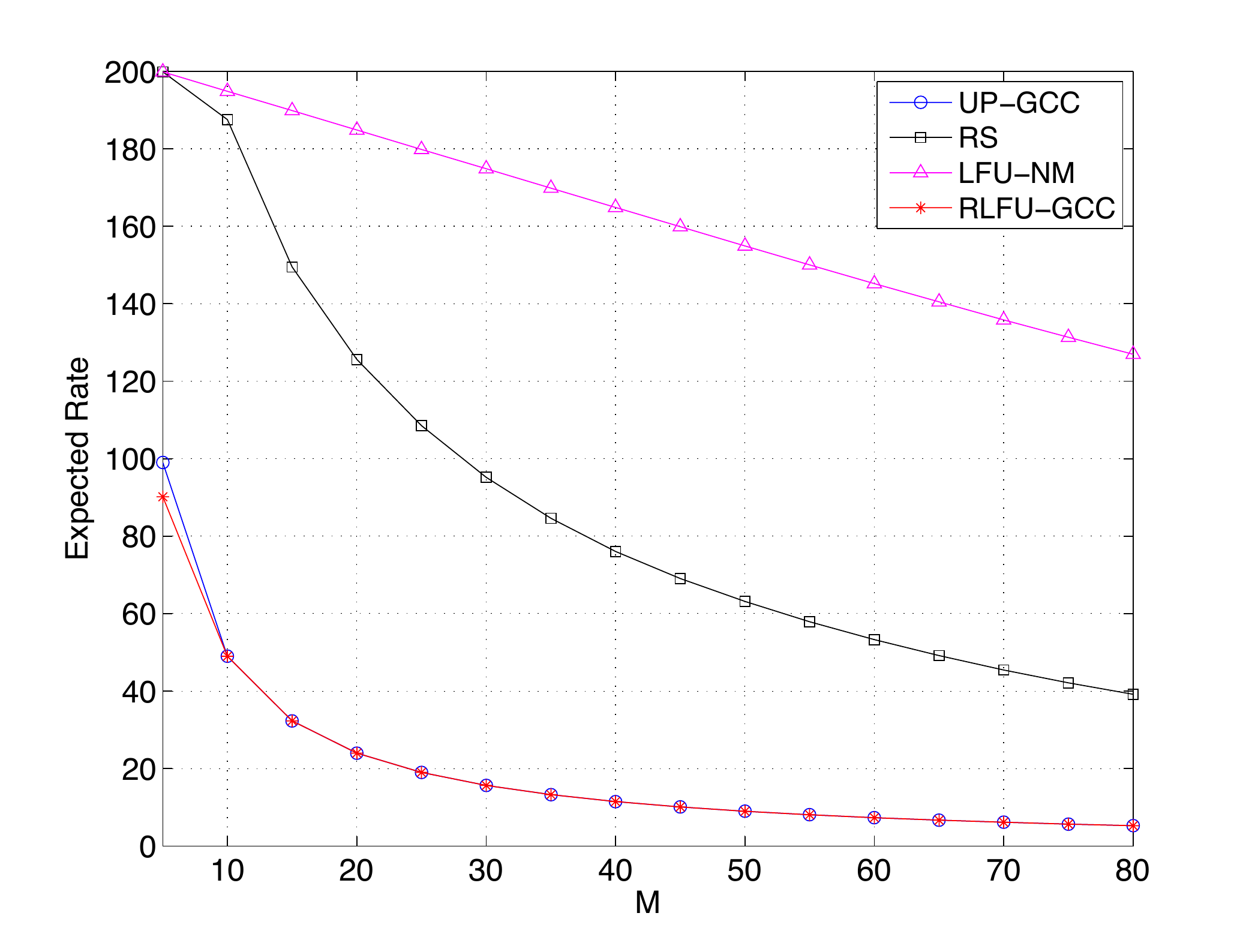}
\label{fig: alpha1p6_m500_n5000_v4}
}
\caption{Simulation results for $\alpha=1.6$. a)~$m=5000,n=50$. b)~$m=5000,n=500$. c)~$m=5000,n=5000$. d)~$m=500,n=5000$. RLFU in this figure corresponds to the RLFU with optimized $\widetilde{m}$ given by (\ref{eqmopt}). %RS denotes the  reference scheme presented in \cite{niesen2013coded}. %Except for the reference scheme, the caching placements are denoted by the legends and the delivery scheme is GCC for all the simulations.
}

\label{fig: result 2}
\end{figure}

In Figs. \ref{fig: result 1} and \ref{fig: result 2},  we plot the  rate achieved by RLFU-GCC, given by $R^{\rm GCC}(n,m,M,\qv,\widetilde \pv)$ with %$\widetilde m$ optimized as
\begin{eqnarray}
\label{eqmopt}
\widetilde m = \arg \! \min \, R^{\rm ub}(n,m,M,\qv,\widetilde m),
\end{eqnarray}
which can be computed via simple one-dimensional search.
For comparison, Figs. \ref{fig: result 1} and \ref{fig: result 2} also show the rate achieved by: 
1) UP-GCC (i.e., letting $\widetilde m =  m$);
2) LFU with naive multicasting (LFU-NM), given by $\sum_{f = M+1}^m \left(1 - \left(1 - q_f \right)^{n} \right)$; 
and 3) the grouping scheme analyzed in \cite{niesen2013coded}, which is referred to  as ``reference scheme" (RS).
The expected rate is shown as a function of the per-user cache capacity $M$ for  $n=\{50, 500,5000\}$, $m=\{500,5000\}$, and $\alpha=\{0.6, 1.6\}$. 
The simulation results agree with the scaling law analysis presented in Section \ref{order opt}. 
In particular,  we observe that, for all scenarios simulated in Figs. \ref{fig: result 1} and \ref{fig: result 2}, 
RLFU-GCC is able to significantly outperform both  LFU-NM and RS. 
For example, when $\alpha=1.6$, $m=500$ and $n=5000$,  Fig. \ref{fig: alpha1p6_m500_n5000_v4} shows that for a cache size equal to just 
$4\%$ of the library ($M=20$), the proposed scheme achieves a factor improvement in expected rate of $5\times$ with respect to the 
reference scheme and $8\times$ with respect to  LFU-NM. 
Interestingly, we notice that the reference scheme (RS) of \cite{niesen2013coded} often yields rate worse than
 UP-GCC, a scheme that does not exploit the knowledge of the demand distribution.

\begin{figure}[ht]
\centering
\subfigure[]{
\centering \includegraphics[width=7.5cm]{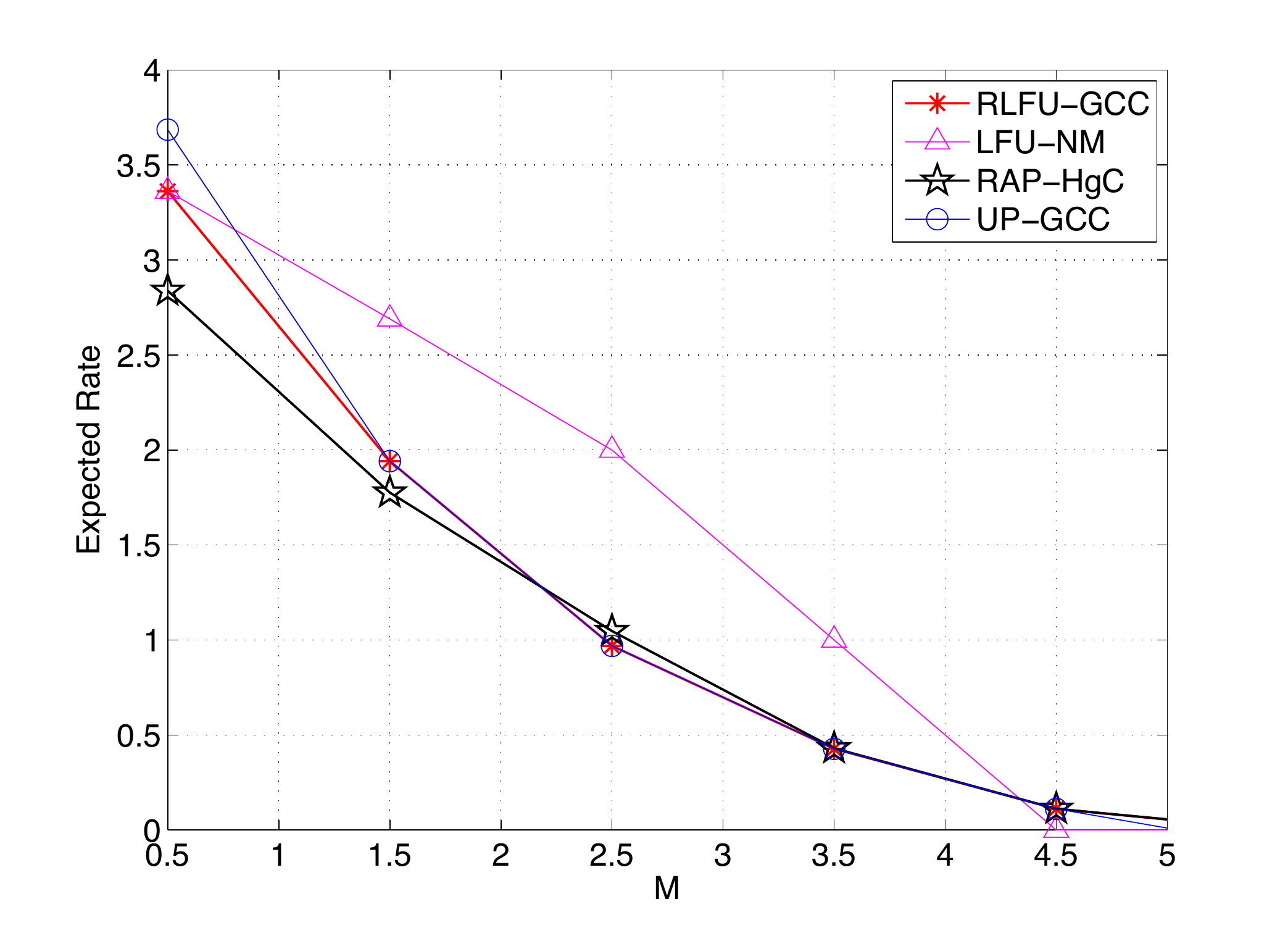}
%\label{fig: }
}
\subfigure[]{
\centering \includegraphics[width=7.5cm]{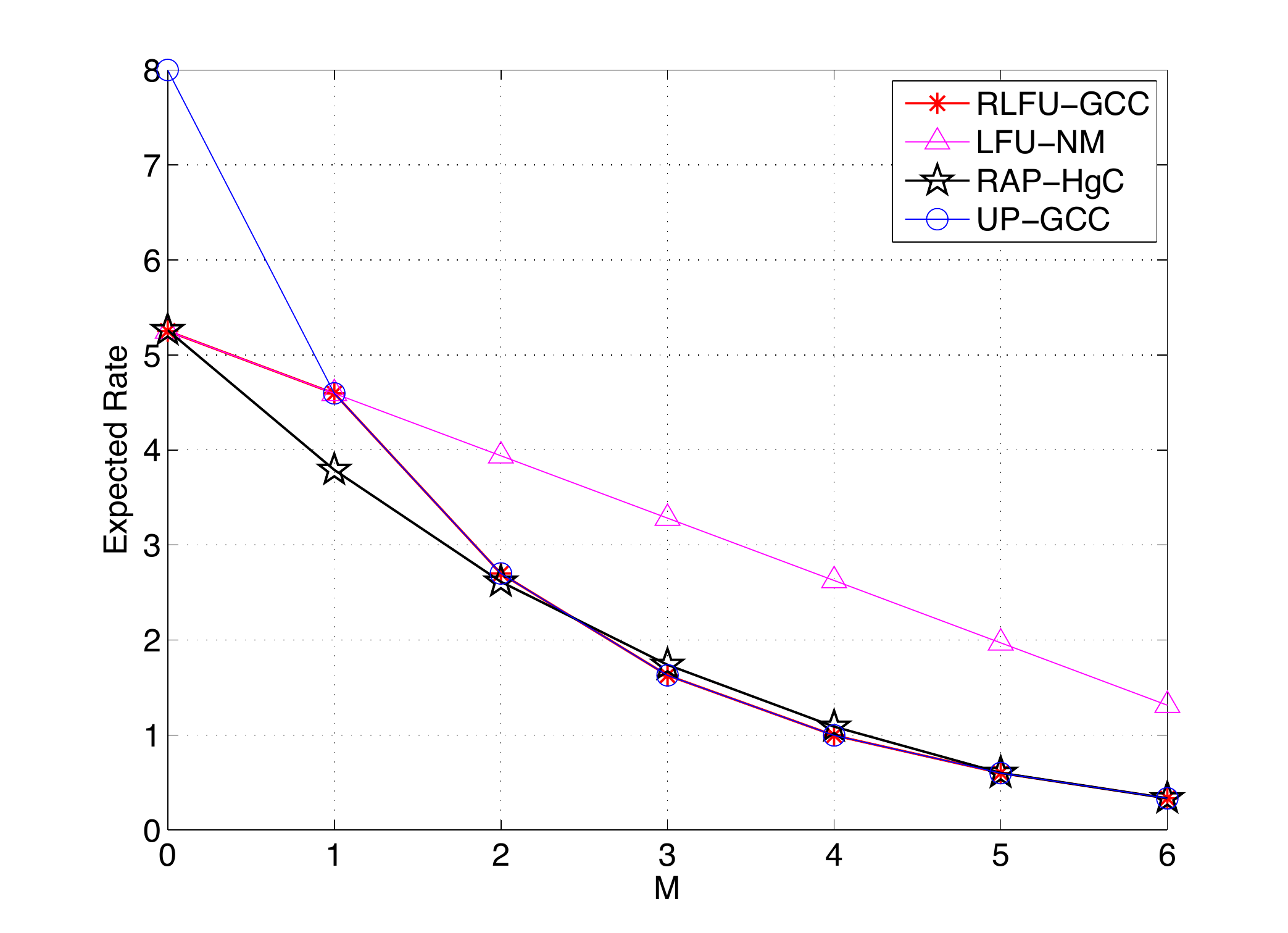}
%\label{fig: }
}
\caption{ Two examples of the simulated expected rate by using RAP-HgC ($B=500$). The comparison includes 
UP-GCC ($B \rightarrow \infty$), RLFU-GCC ($B \rightarrow \infty$ and $\widetilde{m}$ given by (\ref{eqmopt})), 
LFU-NM ($B \rightarrow \infty$). In these simulations, $\alpha=0.6$. a)~$m=n=5$. b)~$m=n=8$.}
\label{fig: result algorithm}
\end{figure}

Computing the chromatic number of a general graph is an NP-hard problem and difficult to approximate \cite{zuckerman2006linear}. 
However, for specific graphs (e.g., Erd\"os-R\'enyi random graphs $G(n,p)$), the chromatic number can be approximated or even computed \cite{bollobas1988chromatic, luczak1991chromatic, luczak1991note, alon1997concentration}. In our case, by using the property of the conflict graph resulting from RAP or RLFU, we  have shown %in Theorem \ref{thm:up}   
that the polynomial time ($O(n^3B^2)$) greedy constrained coloring (GCC) algorithm %given by Algorithm \ref{algorithm: coloring 1} {\BLUE (GCC)} 
can achieve the upper bound of the expected rate given in Theorem \ref{thm:up}. %   (\ref{eq:2}). 
%However, as pointed earlier, one would like to use we remark that
Finally, we remark that, as recently shown in \cite{Shanmugam2014Finite,ji2015efficient}, especially when operating in finite-length 
regimes ($B$ finite), % the performance of GCC can be significantly improved by
one can design improved greedy coloring algorithms that, with the same polynomial-time complexity, further exploit the structure of the conflict graph and the optimized RAP caching distribution to provide significant rate improvements. %This is especially relevant in regimes of finite packetization ($B$)}
%{\BLUE In addition, even a simpler ($O(n^3B^2)$) Greedy Independent-Set based Coloring algorithm (GISC) \cite{luby1986simple} can provide good performance and further improve on the upper bound on the achievable rate in (\ref{eq:2}). 
This is confirmed by the simulation shown in Fig.~\ref{fig: result algorithm}, where in addition to RLFU-GCC, 
UP-GCC, and LFU-NM, we also plot the rate achieved by RAP-HgC, where HgC is the Hierarchical greedy Coloring algorithm proposed in \cite{ji2015efficient}, 
for a network with (a) $m=n=5$, (b) $m=n=8$, $\alpha=0.6$, and a finite number of packets per file $B=500$.  
%{\BLUE Notice that in Fig.~\ref{fig: result algorithm}, we use the RAP caching placement ($\pv^*$ as caching distribution)  with $B=500$, and the greedy independent-set based coloring algorithm \cite{luby1986simple} for the delivery phase. 

%%%%%%%%%%%%%%%%%%%%%%%%%%%%%%%%%%%%%%%%%%
\section{Conclusions}
\label{sec: conclusions}

In this paper, we built on the shared link network with caching, coded delivery, and random demands, firstly considered in \cite{niesen2013coded}. 
We formally defined the problem in an information theoretic sense, giving an operational meaning to the per-user rate averaged over the random demands. 
We analyzed achievability schemes  based on random fractional (packet level) caching and Chromatic-number Index Coding (CIC) delivery, where the latter is defined on a properly constructed conflict graph that involves all the requested packets. 
In particular, any suboptimal (e.g., greedy) technique for coloring such conflict graph yields an achievable rate. 
Our bound (Theorem \ref{thm:up}) considers a particular delivery scheme that we refer to as Greedy Constrained Coloring (GCC), which is polynomial in the system parameters. 
The direct optimization of the bound with respect to the caching distribution yields a caching placement scheme that we refer to as RAndom Popularity-based (RAP). %[[ and an overall scheme referred to as RAP-GCC]]} 
For analytical convenience, we also considered a simpler choice of the caching distribution, where caching is performed with uniform probability up to an optimized file index cut-off value $\widetilde{m}$, and no packets of files with index larger than $\widetilde{m}$ are cached. This placement scheme is referred to as
Random Least Frequencly Used (RLFU), for the obvious resemblance with conventional LFU caching. 
We also provided a general rate lower bound (Theorem \ref{theorem: general lower bound}). 

Then, by analyzing the achievable rate of RLFU-GCC and comparing it with the general rate lower bound, we could establish the order-optimality of
the proposed schemes in the case of Zipf demand distribution, where order-optimality indicates that the ratio between the achievable rate and the best possible rate is upper bounded by a constant as $m,n \rightarrow \infty$ (with the special case of $m$ fixed and $n \rightarrow \infty$ treated apart). 

Beyond the optimal rate scaling laws, we showed the effectiveness of the general RAP-CIC approach with respect to: 1) conventional non-caching approaches such as unicasting or naive multicasting (the default solution in today's wireless networks); 2) local caching policies, such as LFU, with naive multicasting; and 3) a specific embodiment of the general scheme proposed in \cite{niesen2013coded}, which consists of splitting the library  into subsets of files with approximately the same demand probability (in fact, differing at most by a factor of two) and then applying 
greedy coloring index coding separately to the different subsets.

Our scaling results, while seemingly rather cumbersome, point out that the relation between the rate scaling and the various system parameters, 
even restricting to the case of Zipf demand distribution, can be very intricate and non-trivial. 
In particular, we characterized the regimes in which caching is useless (i.e., it provides no order gain with respect to conventional non-caching approaches), as well as the regimes in which caching exhibits multiplicative gains, i.e., the rate decreases (throughput increases) proportionally to a function of the per-user cache size $M$. Specifically, we identified the regions where the multiplicative caching gain is either linear or non-linear in $M$, and how it %as well as cases where the multiplicative gain is non-linear with respect to $M$, and in particular 
depends on the Zipf parameter $\alpha$.  Finally, for the regimes in which caching can provide multiplicative gains, we characterized 
1) the regions in which the order-optimal RAP-CIC converges to conventional LFU with naive multicasting, 
showing when the additional coding complexity is not required,  and 2) the regions in which (cooperative) fractional caching and index coding 
delivery is required for order-optimality. 

\newpage

%%%%%%%%%%%%%%%%%%%%%%%%%%%%%%%%%%%%%%%%%%%%%%%%%%%%%%%%%%%%
\appendices

\section{Proof of Theorem \ref{thm:up}}
\label{sec: Proof of Theorem up}

Let $\Jc(\Csf,\Qsf)$ denote the (random) number of independent sets found by Algorithm \ref{algorithm: coloring 1} applied to the conflict graph
$\Hc_{\Csf,\Qsf}$ defined in Section \ref{sec: Achievable Delivery Scheme}, 
where $\Csf$ is the random cache configuration resulting from the  random caching scheme with caching distribution $\pv$, 
and $\Qsf$ is the packet-level demand vector resulting from the random i.i.d. requests with demand distribution $\qv$. 

Recall that we consider the limit for $F,B \rightarrow \infty$ with fixed packet size $F/B$. Then, since the term $\bar{m}$ in (\ref{eq: m bar}) has been already shown
to upper bound the average rate due to GCC$_2$ (see Remark 3 in Section \ref{sec: Achievable Expected Rate}), Theorem \ref{thm:up} follows by showing that
\begin{equation}  \label{want-to-show}
\lim_{B \rightarrow \infty} \PP \left ( \frac{\EE[\Jc(\Csf,\Qsf) | \Csf]}{B} \leq \psi(\qv,\pv) + \epsilon \right ) = 1,
\end{equation}
for any arbitrarily small $\epsilon > 0$. 

By construction, the independent sets $\Ic$ generated by GCC$_1$ have the same (unordered) label of 
users requesting or caching the packets $\{\rho(v) : v \in \Ic\}$. We shall refer to such unordered label of users as the {\em user label} of the independent set. 
Hence, we count the independent sets by enumerating all possible user labels, and upperbounding how many independent
sets $\Ic$ Algorithm \ref{algorithm: coloring 1} generates for each user label. 

Consider a user label $\Uc_\ell \subseteq \Uc$ of size $\ell$, and let $\Jc_{\Csf,\Qsf}(\Uc_\ell)$ denote the number of independent sets 
generated  by Algorithm \ref{algorithm: coloring 1} with label $\{\mu(v), \eta(v)\} = \Uc_\ell$. 
A necessary condition for the existence of an independent set with user label $\Uc_\ell$ is that, for any user $u \in \Uc_\ell$, there exist a node $v$ 
such that: 1) $\mu(v) = u$ (user $u$ requests the packet corresponding to $v$), and 2) $\eta(v) = \Uc_\ell \setminus \{u\}$ (the packet corresponding to 
$v$ is cached in all users $\Uc_\ell \setminus \{u\}$ and not cached by any other user). Therefore, the following equality  holds with probability 1 (pointwise dominance)
\begin{equation} \label{starting-point}
\Jc_{\Csf,\Qsf}(\Uc_\ell) = \max_{u \in \Uc_\ell} \sum_{v : \rho(v) \ni \fsf_u} 1\left \{ \eta(v) = \Uc_\ell \setminus\{u\} \right \}.
\end{equation}
In (\ref{starting-point}), with a slight abuse of notation, we denote the condition that the packet $\rho(v)$ associated to node $v$
is requested by user $u$ as $\rho(v) \ni \fsf_u$, indicating that the ``file'' field in the packet identifier $\rho(v)$ is equal to 
the $u$-th component of the (random) request vector $ \fsf$. The indicator function captures the necessary condition for 
the existence of an independent set with user label $\Uc_\ell$ expressed (in words) above, and the maximum over $u \in \Uc_\ell$
is necessary to obtain an upper bound. Notice that summing over $u \in \Uc_\ell$ instead of taking the maximum would overcount
the number of independent sets and yield a loose bound. 

Then, using (\ref{starting-point}) and the definition of $\Jc(\Csf,\Qsf)$, we can write
\begin{eqnarray}
\EE[\Jc(\Csf,\Qsf) | \Csf] & = & \EE \left [ \left . \sum_{\ell=1}^n \sum_{\Uc_\ell \subseteq \Uc} \Jc_{\Csf,\Qsf}(\Uc_\ell) \right | \Csf \right ] \nonumber \\
& = & \EE \left [ \left . \sum_{\ell=1}^n \sum_{\Uc_\ell \subseteq \Uc}  \max_{u \in \Uc_\ell} \sum_{v : \rho(v) \ni \fsf_u} 1\left \{ \eta(v) = \Uc_\ell \setminus\{u\} \right \}   \right | \Csf \right ] \nonumber \\
& = & \sum_{\fv \in \Fc^n} \left ( \prod_{i=1}^n q_{f_i} \right ) \left ( \sum_{\ell=1}^n \sum_{\Uc_\ell \subseteq \Uc} 
\max_{u \in \Uc_\ell} \sum_{v : \rho(v) \ni f_u} 1\left \{ \eta(v) = \Uc_\ell \setminus\{u\} \right \} \right ) \label{expectation-q} \\
& = & \sum_{\ell=1}^n {n \choose \ell} \sum_{\dv \in \Fc^\ell} \left ( \prod_{i=1}^\ell q_{d_i} \right ) \left (  
\max_{u = 1,\ldots, \ell}  \sum_{v : \rho(v) \ni d_u} 1\left \{ \eta(v) = \{1,\ldots, \ell\} \setminus\{u\} \right \} \right ) \label{symmetry} \\
& = & \sum_{\ell=1}^n {n \choose \ell} \sum_{\dv \in \Fc^\ell} \left ( \prod_{i=1}^\ell q_{d_i} \right ) \left (  \phantom{\prod_{i=1}^\ell q_{d_i}} \right . \nonumber \\
& &  \sum_{f \in \Fc} 1\left \{ f = \arg\! \max_{j \in \dv}  \sum_{v : \rho(v) \ni j} 1\left \{ \eta(v) = \{1,\ldots, \ell\} \setminus\{u\} \right \} \right \} \label{ziofiga} \\
& & \cdot  \sum_{v : \rho(v) \ni f} 1\left \{ \eta(v) = \{1,\ldots, \ell\} \setminus\{u\} \right \} \label{ziocane} 
\end{eqnarray}
where (\ref{expectation-q}) follows by writing the conditional expectation with respect to the demand vector explicitly in terms of a sum over all 
possible files, after recognizing that the indicator function
\[ 1\left \{ \eta(v) = \Uc_\ell \setminus\{u\} \right \} \]
is a random variable only function of the cache placement $\Csf$ (in fact, this depends only on whether the $\ell-1$ users
in $\Uc_\ell \setminus\{u\}$ have cached or not the packet associated to node $v$), 
and where (\ref{symmetry}) follows by noticing that the term 
\[ \max_{u \in \Uc_\ell} \sum_{v : \rho(v) \ni f_u} 1\left \{ \eta(v) = \Uc_\ell \setminus\{u\} \right \}  \]
depends only on the $\ell$ (possibly repeated) indices $\{f_u : u \in \Uc_\ell\}$. Therefore, after switching the summation order and marginalizing
with respect to all the file indices corresponding to the requests of the users not in $\Uc_\ell$, due to the symmetry of the random caching placement and
the demand distribution (i.i.d. across the users) we can focus on a generic user label of size $\ell$, which without loss of generality can be 
set to be $\{1,\ldots, \ell\}$. At this point, the sum with respect to $\Uc_\ell \subseteq \Uc$ reduces to enumerating all the subsets of size $\ell$ in the user set of size $n$, 
yielding the binomial coefficient ${n \choose \ell}$. Finally, (\ref{ziocane}) follows from replacing the max with a sum over all possible file indices, 
and multiplying by the indicator function that picks the maximum. 

At this point, we need to study the behavior of the random variable
\begin{equation} \label{YY}
\Ysf_{\ell,f} = \sum_{v : \rho(v) \ni f} 1\left \{ \eta(v) = \{1,\ldots, \ell\} \setminus\{u\} \right \},
\end{equation}
where $u \in \{1,\ldots, \ell\}$ and where, by construction, the sum extends to the nodes corresponding to 
file $f$ requested by user $u$, i.e., not present in its cache. By construction of the caching scheme, these nodes are
$B(1 - p_f M)$. 
 Furthermore, the random variable $1\left \{ \eta(v) = \{1,\ldots, \ell\} \setminus\{u\} \right \}$ takes value 1 with probability $(p_f M)^{\ell-1} (1 - p_f M)^{n-\ell}$, corresponding to the fact that 
$\ell-1$ users have cached packet $\rho(v)$ and $n - \ell$ users have not cached it (user $u$ has not cached it by construction, i.e., 
we are conditioning on this event). However, they are not i.i.d. across different $v$. By denoting $P_{\ell,f} \eqdef (p_f M)^{\ell-1} (1 - p_f M)^{n-\ell}$, we can see that
\begin{equation} \label{EY}
\EE[\Ysf_{\ell,f}] = \EE\left[\sum_{v : \rho(v) \ni f} 1\left \{ \eta(v) = \{1,\ldots, \ell\} \setminus\{u\} \right \}\right] = B(1 - p_f M) P_{\ell,f},
\end{equation}
Then, for $\rho(v),\rho(v') \ni f$, 
\begin{eqnarray}
&&\PP\left(1\left \{ \eta(v) = \{1,\ldots, \ell\} \setminus\{u\} \right \} = 1, 1\left \{ \eta(v') = \{1,\ldots, \ell\} \setminus\{u\} \right \} = 1 \right) \notag\\
&& = P_{\ell,f} \PP\left(1\left \{ \eta(v') = \{1,\ldots, \ell\} \setminus\{u\} \right \} = 1 | 1\left \{ \eta(v) = \{1,\ldots, \ell\} \setminus\{u\} \right \} = 1\right) \notag\\
&& = P_{\ell,f} (p_f' M)^{\ell-1} (1 - p_f' M)^{n-\ell},
\end{eqnarray}
where 
\[p_f' = \frac{{B-2 \choose p_fMB - 2}}{{B-1 \choose p_fMB - 1}} = p_f - \frac{p_fM-1}{B-1} = p_f + \delta(B),\] and $\delta(B) \rightarrow 0$ as $B \rightarrow \infty$ independently with $p_f$. Let $P_{\ell,f}' \eqdef (p_f' M)^{\ell-1} (1 - p_f' M)^{n-\ell}$, then we obtain
\begin{eqnarray}
P_{\ell,f}' &=& (p_f' M)^{\ell-1} (1 - p_f' M)^{n-\ell} \notag\\
&=& ((p_f + \delta(B))M)^{\ell-1} (1 - (p_f + \delta(B)) M)^{n-\ell} \notag\\
&=& P_{\ell,f} + \left. \frac{dP_{\ell,f}}{dp_f} \right|_{p_f=p_f'} \cdot \delta(B) + o(\delta(B)) \notag\\
&=& P_{\ell,f} + \left.\left((\ell-1)(p_fM)^{\ell-2}(1-p_fM)^{n-\ell}M - (n-\ell)(p_fM)^{\ell-1}(1-p_fM)^{n-\ell-1} \right)M \right|_{p_f=p_f'} \cdot \delta(B) \notag\\
&& + o(\delta(B)) \notag\\
&=& P_{\ell,f} + \delta'(B),
\end{eqnarray}
where $\delta'(B) \rightarrow 0$ as $B \rightarrow \infty$ independently with $P_{\ell,f}$. Then, we have
\begin{eqnarray}
\EE[\Ysf_{\ell,f}^2] &=& \EE\left[ \left(\sum_{v : \rho(v) \ni f} 1\left \{ \eta(v) = \{1,\ldots, \ell\} \setminus\{u\} \right \}\right)^2 \right] \notag\\
&=& \sum_{v : \rho(v) \ni f} P_{\ell, f}+ \sum_{v : \rho(v) \ni f}\sum_{v' : \rho(v') \ni f, v' \neq v} P_{\ell, f}P_{\ell, f}' \notag\\
&=& \sum_{v : \rho(v) \ni f} P_{\ell, f}+ \sum_{v : \rho(v) \ni f}\sum_{v' : \rho(v') \ni f, v' \neq v} P_{\ell, f}P_{\ell, f} + P_{\ell, f}\delta'(B) \notag\\
&=& \left(\sum_{v : \rho(v) \ni f} P_{\ell, f}+ \sum_{v : \rho(v) \ni f}\sum_{v' : \rho(v') \ni f, v' \neq v} P_{\ell, f}P_{\ell, f}\right) + \sum_{v : \rho(v) \ni f}\sum_{v' : \rho(v') \ni f, v' \neq v}  P_{\ell, f}\delta'(B) \notag\\
&=& \left(\sum_{v : \rho(v) \ni f} P_{\ell, f}+ \sum_{v : \rho(v) \ni f}\sum_{v' : \rho(v') \ni f, v' \neq v} P_{\ell, f}P_{\ell, f}\right) + \frac{\delta'(B)}{P_{\ell,f}}\sum_{v : \rho(v) \ni f}\sum_{v' : \rho(v') \ni f, v' \neq v}  P_{\ell, f}P_{\ell,f} \notag\\
&=& B(1 - p_f M) P_{\ell,f}(1-P_{\ell,f}) + (B(1 - p_f M))^2 P_{\ell,f}^2  + o\left(B^2\right)((1 - p_f M))^2 P_{\ell,f}^2,
\end{eqnarray}
Therefore, by using the fact that $\Var(\Ysf_{\ell,f}) = \EE[\Ysf_{\ell,f}^2] - \EE[\Ysf_{\ell,f}]^2$ and from Chebyshev's inequality \footnote{Here we need only convergence in probability.}, we obtain 
%By the weak law of large numbers\footnote{Here we need only convergence in probability.} 
we have that\footnote{As usual, $\stackrel{p}{\rightarrow}$ indicates limit in probability \cite{grimmett}.}
\[ \frac{\Ysf_{\ell,f}}{B(1 - p_f M)} \stackrel{p}{\rightarrow}  (p_f M)^{\ell-1} (1 - p_f M)^{n-\ell}, \;\;\; \mbox{for} \;\;\; B \rightarrow \infty.  \]

Equivalently, we can write
\begin{equation} \label{lim-in-p}
\lim_{B \rightarrow \infty} \PP \left ( \left | \frac{\Ysf_{\ell,f}}{B}  -  g_\ell(f) \right | \leq \epsilon \right ) = 1,
\end{equation}
for any arbitrarily small $\epsilon > 0$, where we define the function (already introduced in Remark 5 in Section \ref{sec: Achievable Expected Rate}), 
\begin{equation} \label{gfunct}
g_\ell(f) = (p_f M)^{\ell-1} (1 - p_f M)^{n-\ell + 1}. 
\end{equation}
It follows that we can replace the last line of (\ref{ziocane}) by the bound (holding with high probability) $B(g_\ell(f) + \epsilon)$. 
In order to handle the indicator function in (\ref{ziofiga}), we need to consider the concentration (\ref{lim-in-p}) of 
$\frac{\Ysf_{\ell,f}}{B}$ around the values $g_\ell(f)$. Sorting the values $\{g_\ell(f) : f \in \Fc\}$ in increasing order, 
we obtain a grid of at most $m$ discrete values. The limit in probability (\ref{lim-in-p}) states that the random variables
$\Ysf_{\ell,j}/B$ concentrate around their corresponding values $g_\ell(j)$, for any $j \in \Fc$. 
Taking $\epsilon$ sufficiently small, the intervals 
$[g_\ell(j) - \epsilon, g_\ell(j) + \epsilon]$ for different $j$ are mutually disjoint for different values of $j$, 
unless there are some $j \neq j'$ such that $g_\ell(j) = g_\ell(j')$. For the moment we assume that all these values are distinct, and we handle
the case of non-distinct values at the end (we will see that this does not cause any problem). 
Now, we re-write the indicator function in (\ref{ziofiga}) as
\begin{equation} \label{ind1} 
1\left \{ f = \arg\! \max_{j \in \dv}  \sum_{v : \rho(v) \ni j} 1\left \{ \eta(v) = \{1,\ldots, \ell\} \setminus\{u\} \right \} \right \}  = 
1\left \{ f = \arg\! \max_{j \in \dv}  \Ysf_{\ell,j}/B \right \} 
\end{equation}
and compare it with the indicator function
\begin{equation} \label{ind2} 
1\left \{ f = \arg\! \max_{j \in \dv} g_\ell(j) \right \}.
\end{equation}
If $f \notin \dv$ then both indicator functions are equal to 0. 
If $f \in \dv$, suppose that $f = \arg\! \max_{j \in \dv} g_\ell(j)$ such that (\ref{ind2}) is equal to 1. 
Then, (\ref{ind1}) is equal to 0 only if for some $j \in \dv: j \neq f$, $\Ysf_{\ell,j}/B > \Ysf_{\ell,f}/B$. 
Since $\Ysf_{\ell,j}/B \in [g_\ell(j) - \epsilon, g_\ell(j) + \epsilon]$ and $\Ysf_{\ell,f}/B \in [g_\ell(f) - \epsilon, g_\ell(f) + \epsilon]$ with high probability, 
and, by construction, $g_\ell(j) + \epsilon < g_\ell(f) - \epsilon$, it follows that this event has vanishing probability as $B \rightarrow \infty$. 
Similarly, suppose that $f \in \dv$ and that $f \neq \arg\! \max_{j \in \dv} g_\ell(j)$ such that (\ref{ind2}) is equal to 0. 
Then, (\ref{ind1}) is equal to 1 only if $\Ysf_{\ell,f}/B > \Ysf_{\ell,j_{\max}}/B$, where
$j_{\max} =  \arg\! \max_{j \in \dv} g_\ell(j)$.  Again, since $\Ysf_{\ell,f}/B \in [g_\ell(f) - \epsilon, g_\ell(f) + \epsilon]$ and $\Ysf_{\ell,j_{\max}}/B \in [g_\ell(j_{\max}) - \epsilon, g_\ell(j_{\max}) + \epsilon]$ with high probability,  and, by construction, $g_\ell(f) + \epsilon < g_\ell(j_{\max}) - \epsilon$, it follows that this event has vanishing 
probability as $B \rightarrow \infty$. We conclude that 
\begin{equation} \label{ppp}
\lim_{B \rightarrow \infty} \PP \left ( \left | 1\left \{ f = \arg\! \max_{j \in \dv}  \Ysf_{\ell,j}/B \right \}  - 1\left \{ f = \arg\! \max_{j \in \dv} g_\ell(j) \right \} \right | \leq \epsilon \right ) = 1.
\end{equation}
Since convergence in probability implies convergence in the $r$-th mean for uniformly absolutely bounded random variables \cite{grimmett} 
and indicator functions are obviously  bounded by 1, we conclude that
\[ \EE\left [ 1\left \{ f = \arg\! \max_{j \in \Dc}  \Ysf_{\ell,j}/B \right \} \right ] \rightarrow \EE \left [ 1\left \{ f = \arg\! \max_{j \in \Dc} g_\ell(j) \right \} \right ], \]
as $B \rightarrow \infty$, where $\Dc$ is a random subset of $\ell$ elements sampled i.i.d. (with replacement) from $\Fc$ with probability mass function $\qv$. 
Now, replacing the last line of (\ref{ziocane}) with the deterministic bound $B(g_\ell(f) + \epsilon)$ (which holds with high probability as explained before)
and taking expectation of the indicator function using the convergence of the mean said above, we can continue the chain of inequalities after (\ref{ziocane}) and show that the bound
\begin{equation}
\frac{\EE[\Jc(\Csf,\Qsf) | \Csf]}{B} \leq \sum_{\ell=1}^n {n \choose \ell}  \sum_{f \in \Fc}  \PP( f = \arg\! \max_{j \in \Dc} g_\ell(j) ) (g_\ell(f)   + \epsilon) 
\end{equation}
holds with high probability for $B \rightarrow \infty$, for any arbitrary $\epsilon > 0$. In the case where for some distinct $j, j'$ the corresponding values of
$g_\ell(j)$ and $g_\ell(j')$ coincide, we notice that outcome of the indicator functions
(\ref{ind1}) and (\ref{ind2}) are irrelevant to the value of the bound, as long as they pick different indices which yield the same maximum value of
the function $g_\ell(\cdot)$. Hence, the argument can be extended to this case by defining ``equivalent classes'' of indices which yields the same
value in the bound.

Theorem \ref{thm:up} now follows by using (\ref{gfunct}) and by noticing that the probabilities
$\PP( f = \arg\! \max_{j \in \Dc} g_\ell(j) )$ coincide with the terms $\rho_{f,\ell}$ defined in (\ref{rhorho}).

%%%%%%%%%%%%%%%%%%%%%%%%%%%%%%%%%%%%%%%%%%%%%%%%%%%%%%%%%%%%%%%%%%
\section{Proof of Lemma \ref{lemma: achievable}}
\label{sec: Proof of Lemma achievable}

Applying Theorem \ref{thm:up} to the case $\pv = \widetilde{\pv}$, we have that, for all $\epsilon > 0$, 
\begin{equation}  
\lim_{F \rightarrow \infty} \PP \left ( R^{\rm GCC}(n,m,M,\qv,\widetilde{\pv}) \leq \min\{ \psi(\qv, \widetilde{\pv}), \bar{m} \} + \epsilon \right )  = 1.
\end{equation}
Then, we can write
%Since $\psi(\pv^*,\qv) \leq \psi(\widetilde \pv,\qv)$ and $\widetilde \pv$ is given by (\ref{ptilde}), we obtain
\begin{eqnarray}
\psi(\qv,\widetilde{\pv}) 
&=&  \sum_{\ell=1}^n {n \choose \ell}  \sum_{f=1}^m \rho_{f,\ell} (1-p_f M)^{n-\ell+1} (p_f M)^{\ell-1} \notag\\
&=& \EE\left[\sum_{\ell=1}^{\Nsf_{\widetilde m}} {\Nsf_{\widetilde m} \choose \ell} \left(1 - \frac{M}{\widetilde m}\right)^{\Nsf_{\widetilde m}-\ell+1} \left(\frac{M}{\widetilde m}\right)^{\ell-1}\right] + n \sum_{f=\widetilde m + 1}^{m} q_f \notag\\
&=& \EE\left[\left(\frac{\widetilde m}{M}-1\right)\left(1-\left(1-\frac{M}{\widetilde m}\right)^{\Nsf_{\widetilde m}}\right)\right] + n \sum_{f=\widetilde m + 1}^{m} q_f \notag\\
&\buildrel (a) \over \leq& \left(\frac{\widetilde m}{M}-1\right)\left(1-\left(1-\frac{M}{\widetilde m}\right)^{\EE[\Nsf_{\widetilde m}]}\right) + n \sum_{f=\widetilde m + 1}^{m} q_f \notag\\
&\buildrel (b) \over=& \left(\frac{\widetilde m}{M}-1\right)\left(1-\left(1-\frac{M}{\widetilde m}\right)^{nG_{\widetilde m}}\right) + n \left(1-G_{\widetilde m}\right) \notag \\
& = & \widetilde{\psi}(\qv, \widetilde{m}), 
\end{eqnarray}
where $\Nsf_{\widetilde m}$ is the (random) number of users requesting files with index less than or equal to 
$\widetilde m$,  (a) follows from   Jensen's Inequality,  and (b) because $\EE[\Nsf_{\widetilde{m}}] = n \sum_{f=1}^{{\widetilde m}}q_f = n G_{\widetilde m}$.

%%%%%%%%%%%%%%%%%%%%%%%%%%%%%%%%%%%%%%%%%%%%%%%%%%%%%%%%%%%%%%%%%%
\section{Proof of Theorem \ref{theorem: general lower bound}}
\label{sec: proof of theorem: general lower bound}

First, notice that since the users decoder $\lambda_u(\cdot)$ operate independently, the rate of the optimal scheme
$R^*(n,m,M,\qv)$ is non-increasing in $n$. In fact, an admissible scheme for $n$ users is also admissible for any $n' < n$ 
users.\footnote{To see this, simply add $n-n'$ virtual users to the reduced system with $n'$ users, 
generate the corresponding random i.i.d. demands according to $\qv$, and use the code for the system of $n$ users, to achieve the same rate, 
which is clearly larger or equal to the optimal rate for the system with $n'$ users.}

The first step of the proof consists of lower bounding the rate of any admissible scheme with the optimal rate 
of a genie-aided system that eliminates some users. By construction of the genie, we can lower bound
the optimal rate of the genie-aided system by the optimal rate over an ensemble of reduced systems with 
binomially distributed number of users, reduced library size, and uniform demand distribution. 
Finally, we lower bound such ensemble average rate with a lower bound on the optimal rate in the case of 
arbitrary (non-random) demands, by using a result proven in \cite{niesen2013coded}, which we
state here for convenience, expressed in our notation, as Lemma \ref{lemma: uniform}.

Fix $\ell \in \{1, \cdots, m\}$ and consider the following genie-aided system: given the request vector $\fv$, 
all users $u \in \Uc$ such that $f_u > \ell$ are served by a genie at no transmission cost. 
For each $u \in \Uc$ such that $f_u \leq \ell$, the genie flips an independent biased coin and serves user $u$ 
at no transmission cost with probability $1-\frac{q_{\ell}}{q_{f_u}}$, while the system has to serve user $u$ by transmission on the shared link 
with probability $\frac{q_{\ell}}{q_{f_u}}$. 
We let $N$ denote the number of users that require service from the system (i.e., not handled by the genie). It is immediate to see that
$\Nsf \sim$ Binomial$(n, \ell q_\ell)$. In fact, any user $u$ has probability of requiring service from the system with probability
\begin{eqnarray} 
\PP(\mbox{$u$ requires service}) 
& = & \sum_{f=1}^m \PP(\mbox{$u$ requires service} | \fsf_u = f)  q_f \nonumber \\
& = & \sum_{f=1}^\ell \PP(\mbox{$u$ requires service} | \fsf_u = f)  q_f \label{genie1} \\
& = & \sum_{f=1}^\ell \frac{q_\ell}{q_f} q_f  = \ell q_\ell,
\end{eqnarray}
where (\ref{genie1}) follows from the fact that, by construction,  $\PP(\mbox{$u$ requires service} | \fsf_u = f) = 0$ for $f > \ell$. 
Notice also that $\ell q_\ell \leq 1$, since we have assumed a monotonically non-increasing demand distribution $\qv$, 
and if $\ell q_\ell > 1$ then $q_f \geq 1/\ell$ for all $1 \leq f \leq \ell$, such that $\sum_{f=1}^\ell q_f > 1$, which is impossible by the definition of
probability mass function. 

Now, notice that the optimal achievable rate for the genie-aided scheme provides a lower-bound to the optimal achievable rate
$R^*(n,m,M,\qv)$ of the original system. In fact, as argued before, the genie eliminates a random subset of users (which depends on the 
realization of the request vector and on the outcome of the independent coins flipped by the genie). 
We let $R^*_{\rm genie}(n,m,M,\qv)$ denote the optimal rate of the genie-aided scheme. 
Furthermore, we notice that in the genie-aided system the only requests that are handled
by the system are made with uniform independent probability over the reduced library $\{1, \ldots, \ell\}$. In fact, we have
\begin{eqnarray}
\PP(\fsf_u = f | \mbox{$u$ requires service}) & = & \frac{\PP(\fsf_u = f , \; \mbox{$u$ requires service})}{\PP(\mbox{$u$ requires service})} \nonumber \\
& = & \frac{\PP(\mbox{$u$ requires service} | \fsf_u = f) q_f}{\ell q_\ell} \nonumber \\
& = & \left \{ \begin{array}{ll} 
\frac{\frac{q_\ell}{q_f} q_f}{\ell q_\ell} = \frac{1}{\ell} & \;\; \mbox{for} \; f \in \{1, \ldots, \ell\} \\
0 & \;\; \mbox{for} \; f > \ell \end{array} \right .
\end{eqnarray}
It follows that, for a given set of users requiring service, the optimal rate of a system restricted to those users, 
with library size equal to $\ell$, and uniform demand distribution, is not larger than the optimal rate of the genie-aided original system. 
Moreover, by the symmetry of the system with respect to the users, this optimal rate does not depend on the specific set of users requesting service, 
but only on its size, which is given by $\Nsf$, as defined before.  
Consistently with the notation introduced in Section \ref{section: network model}, for any $\Nsf = N$ this optimal rate is denoted by 
$R^*(N, \ell, M, (1/\ell,\ldots, 1/\ell))$.  Then, we can write:
%%%%%%%%%%%%%%%%%%%%%%%%%%%%%%%%%%%%%%%%%%%%%%%%%%
%%%%%%%%%%%%%%%%%%%%%%%%%%%%%%%%%%%%%%%%%%%%%%%%%%
\begin{eqnarray} 
R^*(n,m,M,\qv) & \geq & R^*_{\rm genie}(n,m,M,\qv)  \nonumber \\
& \geq & \EE[ R^*(\Nsf,\ell,M,(1/\ell,\ldots, 1/\ell))]  \nonumber \\
&=& \sum_{N = 1}^n R^*(N, \ell, M, (1/\ell,\ldots, 1/\ell)) \PP\left(\Nsf = N \right) \notag\\
& \geq & \sum_{N = r}^n R^*(N, \ell, M, (1/\ell,\ldots, 1/\ell)) \PP\left(\Nsf = N \right) \label{truncate}\\
& \geq & R^*(r, \ell, M, (1/\ell,\ldots, 1/\ell))  \PP\left(\Nsf \geq r \right) \label{eq: lower bound 3}
\end{eqnarray}
where (\ref{truncate}) holds for any $1 \leq r \leq n$, since the summation contains non-negative terms, 
and  where (\ref{eq: lower bound 3}) follows again by the fact that the optimal rate is non-increasing in the number of users. 

A lower bound on $R^*(r, \ell, M, (1/\ell,\ldots, 1/\ell))$ can be given in terms of the lower bound (converse) result on the optimum rate for a 
shared link network with {\em arbitrary} demands (see Lemma 3 in \cite{niesen2013coded}). This is given by the following:

\begin{lemma}
\label{lemma: uniform}
Any admissible scheme achieving rate $R(r,\ell,M,\{1/\ell, \ldots, 1/\ell\})$ 
for the shared link network with $r$ users, library size $\ell$, cache capacity $M$,  
and uniform demand distribution $\{1/\ell,\ldots, 1/\ell\}$ must satisfy
\begin{equation} \label{lemma-uniform-bound}
R(r,\ell,M,\{1/\ell, \ldots, 1/\ell\}) \geq  z \left (1 -  \frac{M}{\left \lfloor \frac{\ell}{z} \right \rfloor} \right ) \PP(\Zsf \geq z), 
\end{equation}
for any $z = \{1, \ldots, \ell\}$, where $\Zsf$ is a random variable indicating the number of distinct files requested when the random demand vector
is i.i.d. $\sim $Uniform$\{1, \ldots, \ell\}$. \hfill $\square$
\end{lemma}

%\begin{IEEEproof} 
%See Appendix \ref{Proof of Lemma uniform}. 
%\end{IEEEproof}

Using Lemma \ref{lemma: uniform} in (\ref{eq: lower bound 3}) we have
\begin{equation} \label{eq: lower bound 4}
R^*(n,m,M,\qv) \geq   z \PP\left(\Nsf \geq r \right) \PP(\Zsf \geq z) \left (1 -  \frac{M}{\left \lfloor \frac{\ell}{z} \right \rfloor} \right ). 
\end{equation} 
%Therefore, by maximizing over $\ell,r,z$, we get
%\be
%R^*(n,m,M,\qv) \geq  \max_{\ell,r,z} \;\; z \PP\left(\Nsf \geq r \right) \PP(\Zsf \geq z) \left (1 -  \frac{M}{\left \lfloor \frac{\ell}{z} \right \rfloor} \right ). 
%\ee

%The proof is concluded by 
Next, we further lower bound the two probabilities  $\PP\left(\Nsf \geq r \right)$ and $\PP(\Zsf_\ell \geq z)$ and find the range of the corresponding parameters. To this purpose, we 
recall the definition of self-bounding function:

\begin{defn} \label{self-bounding}
Let $\Xc \subseteq \RR$ and consider a nonnegative $\nu$-variate function $g: \Xc^\nu \rightarrow [0, \infty)$. 
We say that $g$ has the self-bounding property 
if there exist functions $g_i: \Xc^{\nu-1} \rightarrow \mathbb{R}$ 
such that, for all $(x_1, \ldots, x_\nu) \in \Xc^\nu$ 
and all $i =1, \ldots, \nu$,
\be
0 \leq g(x_1, \cdots, x_\nu) - g_i(x_1, \cdots, x_{i-1}, x_{i+1}, \cdots, x_\nu) \leq 1,
\ee
and 
\be
\sum_{i=1}^\nu \left( g(x_1, \cdots, x_\nu) - g_i(x_1, \cdots, x_{i-1}, x_{i+1}, \cdots, x_\nu) \right) \leq g(x_1, \cdots, x_\nu).
\ee
\hfill $\lozenge$
\end{defn}

The following lemma \cite{boucheron2013concentration} yields a concentration property of random variables expressed as self-bounding functions of
random vectors. 

\begin{lemma}
\label{lemma: exponential bound}
Consider $\Xc \subseteq \RR$ and the random vector $\Xsf = (\Xsf_1, \ldots, \Xsf_\nu) \in \Xc^\nu$. Let $\Ysf = g(\Xsf)$ 
where $g(\cdot)$ has the self-bounding property of Definition \ref{self-bounding}. Then, for any $0 < \mu \leq \EE[\Ysf]$, 
\be
\label{eq: exponential bound 2}
\PP(\Ysf - \EE[\Ysf] \leq - \mu) \leq \exp\left(- \frac{\mu^2}{2\EE[\Ysf]}\right).  
\ee
\hfill  $\square$
\end{lemma}

Next, we observe that $g(x_1, \ldots, x_\nu) = \sum_{i=1}^\nu x_i$ is self-bounding 
when its argument is a binary vector (i.e., $\Xc = \{0,1\}$). 
Hence, $\Nsf$ satisfies Lemma \ref{lemma: exponential bound} and we can write
\begin{eqnarray} \label{eq: IU 2}
\PP(\Nsf \geq \EE[\Nsf] -\mu) \geq 1- \exp\left(- \frac{\mu^2}{2\EE[\Nsf]}\right),
\end{eqnarray}
with $0 < \mu \leq \EE[\Nsf]$. Since $\Nsf \sim$ Binomial$(n,\ell q_\ell)$ we have $\EE[\Nsf] =  n \ell q_\ell$. Hence, letting
$\mu = \EE[\Nsf] - r$ in (\ref{eq: IU 2}) we obtain
\begin{equation}
\PP\left(\Nsf \geq r \right) \geq 1- \exp\left(- \frac{(n\ell q_\ell - r)^2}{2n \ell q_\ell}\right) \eqdef P_1(\ell, r),  \label{P1}
\end{equation}
for $0 < r \leq n \ell q_\ell$. 

The variable $\Zsf$ defined in Lemma \ref{lemma: uniform} can be written as
\begin{equation}
\Zsf = \sum_{f=1}^{\ell} 1\left \{ \exists \; \mbox{$u$ requesting file $f$} \right \}.
\end{equation} 
Although the binary random variables $1\left \{ \exists \; \mbox{$u$ requesting file $f$} \right \}$ are not mutually independent, 
nevertheless $\Zsf$ is given as the sum of the components of a binary vector and therefore Lemma \ref{lemma: exponential bound} applies. 
In particular, we have 
\[ \EE[\Zsf] = \ell \left ( 1 - \left (1 - \frac{1}{\ell} \right )^{r} \right ) \]
such that, operating as before, we arrive at the lower bound
\begin{equation}
\PP(Z \geq z) \geq 1- \exp\left(- \frac{\left(\EE[\Zsf] - z \right)^2}{2\EE[Z]}\right) \eqdef P_2(\ell, r, z) \label{P2}
\end{equation}
for $0 < z \leq \EE[\Zsf]$.

Then, for some $\ell$, $r$, %using the lower bounds (\ref{P1}) and (\ref{P2}) in (\ref{eq: lower bound 4}) and 
by maximizing the obtained lower bound (\ref{eq: lower bound 4}) 
with respect to the free parameter $z$, we obtain
%\be
%R^*(n,m,M,\qv) \geq  \max_{\ell,r,z} \;\; z \PP\left(\Nsf \geq r \right) \PP(\Zsf \geq z) \left (1 -  \frac{M}{\left \lfloor \frac{\ell}{z} \right \rfloor} \right ). 
%\ee
\begin{eqnarray}
R^*(n,m,M,\qv)
\geq 
\PP(\Nsf \geq r) \max_{z \in \left\{1, \cdots, \left\lceil\min\left\{\EE[\Zsf], r\right\}\right\rceil \right\}}  \PP(\Zsf \geq z) (z-zM/{\lfloor \ell/z \rfloor}) \notag\\
\end{eqnarray}
%For some given $\ell$, $r$, 
Let $\widetilde{z} \in (0,\EE[\Zsf]]$, we consider two cases: $\widetilde{z} \geq 1$ and $0 < \widetilde{z} <1$, %by using the lower bounds (\ref{P1}) and (\ref{P2}) in (\ref{eq: lower bound 4}), we have
\begin{enumerate}
\item if $\widetilde{z} \geq 1$, let $r \geq 1$, then
\begin{eqnarray}
%\label{eq: R lower ground}
R^*(n,m,M,\qv)
&\geq& 
\PP(\Nsf \geq r) \max_{z \in \left\{1, \cdots, \left\lceil \min\left\{\EE[\Zsf], r\right\}\right\rceil \right\}}  \PP(\Zsf \geq z) (z-zM/{\lfloor \ell/z \rfloor}) \notag\\
&\buildrel (a) \over \geq& 
\PP(\Nsf \geq r) \max_{z \in \{1, \cdots, \left\lceil\min\{\widetilde{z}, r\}\right\rceil \}}  \PP(\Zsf \geq z) (z-zM/{\lfloor \ell/z \rfloor}) \notag\\
&\buildrel (b) \over \geq& 
\PP(\Nsf \geq r)\PP(\Zsf \geq \widetilde{z})  \max_{z \in \{1, \cdots, \left\lceil\min\{\widetilde{z}, r\}\right\rceil \}}  (z-zM/{\lfloor \ell/z \rfloor}), %\\
%&\geq& P_1(\ell, r)P_2(\ell, r, z) \max_{z \in \{1, \cdots, \min\{\widetilde{z}, r\}\}}  (z-zM/{\lfloor \ell/z \rfloor}), 
%\label{eq: R lower ground}
\end{eqnarray}
where (a) is because $\widetilde{z} \leq \EE[\Zsf]$ and (b) is because that if $z \leq \widetilde{z}$ then $\PP(\Zsf \geq z) \geq \PP(\Zsf \geq \widetilde{z})$. %Due to the range of $z$, it requires $r \geq 1$.
By using the lower bounds (\ref{P1}) and (\ref{P2}) in (\ref{eq: lower bound 4}), we have
\be
\label{eq: R lower ground}
R^*(n,m,M,\qv) \geq P_1(\ell, r)P_2(\ell, r, \widetilde z) \max_{z \in \{1, \cdots, \left\lceil \min\{\widetilde{z}, r\}\right\rceil \}}  (z-zM/{\lfloor \ell/z \rfloor}).
\ee
\item if $0<\widetilde{z}<1$, then let $r=1$, $z=1$ and, using 
(\ref{P1}) and (\ref{P2}) into (\ref{eq: lower bound 4}), we have
\begin{eqnarray}
\label{eq: R lower ground 1}
R^*(n,m,M,\qv)
&\geq& 
\PP(\Nsf \geq 1) \PP(\Zsf \geq 1) (1-M/{\ell}) \notag\\
&\buildrel (a) \over =& 
\PP(\Nsf \geq 1) \PP(\Zsf \geq \widetilde z) (1-M/{\ell}) \notag\\
&\geq& \PP(\Nsf \geq 1) P_2(\ell,1,\widetilde z) (1-M/{\ell}) \notag\\
& \buildrel (b) \over \geq & P_1(\ell,r) P_2(\ell,1,\widetilde z) (1-M/{\ell}), 
\end{eqnarray}
where (a) follows from observing that $\Zsf$ is an integer, so that 
$\PP(\Zsf \geq 1) = \PP(\Zsf \geq \widetilde z)$ when $\widetilde z \in (0,1)$.  
Similarly, (b) holds also for $r > 1$, since in this case $\PP(\Nsf \geq 1) \geq \PP(\Nsf \geq r)$. 
\end{enumerate}

Therefore, taking the maximum of (\ref{eq: R lower ground}) and (\ref{eq: R lower ground 1}), we obtain
\begin{eqnarray}
R^*(n,m,M,\qv) \geq && \max\left\{ P_1(\ell, r)P_2(\ell, r, z)  \max_{z \in \{1, \cdots, \left\lceil \min\{\widetilde{z}, r\}\right\rceil \}}  (z-zM/{\lfloor \ell/z \rfloor})1\{\widetilde z, r \geq 1\},  \right. \notag\\
&&\left. \phantom{\max_{z \in \{1, \cdots, \left\lceil \min\{\widetilde{z}, r\}\right\rceil \}} } P_1(\ell, r) P_2(\ell,1,\widetilde z)  (1-M/{\ell})1\{\widetilde z \in (0,1)\}\right\}. 
\end{eqnarray}
Maximizing over $\ell, r$, and $\widetilde z$, we obtain  (\ref{eq: general lower bound}) in Theorem \ref{theorem: general lower bound}.

%%%%%%%%%%%%%%%%%%%%%%%%%%%%%%%%%%%%%%%%%%%%%%%%%%%%%%%%%%%%%%%%%
%%%%%%%%%%%%%%%%%%%%%%%%%%%%%%%%%%%%%%%%%%%%%%%%%%%%%%%%%%%%%%%%%
\section{Proof of Theorem \ref{theorem: gamma < 1}}
\label{sec: Proof of Theorem gamma < 1}

Letting  $\widetilde m = m$ and using Lemma \ref{lemma: achievable} we obtain
\begin{eqnarray}
\label{eq: achievable alpha<1 1}
R^{\rm ub}(n,m,M, \qv, m) &=& \min\left\{\left(\frac{m}{M}-1\right)\left(1-\left(1-\frac{M}{m}\right)^n\right), m\right\} \notag\\
&\buildrel (a) \over \leq& \min\left\{\left(\frac{m}{M}-1\right)\left(1-\left(1-\frac{nM}{m}\right)\right), \frac{m}{M}-1, m\right\} \notag\\
&\leq&  \min\left\{n\left(1-\frac{M}{m}\right), \frac{m}{M}-1, m\right\} \notag\\
&\leq& \min\left\{\frac{m}{M}-1, m,n\right\},
\end{eqnarray}
where in (a) is because that $\left(1-x\right)^n \geq 1-nx$ for $x \leq 1$.
The proof of Theorem \ref{theorem: gamma < 1} follows by showing that $\min\left\{\frac{m}{M}-1, m,n\right\}$ is order-optimal. 
%Besides letting $n \rightarrow \infty$, we also let $m \rightarrow \infty$ for simplicity of the computation and notations. 
%For finite $m$, we can prove the result by using a similar method, which is omitted here for brevity. 

%%%%%%%%%%%%%%%%%%%%%%%%%%%%%%%%%%%%%%%%%%%%%%%%%%%%%%%%%%%%%%%% HERE HERE 

In the following, we will evaluate the converse shown in Theorem \ref{theorem: general lower bound},
and compute the gap between $R^{\rm lb}(n,m,M,\qv)$, given in (\ref{eq: general lower bound}),  and $R^{\rm ub}(n,m,M,\qv, \widetilde m)$ to show the order-optimality of RLFU-GCC by appropriately choosing the parameters $\ell$, $r$, $\widetilde z$, $z$. 
Specifically we choose: 
\begin{align}
\ell &= m, \label{assigmnet1}\\
r &= \delta(1-\alpha)n, \label{assigmnet2}\\
\widetilde{z} 
&=  \sigma m \left(1 - \exp\left(-\delta(1-\alpha)\frac{n}{m}\right)\right), 
\label{assigmnet3}
\end{align}
 where $\delta \in (0,1) $   and  $ \sigma \in (0,1) $ 
are positive constant independent of the system parameters $m,n,M$, and determined in the following while  $z$ will be determined later according to the different value of $m,n,M$.  Note that  $  \sigma m \left(1 - \exp\left(-\delta(1-\alpha)\frac{n}{m}\right)\right) \leq  \delta(1-\alpha)n $ hence by definition $\widetilde{z}  \leq r$.  
We now compute each term in (\ref{eq: general lower bound}) individually.\footnote{\label{fnm:1}In evaluating (\ref{eq: general lower bound}), anytime that the value of $z$  or  $  \lceil \widetilde{z} \rceil$ diverges as  $m \rightarrow \infty$,  we ignore the non-integer effect without mentioning.}
%For example, $\left\lceil \delta(1-\alpha)n \right\rceil = \delta(1-\alpha)n + o(n)$.
To this end, using  \eqref{assigmnet1} and \eqref{assigmnet2} we first find an expression  for 
$n\ell q_\ell$ and $\ell\left(1-\left(1-\frac{1}{\ell}\right)^r\right)$ in terms of  $\delta, \sigma,m,n$ and $\alpha$. 

%From Theorem \ref{theorem: general lower bound},  we can see that it requires $n\ell q_\ell$ and $\ell\left(1-\left(1-\frac{1}{\ell}\right)^r\right)$. 
%(see the definitions of $\Nsf$ and $\Zsf$ in Appendix \ref{sec: proof of theorem: general lower bound}). 
%First, we need to introduce the following lemma.
%\begin{lemma}
%\label{lemma: H}
%Let $H(\alpha,x,y) = \sum_{i=x}^y i^{-\alpha}$, for $\alpha \neq 1$, then 
%\begin{align}
%\frac{1}{1-\alpha}(y+1)^{1-\alpha} - \frac{1}{1-\alpha}x^{1-\alpha} \leq H(\alpha,x,y) \leq 
%\frac{1}{1-\alpha}y^{1-\alpha} - \frac{1}{1-\alpha}x^{1-\alpha} + \frac{1}{x^{\alpha}}.
%\end{align}
%\hfill$\square$
%\end{lemma}
%The proof of Lemma \ref{lemma: H} can be found in \cite{ji2013throughput}.
Specifically,  
%Let $\ell = m$, 
%by using (\ref{eq: EIU}) and 
by using Lemma \ref{lemma: H}, we can write
\begin{eqnarray}
\label{eq: alpha<1 EIU 1}
n\ell q_\ell &=& n m \frac{m^{-\alpha}}{H(\alpha, 1, m)} \notag\\
&\geq& \frac{nm^{1-\alpha}}{\frac{1}{1-\alpha}m^{1-\alpha} - \frac{1}{1-\alpha} + 1} \notag\\
&\geq& (1-\alpha)n + o(n),
\end{eqnarray}
and
\begin{eqnarray}
\label{eq: alpha<1 EIU 2}
n\ell q_\ell &=& n m \frac{m^{-\alpha}}{H(\alpha, 1, m)} \notag\\
&\leq& \frac{nm^{1-\alpha}}{\frac{1}{1-\alpha}(m+1)^{1-\alpha} - \frac{1}{1-\alpha}} \notag\\
&=& (1-\alpha)n + o(n).
\end{eqnarray}
from which it follows that  
%Thus, by using (\ref{eq: alpha<1 EIU 1}) and (\ref{eq: alpha<1 EIU 2}), we have
\be
\label{eq: IU 5}
n\ell q_\ell = (1-\alpha)n + o(n).
\ee

Furthermore, using \eqref{assigmnet2}, 
%We let $r = \delta(1-\alpha)n$, where $\delta \in (0,1)$ is a constant independent of all the system parameters $m,n,M$, then 
we have %by using (\ref{eq: EZb}), we can write
\begin{eqnarray}
\ell\left(1-\left(1-\frac{1}{\ell}\right)^r\right) %&=& \ell\left(1-\left(1-\frac{1}{\ell}\right)^r\right) \notag\\
&=& m \left(1 - \left(1 - \frac{1}{m}\right)^{\delta(1-\alpha)n} \right) \notag\\
&=& m \left(1 - \exp\left(-\delta(1-\alpha)\frac{n}{m}\right)\right) + o\left( m \left(1 - \exp\left(-\delta(1-\alpha)\frac{n}{m}\right)\right) \right).
\label{eq: IU6}
\end{eqnarray} 

Then, by using   \eqref{assigmnet1}-\eqref{assigmnet3}, \eqref{eq: IU6}  and  \eqref{eq: IU 5}   in  (\ref{eq: P1}) and  (\ref{eq: P2}), we obtain
\begin{eqnarray}
%\label{eq: IU 5}
%P_1(\ell,r) 
P_1(\ell,r) &=& 1- \exp\left(- \frac{(n\ell q_\ell - \delta(1-\alpha)n)^2}{2n\ell q_\ell}\right) \notag\\
&=& 1- \exp\left(- \frac{(((1-\alpha)n + o(n)) - \delta(1-\alpha)n)^2}{2((1-\alpha)n + o(n))}\right) \notag\\
&=& 1- \exp\left(- \frac{((1-\delta)(1-\alpha)n + o(n))^2}{2(\delta(1-\alpha)n + o(n))}\right) \notag\\
&=& 1 - o(1),
\end{eqnarray}
and 
\begin{eqnarray}
\label{eq: Zb 6}
%&&\PP(\Zsf \geq \widetilde{z}) \notag \\
&& P_2(\ell, r, \widetilde z) \notag\\
&& = 1- \exp\left(- \frac{\left(\ell\left(1-\left(1-\frac{1}{\ell}\right)^r\right) - \widetilde{z}\right)^2}{2\ell\left(1-\left(1-\frac{1}{\ell}\right)^r\right)}\right) \notag\\
&& \buildrel (a) \over \geq 
%1- \notag\\
%&& \exp\left(- \frac{\left(m \left(1 - \exp\left(-\delta(1-\alpha)\frac{n}{m}\right)\right) + o\left( m \left(1 - \exp\left(-\delta(1-\alpha)\frac{n}{m}\right)\right) \right) - \sigma m \left(1 - \exp\left(-\delta(1-\alpha)\frac{n}{m}\right)\right) \right)^2}{2\left( m \left(1 - \exp\left(-\delta(1-\alpha)\frac{n}{m}\right)\right) + o\left( m \left(1 - \exp\left(-\delta(1-\alpha)\frac{n}{m}\right)\right) \right) \right)}\right) \notag\\
%&& = 1 - \exp\left(- \frac{\left((1-\sigma)m\left(1 - \exp\left(-\delta(1-\alpha)\frac{n}{m}\right)\right)  + o\left( m \left(1 - \exp\left(-\delta(1-\alpha)\frac{n}{m}\right)\right) \right)\right)^2}{2\left( m \left(1 - \exp\left(-\delta(1-\alpha)\frac{n}{m}\right)\right) + o\left( m \left(1 - \exp\left(-\delta(1-\alpha)\frac{n}{m}\right)\right) \right) \right)} \right)\notag
%\\
%&& = 
1- o(1)
\end{eqnarray} 
where (a) follows from the fact that 
%$$
%\ell\left(1-\left(1-\frac{1}{\ell}\right)^r\right) - \widetilde{z}= \Theta \left(m \left(1 - \exp\left(-\delta(1-\alpha)\frac{n}{m}\right)\right) \right) 
%$$
%or 
$$
\frac{\left(\ell\left(1-\left(1-\frac{1}{\ell}\right)^r\right) - \widetilde{z}\right)^2}{2\ell\left(1-\left(1-\frac{1}{\ell}\right)^r\right)}=
\Theta \left(m \left(1 - \exp\left(-\delta(1-\alpha)\frac{n}{m}\right)\right) \right) 
$$
%where in \eqref{p2step} we have used the fact that 
%\be
%\label{eq: gamma < 1 zb 1}
%\widetilde{z} = \min\left\{\sigma m \left(1 - \exp\left(-\delta(1-\alpha)\frac{n}{m}\right)\right), \delta(1-\alpha)n\right\} = \sigma m \left(1 - \exp\left(-\delta(1-\alpha)\frac{n}{m}\right)\right).
%\ee 

Thus, by using Theorem \ref{theorem: general lower bound}, we obtain
\begin{eqnarray}
\label{eq: R lower 1}
&&R^{\rm lb}(n,m,M,\qv) \notag\\
&&\buildrel (a) \over \geq 
P_1(\ell,r) P_2(\ell, r, \widetilde z)\max_{z \in \{1, \cdots,  \lceil \widetilde{z} \rceil\}} (z-zM/{\lfloor \ell/z \rfloor}) \notag\\
&& \geq (1-o(1)) (1-o(1)) \max_{z \in \{1, \cdots,   \lceil \widetilde{z} \rceil\}} (z-zM/{\lfloor \ell/z \rfloor}) \notag\\
&& \geq (1-o(1))^2\max_{z \in \{1, \cdots,  \lceil \widetilde{z} \rceil\}} (z-zM/{\lfloor \ell/z \rfloor}),
\end{eqnarray}
where (a) is because $\widetilde{z} \leq r$. 

\begin{figure}[ht]
\centerline{\includegraphics[width=10cm]{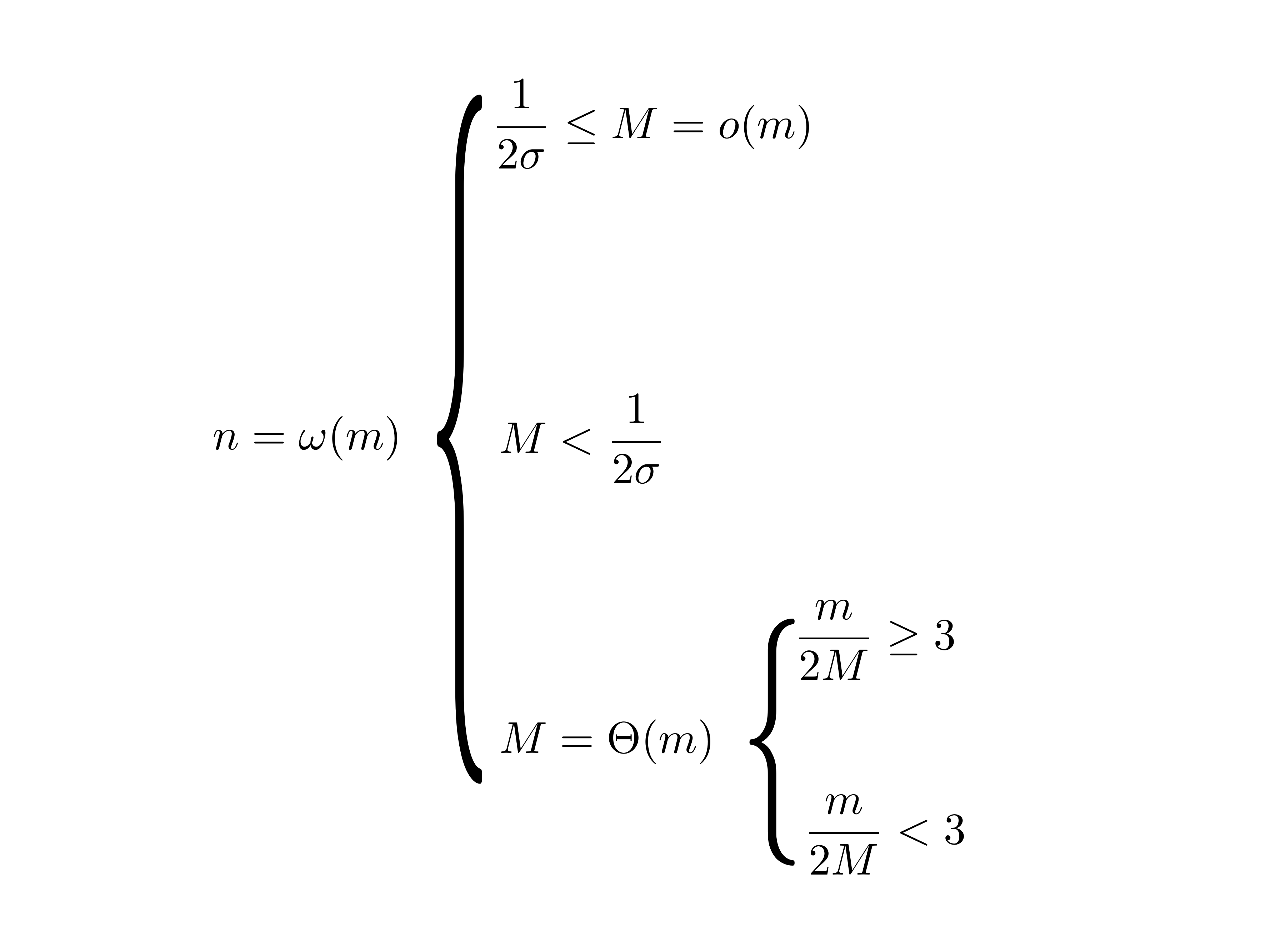}}
\caption{The sub-cases of the regimes of $M$ when $n = \omega(m)$. }
\label{fig: regimes_M_Thm3_1}
\end{figure}

\begin{figure}[ht]
\centerline{\includegraphics[width=10cm]{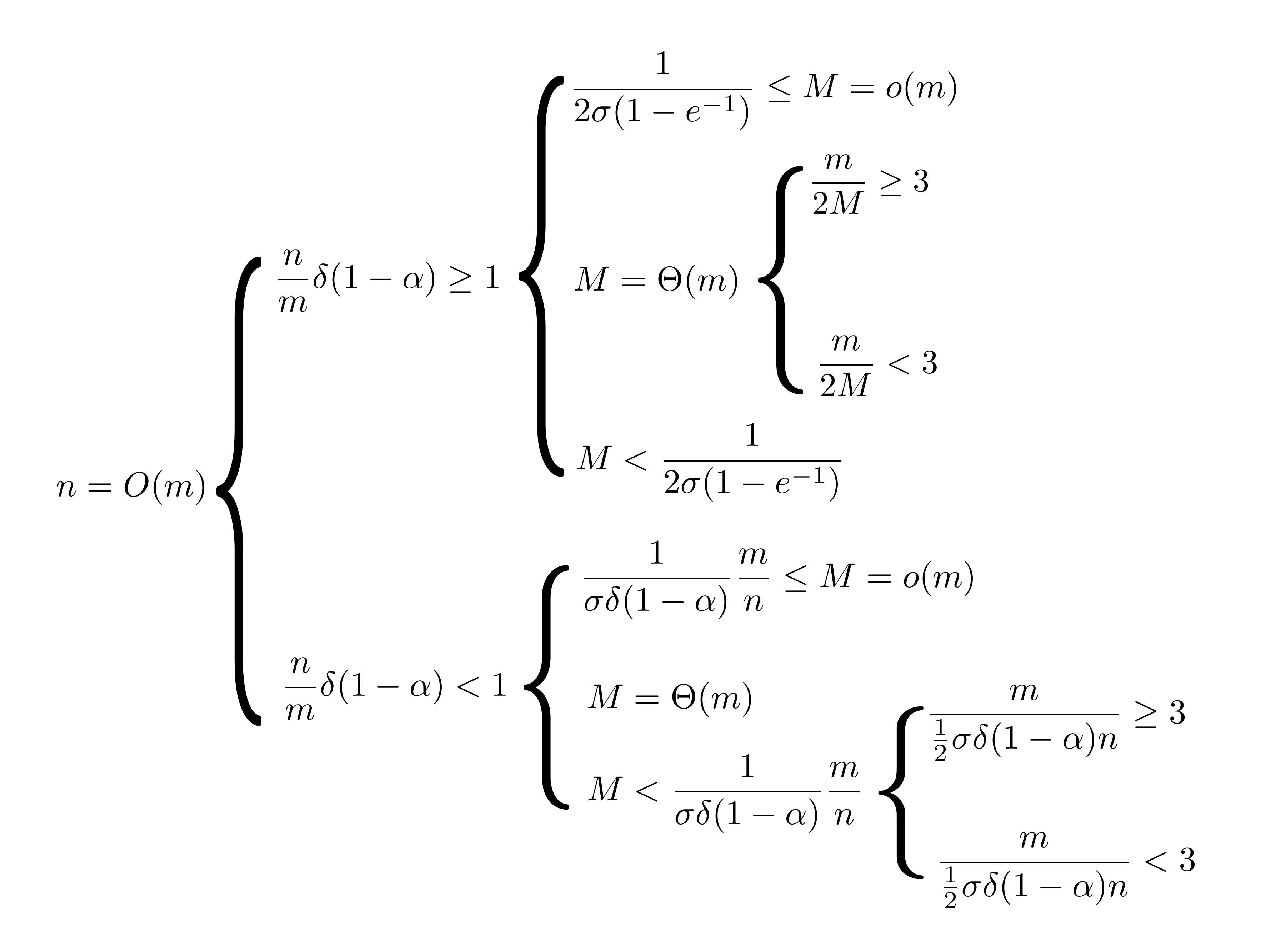}}
\caption{The sub-cases of the regimes of $M$ when $n = O(m)$. }
\label{fig: regimes_M_Thm3_2}
\end{figure}

In the following, we consider two cases, namely $n = \omega(m)$ and $n=O(m)$.  For each of these two cases, we treat separately the sub-regions of $M$ illustrated in Fig. \ref{fig: regimes_M_Thm3_1} and Fig. \ref{fig: regimes_M_Thm3_2}, respectively.

\subsection{Region of $n = \omega(m)$}
\label{subsectioRegion of n = omega(m)}
In this regime, using the fact that $\frac{n}{m} = \omega(1)$, 
the expression of $\widetilde{z}$, given in (\ref{assigmnet3}), reduces to: 
\begin{eqnarray}
%\label{eq: gamma < 1 zb 1}
\widetilde{z} %&=& \min\left\{\sigma m \left(1 - \exp\left(- \frac{n}{m}\delta(1-\alpha)\right)\right), \delta(1-\alpha)n\right\} \notag\\
%\buildrel (a) \over 
= \sigma m \left(1 - o(1)\right).
\end{eqnarray}
%where (a) is because that $\frac{n}{m} = \omega(1)$. 
Consequently,  (\ref{eq: R lower 1}) can be rewritten as:
\begin{eqnarray}
\label{eq: converse alpha<1 1}
R^{\rm lb}(n,m,M,\qv)  \geq (1-o(1))^2\max_{z \in \{1, \cdots, \lceil\sigma m \left(1 - o(1)\right) \rceil\}} (z-zM/{\lfloor \ell/z \rfloor}).
\end{eqnarray}

\subsubsection{When $\frac{1}{2\sigma (1-o(1))} \leq M = o(m)$}

Letting $z = \left\lfloor \frac{m}{2M} \right\rfloor$,  from   (\ref{eq: converse alpha<1 1}) we obtain
\begin{eqnarray}
\label{eq: converse sl 1}
R^{\rm lb}(n,m,M,\qv)  &\geq& (1-o(1))^2\left(z - \frac{z}{\left\lfloor \frac{m}{z} \right\rfloor} M\right) \notag\\
&=& (1-o(1))^2\left(\left\lfloor \frac{m}{2M} \right\rfloor - \frac{\left\lfloor \frac{m}{2M} \right\rfloor}{\left\lfloor \frac{m}{\left\lfloor \frac{m}{2M} \right\rfloor} \right\rfloor} M\right) \notag\\
&=& (1-o(1))^2\left(\frac{m}{4M} + o\left(\frac{m}{4M}\right)\right),
\end{eqnarray}
from which  using (\ref{eq: achievable alpha<1 1})  we have:
\be
\frac{R^{\rm ub}(n,m,M,\qv, \widetilde m)}{R^{\rm lb}(n,m,M,\qv) } \leq \frac{\frac{m}{M} - 1 + o\left(\frac{m}{M}\right)}{(1-o(1))^2\left(\frac{m}{4M} + o\left(\frac{m}{4M}\right)\right)} = 4 + o(1).
\ee

\subsubsection{When $M = \Theta(m)$}

\begin{itemize}
\item If $\frac{m}{2M} \geq 3$, letting  $z = \left\lfloor \frac{m}{2M} \right\rfloor$, from  (\ref{eq: converse alpha<1 1}) we get
\begin{eqnarray}
\label{eq: converse sl 2}
R^{\rm lb}(n,m,M,\qv)  
&\geq& (1-o(1))^2\left(\left\lfloor \frac{m}{2M} \right\rfloor - \frac{\left\lfloor \frac{m}{2M} \right\rfloor}{\left\lfloor \frac{m}{\left\lfloor \frac{m}{2M} \right\rfloor} \right\rfloor} M\right) \notag\\
&\geq& (1-o(1))^2\left(\frac{m}{2M} - 1\right)\left(1 - \frac{M}{2M} + o(1)\right) \notag\\
&\geq&(1-o(1))^2\left(\frac{\frac{m}{2M} - 1}{2} + o(1)\right),
\end{eqnarray}
from which, using (\ref{eq: achievable alpha<1 1}), we obtain
\be
\label{eq: gap sl 1}
\frac{R^{\rm ub}(n,m,M,\qv, \widetilde m)}{R^{\rm lb}(n,m,M,\qv) } \leq \frac{\frac{m}{M} - 1 + o\left(\frac{m}{M}\right)}{(1-o(1))^2\left(\frac{\frac{m}{2M} - 1}{2} + o(1)\right)} \leq 6 + o(1).
\ee
\item If $\frac{m}{2M} < 3$, letting $z = 1$, from (\ref{eq: converse alpha<1 1}), we obtain
\be
\label{eq: converse sl 3}
R^{\rm lb}(n,m,M,\qv)   \geq (1-o(1))^2\left(1-\frac{M}{m}\right),
\ee
from  which, using (\ref{eq: achievable alpha<1 1}), we have
\be
\label{eq: gap sl 2}
\frac{R^{\rm ub}(n,m,M,\qv, \widetilde m)}{R^{\rm lb}(n,m,M,\qv) } \leq \frac{\frac{m}{M} - 1 + o\left(\frac{m}{M}\right)}{(1-o(1))^2\left(1-\frac{M}{m}\right)} \leq \frac{m}{M} + o(1) \leq 6 + o(1). 
\ee
\end{itemize}

\subsubsection{When $M < \frac{1}{2\sigma (1-o(1))}$}

letting $z = \sigma m \left(1 - o(1)\right)$,  from (\ref{eq: converse alpha<1 1}), we get
\be
\label{eq: converse sl 4}
R^{\rm lb}(n,m,M,\qv)  \geq (1-o(1))^2 \sigma m\left(1-\frac{M}{\left\lfloor \frac{1}{\sigma} \right\rfloor}\right),
\ee
from which, using (\ref{eq: achievable alpha<1 1}), we obtain
\be
\label{eq: gap alpha<1 2}
\frac{R^{\rm ub}(n,m,M,\qv, \widetilde m)}{R^{\rm lb}(n,m,M,\qv) } \leq \frac{m}{(1-o(1))^2 \sigma m\left(1-\frac{M}{\left\lfloor \frac{1}{\sigma} \right\rfloor}\right)} \leq \frac{1}{\sigma \left(1-\frac{\frac{1}{\sigma (1-o(1))}}{2\left\lfloor \frac{1}{\sigma} \right\rfloor}\right)} + o(1).
\ee

\subsection{Region of  $n=O(m)$}

\subsubsection{When $\frac{n}{m}\delta(1-\alpha) \geq 1$}
$\widetilde{z}$, given  in (\ref{assigmnet3}), boils down to: 
\begin{eqnarray}
\label{eq: gamma < 1 zb 3}
\widetilde{z} &=& \sigma m \left(1 - \exp\left(-\delta(1-\alpha)\frac{n}{m}\right)\right) \notag\\
&\geq& \sigma m \left(1 - e^{-1}\right).
\end{eqnarray} 

\begin{itemize}
\item If  $\frac{1}{2(\sigma (1-e^{-1}))} \leq M = o(m)$,  
by letting $z = \left\lfloor \frac{m}{2M} \right\rfloor$, $R^{\rm lb}(n,m,M,\qv) $ is given by (\ref{eq: converse sl 1}) and consequently using (\ref{eq: achievable alpha<1 1}), we have 

\be
\label{eq: gap alpha<1 11}
\frac{R^{\rm ub}(n,m,M,\qv, \widetilde m)}{R^{\rm lb}(n,m,M,\qv) } \leq \frac{ \frac{m}{M} - 1 + o\left(\frac{m}{M}\right)}{(1-o(1))^2\left(\frac{m}{4M} + o\left(\frac{m}{4M}\right)\right)} = 4 + o(1).
\ee

%\begin{enumerate}
%
%\item If $t = \frac{nM}{m} = \omega(1)$, by using (\ref{eq: achievable alpha<1 1}) and (\ref{eq: converse sl 1}),  we obtain
%\be
%\label{eq: gap alpha<1 11}
%\frac{R^{\rm ub}(n,m,M,\qv, \widetilde m)}{R^{\rm lb}(n,m,M,\qv) } \leq \frac{ \frac{m}{M} - 1 + o\left(\frac{m}{M}\right)}{(1-o(1))^2\left(\frac{m}{4M} + o\left(\frac{m}{M}\right)\right)} = 4 + o(1).
%\ee
%
%\item If $t = \frac{nM}{m} = \Theta(1)$, then we have
%
%\begin{itemize}
%
%\item If $t = \frac{nM}{m} \geq 1$, by using (\ref{eq: achievable alpha<1 1}) and (\ref{eq: converse sl 1}), then
%\be
%\label{eq: gap alpha<1 12}
%\frac{R^{\rm ub}(n,m,M,\qv, \widetilde m)}{R^{\rm lb}(n,m,M,\qv) } \leq \frac{ \frac{m}{M} - 1 + o\left(\frac{m}{M}\right)}{(1-o(1))^2\left(\frac{m}{4M} + o\left(\frac{m}{M}\right)\right)} = 4 + o(1).
%\ee
%
%\item If $t = \frac{nM}{m} < 1$, by using (\ref{eq: achievable alpha<1 1}) and (\ref{eq: converse sl 1}), 
%then we can obtain
%\be
%\label{eq: gap sl 3}
%\frac{R^{\rm ub}(n,m,M,\qv, \widetilde m)}{R^{\rm lb}(n,m,M,\qv) } \leq \frac{n}{(1-o(1))^2\left(\frac{m}{4M} + o\left(\frac{m}{M}\right)\right)} = 4 + o(1).
%\ee
%
%\end{itemize}
%
%\end{enumerate}

\item If $M = \Theta(m)$, by letting $z = \left\lfloor \frac{m}{2M} \right\rfloor$ when $\frac{m}{2M} \geq 3$ and  letting $z = 1$ when $\frac{m}{2M} < 3$, and by using (\ref{eq: gap sl 1}) and (\ref{eq: gap sl 2}), respectively,  we conclude that 
\be
\frac{R^{\rm ub}(n,m,M,\qv, \widetilde m)}{R^{\rm lb}(n,m,M,\qv) } \leq 6 + o(1).
\ee

\item If  $M < \frac{1}{2(\sigma (1-e^{-1}))}$, 
letting $z =\sigma m \left(1 - e^{-1}\right)$, by using  \eqref{assigmnet1}-\eqref{assigmnet3} and (\ref{eq: R lower 1}), we have
\be
\label{eq: converse sl 5}
R^{\rm lb}(n,m,M,\qv)  \geq (1-o(1))^2 \sigma \left(1 - e^{-1}\right) m\left(1-\frac{M}{\left\lfloor \frac{1}{\sigma\left(1 - e^{-1}\right)} \right\rfloor}\right),
\ee
from which  using (\ref{eq: achievable alpha<1 1}), we obtain
\begin{eqnarray}
\label{eq: gap alpha<1 2 1}
\frac{R^{\rm ub}(n,m,M,\qv, \widetilde m)}{R^{\rm lb}(n,m,M,\qv) } &\leq& \frac{m}{(1-o(1))^2 \sigma \left(1 - e^{-1}\right) m\left(1-\frac{M}{\left\lfloor \frac{1}{\left(1 - e^{-1}\right)\sigma} \right\rfloor}\right)} \notag\\
&\leq& \frac{1}{\sigma \left(1 - e^{-1}\right)\left(1-\frac{\frac{1}{\sigma (1-e^{-1})}}{2\left\lfloor \frac{1}{\left(1 - e^{-1}\right)\sigma} \right\rfloor}\right)} + o(1).
\end{eqnarray}

\end{itemize}

\subsubsection{When $\frac{n}{m}\delta(1-\alpha) < 1$}
$\widetilde{z}$ boils down to: 
\begin{eqnarray}
\label{eq: gamma < 1 zb 4}
\widetilde{z} &=& \sigma m \left(1 - \exp\left(- \frac{n}{m}\delta(1-\alpha)\right)\right) \notag\\
&\buildrel (a) \over \geq& \sigma m \left(\frac{n}{m}\delta(1-\alpha) - \frac{1}{2}\left(\frac{n}{m}\delta(1-\alpha)\right)^2 \right) \notag\\
&=& \sigma \delta(1-\alpha)n - \frac{1}{2} \sigma\frac{n^2}{m} \delta^2(1-\alpha)^2 \notag\\
&\buildrel (b) \over \geq& \sigma \delta(1-\alpha)n - \frac{1}{2} \sigma \frac{1}{\delta(1-\alpha)} \delta^2(1-\alpha)^2n \notag\\
&=& \sigma \delta(1-\alpha)n - \frac{1}{2} \sigma\delta(1-\alpha)n \notag\\
&=& \frac{1}{2} \sigma\delta(1-\alpha)n,
\end{eqnarray}
where (a) follows from  $1-e^{-x} \geq x - \frac{x^2}{2}$,  (b) is due to  the fact that $\frac{n}{m}\delta(1-\alpha) < 1$.

\begin{itemize}

\item If $\frac{1}{\sigma\delta(1-\alpha)}\frac{m}{n} \leq M = o(m)$, by letting $z = \left\lfloor \frac{m}{2M} \right\rfloor$, $R^{\rm lb}(n,m,M,\qv) $ is given by (\ref{eq: converse sl 1}) and finally, using (\ref{eq: gap alpha<1 11}),  we obtain 
\be
\frac{R^{\rm ub}(n,m,M,\qv, \widetilde m)}{R^{\rm lb}(n,m,M,\qv) } \leq 4 + o(1).
\ee
\item If  $M = \Theta(m)$, letting $z = \left\lfloor \frac{m}{2M} \right\rfloor$  when $\frac{m}{2M} \geq 3$ and letting $z = 1$ when $\frac{m}{2M} < 3$, and by using (\ref{eq: gap sl 1}) and (\ref{eq: gap sl 2}), respectively, we conclude that 
\be
\frac{R^{\rm ub}(n,m,M,\qv, \widetilde m)}{R^{\rm lb}(n,m,M,\qv) } \leq 6 + o(1).
\ee

\item If $M < \frac{1}{\sigma\delta(1-\alpha)}\frac{m}{n}$, letting $z = \frac{1}{2} \sigma\delta(1-\alpha)n$, and  using 
\eqref{assigmnet1}-\eqref{assigmnet3} and (\ref{eq: R lower 1}), we have
\be
\label{eq: converse sl 6}
R^{\rm lb}(n,m,M,\qv)  \geq (1-o(1))^2 \frac{1}{2} \sigma\delta(1-\alpha)n\left(1-\frac{M}{\left\lfloor \frac{m}{\frac{1}{2} \sigma\delta(1-\alpha)n} \right\rfloor}\right). 
\ee

Recalling that by assumption $\frac{n}{m}\delta(1-\alpha) < 1$ and 
 $\sigma <1$,  we have  that  $\frac{m}{\frac{1}{2} \sigma\delta(1-\alpha)n} \geq 2$.  We hence consider  two cases. 
For $\frac{m}{\frac{1}{2} \sigma\delta(1-\alpha)n} \geq 3$,  using (\ref{eq: achievable alpha<1 1}) and (\ref{eq: converse sl 6}),  we obtain
\begin{eqnarray}
\label{eq: gap alpha<1 3}
\frac{R^{\rm ub}(n,m,M,\qv, \widetilde m)}{R^{\rm lb}(n,m,M,\qv) } &\leq& \frac{n}{(1-o(1))^2 \frac{1}{2} \sigma\delta(1-\alpha)n\left(1-\frac{M}{\left\lfloor \frac{m}{\frac{1}{2} \sigma\delta(1-\alpha)n} \right\rfloor}\right)} \notag\\
&\leq& \frac{2}{(1-o(1))^2\sigma\delta(1-\alpha)\left(1-\frac{\frac{1}{\sigma\delta(1-\alpha)}\frac{m}{n}}{\frac{m}{\frac{1}{2} \sigma\delta(1-\alpha)n}-1}\right)} \notag\\
&=& \frac{2}{(1-o(1))^2\sigma\delta(1-\alpha)\left(1- \frac{1}{2 - \frac{\sigma\delta(1-\alpha)n}{m}}\right)} \notag\\
&\leq& \frac{2}{(1-o(1))^2\sigma\delta(1-\alpha)\left(1- \frac{1}{2 - \frac{2}{3}}\right)} \notag\\
&=& \frac{8}{(1-o(1))^2\sigma\delta(1-\alpha) },
\end{eqnarray}
while for  $2 \leq \frac{m}{\frac{1}{2} \sigma\delta(1-\alpha)n} < 3$, using (\ref{eq: achievable alpha<1 1}) and (\ref{eq: converse sl 6}),  we obtain
\begin{eqnarray}
\label{eq: gap alpha<1 4}
\frac{R^{\rm ub}(n,m,M,\qv, \widetilde m)}{R^{\rm lb}(n,m,M,\qv) } &\leq& \frac{n}{(1-o(1))^2 \frac{1}{2} \sigma\delta(1-\alpha)n\left(1-\frac{M}{\left\lfloor \frac{m}{\frac{1}{2} \sigma\delta(1-\alpha)n} \right\rfloor}\right)} \notag\\
&\leq& \frac{2}{(1-o(1))^2\sigma\delta(1-\alpha)\left(1-\frac{\frac{1}{\sigma\delta(1-\alpha)}\frac{m}{n}}{2}\right)} \notag\\
&\leq& \frac{2}{(1-o(1))^2\sigma\delta(1-\alpha)\left(1- \frac{3}{4}\right)} \notag\\
&=& \frac{8}{(1-o(1))^2\sigma\delta(1-\alpha) }
\end{eqnarray}

\end{itemize}

\section{Proof of Theorem \ref{theorem: gamma > 1 achievable 2}}
\label{Proof of Theorem gamma > 1 achievable 2}
In the regime considered by  Theorem \ref{theorem: gamma > 1 achievable 2},  the number of users is much larger than the library size. For simplicity, we write $n = \rho m^{\alpha}$. 
%The order-optimality result when $m$ is finite can be proved in the same procedure of the proof of Theorem \ref{theorem: gamma < 1} with minor changes. 

From Lemma \ref{lemma: achievable}, we have that:
\begin{eqnarray}
\label{eq: general achievable 1}
%R^{\rm GCC}(m,n,M,\widetilde \pv,\qv) &\leq& 
R^{\rm ub}(n,m,M,\qv, \widetilde m) %\notag\\
&=& \min \left \{ \left(\frac{{\widetilde m}} {M}-1\right)\left( 1 -\left(1-\frac{M}{\widetilde m} \right)^{ n \, G_{\widetilde m}} \right) +  (1-G_{\widetilde m})\,n,  \,  \bar m \right\} 
\end{eqnarray}
from which, under the condition that  $\rho \geq 1$,  letting  $\widetilde m = m$, we have 
\begin{eqnarray}
\label{eq: general achievable 1bis}
R^{\rm ub}(n,m,M,\qv, \widetilde m) 
&\leq& 
 \frac{m}{M} -1 + o\left(\frac{m}{M} \right)\notag .
\end{eqnarray}

\begin{figure}[ht]
\centerline{\includegraphics[width=10cm]{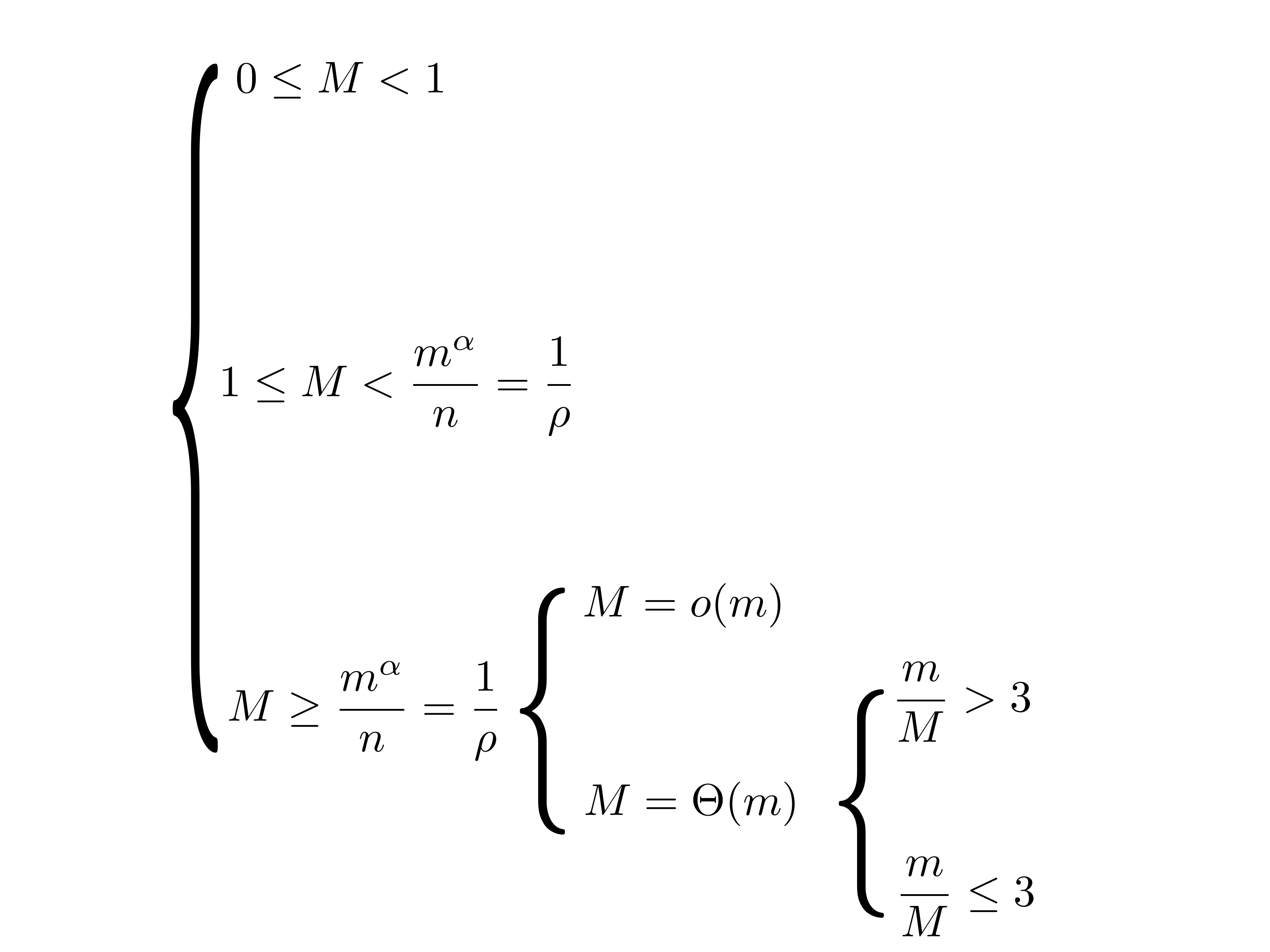}}
\caption{The sub-cases of the regimes of $M$ when $n = \rho m^{\alpha}$. }
\label{fig: regimes_M_Thm5}
\end{figure}

%\begin{eqnarray}
%R^{\rm ub}(n,m,M,\qv, \widetilde m)
%&\leq& \left(\frac{{\widetilde m}} {M}-1\right)\left( 1 -\left(1-\frac{M}{\widetilde m} \right)^{ n \, G_{\widetilde m}} \right) +  (1-G_{\widetilde m})\,n \notag\\
%&\leq& \frac{m}{M} -1 + o\left(\frac{m}{M} \right).
%\end{eqnarray}
In this case, the converse and order-optimal results follow, with minor changes, the same procedure as in the case of $n = \omega(m^\alpha)$ shown in the proof 
%{\BLUE of Theorem \ref{theorem: gamma > 1 achievable 1} and the proof }
of Theorem \ref{theorem: gamma < 1}  (see Appendix \ref{subsectioRegion of n = omega(m)}).
The non-trivial case is when $0<\rho < 1$, where we have the following regions: $0 \leq M < 1$, $1 \leq M < \frac{m^{\alpha}}{n} = \frac{1}{\rho}$, $M \geq \frac{1}{\rho}$.  All the sub-regions of $M$ are illustrated in Fig. \ref{fig: regimes_M_Thm5}, and will be treated separately in the following proofs.  

For the remainder of this section, in evaluating (\ref{eq: general lower bound}), anytime that the value of either $z$  or  $ \left\lceil \min\{\widetilde{z}, r\}\right\rceil$ diverges as  $m \rightarrow \infty$,  we ignore the non-integer effect without mentioning.

\subsection{Region of  $0 \leq M < 1$}
\label{subsectionMsmal1}
In this case, %we let $\widetilde m = \rho^\frac{1}{\alpha}m$,  
by using the second term of (\ref{eq: general achievable 1}), we can obtain 
%{\BLUE  how do we know that that is the min between the two??}
\begin{eqnarray}\label{eq: up M<1}
 %R^{\rm ub}(n,m,M,\qv, \widetilde m) \leq \bar m \leq \widetilde m + (1-G_{\widetilde m})\,n = 2 \rho^\frac{1}{\alpha}m + o(\rho^\frac{1}{\alpha}m).
R^{\rm ub}(n,m,M,\qv, \widetilde m) \leq \bar m \leq \rho^\frac{1}{\alpha}m+ (1-G_{\widetilde m})\,n = 2 \rho^\frac{1}{\alpha}m + o(\rho^\frac{1}{\alpha}m).
\end{eqnarray}

As in the proof of Theorem \ref{theorem: gamma < 1}, we compute the converse using  Theorem \ref{theorem: general lower bound}
and appropriately choosing the parameters $\ell$, $r$, $\widetilde z$, $z$. Specifically, we choose:
\begin{align}
\label{eq:commy1}
\ell &= \widetilde m = \rho^{\frac{1}{\alpha}}m, \\
r &= \frac{\delta(\alpha-1)}{\alpha}\rho^\frac{1}{\alpha}m,\label{eq:commy2}\\
\widetilde{z} 
&=   \sigma m \left(1 -  \exp\left(- \frac{\delta(\alpha-1)}{\alpha}\rho^\frac{1}{\alpha}\right) \right), \label{eq:commy3} \\ %\frac{\sigma\delta(\alpha-1)}{\alpha}\rho^\frac{1}{\alpha}m, \label{eq:commy3}\\
z &= \lfloor \widetilde{z} \rfloor, \label{eq:commy4}
\end{align}
with   $0 < \delta < 1$ and $ 0<\sigma <1$ being positive constants to be determined in the following, and such that 
$\widetilde{z} \leq r$.  Next, we compute each term in (\ref{eq: general lower bound}) individually. To this end, using  \eqref{eq:commy1} and
\eqref{eq:commy2}   we first find an expression  for 
$n\ell q_\ell$ and $\ell\left(1-\left(1-\frac{1}{\ell}\right)^r\right)$ in terms of  $\delta, \sigma,m,n$ and $\alpha$. 

Using \eqref{eq:commy1}, %(\ref{eq: EIU}) and 
Lemma \ref{lemma: H}, and the fact that $n = \rho m^{\alpha}$,  
we can write:
\begin{eqnarray}
\label{eq: IU M<1 1}
n\ell q_\ell &=& n \cdot \widetilde{m} \cdot \frac{{ \left(\widetilde{m}\right)}^{-\alpha}}{H(\alpha, 1, m)} \notag\\
&\geq& \frac{\rho^{\frac{1}{\alpha} }{m}}{\frac{1}{1-\alpha}m^{1-\alpha}  - \frac{1}{1-\alpha} + 1} \notag\\
&=& \frac{\alpha-1}{\alpha} \rho^{\frac{1}{\alpha} }m + o\left(m\right).
\end{eqnarray}
and
\begin{eqnarray}
\label{eq: IU M<1 2}
n\ell q_\ell &=& n \cdot \widetilde{m} \cdot \frac{{ \left(\widetilde{m}\right)}^{-\alpha}}{H(\alpha, 1, m)} \notag\\
&\leq& \frac{\rho^{\frac{1}{\alpha} }{m}}{\frac{1}{1-\alpha}(m+1)^{1-\alpha} - \frac{1}{1-\alpha}} \notag\\
&=&  (\alpha-1)\rho^{\frac{1}{\alpha} } m+ o\left(m\right),
\end{eqnarray}
from which we have
\begin{eqnarray}
\label{eq: IU M<1 30}
&&\frac{\alpha-1}{\alpha}\rho^{\frac{1}{\alpha} } m + o\left(m\right) \leq n\ell q_\ell\leq (\alpha-1)\rho^{\frac{1}{\alpha} } m+ o\left(m\right),
\end{eqnarray}
where \eqref{eq: IU M<1 1} and  \eqref{eq: IU M<1 2} follow from the fact that $\alpha>1$. 

Using \eqref{eq:commy2},  and the fact that $(1-1/x)^{\gamma x} \rightarrow e^{-\gamma}$ as $x \rightarrow \infty$, we have:
\begin{eqnarray}
&& \ell\left(1-\left(1-\frac{1}{\ell}\right)^r\right) \notag\\ 
&& = m \left(1 - \left(1 - \frac{1}{m}\right)^{\frac{\delta(\alpha-1)}{\alpha}\rho^\frac{1}{\alpha}m} \right) \notag\\
&& = m \left(1 -  \exp\left(- \frac{\delta(\alpha-1)}{\alpha}\rho^\frac{1}{\alpha}\right) \right) + o(m), %\notag\\
%&& = \frac{\delta(\alpha-1)}{\alpha}\rho^\frac{1}{\alpha}m + o(m).
\label{eq:Antonia}
\end{eqnarray}  
from which, by using (\ref{eq: P1}),  (\ref{eq: P2}),  (\ref{eq: IU M<1 30}), and (\ref{eq:Antonia}), we obtain
\begin{eqnarray}
\label{eq: P IU M<1}
P_1(\ell,r,z) &=& 1- \exp\left(- \frac{\left(n\ell q_\ell - \frac{\delta(\alpha-1)}{\alpha}\rho^\frac{1}{\alpha}m\right)^2}{2n\ell q_\ell}\right) \notag\\
&\geq& 1 -\exp \notag\\
&&\left(- \frac{\left(\frac{\alpha-1}{\alpha}\rho^{\frac{1}{\alpha}}m + o\left(m\right) - \frac{\delta(\alpha-1)}{\alpha}\rho^\frac{1}{\alpha}m\right)^2}{2\left((\alpha-1)\rho^{\frac{1}{\alpha}}m + o\left(m\right)\right)}\right) \notag\\
&=& 1 - o(1),
\end{eqnarray}
%Let  
%\begin{eqnarray}
%\widetilde{z} &=& \min\left\{\sigma_1\left(\frac{\delta_1(\alpha-1)}{\alpha}n^{\frac{1}{\alpha}} \right), \frac{\delta_1(\alpha-1)}{\alpha}n^{\frac{1}{\alpha}} \right\} \notag\\
%&=&  \frac{\sigma_1\delta_1(\alpha-1)}{\alpha}n^{\frac{1}{\alpha}},
%\end{eqnarray}
%where $0<\sigma_1 <1$ is a constant to guarantee $\widetilde{z} \leq r$, 
and 
\begin{eqnarray}
\label{eq: P Zb M<1}
&&P_2(\ell,r,\widetilde z) \notag \\
&& = 1- \exp\left(- \frac{\left(\ell\left(1-\left(1-\frac{1}{\ell}\right)^r\right) - \widetilde{z}\right)^2}{2 \ell \left(1 - \left(1 - \frac{1}{\ell}\right)^{r} \right)}\right) \notag\\
&& = 1 - \exp\left( \frac{\left((1-\sigma)\left(m \left(1 -  \exp\left(- \frac{\delta(\alpha-1)}{\alpha}\rho^\frac{1}{\alpha}\right) \right) + o(m)\right) \right)^2}{2\left(\left(1 -  \exp\left(- \frac{\delta(\alpha-1)}{\alpha}\rho^\frac{1}{\alpha}\right) \right)m+ o(m) \right)} \right). \notag\\
&& = 1- o(1).
\end{eqnarray} 

Hence, replacing  \eqref{eq:commy1}-\eqref{eq:commy4}, (\ref{eq: P IU M<1}) and (\ref{eq: P Zb M<1})  in   (\ref{eq: general lower bound}), we obtain:
\begin{eqnarray}
\label{eq: converse M<1}
&&R^{\rm lb}(n,m,M,\qv)  \notag\\
&& \geq 
P_1(\ell,r) P_2(\ell,r,\widetilde z) \max_{z \in \{1, \cdots, \lceil \widetilde{z} \rceil \}} (z-zM/{\lfloor \ell/z \rfloor}) \notag\\
&& \geq (1-o(1)) (1-o(1)) \max_{z \in \{1, \cdots, \lceil \widetilde{z} \rceil\}} (z-zM/{\lfloor \ell/z \rfloor}) \notag\\
&& \geq (1-o(1))^2 \left(\widetilde{z} - \frac{\widetilde{z}M}{\left\lfloor \frac{\widetilde{m}}{\widetilde{z}} \right\rfloor}\right) \notag\\
&& \geq (1-o(1))^2 \sigma m \left(1 -  \exp\left(- \frac{\delta(\alpha-1)}{\alpha}\rho^\frac{1}{\alpha}\right) \right) \left(1 - \frac{M}{\left\lfloor \frac{1}{\frac{\sigma\delta(\alpha-1)}{\alpha}}\right\rfloor} \right) \notag\\
&&\buildrel (a) \over \geq (1-o(1))^2 \sigma \left( \frac{\rho^\frac{1}{\alpha}\delta(\alpha-1)}{\alpha} - \frac{1}{2}\left(\frac{\rho^\frac{1}{\alpha}\delta(\alpha-1)}{\alpha}\right)^2 \right)\left(1 - \frac{1}{\left\lfloor \frac{1}{\frac{\sigma\delta(\alpha-1)}{\alpha}}\right\rfloor}\right)m,
\end{eqnarray}
where (a) is because $1 - \exp(-x) \geq x - \frac{x^2}{2}$ for $x \geq 0$. 

Using \eqref{eq: up M<1} and \eqref{eq: converse M<1},
%from which, 
we obtain
\begin{eqnarray}
\label{eq: gap M<1 1}
\frac{R^{\rm ub}(n,m,M,\qv, \widetilde m)}{R^{\rm lb}(n,m,M,\qv) } &\leq& \frac{2\rho^{\frac{1}{\alpha}}m}{(1-o(1))^2 \sigma \left( \frac{\rho^\frac{1}{\alpha}\delta(\alpha-1)}{\alpha} - \frac{1}{2}\left(\frac{\rho^\frac{1}{\alpha}\delta(\alpha-1)}{\alpha}\right)^2 \right)\left(1 - \frac{1}{\left\lfloor \frac{1}{\frac{\sigma\delta(\alpha-1)}{\alpha}}\right\rfloor}\right)m} \notag\\
&=& \frac{2}{(1-o(1))^2 \left(1 - \frac{1}{\left\lfloor \frac{1}{\frac{\sigma\delta(\alpha-1)}{\alpha}}\right\rfloor}\right) \left( \frac{\sigma\delta(\alpha-1)}{\alpha} - \frac{1}{2} \sigma\rho^\frac{1}{\alpha} \left(\frac{\delta(\alpha-1)}{\alpha}\right)^2 \right)} \notag\\
&\buildrel (a) \over \leq& \frac{2}{(1-o(1))^2 \left(1 - \frac{1}{\left\lfloor \frac{1}{\frac{\sigma\delta(\alpha-1)}{\alpha}}\right\rfloor}\right) \left( \frac{\sigma\delta(\alpha-1)}{\alpha} - \frac{1}{2} \sigma \left(\frac{\delta(\alpha-1)}{\alpha}\right)^2 \right)},
\end{eqnarray}
where (a) is because $\rho < 1$.
(\ref{eq: gap M<1 1}) shows the order-optimality of the achievable expected rate. In fact, it is easy to find values for the  parameters $\sigma, \delta, \in (0,1)$ such that the  right-hand side of  (\ref{eq: gap M<1 1}) is uniformly bounded with respect to $n , m $ and $M$, as shown in the  following example: Let $\sigma=\delta = \left(\frac{1}{2}\right)^{\frac{1}{2}}$. Then $\sigma \delta = \frac{1}{2}$, which replaced in (\ref{eq: gap M<1 1}) yields
\be
\label{eq: gap alpha>1 M<1}
\frac{R^{\rm ub}(n,m,M,\qv, \widetilde m)}{R^{\rm lb}(n,m,M,\qv) } \leq \frac{2}{\left(\frac{(\alpha-1)}{2\alpha} - \frac{1}{4}\left(\frac{1}{2}\right)^{\frac{1}{2}}\left(\frac{(\alpha-1)}{2\alpha}\right)^2\right) \left(1 - \frac{1}{\lfloor\frac{2\alpha}{(\alpha-1)}\rfloor}\right)}.
\ee
Since $\frac{2\alpha}{(\alpha-1)} > 2$ for $\alpha>1$,  then the right hand side of (\ref{eq: gap alpha>1 M<1}) is a positive constant. In the following proofs, for brevity, we do not illustrate the values of the constant parameters.  %DO WE NEED THIS ONE???

\subsection{Region of  $1 \leq M < \frac{m^{\alpha}}{n} = \frac{1}{\rho}$}
%M^{\frac{1}{\alpha}}n^{\frac{1}{\alpha}} = 
Let $\widetilde m = \rho^{\frac{1}{\alpha}}M^{\frac{1}{\alpha}}m$, by using (\ref{eq: general achievable 1}) and Lemma \ref{lemma: H}, we obtain
\begin{eqnarray}
\label{eq: achievable M>1 14}
R^{\rm ub}(n,m,M,\qv, \widetilde m)
&\leq& \left(\frac{{\widetilde m}} {M}-1\right)\left( 1 -\left(1-\frac{M}{\widetilde m} \right)^{ n \, G_{\widetilde m}} \right) +  (1-G_{\widetilde m})\,n \notag\\
%&\leq& \frac{2M^{\frac{1}{\alpha}}n^{\frac{1}{\alpha}}}{M} + o\left(\frac{M^{\frac{1}{\alpha}}n^{\frac{1}{\alpha}}}{M} \right) \notag\\
&=& \frac{2\rho^{\frac{1}{\alpha}}m}{M^{1-\frac{1}{\alpha}}} + o\left(\frac{\rho^{\frac{1}{\alpha}}m}{M^{1-\frac{1}{\alpha}}}\right).
\end{eqnarray}
%Next, we prove the order-optimality of the expected rate achieved by RLFU. 
Following the same procedure as in Appendix \ref{subsectionMsmal1}, we use Theorem \ref{theorem: general lower bound} to compute the converse. 
All the parameters of Theorem \ref{theorem: general lower bound} are summarized in the following:
\begin{align}
\ell &= \widetilde m = \rho^{\frac{1}{\alpha}}M^{\frac{1}{\alpha}}m, \label{ming1}\\
r &=   \frac{\delta(\alpha-1)}{\alpha}M^{\frac{1-\alpha}{\alpha}}\rho^{\frac{1}{\alpha}}m, \label{ming2}\\
\widetilde z &=   \sigma\left(1 - \exp\left(-\frac{\delta(\alpha-1)}{\alpha}M^{\frac{1-\alpha}{\alpha}}\rho^{\frac{1}{\alpha}}\right)\right)m \label{ming3}\\
z &= \lfloor \widetilde z \rfloor,\label{ming4}
\end{align}
with the constant parameters $\delta \in (0, 1)$ and  $\sigma \in (0, 1)$  to be determined in the following,  and such that 
$\widetilde{z} \leq r$. Based on \eqref{ming1}-\eqref{ming4},  we now compute each term in (\ref{eq: general lower bound}) individually in detail.

 Using (\ref{ming1}) and 
Lemma \ref{lemma: H}, 
and following the same procedure as in \eqref{eq: IU M<1 1} and \eqref{eq: IU M<1 2}, we can write
\begin{eqnarray}
\label{eq: EIU M>1 14}
&&\frac{(\alpha-1)}{\alpha}M^{\frac{1-\alpha}{\alpha}}\rho^{\frac{1}{\alpha}}m + o(m) \leq n\ell q_\ell \leq (\alpha-1) M^{\frac{1-\alpha}{\alpha}}\rho^{\frac{1}{\alpha}}m + o(m). 
\end{eqnarray}
which replaced in (\ref{eq: P1}), similar to  \eqref{eq: P IU M<1},  gives: 
\begin{eqnarray}
\label{eq: P IU M>1 14}
P_1(\ell,r,z) 
\geq 1 - o(1).
\end{eqnarray}

Furthermore, using (\ref{ming2}),  we obtain
\begin{eqnarray}
\label{eq: EZb M>1 14}
\ell\left(1-\left(1-\frac{1}{\ell}\right)^r\right) 
&=& m \left(1 - \left(1 - \frac{1}{m}\right)^{\frac{\delta(\alpha-1)}{\alpha}M^{\frac{1-\alpha}{\alpha}}\rho^{\frac{1}{\alpha}}m} \right) \notag\\
&=& m \left(1 - \exp\left(-\frac{\delta(\alpha-1)}{\alpha}M^{\frac{1-\alpha}{\alpha}}\rho^{\frac{1}{\alpha}}\right)\right) + o\left( m \right).
\end{eqnarray} 
Replacing  (\ref{ming3})  and \eqref{eq: EZb M>1 14} in (\ref{eq: P2}), as in \eqref{eq: P Zb M<1}, we obtain: 
%\begin{eqnarray}
%\widetilde{z} 
%=\sigma_{14}\left(1 - \exp\left(-\frac{\delta_{14}\rho(\alpha-1)}{\alpha}M^{\frac{1-\alpha}{\alpha}}\rho^{\frac{1}{\alpha}}\right)\right)m, 
%\end{eqnarray}
%where $0<\sigma_{14} <1$ is a constant, then by using (\ref{eq: P2}) and (\ref{eq: EZb M>1 14}), we have
\begin{eqnarray}
\label{eq: P Zb M>1 14}
&&P_2(\ell, r, \widetilde z)
= 1- o(1).
\end{eqnarray} 

Finally, replacing  \eqref{ming1}-\eqref{ming4}, (\ref{eq: EZb M>1 14}), and (\ref{eq: P Zb M>1 14})  in   (\ref{eq: general lower bound}), we obtain:
\begin{eqnarray}
\label{eq: converse 14}
&&R^{\rm lb}(n,m,M,\qv)  \notag\\
&& \geq  P_1(\ell,r) P_2(\ell, r, \widetilde z)\max_{z \in \{1, \cdots, \lceil \widetilde{z} \rceil\}} (z-zM/{\lfloor \ell/z \rfloor}) \notag\\
&& \geq (1-o(1)) (1-o(1)) \max_{z \in \{1, \cdots, \lceil \widetilde{z} \rceil\}} (z-zM/{\lfloor \ell/z \rfloor}) \notag\\
&& \geq (1-o(1))^2 \left(\widetilde{z} - \frac{\widetilde{z}M}{\left\lfloor \frac{\widetilde{m}}{\widetilde{z}} \right\rfloor}\right) \notag\\
&& \geq (1-o(1))^2 \widetilde{z}\left(1 -  \frac{M}{\frac{\widetilde{m}}{\widetilde{z}} - 1}\right) \notag\\
&& =  (1-o(1))^2 \sigma\left(1 - \exp\left(-\frac{\delta(\alpha-1)}{\alpha}M^{\frac{1-\alpha}{\alpha}}\rho^{\frac{1}{\alpha}}\right)\right)m \notag\\
&& \cdot \left(1 -  \frac{M}{\frac{M^{\frac{1}{\alpha}}\rho^{\frac{1}{\alpha}} m}{\sigma\left(1 - \exp\left(-\frac{\delta(\alpha-1)}{\alpha}M^{\frac{1-\alpha}{\alpha}}\rho^{\frac{1}{\alpha}}\right)\right)m} - 1}\right) \notag\\
&& = (1-o(1))^2 \sigma\left(1 - \exp\left(-\frac{\delta(\alpha-1)}{\alpha}M^{\frac{1-\alpha}{\alpha}}\rho^{\frac{1}{\alpha}}\right)\right)m \notag\\
&& \cdot \left(1 -  \frac{M}{\frac{M^{\frac{1}{\alpha}}}{\sigma\left(1 - \exp\left(-\frac{\delta(\alpha-1)}{\alpha}M^{\frac{1-\alpha}{\alpha}}\rho^{\frac{1}{\alpha}} \right)\right)} - 1} \right) \notag\\
&& \geq (1-o(1))^2 \sigma\left(1 - \exp\left(-\frac{\delta(\alpha-1)}{\alpha}M^{\frac{1-\alpha}{\alpha}} \rho^{\frac{1}{\alpha}} \right)\right)m \notag\\
&& \cdot \left( 1 - \frac{M}{\frac{\left( M\rho \right)^{\frac{1}{\alpha}}}{\frac{\sigma\delta(\alpha-1)}{\alpha}M^{\frac{1-\alpha}{\alpha}} \rho^{\frac{1}{\alpha}} } - 1} \right)  \label{barca}\\
&& \geq (1-o(1))^2 \sigma\left(1 - \exp\left(-\frac{\delta(\alpha-1)}{\alpha}M^{\frac{1-\alpha}{\alpha}} \rho^{\frac{1}{\alpha}} \right)\right)m \cdot \left( 1 - \frac{M}{\frac{M}{\frac{\sigma\delta(\alpha-1)}{\alpha}} - 1}\right) \notag\\
&& = (1-o(1))^2 \sigma\left(1 - \exp\left(-\frac{\delta(\alpha-1)}{\alpha}M^{\frac{1-\alpha}{\alpha}} \rho^{\frac{1}{\alpha}} \right)\right)m \left( 1 - \frac{1}{\frac{\alpha}{\sigma\delta(\alpha-1)} - \frac{1}{M}} \right) \notag \\
%\end{eqnarray}
%\begin{eqnarray}
&& \geq (1-o(1))^2 \sigma\left(1 - \exp\left(-\frac{\delta(\alpha-1)}{\alpha} M^{\frac{1-\alpha}{\alpha}} \rho^{\frac{1}{\alpha}} \right)\right)m \left( 1 - \frac{1}{\frac{\alpha}{\sigma\delta(\alpha-1)} - 1} \right) %\label{eq: converse 14},
\end{eqnarray}
where \eqref{barca} follows from the fact that since $ 1<M $, $1 <\alpha$, $\delta \in (0, 1)$, and  $\sigma \in (0, 1)$, we have
$$
1 - \exp\left(-\frac{\delta(\alpha-1)}{\alpha}M^{\frac{1-\alpha}{\alpha}} \rho^{\frac{1}{\alpha}} \right) < \frac{\sigma\delta(\alpha-1)}{\alpha}M^{\frac{1-\alpha}{\alpha}} \rho^{\frac{1}{\alpha}}  < \left( M\rho \right)^{\frac{1}{\alpha}} .$$

On the other hand, since 
\begin{eqnarray}
\label{eq: converse 14 1}
&& 1 - \exp\left(-\frac{\delta(\alpha-1)}{\alpha}M^{\frac{1-\alpha}{\alpha}} \rho^{\frac{1}{\alpha}} \right) \notag\\
&& \geq \frac{\delta(\alpha-1)}{\alpha}M^{\frac{1-\alpha}{\alpha}}  \rho^{\frac{1}{\alpha}} - \frac{1}{2}\left( \frac{\delta(\alpha-1)}{\alpha}M^{\frac{1-\alpha}{\alpha}} \rho^{\frac{1}{\alpha}} \right)^2 \notag\\
&& \geq \frac{1}{2}\left( \frac{\delta(\alpha-1)}{\alpha}M^{\frac{1-\alpha}{\alpha}}  \rho^{\frac{1}{\alpha}} \right),
\end{eqnarray}
using (\ref{eq: converse 14}), we obtain
\be
R^{\rm lb}(n,m,M,\qv)  \geq (1-o(1))^3 \sigma\frac{1}{2}\left( \frac{\delta(\alpha-1)}{\alpha} M^{\frac{1-\alpha}{\alpha}} \rho^{\frac{1}{\alpha}}  \right)  \left( 1 - \frac{1}{\frac{\alpha}{\sigma\delta(\alpha-1)} - 1} \right)m,
\ee
from which, using (\ref{eq: achievable M>1 14}), it follows: 
\begin{eqnarray}
\frac{R^{\rm ub}(n,m,M, \qv, \widetilde m)}{R^{\rm lb}(n,m,M,\qv) }  &\leq& \frac{ 2 M^{\frac{1-\alpha}{\alpha}} \rho^{\frac{1}{\alpha}}  m }{(1-o(1))^3 \sigma\frac{1}{2}\left( \frac{\delta (\alpha-1)}{\alpha} M^{\frac{1-\alpha}{\alpha}} \rho^{\frac{1}{\alpha}}  \right)  \left( 1 - \frac{1}{\frac{\alpha}{\sigma\delta(\alpha-1)} - 1} \right)m} \notag\\
&=& \frac{4\alpha}{(1-o(1))^3 \sigma\delta(\alpha-1)\left( 1 - \frac{1}{\frac{\alpha}{\sigma\delta(\alpha-1)} - 1} \right)},
\end{eqnarray}
which shows the order-optimality of the achievable expected rate of RLFU-GCC.

\subsection{Region of  $M \geq \frac{m^{\alpha}}{n} = \frac{1}{\rho}$}

In this regime, let $\widetilde m = m$. Using (\ref{eq: general achievable 1}),  we obtain
\begin{eqnarray}
\label{eq: achievable M>1 15}
R^{\rm ub}(n,m,M,\qv, \widetilde m)
\leq \frac{m}{M} - 1 + o\left(\frac{m}{M}\right).
\end{eqnarray}

In order to prove the order-optimality of the expected rate achieved by RLFU-GCC, as before, we use Theorem \ref{theorem: general lower bound} to compute the converse.  All the parameters of Theorem \ref{theorem: general lower bound} are summarized in the following: 
\begin{align}
\ell &= m, \label{Giusep1}\\
r &=  \frac{\delta\rho (\alpha-1)}{\alpha}  m, \label{Giusep2}\\
\widetilde z &=   \sigma m \left(1 - \exp\left(-\frac{\delta \rho(\alpha-1)}{\alpha} \right)\right), \label{Giusep3}\\
z &=  \left\{\begin{array}{cc}  \lfloor \frac{\delta\sigma(\alpha-1)}{2\alpha}\frac{m}{M} \rfloor , & M = o(m) \\
\max\{\left\lfloor \frac{m}{2M} \right\rfloor, 1\}, & \rm{otherwise} 
 \end{array} \right., \label{Giusep4} 
\end{align}
with the constant parameters $\delta \in (0, 1)$ and  $\sigma \in (0, 1)$  to be determined in the following,  and such that 
$\widetilde{z} \leq r$.  We now compute each term in (\ref{eq: general lower bound}) individually in detail.

Using \eqref{Giusep1} %by using (\ref{eq: EIU}), 
and following the same steps as in \eqref{eq: alpha<1 EIU 1}  and \eqref{eq: alpha<1 EIU 2}, we have
\be
\label{eq: EIU M>1 15}
\frac{\alpha-1}{\alpha}\rho m+ o\left( m \right) \leq n\ell q_\ell \leq (\alpha-1)\rho m + o\left(m\right).
\ee
from which, using (\ref{eq: P1}),  similar as in \eqref{eq: P Zb M<1},  we obtain
\begin{eqnarray}
\label{eq: P IU M>1 15}
P_1(\ell,r) = 1 - o(1).
\end{eqnarray}

Furthermore using  \eqref{Giusep2}, we can write 
\begin{eqnarray}
\label{eq: EZb M>1 15}
\ell\left(1-\left(1-\frac{1}{\ell}\right)^r\right) &=& m \left(1 - \left(1 - \frac{1}{m}\right)^{\frac{\delta\rho(\alpha-1)}{\alpha} m} \right) \notag\\
&=& m \left(1 - \exp\left(-\frac{\delta \rho(\alpha-1)}{\alpha} \right)\right) + o\left( m \right) \notag\\
\end{eqnarray} 
from which, using (\ref{eq: P2})  and \eqref{Giusep3}, we have
\begin{eqnarray}
\label{eq: P Zb M>1 15}
P_2(\ell, r, \widetilde z)= 1- o(1).
\end{eqnarray} 

Replacing \eqref{Giusep1}-\eqref{Giusep4},   \eqref{eq: P IU M>1 15}  and \eqref{eq: P Zb M>1 15}  in  (\ref{eq: general lower bound}),   we obtain
\begin{eqnarray}
\label{eq: converse M>1 15}
R^{\rm lb}(n,m,M,\qv)  &\geq&  P_1(\ell,r) P_2(\ell, r, \widetilde z)\max_{z \in \{1, \cdots, \lceil \widetilde{z} \rceil \}} (z-zM/{\lfloor \ell/z \rfloor}) \notag\\
&\geq& (1-o(1))^2 \max_{z \in \{1, \cdots, \lceil \widetilde{z} \rceil \}} (z-zM/{\lfloor \ell/z \rfloor}),
\end{eqnarray}
with $\widetilde z$ and $z$ given in  \eqref{Giusep3} and \eqref{Giusep4}, respectively. 
Note that, from   \eqref{Giusep3} , it follows that $\widetilde{z}$ is lower bounded by: 
\begin{eqnarray}
\widetilde{z} &=& \sigma m \left(1 - \exp\left(-\frac{\delta \rho (\alpha-1)}{\alpha}\right)\right) \notag\\
&\geq& \sigma m \left(\frac{\delta \rho (\alpha-1)}{\alpha} - \frac{1}{2}\left(\frac{\delta \rho (\alpha-1)}{\alpha}\right)^2\right) \notag\\
&\geq& \frac{\sigma \delta \rho (\alpha-1)}{2\alpha}m \notag\\
&\geq& \frac{\sigma \delta  (\alpha-1)}{2\alpha} \frac{m}{M},
\label{ztildelowbound}
\end{eqnarray}
where \eqref{ztildelowbound} follows from the fact that,  in this regime, by assumption $M \geq \frac{m^{\alpha}}{n} = \frac{1}{\rho}$ and consequently   $\rho m \geq \frac{m}{M}$.

Now we consider two cases of $M$: $M=o(m)$ and $M=\Theta(m)$ (see Fig. \ref{fig: regimes_M_2}).
%\begin{itemize}
\subsubsection {When $M = o(m)$}  From  \eqref{Giusep4},  
$$z = \left \lfloor  \frac{\delta \sigma(\alpha-1)}{2\alpha}\frac{m}{M} \right \rfloor ,$$ 
from which, using \eqref{ztildelowbound}, it follows that $z \leq \widetilde z$. Hence,  using 
 \eqref{Giusep1},  (\ref{eq: converse M>1 15}),   and footnote \ref{fnm:1}, we obtain
\begin{eqnarray}
\label{eq: R > 1 f1}
R^{\rm lb}(n,m,M,\qv) &\buildrel (a) \over\geq& (1-o(1))^2 \frac{\delta \sigma(\alpha-1)}{2\alpha}\frac{m}{M} \left(1 - \frac{M}{\left\lfloor \frac{m}{\frac{\delta \sigma(\alpha-1)}{2\alpha}\frac{m}{M}} \right\rfloor}\right) \notag\\
&=& (1-o(1))^2 \frac{\delta \sigma(\alpha-1)}{2\alpha}\frac{m}{M} \left(1 - \frac{M}{\left\lfloor \frac{2\alpha}{\delta \sigma(\alpha-1)} M\right\rfloor}\right) \notag\\
&\geq& (1-o(1))^2 \frac{\delta \sigma(\alpha-1)}{2\alpha}\frac{m}{M} \left(1 - \frac{1}{ \frac{2\alpha}{\delta \sigma(\alpha-1)} - \frac{1}{M}}\right) \notag\\
&\buildrel (b) \over\geq& (1-o(1))^2 \frac{\delta \sigma(\alpha-1)}{2\alpha}\frac{m}{M} \left(1 - \frac{1}{ \frac{2\alpha}{\delta \sigma(\alpha-1)} - 1}\right),
\end{eqnarray}
where in (a)  we have used the fact that  $\left \lfloor \frac{\delta \sigma(\alpha-1)}{2\alpha}\frac{m}{M} \right \rfloor =\frac{\delta \sigma(\alpha-1)}{2\alpha}\frac{m}{M} + o(m)$ and that  
$\frac{\delta \sigma(\alpha-1)}{2\alpha}\frac{m}{M} \rightarrow \infty $, and in (b) the fact that  $M > 1$.
Then,  using (\ref{eq: achievable M>1 15}), we obtain:
\begin{eqnarray}
\label{eq: gap 5 1}
\frac{R^{\rm ub}(n,m,M, \qv, \widetilde m)}{R^{\rm lb}(n,m,M,\qv) } &=& \frac{\frac{m}{M}-1 + o(1)}{(1-o(1))^2 \frac{\delta \sigma(\alpha-1)}{2\alpha}\frac{m}{M} \left(1 - \frac{1}{ \frac{2\alpha}{\delta \sigma(\alpha-1)} - 1}\right)} \notag\\
& \leq & \frac{1}{(1-o(1))^2 \frac{\delta \sigma(\alpha-1)}{2\alpha} \left(1 - \frac{1}{ \frac{2\alpha}{\delta \sigma(\alpha-1)} - 1}\right)} + o(1).
\end{eqnarray}

\subsubsection{ When $M = \Theta(m)$} Then 
{
\begin{itemize}
\item If $\frac{m}{M} \leq 3$,  from  \eqref{Giusep4}, we have that $z = 1$. Hence  using  \eqref{Giusep1} and   (\ref{eq: converse M>1 15}), we have:
\be
R^{\rm lb}(n,m,M,\qv)  \geq (1-o(1))^2 \left(1 - \frac{M}{m }\right),
\ee
from which
\begin{eqnarray}
\frac{R^{\rm ub}(n,m,M, \qv, \widetilde m)}{R^{\rm lb}(n,m,M,\qv) } &=& \frac{\frac{m}{M}-1 + o(1))}{(1-o(1))^2\left(1 - \frac{M}{m   }\right)} \notag\\
&\leq& \frac{m}{M} + o(1) \notag\\
&\leq& 3 + o(1). 
\end{eqnarray}
\item If  $\frac{m}{M} > 3$,  from  \eqref{Giusep4}, we have that  $z = \left\lfloor \frac{m}{2M} \right\rfloor$. Hence using \eqref{Giusep4} and  (\ref{eq: converse M>1 15}) and the fact that $M = \Theta(m)$, we obtain
\begin{eqnarray}
\label{eq: converse M>1 15 1}
R^{\rm lb}(n,m,M,\qv)  
&& \geq P_1(\ell,r) P_2(\ell,r,\widetilde z) \max_{z \in \{1, \cdots,  \lceil \widetilde{z} \rceil \}} (z-zM/{\lfloor \ell/z \rfloor}) \notag\\
%&& \geq (1-o(1)) (1-o(1)) \max_{z \in \{1, \cdots, \lceil \widetilde{z} \rceil\}} (z-zM/{\lfloor \ell/z \rfloor}) \notag\\
%&& \geq (1-o(1))^2 z \left(1 - \frac{M}{\left\lfloor \frac{ {m} }{z} \right\rfloor } \right) \notag\\
&& \geq (1-o(1))^2 \left\lfloor \frac{m}{2M} \right\rfloor \left(1 - \frac{M}{\left\lfloor\frac{m}{\left\lfloor \frac{m}{2M} \right\rfloor}\right\rfloor} \right) \notag\\
&& \geq (1-o(1))^2 \left\lfloor \frac{m}{2M} \right\rfloor \left(1 - \frac{M}{\frac{m}{ \frac{m}{2M}} - 1} \right) \notag\\
&& = (1-o(1))^2 \left\lfloor \frac{m}{2M} \right\rfloor \left(1 - \frac{1}{2-\frac{1}{M}}\right) \notag\\
&& \geq (1-o(1))^3 \frac{1}{2}\left( \frac{m}{2M} - 1 \right),
\end{eqnarray}
from which  using (\ref{eq: achievable M>1 15}), we have
\begin{eqnarray}
\frac{R^{\rm ub}(n,m,M, \qv, \widetilde m)}{R^{\rm lb}(n,m,M,\qv) } &=& \frac{\frac{m}{M}-1 + o(1))}{(1-o(1))^3 \frac{1}{2}\left( \frac{m}{2M} - 1 \right)} \notag\\
&\leq& \frac{1 + o(1))}{(1-o(1))^3 \frac{1}{2}\left(\frac{1}{2} - \frac{M}{m}\right)} \notag\\
&\leq& 12 + o(1). 
\end{eqnarray}

\end{itemize}
}

Thus, we finish the proof of Theorem \ref{theorem: gamma > 1 achievable 2}.

\section{Proof of Table \ref{table: table_1_1} in Theorem \ref{theorem: gamma > 1 achievable 3}} 
\label{Proof of Theorem gamma > 1 3}

\begin{figure}[ht]
\centerline{\includegraphics[width=10cm]{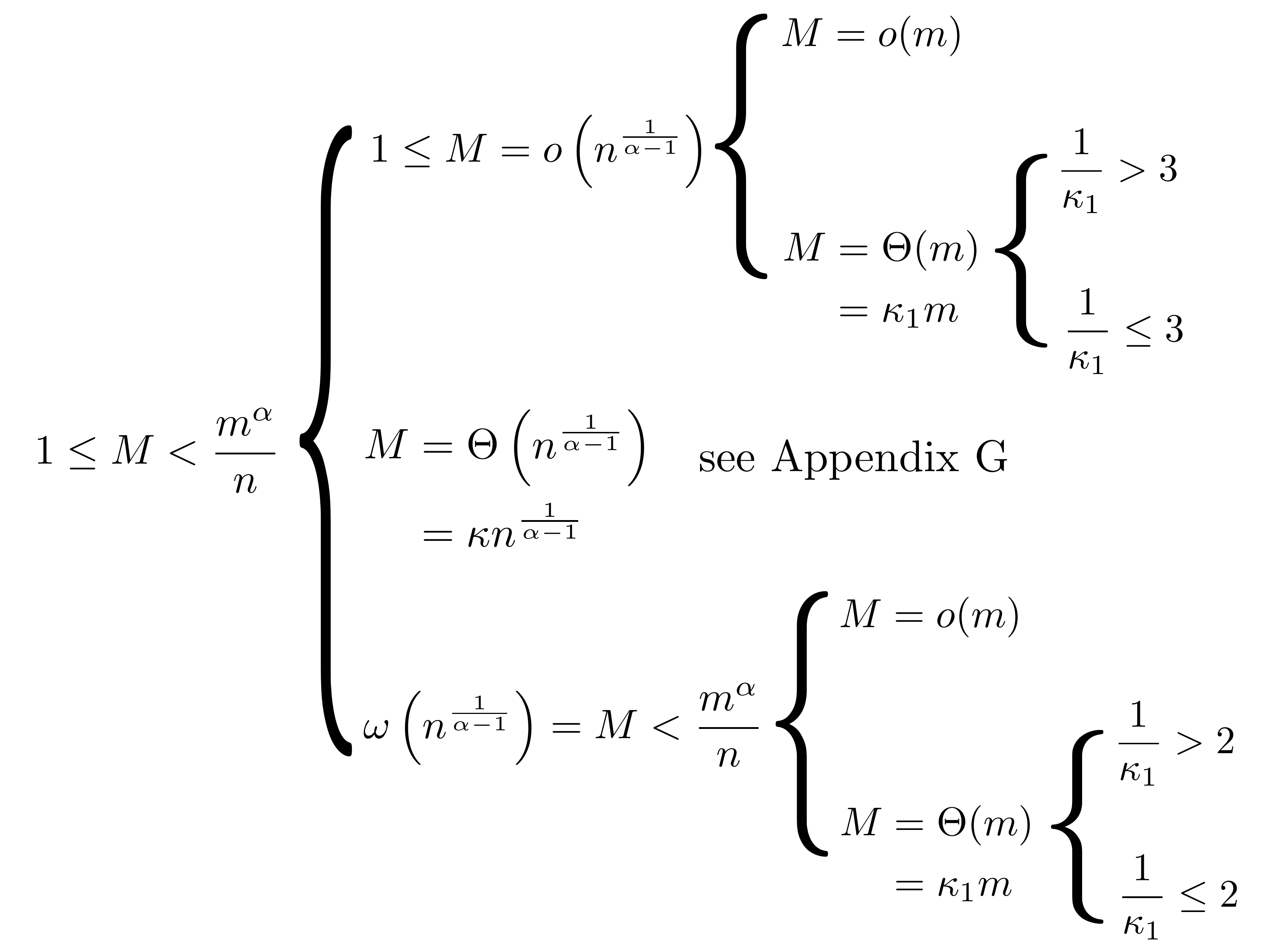}}
\caption{The sub-cases of the regimes of $M$ when $1 \leq M < \frac{m^{\alpha}}{n}$, where $\kappa$, $\kappa_1$ 
are some constants which will be given later.}
\label{fig: regimes_M_2}
\end{figure}

\begin{figure}[ht]
\centerline{\includegraphics[width=12cm]{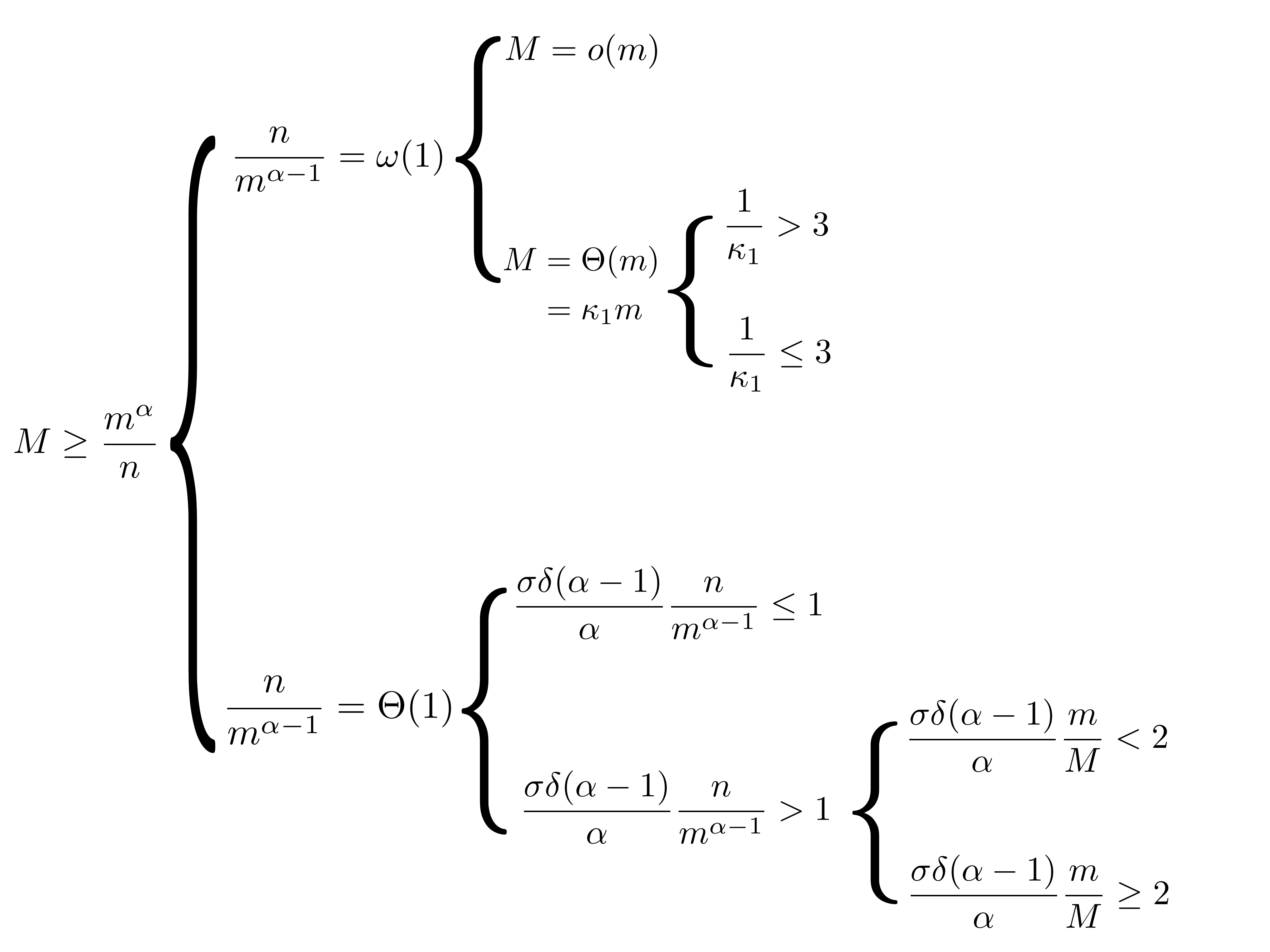}}
\caption{The sub-cases of the regimes of $M$ when $M \geq \frac{m^{\alpha}}{n}$, where $\kappa_1$, $\sigma$, $\delta$ are  some constants which will be given later.}
\label{fig: regimes_M_1}
\end{figure}

In this section, we provide the proof of Table \ref{table: table_1_1} in  Theorem \ref{theorem: gamma > 1 achievable 3},
where we assume $n,m \rightarrow \infty $ and $n = o\left(m^{\alpha}\right)$. Table \ref{table: table_1_1}  considers the regions $0 \leq M<1$, $1 \leq M < \frac{m^{\alpha}}{n}$, and $M \geq \frac{m^{\alpha}}{n}$, excluding the sub-region
$\{ 1 \leq M < \frac{m^{\alpha}}{n} \} \cap  \{  M = \kappa n^{\frac{1}{\alpha-1}}\} $ whose order-optimality is analyzed in Appendix \ref{sec: proof of table II} and whose corresponding order-optimal results are provided in Table \ref{table: table_1_2}. Except for
the region  $0 \leq M<1$, we further consider the subregions illustrated 
in Figs. \ref{fig: regimes_M_2} and \ref{fig: regimes_M_1}, treated separately in the following proofs.

\subsection{Region of  $0 \leq M < 1$}

In this case, we want to prove that RLFU-GCC with  $\widetilde m = n^{\frac{1}{\alpha}}$ is order optimal. To this end,  using (\ref{eq: general achievable 1})  {and Lemma \ref{lemma: H} },  we can write the rate for  RLFU-GCC with  $\widetilde m = n^{\frac{1}{\alpha}}$ 
as: 
\be
\label{eq: achievable M>1 1}
 R^{\rm ub}(n,m,M,\qv, \widetilde m) \leq \widetilde m + (1-G_{\widetilde m})\,n = 2 n^{\frac{1}{\alpha}} + o( n^{\frac{1}{\alpha}}).
\ee
Next, similarly as in Section \ref{subsectionMsmal1}, we evaluate  the converse replacing in   Theorem \ref{theorem: general lower bound}, the following  parameters: 
\begin{align}
\ell &= \widetilde m = n^{\frac{1}{\alpha}}, \notag\\
r &= \frac{\delta(\alpha-1)}{\alpha} n^{\frac{1}{\alpha}}, \notag\\
\widetilde{z} 
&=  \frac{\sigma\delta(\alpha-1)}{\alpha}n^{\frac{1}{\alpha}}, \notag\\
z &= \lfloor \widetilde{z} \rfloor, \notag
\end{align}
with   $0 < \delta < 1$, and $ 0<\sigma <1$  positive constants determined in the following and such that 
$\widetilde{z} \leq r$.   
After some algebraic manipulations similar to the ones conducted in Section \ref{subsectionMsmal1}
we  obtain
\begin{eqnarray}
\label{eq: gap M<1}
\frac{R^{\rm ub}(n,m,M,\qv, \widetilde m)}{R^{\rm lb}(n,m,M,\qv) } &\leq& \frac{2n^{\frac{1}{\alpha}}}{(1-o(1))^2 \frac{\sigma\delta(\alpha-1)}{\alpha}\left(1 - \frac{1}{\left\lfloor \frac{1}{\frac{\sigma\delta(\alpha-1)}{\alpha}}\right\rfloor}\right) n^{\frac{1}{\alpha}}} \notag\\
&=& \frac{2}{(1-o(1))^2 \frac{\sigma\delta(\alpha-1)}{\alpha}\left(1 - \frac{1}{\left\lfloor \frac{1}{\frac{\sigma\delta(\alpha-1)}{\alpha}}\right\rfloor}\right)},
\end{eqnarray}
which shows the order-optimality of the achievable expected rate. 
Eq. \eqref{eq: gap M<1} proves that in this regime ($0 \leq M < 1$ small enough), caching cannot provide large gain, or the gain of caching is at most additive so that $M$ cannot affect the order of the expected rate.  

\subsection{Region of $1 \leq M < \frac{m^{\alpha}}{n}$}

In this regime of $M$, we have three cases to consider, which are $1 \leq M = o\left(n^{\frac{1}{\alpha-1}}\right)$, $M = \Theta\left(n^{\frac{1}{\alpha-1}}\right)$ and $\omega\left(n^{\frac{1}{\alpha-1}}\right) = M < \frac{m^{\alpha}}{n}$ (see Fig. \ref{fig: regimes_M_2}).

\subsubsection{When $1 \leq M = o\left(n^{\frac{1}{\alpha-1}}\right)$}

 This case further splits in two scenarios: $M = o(m)$ and $M = \Theta(m)$ (see Fig. \ref{fig: regimes_M_2}).

\begin{itemize}

\item If $M = o(m)$, letting $\widetilde m = M^{\frac{1}{\alpha}}n^{\frac{1}{\alpha}}$,  using (\ref{eq: general achievable 1})  {and Lemma \ref{lemma: H} }, after some algebraic manipulations, we obtain
\begin{eqnarray}
\label{eq: achievable M>1 2}
R^{\rm ub}(n,m,M,\qv, \widetilde m)
&\leq& \left(\frac{{\widetilde m}} {M}-1\right)\left( 1 -\left(1-\frac{M}{\widetilde m} \right)^{ n \, G_{\widetilde m}} \right) +  (1-G_{\widetilde m})\,n \notag\\
&\leq& \frac{2M^{\frac{1}{\alpha}}n^{\frac{1}{\alpha}}}{M} + o\left(\frac{M^{\frac{1}{\alpha}}n^{\frac{1}{\alpha}}}{M} \right).
\end{eqnarray}

Next, we prove the order-optimality of the expected rate achieved by the proposed scheme. Following the similar steps as before, we use   Theorem \ref{theorem: general lower bound} to compute the converse. The parameters required in Theorem \ref{theorem: general lower bound} are summarized in the following.
\begin{align}
\ell &= \widetilde m =M^{\frac{1}{\alpha}}n^{\frac{1}{\alpha}}, \label{max1}\\
r &= \frac{\delta(\alpha-1)}{\alpha}M^{\frac{1-\alpha}{\alpha}}n^{\frac{1}{\alpha}}, \label{max2}\\
\widetilde z &=  \frac{\sigma\delta(\alpha-1)}{2\alpha}M^{\frac{1-\alpha}{\alpha}}n^{\frac{1}{\alpha}}, \label{max3}\\
z &= \lfloor \widetilde z \rfloor , \label{max4}
\end{align}
with   $0 < \delta < 1$, and $ 0<\sigma <1$  positive constants determined in the following.
Note that by definition 
$\widetilde{z} < r < \widetilde m$. Next  we  compute each term in (\ref{eq: general lower bound}) individually.
To this end, using  \eqref{max1} and
\eqref{max2},    we first find an expression  for 
$n\ell q_\ell$ and $\ell\left(1-\left(1-\frac{1}{\ell}\right)^r\right)$ in terms of  $\delta, \sigma,m,n$ and $\alpha$. 
Specifically, using   \eqref{max1}  and Lemma \ref{lemma: H}, we have
\begin{eqnarray}
\label{eq: IU M>1 1}
n\ell q_\ell &=& n \cdot \widetilde{m} \cdot \frac{{ \left(\widetilde{m}\right)}^{-\alpha}}{H(\alpha, 1, m)} \notag\\
&\geq& \frac{n\widetilde{m}^{1-\alpha}}{\frac{1}{1-\alpha}m^{1-\alpha} - \frac{1}{1-\alpha} + 1} \notag\\
&=& \frac{n\left(M^{\frac{1}{\alpha}}n^{\frac{1}{\alpha}}\right)^{1-\alpha}m^{\alpha-1}}{\frac{1}{1-\alpha} - \frac{1}{1-\alpha}m^{\alpha - 1} + m^{\alpha - 1}} \notag\\
&=& \frac{nM^{\frac{1-\alpha}{\alpha}}n^{\frac{1-\alpha}{\alpha}}m^{\alpha-1}}{\frac{\alpha}{\alpha-1}m^{\alpha - 1} - \frac{1}{\alpha - 1}} \notag\\
&=& \frac{\alpha-1}{\alpha}M^{\frac{1-\alpha}{\alpha}}n^{\frac{1}{\alpha}} + o\left(M^{\frac{1-\alpha}{\alpha}}n^{\frac{1}{\alpha}}\right),
\end{eqnarray}
and  
\begin{eqnarray}
\label{eq: IU M>1 2}
n\ell q_\ell &=& n \cdot \widetilde{m} \cdot \frac{{ \left(\widetilde{m}\right)}^{-\alpha}}{H(\alpha, 1, m)} \notag\\
&\leq& \frac{n\widetilde{m}^{1-\alpha}}{\frac{1}{1-\alpha}(m+1)^{1-\alpha} - \frac{1}{1-\alpha}} \notag\\
&=& \frac{\left(M^{\frac{1}{\alpha}}n^{\frac{1}{\alpha}}\right)^{1-\alpha}nm^{\alpha-1}}{\frac{1}{\alpha-1} m^{\alpha - 1} - \frac{1}{\alpha-1} \left(\frac{m}{m+1}\right)^{\alpha-1}} \notag\\
&=&  (\alpha-1)M^{\frac{1-\alpha}{\alpha}}n^{\frac{1}{\alpha}} + o\left(M^{\frac{1-\alpha}{\alpha}}n^{\frac{1}{\alpha}}\right),
\end{eqnarray}
from which 
\begin{eqnarray}
\label{eq: IU M>1 3}
&\frac{\alpha-1}{\alpha}M^{\frac{1-\alpha}{\alpha}}n^{\frac{1}{\alpha}} + o\left(M^{\frac{1-\alpha}{\alpha}}n^{\frac{1}{\alpha}}\right) \leq n\ell q_\ell \leq (\alpha-1)M^{\frac{1-\alpha}{\alpha}}n^{\frac{1}{\alpha}} + o\left(M^{\frac{1-\alpha}{\alpha}}n^{\frac{1}{\alpha}}\right). \notag \\
\end{eqnarray}

Next using \eqref{max1} and \eqref{max2} , we have
\begin{eqnarray}
\label{eq: Zb M>1 1}
\ell\left(1-\left(1-\frac{1}{\ell}\right)^r\right) 
&=& \widetilde m \left(1 - \left(1 - \frac{1}{\widetilde m}\right)^{\frac{\delta(\alpha-1)}{\alpha}M^{\frac{1-\alpha}{\alpha}}n^{\frac{1}{\alpha}}} \right) \notag\\
&= &  M^{\frac{1}{\alpha}}n^{\frac{1}{\alpha}} \left(1 - \exp\left(-\frac{\delta (\alpha-1)}{\alpha M} \right)\right) + o\left(  M^{\frac{1}{\alpha}}n^{\frac{1}{\alpha}} \left(1 - \exp\left(-\frac{\delta (\alpha-1)}{\alpha M} \right)\right)  \right).   \notag \\
%\frac{\delta(\alpha-1)}{\alpha}M^{\frac{1-\alpha}{\alpha}}n^{\frac{1}{\alpha}} + o(M^{\frac{1-\alpha}{\alpha}}n^{\frac{1}{\alpha}}).
\end{eqnarray}  

Then, by using (\ref{eq: P1}),  (\ref{max2}), and  (\ref{eq: IU M>1 3}), we obtain
\begin{eqnarray}
\label{eq: P IU M>1 1}
P_1(\ell,r) &=& 1- \exp\left(- \frac{\left(n\ell q_\ell - \frac{\delta(\alpha-1)}{\alpha}M^{\frac{1-\alpha}{\alpha}}n^{\frac{1}{\alpha}}\right)^2}{2n\ell q_\ell}\right) \notag\\
&\geq& 1 -\exp \notag\\
&&\left(- \frac{\left(\frac{\alpha-1}{\alpha}M^{\frac{1-\alpha}{\alpha}}n^{\frac{1}{\alpha}} + o\left(M^{\frac{1-\alpha}{\alpha}}n^{\frac{1}{\alpha}}\right) - \frac{\delta(\alpha-1)}{\alpha}M^{\frac{1-\alpha}{\alpha}}n^{\frac{1}{\alpha}}\right)^2}{2\left((\alpha-1)M^{\frac{1-\alpha}{\alpha}}n^{\frac{1}{\alpha}} + o\left(M^{\frac{1-\alpha}{\alpha}}n^{\frac{1}{\alpha}}\right)\right)}\right) \notag\\
&=& 1 - o(1).
\end{eqnarray}
while using (\ref{eq: P2}), (\ref{max3})
%\begin{eqnarray}
%\widetilde{z} &=& \min\left\{\sigma_2\left(\frac{\delta_2(\alpha-1)}{\alpha}M^{\frac{1-\alpha}{\alpha}}n^{\frac{1}{\alpha}} \right), \frac{\delta_2(\alpha-1)}{\alpha}M^{\frac{1-\alpha}{\alpha}}n^{\frac{1}{\alpha}} \right\} \notag\\
%&=&  \frac{\sigma_2\delta_2(\alpha-1)}{\alpha}M^{\frac{1-\alpha}{\alpha}}n^{\frac{1}{\alpha}},
%\end{eqnarray}
and (\ref{eq: Zb M>1 1}), we have
\begin{eqnarray}
\label{eq: P Zb M>1 1}
&&P_2(\ell,r,\widetilde z) \notag \\
&& = 1- \exp\left(- \frac{\left(\ell\left(1-\left(1-\frac{1}{\ell}\right)^r\right) - \widetilde{z}\right)^2}{2\ell\left(1-\left(1-\frac{1}{\ell}\right)^r\right)}\right) \notag\\
%&& = 1 - \exp\left( \frac{\left((1-\sigma)\left(\frac{\delta(\alpha-1)}{\alpha}M^{\frac{1-\alpha}{\alpha}}n^{\frac{1}{\alpha}} + o(M^{\frac{1-\alpha}{\alpha}}n^{\frac{1}{\alpha}})\right) \right)^2}{2\left( \frac{\delta\alpha-1)}{\alpha}M^{\frac{1-\alpha}{\alpha}}n^{\frac{1}{\alpha}} + o(M^{\frac{1-\alpha}{\alpha}}n^{\frac{1}{\alpha}}) \right)} \right). \notag\\
&& \buildrel (a) \over = 1- o(1),
\end{eqnarray} 
where  (a) follows from  
\begin{eqnarray}
\label{eq: special 1}
\frac{\delta(\alpha-1)}{2\alpha}M^{\frac{1-\alpha}{\alpha}}n^{\frac{1}{\alpha}} <
\ell\left(1-\left(1-\frac{1}{\ell}\right)^r\right)   < 
 \frac{\delta(\alpha-1)}{\alpha}M^{\frac{1-\alpha}{\alpha}}n^{\frac{1}{\alpha}}
\end{eqnarray}
with \eqref{eq: special 1} derived from   \eqref{eq: Zb M>1 1}  using $1 - \exp(-x) \geq x - \frac{x^2}{2}$ for $x>0$, and $1 - \exp(-x) \leq x$ for $x>0$.

Finally,  replacing 
Eqs. \eqref{max1}-\eqref{max4},  \eqref{eq: P IU M>1 1}, and \eqref{eq: P Zb M>1 1} in Theorem \ref{theorem: general lower bound}  and using footnote \ref{fnm:1}, we obtain

\begin{eqnarray}
\label{eq: converse M>1 1}
&&R^{\rm lb}(n,m,M,\qv)  \notag\\
&& \geq P_1(\ell,r) P_2(\ell,r,\widetilde z) \max_{z \in \{1, \cdots,  \lceil \widetilde{z} \rceil\}} (z-zM/{\lfloor \ell/z \rfloor}) \notag\\
&& \geq (1-o(1)) (1-o(1)) \max_{z \in \{1, \cdots, \lceil \widetilde{z} \rceil \}} (z-zM/{\lfloor \ell/z \rfloor}) \notag\\
&& \geq (1-o(1))^2 \widetilde{z} \left(1 - \frac{M}{\left\lfloor \frac{ \widetilde{m} }{\widetilde{z}} \right\rfloor } \right) \notag\\
&& \geq (1-o(1))^2 \frac{\sigma\delta(\alpha-1)}{2\alpha}M^{\frac{1-\alpha}{\alpha}}n^{\frac{1}{\alpha}} \left(1 - \frac{M}{\left\lfloor\frac{M^{\frac{1}{\alpha}}n^{\frac{1}{\alpha}}}{\frac{\sigma\delta(\alpha-1)}{2\alpha}M^{\frac{1-\alpha}{\alpha}}n^{\frac{1}{\alpha}}}\right\rfloor} \right) \notag\\
&& \geq (1-o(1))^2 \frac{\sigma\delta(\alpha-1)}{2\alpha}M^{\frac{1-\alpha}{\alpha}}n^{\frac{1}{\alpha}} \left(1 - \frac{M}{\frac{M^{\frac{1}{\alpha}}n^{\frac{1}{\alpha}}}{\frac{\sigma\delta(\alpha-1)}{2\alpha}M^{\frac{1-\alpha}{\alpha}}n^{\frac{1}{\alpha}}} - 1} \right) \notag\\
%&& = (1-o(1))^2 \frac{\sigma\delta(\alpha-1)}{2\alpha}M^{\frac{1-\alpha}{\alpha}}n^{\frac{1}{\alpha}} \left(1 - \frac{M}{\frac{2\alpha}{\sigma\delta(\alpha-1)}M - 1} \right) \notag\\
%&& = (1-o(1))^2 \frac{\sigma\delta(\alpha-1)}{2\alpha}M^{\frac{1-\alpha}{\alpha}}n^{\frac{1}{\alpha}} \left(1 - \frac{1}{\frac{2\alpha}{\sigma\delta(\alpha-1)} - \frac{1}{M}} \right) \notag\\
&&=  (1-o(1))^2 \frac{\sigma\delta(\alpha-1)}{2\alpha}M^{\frac{1-\alpha}{\alpha}}n^{\frac{1}{\alpha}} \left(1 - \frac{1}{\frac{2\alpha}{\sigma\delta(\alpha-1)} - \frac{1}{M}} \right) \notag\\
&& \geq (1-o(1))^2 \frac{\sigma\delta(\alpha-1)}{2\alpha} \left(1 - \frac{1}{\frac{2\alpha}{\sigma\delta(\alpha-1)} - 1} \right)M^{\frac{1-\alpha}{\alpha}}n^{\frac{1}{\alpha}},
\end{eqnarray}
from which, using  \eqref{eq: achievable M>1 2}, we obtain
\begin{eqnarray}
\frac{R^{\rm ub}(n,m,M, \qv, \widetilde m)}{R^{\rm lb}(n,m,M,\qv) } &\leq& \frac{\frac{2M^{\frac{1}{\alpha}}n^{\frac{1}{\alpha}}}{M} + o\left(\frac{M^{\frac{1}{\alpha}}n^{\frac{1}{\alpha}}}{M} \right)}{ (1-o(1))^2 \frac{\sigma\delta(\alpha-1)}{2\alpha} \left(1 - \frac{1}{\frac{2\alpha}{\sigma\delta(\alpha-1)} - 1} \right)M^{\frac{1-\alpha}{\alpha}}n^{\frac{1}{\alpha}}} \notag\\
&\leq& \frac{2+ o\left(1\right)}{ (1-o(1))^2 \frac{\sigma\delta(\alpha-1)}{2\alpha} \left(1 - \frac{1}{\frac{2\alpha}{\sigma\delta(\alpha-1)} - 1} \right)},
\end{eqnarray}
{ where $\sigma, \delta \in (0,1)$ can be chosen accordingly (possibly a function of $\alpha$) such that $\frac{2}{\frac{\sigma\delta(\alpha-1)}{2\alpha} \left(1 - \frac{1}{\frac{2\alpha}{\sigma\delta(\alpha-1)} - 1} \right)}$ is a positive constant, which shows the order-optimality of the expected rate.
}

\item If $M = \Theta(m) = \kappa_1 m + o(m)$, where $0 < \kappa_1 < 1$ is a given constant such that $1 \leq M < \frac{m^\alpha}{n}$ and $1 \leq M = o(n^{\frac{1}{\alpha-1}})$, 
using (\ref{eq: general achievable 1}) and letting $\widetilde{m} = m$, we  obtain
\begin{eqnarray}
\label{eq: achievable M>1 2 1}
R^{\rm ub}(n,m,M,\qv, \widetilde m)
&\leq& \left(\frac{{\widetilde m}} {M}-1\right)\left( 1 -\left(1-\frac{M}{\widetilde m} \right)^{ n \, G_{\widetilde m}} \right) +  (1-G_{\widetilde m})\,n \notag\\
&\leq& \frac{m}{M} -1 + o\left(\frac{m}{M} \right).
\end{eqnarray}

Following the similar steps as before, we use  Theorem \ref{theorem: general lower bound} to compute the converse. 
The parameters appeared in Theorem \ref{theorem: general lower bound} are summarized in the following.
\begin{align}
\ell &= \widetilde m=m, \label{leo1}\\
r &=  \frac{\delta(\alpha-1)}{\alpha}\frac{n}{m^{\alpha-1}}, \label{leo2}\\
\widetilde z &=  \frac{\sigma\delta(\alpha-1)}{\alpha}\frac{n}{m^{\alpha-1}} \label{leo3}\\ 
z &= \max\left\{\left\lfloor \frac{m}{2M} \right\rfloor, 1\right\}, \label{leo4}
\end{align}
with   $0 < \delta < 1$, and $ 0<\sigma <1$  positive constants determined in the following.
Note that by definition 
$\widetilde{z} < r $.
As before, we  compute each term in (\ref{eq: general lower bound}) individually.
To this end, using  \eqref{leo1} and
\eqref{leo2},    we first find an expression  for 
$n\ell q_\ell$ and $\ell\left(1-\left(1-\frac{1}{\ell}\right)^r\right)$ in terms of  $\delta, \sigma,m,n$ and $\alpha$. 
Specifically, using   \eqref{leo1}  and Lemma \ref{lemma: H}, 
{  following similar steps as in \eqref{eq: IU M<1 1} and \eqref{eq: IU M<1 2}}, we obtain
\begin{eqnarray}
\label{eq: IU M>1 21} \frac{(\alpha-1)}{\alpha}\frac{n}{m^{\alpha-1}} + o\left(\frac{n}{m^{\alpha-1}}\right) \leq n\ell q_\ell 
\leq  (\alpha-1)\frac{n}{m^{\alpha-1}} + o\left(\frac{n}{m^{\alpha-1}}\right),
\end{eqnarray}
from which, using (\ref{eq: P1}) and \eqref{leo2},  we obtain
\begin{eqnarray}
\label{eq: P IU 2}
P_1(\ell,r,z) &=& 1- \exp\left(- \frac{\left(n\ell q_\ell - \frac{\delta(\alpha-1)}{\alpha}\frac{n}{m^{\alpha-1}}\right)^2}{2n\ell q_\ell }\right) \notag\\
&\geq& 1- \exp\left(- \frac{\left(\frac{\alpha-1}{\alpha}\frac{n}{m^{\alpha-1}} + o\left(\frac{n}{m^{\alpha-1}}\right) - \frac{\delta(\alpha-1)}{\alpha}\rho_2\frac{n}{m^{\alpha-1}}\right)^2}{2\left(\alpha-1)\frac{n}{m^{\alpha-1}} + o\left(\frac{n}{m^{\alpha-1}}\right)\right)}\right) \notag\\
&=& 1 - o(1).
\end{eqnarray}
On the other hand, using  \eqref{max1}  and  \eqref{max2}, via Taylor expansion,
we obtain
\begin{eqnarray}
\label{eq: Zb M>1 2}
\ell\left(1-\left(1-\frac{1}{\ell}\right)^r\right)  &=& m \left(1 - \left(1 - \frac{1}{m}\right)^{\frac{\delta(\alpha-1)}{\alpha}\frac{n}{m^{\alpha-1}}} \right) \notag\\
&=& \frac{\delta(\alpha-1)}{\alpha}\frac{n}{m^{\alpha-1}} + o\left(\frac{n}{m^{\alpha-1}}\right),
\end{eqnarray}  
from which,  using  (\ref{eq: P2}) and   \eqref{leo3},  after some algebra, we have
\begin{eqnarray}
\label{eq: P Zb M>1 2}
&&P_2(\ell,r,\widetilde z) \notag \\
&& = 1- \exp\left(- \frac{\left(\ell\left(1-\left(1-\frac{1}{\ell}\right)^r\right) - \widetilde{z}\right)^2}{2\ell\left(1-\left(1-\frac{1}{\ell}\right)^r\right)}\right) \notag\\
&& \buildrel (a) \over = 1- o(1),
\end{eqnarray}
 where (a)  follows from the fact that  
$$
\frac{\left(\ell\left(1-\left(1-\frac{1}{\ell}\right)^r\right) - \widetilde{z}\right)^2}{2\ell\left(1-\left(1-\frac{1}{\ell}\right)^r\right)}= \Theta \left (\frac{n}{m^{\alpha-1}} \right ).
$$

In the following, we distinguish between two cases: $\frac{m}{M} = \frac{1}{\kappa_1} >  3$ and $\frac{m}{M} = \frac{1}{\kappa_1} \leq 3$ (see Fig. \ref{fig: regimes_M_2}).
\begin{itemize}
\item If $\frac{m}{M} = \frac{1}{\kappa_1} >  3$,  from   \eqref{leo4}, we have: 
\begin{eqnarray}
\label{leo5}
z &=&  \left\lfloor \frac{m}{2M} \right\rfloor,
\end{eqnarray}
from which replacing  Eqs.  \eqref{leo1}-\eqref{leo3}, \eqref{eq: P IU 2},  \eqref {eq: P Zb M>1 2}  and  \eqref{leo5} in  \eqref{eq: general lower bound}, we have
\begin{eqnarray}
\label{eq: converse M>1 2}
&&R^{\rm lb}(n,m,M,\qv)  \notag\\
&& \geq P_1(\ell,r) P_2(\ell,r,\widetilde z) \max_{z \in \{1, \cdots,  \lceil \widetilde{z} \rceil \}} (z-zM/{\lfloor \ell/z \rfloor}) \notag\\
&& \geq (1-o(1)) (1-o(1)) \max_{z \in \{1, \cdots, \lceil \widetilde{z} \rceil\}} (z-zM/{\lfloor \ell/z \rfloor}) \notag\\
&& \geq (1-o(1))^2 z \left(1 - \frac{M}{\left\lfloor \frac{ {m} }{z} \right\rfloor } \right) \notag\\
&& \geq (1-o(1))^2 \left\lfloor \frac{m}{2M} \right\rfloor \left(1 - \frac{M}{\left\lfloor\frac{m}{\left\lfloor \frac{m}{2M} \right\rfloor}\right\rfloor} \right) \notag\\
&& \geq (1-o(1))^2 \left\lfloor \frac{m}{2M} \right\rfloor \left(1 - \frac{M}{\frac{m}{ \frac{m}{2M}} - 1} \right) \notag\\
&& = (1-o(1))^2 \left\lfloor \frac{m}{2M} \right\rfloor \left(1 - \frac{1}{2-\frac{1}{M}}\right) \notag\\
&& \geq (1-o(1))^3 \frac{1}{2}\left( \frac{m}{2M} - 1 \right).
\end{eqnarray}
Thus, we obtain
\begin{eqnarray}
\frac{R^{\rm ub}(n,m,M,\qv, \widetilde m)}{R^{\rm lb}(n,m,M,\qv) } &\leq& \frac{\frac{m}{M}}{(1-o(1))^4 \frac{1}{2}\left( \frac{m}{2M} - 1 \right)} \notag\\
&=& \frac{2}{\frac{1}{2} - \frac{M}{m}} \notag\\
&\leq& \frac{2}{\frac{1}{2} - \frac{1}{3}} = 12,
\end{eqnarray}
proving the order-optimality of the  achievable expected rate.

\item If $\frac{m}{M} = \frac{1}{\kappa_1} \leq 3$, 
from   \eqref{leo4}, we have: 
\begin{eqnarray}
\label{leo6}
z =  1, 
\end{eqnarray}
from which replacing  Eqs.  \eqref{leo1}-\eqref{leo3}, \eqref{eq: P IU 2},  \eqref {eq: P Zb M>1 2}  and  \eqref{leo6} in  \eqref{eq: general lower bound}, we have
\begin{eqnarray}
\label{eq: converse M>1 2 1}
&&R^{\rm lb}(n,m,M,\qv)  \notag\\
&& \geq  P_1(\ell,r) P_2(\ell,r,\widetilde z) \max_{z \in \{1, \cdots,  \lceil \widetilde{z} \rceil \}} (z-zM/{\lfloor \ell/z \rfloor}) \notag\\
%&& \geq (1-o(1)) (1-o(1)) \max_{z \in \{1, \cdots, \lceil \widetilde{z} \rceil\}} (z-zM/{\lfloor \ell/z \rfloor}) \notag\\
%&& \geq (1-o(1))^2 \widetilde{z} \left(1 - \frac{M}{\left\lfloor \frac{ \widetilde{m} }{\widetilde{z}} \right\rfloor } \right) \notag\\
%&& \geq (1-o(1))^2 \left(1 - \frac{M}{\left\lfloor\frac{m}{1}\right\rfloor} \right) \notag\\
&& \geq (1-o(1))^2 \left(1 - \frac{M}{m}\right).
\end{eqnarray}
Thus, we obtain
\begin{eqnarray}
\frac{R^{\rm ub}(n,m,M,\qv, \widetilde m)}{R^{\rm lb}(n,m,M,\qv) } &\leq& \frac{\frac{m}{M}-1}{(1-o(1))^3  \left(1 - \frac{M}{m}\right)} \notag\\
&=& \frac{\frac{m}{M}\left(1-\frac{M}{m}\right)}{(1-o(1))^3  \left(1 - \frac{M}{m}\right)} \notag\\
&=& \frac{m}{M} \leq 3.
\end{eqnarray}
\end{itemize}

\end{itemize}

%{\BLUE I SUGGEST TO DELATE: JAIME WHAT DO YOU THINK??? {\BLUE Maybe we can put it as a footnote to remind the user that we do not forget this case, MJ. ANTONIA SAYS: "WE MENTION THIS AT THE BEGINNING OF THE SECTION"}
%\subsubsection{When $M = \Theta(n^{\frac{1}{\alpha-1}}) = \kappa n^{\frac{1}{\alpha-1}} + o(n^{\frac{1}{\alpha-1}})< \frac{m^{\alpha}}{n}$ for some positive constant $\kappa$} due to the complexity, this case is proved in Appendix \ref{sec: proof of table II}.  }

\subsubsection{When $\omega(n^{\frac{1}{\alpha-1}}) = M < \frac{m^{\alpha}}{n}$ (see Fig. \ref{fig: regimes_M_2} for a reminder)} 
In this case, letting $\widetilde{m} = M$,  we have  
\begin{eqnarray}
\label{eq: achievable M>1 90}
R^{\rm ub}(n,m,M,\qv, \widetilde m) &\leq& (1-G_{\widetilde m})\,n.
%&\leq& nM^{1-\alpha} + o\left(nM^{1-\alpha}\right). 
\end{eqnarray}
Next we  consider two regimes: $M = o(m)$ and $M = \Theta(m)$.

\begin{itemize}

\item If $M = o(m)$, using (\ref{eq: achievable M>1 90}) and Lemma \ref{lemma: H}, we have
\begin{eqnarray}
\label{eq: achievable M>1 9}
R^{\rm ub}(n,m,M,\qv, \widetilde m) 
&\leq& nM^{1-\alpha} + o\left(nM^{1-\alpha}\right). 
\end{eqnarray}

To compute the converse, in this case  we use the second term of (\ref{eq: general lower bound}), with the parameters given by:
\begin{align}
\ell &= cM, \label{messi1}\\
r &= \frac{\delta(\alpha-1)c^{1-\alpha}}{\alpha}nM^{1-\alpha}, \label{messi2}\\
\widetilde z &=   \sigma, \label{messi3} %\notag\\
%z &= \widetilde z, \notag
\end{align}
and  $ c >1 $, $0 < \delta < 1$, and $ 0<\sigma <1$  positive constants determined in the following. Next 
we compute each term in (\ref{eq: general lower bound}) individually. To this end, using  \eqref{messi1} and
\eqref{messi2}, and recalling that  $\widetilde{m}=M$  we first find an expression  for 
$n\ell q_\ell$ and $\ell\left(1-\left(1-\frac{1}{\ell}\right)^r\right)$ in terms of  $\delta, \sigma,m,n$ and $\alpha$. 
Specifically, using   \eqref{max1}  and Lemma \ref{lemma: H}, we have
%Let $\ell=cM$, where $c > 1$ is a constant,  by using %(\ref{eq: EIU}) and  Lemma \ref{lemma: H}, following the same steps as (\ref{eq: IU M<1 3}), then we can write
\begin{eqnarray}
\label{eq: EIU M>1 9 1}
n\ell q_\ell &=& n \cdot cM \cdot \frac{{ \left(cM\right)}^{-\alpha}}{H(\alpha, 1, m)} \notag\\
&\geq& \frac{n( c M)^{1-\alpha}}{\frac{1}{1-\alpha}m^{1-\alpha} - \frac{1}{1-\alpha} + 1} \notag\\
%&=& \frac{c^{1-\alpha}nM^{1-\alpha}}{\frac{\alpha}{\alpha-1} - \frac{1}{\alpha-1}\frac{1}{m^{\alpha-1}}} \notag\\
&=& \frac{(\alpha-1)c^{1-\alpha}}{\alpha}nM^{1-\alpha} + o(nM^{1-\alpha}),
\end{eqnarray}
and
\begin{eqnarray}
\label{eq: EIU M>1 9 2}
n\ell q_\ell &=& n \cdot c M \cdot \frac{{ \left(cM\right)}^{-\alpha}}{H(\alpha, 1, m)} \notag\\
&\leq& \frac{n ( c M)^{1-\alpha}}{\frac{1}{1-\alpha}(m+1)^{1-\alpha} - \frac{1}{1-\alpha}} \notag\\
%&=& \frac{c^{1-\alpha}nM^{1-\alpha}}{\frac{1}{\alpha-1} - \frac{1}{\alpha-1} \frac{1}{\left(m+1\right)^{\alpha-1}}} \notag\\
&=&  (\alpha-1)c^{1-\alpha}  nM^{1-\alpha} + o\left(nM^{1-\alpha}\right),
\end{eqnarray}
from which  
%Thus, by using (\ref{eq: EIU M>1 9 1}) and (\ref{eq: EIU M>1 9 2}), we have
\be
\label{eq: EIU M>1 9}
\frac{(\alpha-1)c^{1-\alpha} }{\alpha}nM^{1-\alpha} + o(nM^{1-\alpha}) \leq n\ell q_\ell
\leq (\alpha-1)c^{1-\alpha} nM^{1-\alpha} + o\left(nM^{1-\alpha}\right).
\ee
%
%We let $r = 1$, %by using (\ref{eq: Zb}), 
%then we can write
%\begin{eqnarray}
%\label{eq: EZb M>1 9}
%&& \ell\left(1-\left(1-\frac{1}{\ell}\right)^r\right) \notag\\ 
%&& = m \left(1 - \left(1 - \frac{1}{m}\right)^1 \right) \notag\\
%&& = 1.
%\end{eqnarray}  

Using (\ref{eq: P1}),  \eqref{messi2}, and  \eqref{eq: EIU M>1 9},
%Then, let $r = \frac{\delta_9(\alpha-1)c^{1-\alpha}}{\alpha}nM^{1-\alpha}$, where $0 < \delta_9  < 1$, then we observe $r<1$. By using (\ref{eq: P1}), 
we obtain %and (\ref{eq: EIU M>1 9}), by using the definition of $\Nsf$ given in Appendix \ref{sec: proof of theorem: general lower bound}, we obtain
\begin{eqnarray}
\label{eq: IU 9}
%&& \PP\left(\Nsf \geq 1\right) \notag\\
%&& \buildrel (a) \over = \PP\left(\Nsf \geq \widetilde r\right) \geq 
&& P_1(\ell, r) \notag\\
&& = 1- \exp\left(- \frac{\left(n\ell q_\ell - \frac{\delta(\alpha-1)c^{1-\alpha}}{\alpha}nM^{1-\alpha}\right)^2}{2n\ell q_\ell }\right) \notag\\
&& \geq 1- \exp\left(- \frac{\left(\frac{(\alpha-1)c^{1-\alpha}}{\alpha}nM^{1-\alpha} + o\left(nM^{1-\alpha}\right) - \frac{\delta(\alpha-1)c^{1-\alpha}}{\alpha}nM^{1-\alpha}\right)^2}{2\left((\alpha-1)c^{1-\alpha} nM^{1-\alpha} + o\left(nM^{1-\alpha}\right)\right)}\right) \notag\\
&&\geq \frac{(1-\delta)^2(\alpha - 1)c^{1-\alpha}}{2\alpha^2}nM^{1-\alpha} + o(nM^{1-\alpha}),
\end{eqnarray}
%where (a) is because that $\Nsf$ is an integer such that $\PP\left(\Nsf \geq 1\right) = \PP\left(\Nsf \geq \widetilde r\right)$ for $\widetilde r<1$.
while using (\ref{eq: P2})  and   \eqref{messi3}
%and (\ref{eq: EZb M>1 9}) and the definition of $\Zsf$ given in Appendix \ref{sec: proof of theorem: general lower bound}, 
we have
\begin{eqnarray}
\label{eq: Zb 9}
 %\PP(\Zsf \geq \widetilde{z}) &=& \PP(\Zsf \geq \sigma_9) \notag \\
%&\geq& 
P_2(\ell,1,\widetilde z)% \notag\\ 
%&\geq& 1- \exp\left(- \frac{\left(\ell\left(1-\left(1-\frac{1}{\ell}\right)^r\right) - \sigma_9\right)^2}{2\EE[\Zsf]}\right) \notag\\
&=& 1 - \exp\left( \frac{\left(1-\sigma \right)^2}{2} \right). 
\end{eqnarray} 

Then,  replacing  \eqref{messi1}-\eqref{messi3},   \eqref{eq: IU 9} , and   \eqref{eq: Zb 9} in  the second term of \eqref{eq: general lower bound}, 
 %let $z = 1$ and given all parameters $\ell, r, \widetilde{z}$, by using (\ref{eq: R lower ground 1}), 
we obtain
\begin{eqnarray}
\label{eq: converse M>1 9}
&&R^{\rm lb}(n,m,M,\qv)  \notag\\
&& \geq  P_1(\ell,r) P_2(\ell,1,\widetilde z) (1-M/{\ell}) \notag\\
&& \geq (1-o(1))\left(\frac{(1-\delta)^2(\alpha-1)c^{1-\alpha}}{2\alpha^2}nM^{1-\alpha} + o(nM^{1-\alpha})\right) \left(1 - \exp\left( \frac{\left(1-\sigma \right)^2}{2} \right)\right)\left(1-\frac{M}{cM}\right) \notag\\
%&& \geq (1-o(1))\left(\frac{(1-\delta)^2c^{1-\alpha}}{2\alpha}nM^{1-\alpha} + o(nM^{1-\alpha})\right) \left(1 - \exp\left( \frac{\left(1-\sigma \right)^2}{2} \right)\right)\left(1 - \frac{1}{c} \right) \notag\\
&& \geq (1-o(1))\left(1 - \exp\left( \frac{\left(1-\sigma \right)^2}{2} \right)\right)\left(1 - \frac{1}{c} \right)\frac{(1-\delta)^2(\alpha-1)c^{1-\alpha}}{2\alpha^2}nM^{1-\alpha} + o(nM^{1-\alpha}), \notag\\
\end{eqnarray}
from which using 
 (\ref{eq: achievable M>1 9}), we obtain
\begin{eqnarray}
\label{eq: gap M>1 9}
\frac{R^{\rm ub}(n,m,M,\qv, \widetilde m)}{R^{\rm lb}(n,m,M,\qv) } &\leq&
 \frac{n M^{1-\alpha} + o (n M^{1-\alpha} )}{(1-o(1))\left(1 - \exp\left( \frac{\left(1-\sigma \right)^2}{2} \right)\right)\left(1 - \frac{1}{c} \right)\frac{(1-\delta)^2(\alpha-1)c^{1-\alpha}}{2\alpha^2}nM^{1-\alpha} + o(nM^{1-\alpha})} \notag\\
&\leq& \frac{1}{\left(1 - \exp\left( \frac{\left(1-\sigma\right)^2}{2} \right)\right)\left(1 - \frac{1}{c} \right)\frac{(1-\delta)^2(\alpha-1)c^{1-\alpha}}{2\alpha^2}} + o(1),
\end{eqnarray}
which shows the order-optimality of the achievable expected rate for RLFU and RAP.

\item If $M = \Theta(m) = \kappa_1 m + o(m)$, with $ 0<  \kappa_1 < 1$,
%(under the case that $\omega(n^{\frac{1}{\alpha-1}}) = M < \frac{m^\alpha}{n}$, see Fig. \ref{fig: regimes_M_2} for a reminder), where $0 < \kappa_1 \leq 1$, let $\widetilde m = M$, by 
using \eqref{eq: achievable M>1 90} and Lemma \ref{lemma: H}, we have
\begin{eqnarray}
\label{eq: achievable M>1 10}
R^{\rm ub}(n,m,M,\qv, \widetilde m) %&\leq& (1-G_{\widetilde m})\,n \notag\\ 
&\leq& \left(\kappa_1^{1-\alpha} - 1\right)nm^{1-\alpha} + o\left(1\right) \notag\\
%&=& \left(\kappa^{1-\alpha} - 1\right)\left(\frac{1}{\kappa_1}\right)^{1-\alpha}nM^{1-\alpha} + o\left(1\right) \notag\\
&=& \left(1 - \kappa_1^{\alpha-1}\right)nM^{1-\alpha} + o(1).
\end{eqnarray}

%{\BLUE MING WHY DO YOU REMIND TO THE READER THAT
%$\omega(n^{\frac{1}{\alpha-1}}) = M < \frac{m^\alpha}{n}$  , I DO NOT SEE WHERE DO YOU USE THIS!!  also for $k_1$ do we need to say that is smaller than 1??}

To compute the converse, similar as before, we use the second term of (\ref{eq: general lower bound}). The values of the parameters
$\ell$, $r$, and $\widetilde z$  in  (\ref{eq: general lower bound}) will be different based on the fact that  
 $\frac{m}{M} = \frac{1}{\kappa_1} > 2$ and $\frac{m}{M} = \frac{1}{\kappa_1} \leq 2$ (see Fig. \ref{fig: regimes_M_2}).

\begin{itemize}

\item When $\frac{m}{M} = \frac{1}{\kappa_1} > 2$, the parameters $\ell$, $r$, and $\widetilde z$  in  (\ref{eq: general lower bound})  are given as in 
\eqref{messi1}-\eqref{messi3} with the additional constraint that  $1 < c < 2$ to guarantee that $\ell \leq m$.
Following the same steps  as  in the case of $M = o(m)$, we obtain  (\ref{eq: converse M>1 9}),  from which, using (\ref{eq: achievable M>1 10}) we obtain:
\begin{eqnarray}
\label{eq: gap M>1 9 1}
\frac{R^{\rm ub}(n,m,M, \qv, \widetilde m)}{R^{\rm lb}(n,m,M,\qv) } 
%&\leq& \frac{\left(1 - \kappa_1^{\alpha-1}\right)n M^{1-\alpha} + o (n M^{1-\alpha} )}{(1-o(1))\left(1 - \exp\left( \frac{\left(1-\sigma \right)^2}{2} \right)\right)\left(1 - \frac{1}{c} \right)\frac{(1-\delta)^2c^{1-\alpha}}{2\alpha}nM^{1-\alpha} + o(nM^{1-\alpha})} \notag\\
&\leq& \frac{1}{\left(1 - \exp\left( \frac{\left(1-\sigma \right)^2}{2} \right)\right)\left(1 - \frac{1}{c} \right)\frac{(1-\delta)^2(\alpha-1)c^{\alpha-1}}{2\alpha}} + o(1),
\end{eqnarray}
which shows the order-optimality of the achievable expected rate for RLFU and RAP.
\item When $\frac{m}{M} = \frac{1}{\kappa_1} \leq 2$, 
the parameters $\ell$, $r$, and $\widetilde z$  in  (\ref{eq: general lower bound})  are given by: 
\begin{align}
\ell &= m, \label{ronaldo1}\\
r &= \frac{\alpha-1}{\alpha }\kappa_1^{\alpha-1}\delta nM^{1-\alpha}, \label{ronaldo2}\\ %1, \notag\\
\widetilde z &=   \sigma, \label{ronaldo3} %\notag\\
%z &= \widetilde z, \notag
\end{align}

with  $0 < \delta < 1$, and $ 0<\sigma <1$  positive constants determined in the following. Next 
we compute each term in (\ref{eq: general lower bound}) individually.  
Specifically, using   \eqref{ronaldo1}  and Lemma \ref{lemma: H},  recalling that $\widetilde m=M$ 
we have
\begin{eqnarray}
\label{eq: EIU M>1 {10} 1}
n\ell q_\ell &=& n \cdot m \cdot \frac{{ \left({m}\right)}^{-\alpha}}{H(\alpha, 1, m)} \notag\\
&\geq& \frac{n {m}^{1-\alpha}}{\frac{1}{1-\alpha}m^{1-\alpha} - \frac{1}{1-\alpha} + 1} \notag\\
&=& \frac{\alpha-1}{\alpha}n \kappa_1^{\alpha-1} M^{1-\alpha} + o(nM^{1-\alpha}),
\end{eqnarray}
and
\begin{eqnarray}
\label{eq: EIU M>1 {10} 2}
n\ell q_\ell &=& n \cdot {m} \cdot \frac{{ \left({m}\right)}^{-\alpha}}{H(\alpha, 1, m)} \notag\\
&\leq& \frac{n {M}^{1-\alpha}}{\frac{1}{1-\alpha}(m+1)^{1-\alpha} - \frac{1}{1-\alpha}} \notag\\
&=& (\alpha-1)n \kappa_1^{\alpha-1}  M^{1-\alpha} + o(nM^{1-\alpha}).
\end{eqnarray}
Thus, by using (\ref{eq: EIU M>1 {10} 1}) and (\ref{eq: EIU M>1 {10} 2}), we obtain
\begin{eqnarray}
\label{eq: EIU M>1 {10} 3}
\frac{\alpha-1}{\alpha}n \kappa_1^{\alpha-1}  M^{1-\alpha} + o(nM^{1-\alpha}) \leq n\ell q_\ell \leq(\alpha-1)n\kappa_1^{\alpha-1}  M^{1-\alpha} + o(nM^{1-\alpha}),
\end{eqnarray}
from which, using \eqref{ronaldo2} we have:
\begin{eqnarray}
\label{eq: P IU M>1 {10}}
&& P_1(\ell,r) \notag\\
%&& \buildrel (a) \over = \PP\left(\Nsf \geq \widetilde r \right) \notag\\
&& = 1- \exp\left(- \frac{\left(n\ell q_\ell - \frac{\alpha-1}{\alpha }\kappa_1^{\alpha-1}  \delta n  M^{1-\alpha}\right)^2}{2n\ell q_\ell}\right) \notag\\
&&\geq 1 - \exp\left(- \frac{(1-\delta)^2(\alpha-1)\kappa_1^{\alpha-1} }{2\alpha^2 } nM^{1-\alpha}\right) \notag\\
&&\geq \frac{(1-\delta)^2(\alpha-1)\kappa_1^{\alpha-1} }{2\alpha^2  } nM^{1-\alpha} - \frac{1}{2}\left(\frac{(1-\delta)^2(\alpha-1)\kappa_1^{\alpha-1} }{2\alpha^2 } nM^{1-\alpha}\right)^2. \notag\\
\end{eqnarray}

%Let $r = 1$, then %by using (\ref{eq: EZb}), 
%we obtain
%\begin{eqnarray}
%\label{eq: EZb M>1 {10}}
%\ell\left(1-\left(1-\frac{1}{\ell}\right)^r\right) = 1.
%\end{eqnarray}  
%Using 
%Then, let $r = \frac{\alpha-1}{\alpha}\delta_{10}\kappa_1^{\alpha-1}nM^{1-\alpha}$, where $0 < \delta_{10} < 1$ is some positive constant determined later, then we can observe that $r<1$. By using (\ref{eq: P1}) and (\ref{eq: EIU M>1 {10} 3}), % and the definition of $\Nsf$ given in Appendix \ref{sec: proof of theorem: general lower bound}, 
%we obtain
%\begin{eqnarray}
%\label{eq: P IU M>1 {10}}
%&& P_1(\ell,r) \notag\\
%%&& \buildrel (a) \over = \PP\left(\Nsf \geq \widetilde r \right) \notag\\
%&& = 1- \exp\left(- \frac{\left(n\ell q_\ell - \frac{\alpha-1}{\alpha}\delta_{10}\kappa_1^{\alpha-1}nM^{1-\alpha}\right)^2}{2n\ell q_\ell}\right) \notag\\
%&&\geq 1 - \exp\left(- \frac{(1-\delta_{10})^2(\alpha-1)}{2\alpha^2}\kappa_1^{\alpha-1}nM^{1-\alpha}\right) \notag\\
%&&\geq \frac{(1-\delta_{10})^2(\alpha-1)}{2\alpha^2}\kappa_1^{\alpha-1}nM^{1-\alpha} - \frac{1}{2}\left(\frac{(1-\delta_{10})^2(\alpha-1)}{2\alpha^2}\kappa_1^{\alpha-1}nM^{1-\alpha}\right)^2, \notag\\
%\end{eqnarray}
%where (a) is because that $\Nsf$ is an integer such that $P_1(\ell,r) = \PP\left(\Nsf \geq \widetilde r\right)$ for $\widetilde r<1$.
Furthermore using  using (\ref{eq: P2})  and \eqref{ronaldo3}, %and (\ref{eq: EZb M>1 {10}}), 
we have
\begin{eqnarray}
\label{eq: P Zb M>1 {10}}
%\PP(\Zsf \geq \widetilde{z}) &=& \PP(\Zsf \geq \sigma_{10}) \notag \\
P_2(\ell, 1, \widetilde z) %&=& %1- \exp\left(- \frac{\left(\ell\left(1-\left(1-\frac{1}{\ell}\right)^r\right) - \sigma_{10}\right)^2}{2\EE[\Zsf]}\right) \notag\\
&=& 1 - \exp\left( \frac{\left(1-\sigma\right)^2}{2} \right),
\end{eqnarray} 
from which replacing   \eqref{ronaldo1}-\eqref{ronaldo3}, \eqref {eq: P IU M>1 {10}} and \eqref{eq: P Zb M>1 {10}} in the second term of 
 (\ref{eq: general lower bound}) 
%Then, %let $z=1$ and 
%given all parameters $\ell, r,\widetilde{z}$, by using Theorem \ref{theorem: general lower bound}, 
we obtain
\begin{eqnarray}
\label{eq: converse M>1 {10}}
&&R^{\rm lb}(n,m,M,\qv)  \notag\\
&& \geq  P_1(\ell,r) P_2(\ell, 1, \widetilde z) (1-M/{\ell}) \notag\\
&& \geq \left(\frac{(1-\delta)^2(\alpha-1)\kappa_1^{\alpha-1} }{2\alpha^2 } nM^{1-\alpha} - \frac{1}{2}\left(\frac{(1-\delta)^2(\alpha-1)\kappa_1^{\alpha-1} }{2\alpha^2 }nM^{1-\alpha}\right)^2\right) \notag\\
&& \quad  \left(1 - \exp\left( \frac{\left(1-\sigma \right)^2}{2} \right)\right) \left(1-\frac{M}{m}\right).
\end{eqnarray}

Thus, by using (\ref{eq: achievable M>1 10}) and (\ref{eq: converse M>1 {10}}), we obtain
\begin{eqnarray}
\label{eq: gap M>1 {10}}
&& \frac{R^{\rm ub}(n,m,M, \qv, \widetilde m)}{R^{\rm lb}(n,m,M,\qv) } \leq \notag\\
&& \frac{\left(1 - \kappa_1^{\alpha-1}\right)nM^{1-\alpha} + o(1)}{\left(\frac{(1-\delta)^2(\alpha-1)\kappa_1^{\alpha-1} }{2\alpha^2 }nM^{1-\alpha} - \frac{1}{2}\left(\frac{(1-\delta)^2(\alpha-1)\kappa_1^{\alpha-1} }{2\alpha^2 }nM^{1-\alpha}\right)^2\right) \left(1 - \exp\left( \frac{\left(1-\sigma \right)^2}{2} \right)\right)\left(1-\frac{M}{m}\right)} \notag\\
&&\leq \frac{1}{\frac{(1-\delta)^2(\alpha-1)\kappa_1^{\alpha-1} }{2\alpha^2 }\left(1 - \exp\left( \frac{\left(1-\sigma \right)^2}{2} \right)\right)} \frac{1 - \left(\frac{M}{m}\right)^{\alpha-1}}{1-\frac{M}{m}} + o(1)\notag\\
&& \leq \frac{1}{\frac{(1-\delta_{10})^2(\alpha-1)}{2\alpha^2}\frac{1}{2^{\alpha-1}} \left(1 - \exp\left( \frac{\left(1-\sigma_{10} \right)^2}{2} \right)\right)} \frac{1 - \left(\frac{M}{m}\right)^{\alpha-1}}{1-\frac{M}{m}} + o(1). 
\end{eqnarray}

Note that 
\begin{subequations}
\begin{empheq}[left={\frac{1 - \frac{M}{m}^{\alpha-1}}{1-\frac{M}{m}} \leq }\empheqlbrace]{align}
&1 ,\,\, \,\,  \,\, \,\,\alpha \leq 2  \label{rotta1} \\
& \alpha-1 , \,\, \,\, \,\, \,\, \,\, \,\,\alpha > 2 \label{rotta2} 
 \end{empheq}
\label{rotta0} 
\end{subequations}
from which 
%When $\alpha \leq 2$, then 
%\be
%\label{eq: gap M>1 {10} 1}
%\frac{1 - \frac{M}{m}^{\alpha-1}}{1-\frac{M}{m}} \leq 1.
%\ee
%When $\alpha > 2$, then
%\be
%\label{eq: gap M>1 {10} 2}
%\frac{1 - \frac{M}{m}^{\alpha-1}}{1-\frac{M}{m}} \leq \alpha-1.
%\ee
using (\ref{eq: gap M>1 {10}}), and (\ref{rotta0}), we obtain
\be
 \frac{R^{\rm ub}(n,m,M, \qv, \widetilde m)}{R^{\rm lb}(n,m,M,\qv) } \leq \frac{\max\{1, \alpha-1\}}{\frac{(1-\delta)^2(\alpha-1)}{2\alpha^2}\frac{1}{2^{\alpha-1}} \left(1 - \exp\left( \frac{\left(1-\sigma \right)^2}{2} \right)\right)},
\ee
which shows the order-optimality of the achievable expected rate for RLFU and RAP.

\end{itemize}

\end{itemize}

\subsection{Region of  $M \geq \frac{m^{\alpha}}{n}$} 
In this case, we want to prove that RLFU-GCC with  $\widetilde m = m$ is order optimal. To this end,  using (\ref{eq: general achievable 1}),  we can write the rate for  RLFU-GCC with  $\widetilde m =m$  as:
\begin{eqnarray}
\label{eq: achievable M>1 11}
R^{\rm ub}(n,m,M,\qv, \widetilde m)
&\leq& \left(\frac{{\widetilde m}} {M}-1\right)\left( 1 -\left(1-\frac{M}{\widetilde m} \right)^{ n \, G_{\widetilde m}} \right) +  (1-G_{\widetilde m})\,n \notag\\
&\leq& \frac{m}{M} -1 + o\left(\frac{m}{M} \right).
\end{eqnarray}

At this point, we distinguish between two cases: $\frac{n}{m^{\alpha-1}} = \omega(1)$ and $\frac{n}{m^{\alpha-1}} = \Theta(1)$ (see Fig. \ref{fig: regimes_M_2}).

\subsubsection{When $\frac{n}{m^{\alpha-1}} = \omega(1)$} 
To compute the converse,  again,  we evaluate the first term of  (\ref{eq: general lower bound})  in  Theorem \ref{theorem: general lower bound},  with  the parameters defined as:
\begin{align}
\ell &= m,  \label{babbo1} \\\
r &=  \frac{\delta(\alpha-1)}{\alpha}\frac{n}{m^{\alpha-1}}, \label{babbo2}\\
\widetilde z &=   \frac{\sigma\delta(\alpha-1)}{\alpha}\frac{n}{m^{\alpha-1}}, \label{babbo3}\\
z &=  \left\{\begin{array}{cc}  \left \lfloor \frac{\sigma \delta(\alpha-1)}{\alpha}\frac{m}{M}  \right \rfloor, & M=o(m) \\
\max\{\left\lfloor \frac{m}{2M} \right\rfloor, 1\}, & M = \Theta(m) \end{array}\right., \label{babbo4}
\end{align}
with $0 < \delta < 1$, and $ 0<\sigma <1$  positive constants determined in the following. Note that by definition 
$\widetilde{z}<r$. Next 
we compute each term in (\ref{eq: general lower bound}) individually. To this end, using  \eqref{babbo1} and
\eqref{babbo2}, and recalling that  $\widetilde{m}=m$  we first find an expression  for 
$n\ell q_\ell$ and $\ell\left(1-\left(1-\frac{1}{\ell}\right)^r\right)$ in terms of  $\delta, \sigma,m,n$ and $\alpha$. 
Specifically, using   \eqref{babbo1}  and Lemma \ref{lemma: H}, 
{following similar steps as in \eqref{eq: alpha<1 EIU 1} and \eqref{eq: alpha<1 EIU 2}}, we have
\be
\label{eq: EIU M>1 11}
\frac{\alpha-1}{\alpha}\frac{n}{m^{\alpha-1}} + o\left( \frac{n}{m^{\alpha-1}} \right) \leq n\ell q_\ell  \leq (\alpha-1)\frac{n}{m^{\alpha-1}} + o\left(\frac{n}{m^{\alpha-1}}\right),
\ee
from which, replacing   \eqref{babbo2} and  \eqref {eq: EIU M>1 11}  in (\ref{eq: P1}),    we have:
\begin{eqnarray}
\label{eq: P IU M>1 11}
P_1(\ell,r) 
%&=& 
%1- \exp\left(- \frac{\left(n\ell q_\ell - \frac{\delta_{11}(\alpha-1)}{\alpha}\frac{n}{m^{\alpha-1}}\right)^2}{2n\ell q_\ell}\right) \notag\\
&\geq& 1- \exp\left(- \frac{\left(\frac{(\alpha-1)}{\alpha}\frac{n}{m^{\alpha-1}} - \frac{\delta(\alpha-1)}{\alpha}\frac{n}{m^{\alpha-1}} \right)^2}{2\left( (\alpha-1)\frac{n}{m^{\alpha-1}} + o\left(\frac{n}{m^{\alpha-1}}\right) \right)}\right) \notag\\
%&=& 1- \exp\left(- \frac{\left( (1-\delta_{11})\frac{(\alpha-1)}{\alpha}\frac{n}{m^{\alpha-1}} \right)^2}{2\left( (\alpha-1)\frac{n}{m^{\alpha-1}} + o\left(\frac{n}{m^{\alpha-1}}\right) \right)}\right)  
&& \buildrel (a) \over = 1- o(1).
\end{eqnarray}

Furthermore, using  \eqref{babbo1} and  \eqref{babbo2},  via Taylor expansion,  we obtain
\begin{eqnarray}
\label{eq: EZb M>1 11}
\ell\left(1-\left(1-\frac{1}{\ell}\right)^r\right) 
&=& m \left(1 - \left(1 - \frac{1}{m}\right)^{\frac{\delta(\alpha-1)}{\alpha}\frac{n}{m^{\alpha-1}}} \right) \notag\\
&=& \frac{\delta(\alpha-1)}{\alpha}\frac{n}{m^{\alpha-1}} + o\left(\frac{n}{m^{\alpha-1}}\right),
\end{eqnarray} 
from which, replacing   \eqref{babbo3}, and  (\ref{eq: EZb M>1 11}) in   (\ref{eq: P2}), we have:
%Let  $\widetilde{z} = \min\left\{\frac{\sigma_{11}\delta_{11}(\alpha-1)}{\alpha}\frac{n}{m^{\alpha-1}}, \frac{\delta_{11}(\alpha-1)}{\alpha}\frac{n}{m^{\alpha-1}}\right\} = \frac{\sigma_{11}\delta_{11}(\alpha-1)}{\alpha}\frac{n}{m^{\alpha-1}}$, where $0<\sigma_{11}<1$ is a constant such that $\widetilde{z}<r$, then by , we have
\begin{eqnarray}
\label{eq: P Zb M>1 11}
&&P_2(\ell, r, \widetilde z) \notag \\
%&& = 1- \exp\left(- \frac{\left(\ell\left(1-\left(1-\frac{1}{\ell}\right)^r\right) - \widetilde{z}\right)^2}{2\ell\left(1-\left(1-\frac{1}{\ell}\right)^r\right) }\right) \notag\\
&& \geq 1- \exp\left(- \frac{\left(\frac{\delta(\alpha-1)}{\alpha}\frac{n}{m^{\alpha-1}} + o\left(\frac{n}{m^{\alpha-1}}\right) - \frac{\sigma\delta(\alpha-1)}{\alpha}\frac{n}{m^{\alpha-1}} \right)^2}{2\left( \frac{\delta(\alpha-1)}{\alpha}\frac{n}{m^{\alpha-1}} + o\left(\frac{n}{m^{\alpha-1}}\right) \right)}\right) \notag\\
%&& = 1 - \exp\left( \frac{\left((1-\sigma_{11})\frac{\delta_{11}(\alpha-1)}{\alpha}\frac{n}{m^{\alpha-1}}  + o(\frac{n}{m^{\alpha-1}} )\right)^2}{2\left( \frac{\delta_{11}(\alpha-1)}{\alpha}\frac{n}{m^{\alpha-1}} + o\left(\frac{n}{m^{\alpha-1}}\right) \right)} \right). \notag\\
&& = 1- o(1).
\end{eqnarray} 
Thus,  replacing   \eqref{eq: P IU M>1 11} and \eqref{eq: P Zb M>1 11} in the second term of 
 (\ref{eq: general lower bound}), we obtain
\begin{eqnarray}
\label{eq: converse 11 1}
&&R^{\rm lb}(n,m,M,\qv)  \notag\\
&& \geq  P_1(\ell,r) P_2(\ell, r, \widetilde z) \max_{z \in \{1, \cdots, \lceil \widetilde{z} \rceil \}} (z-zM/{\lfloor \ell/z \rfloor}) \notag\\
&& \geq (1-o(1)) (1-o(1)) \max_{z \in \{1, \cdots,  \lceil \widetilde{z} \rceil \}} (z-zM/{\lfloor \ell/z \rfloor}) \notag\\
&& \geq (1-o(1))^2\max_{z \in \{1, \cdots,  \lceil \widetilde{z} \rceil \}} z (1-M/{\lfloor \ell/z \rfloor}).
\end{eqnarray}

Next, we find an explicit expression for $\max_{z \in \{1, \cdots,  \lceil \widetilde{z} \rceil \}} z \left(1-M/{\lfloor \ell/z \rfloor} \right)$.
Using  \eqref{babbo4},  after some algebraic manipulations,  we have  
\begin{subequations}
\begin{empheq}[left={\vspace{-5cm}\displaystyle \max_{z \in \{1, \cdots,  \lceil \widetilde{z} \rceil \}} z  \left(1-\frac{M}{\lfloor \ell/z \rfloor} \right ) > }\empheqlbrace]{align}
& \frac{\sigma\delta(\alpha-1)}{\alpha} \left(1 - \frac{\sigma\delta(\alpha-1)}{\alpha}\right)\frac{m}{M} + o\left(\frac{m}{M} \right),\,\, \,\,  \,\, \,\,M=o(m)  \label{Estefan1} \\
&\left(1 - \frac{M}{m}\right) , \,\, \,\, \,\, \,\, \,\, \,\,M = \Theta(m)  \mbox{\,\,\,\,and\,\,\,\,}  \frac{m}{M} \leq  3  \label{Estefan2}  \\
&\frac{1}{2} \left(\frac{m}{2M} - 1\right) + o(1),   \,\, \,\, \,\, \,\,\,\, \,\,M = \Theta(m)  \mbox{\,\,\,\,and\,\,\,\,}   \frac{m}{M}  > 3 \label{Estefan3} , 
 \end{empheq}
\label{Estefan0}  
\end{subequations}
where  \eqref{Estefan1} follows from footnote \ref{fnm:1} and from 
the fact that $M \geq \frac{m^{\alpha}}{n}$, %(which implies $ \frac{n}{m^{\alpha-1}} \geq \frac{m}{M}$)
while  \eqref{Estefan3} follows from the fact that  
 $z = \left\lfloor \frac{m}{2M} \right\rfloor >\frac{m}{2M}-1 $ and  ${\lfloor m/z \rfloor} >   \frac{m}{\frac{m}{2M}}-1$.

%\begin{eqnarray}
%z  (1-M/{\lfloor \ell/z \rfloor}) =  \left\{
%\begin{array}{cll}  
%\frac{\sigma\delta(\alpha-1)}{\alpha} \left(1 - \frac{\sigma\delta(\alpha-1)}{\alpha}\right)\frac{m}{M} + o\left(\frac{m}{M} \right), & M=o(m)  \label{Estefan1} \\
%\left(1 - \frac{M}{m}\right) , & M = \Theta(m)  \mbox{\,and\,}  \frac{m}{M} \leq  2  \label{Estefan2}  \\
%\frac{1}{4} \left(\frac{m}{M} - 1\right) + o(1),  & M = \Theta(m)  \mbox{\,and\,}   \frac{m}{M}  > 2 \label{Estefan3}  \\
% \end{array}
% \right., 
%\end{eqnarray}

Replacing  \ref{Estefan0}   in  \eqref{eq: converse 11 1} and using (\ref{eq: achievable M>1 11}), after simple algebraic manipulations we have:
%\begin{subequations}
%\begin{empheq}[left={\frac{R^{\rm ub}(n,m,M,\widetilde \pv, \qv)}{R^{\rm lb}(n,m,M,\qv) } < }\empheqlbrace]{align}
%&\frac{1}{\frac{\sigma\delta(\alpha-1)}{\alpha} \left(1 - \frac{\sigma\delta(\alpha-1)}{\alpha}\right)} + o(1),\,\, \,\,  \,\, \,\,M=o(m)  \notag \\
%& \frac{m}{M} + o(1)  , \,\, \,\, \,\, \,\, \,\, \,\,M = \Theta(m)  \mbox{\,\,and\,\,}  \frac{m}{M} \leq  2 \notag  \\
%&{(1-o(1))^2 \frac{1}{{2}} \left(\frac{m}{2M} - 1\right) + o(1)} < 4+o(1),    \,\, \,\, \,\, \,\,\,\, \,\,M = \Theta(m)  \mbox{\,\,and\,\,}   \frac{m}{M}  > 2 \notag
% \end{empheq}
%\notag 
%\end{subequations}
\begin{subequations}
\begin{empheq}[left={\frac{R^{\rm ub}(n,m,M,\qv, \widetilde m)}{R^{\rm lb}(n,m,M,\qv) } < }\empheqlbrace]{align}
&\frac{1}{\frac{\sigma\delta(\alpha-1)}{\alpha} \left(1 - \frac{\sigma\delta(\alpha-1)}{\alpha}\right)} + o(1),\,\, \,\,  \,\, \,\,M=o(m)  \notag \\
& \frac{m}{M} + o(1) \leq 3 + o(1)  , \,\, \,\, \,\, \,\, \,\, \,\,M = \Theta(m)  \mbox{\,\,and\,\,}  \frac{m}{M} \leq  3 \notag  \\
&{(1-o(1))^2 \frac{2}{(\frac{1}{2} - \frac{M}{m})} + o(1)} < 12+o(1),    \,\, \,\, \,\, \,\,\,\, \,\,M = \Theta(m)  \mbox{\,\,and\,\,}   \frac{m}{M}  > 3 \notag
 \end{empheq}
\notag 
\end{subequations}

\subsubsection{ When $\frac{n}{m^{\alpha-1}} = \Theta(1)$ (see Fig. \ref{fig: regimes_M_2})} Since $M \geq \frac{m^{\alpha}}{n}$, we have that $\frac{n}{m^{\alpha-1}} \geq \frac{m}{M}$ and  $M = \Theta(m)$. To compute the converse, similar as before, we use   (\ref{eq: general lower bound}) in  Theorem \ref{theorem: general lower bound}, with  all the parameters given by: 
\begin{align}
\ell &= m, \label{capodanno1}\\
r &=  \frac{\delta(\alpha-1)}{\alpha}\frac{n}{m^{\alpha-1}}, \label{capodanno2}\\
\widetilde z &=   \left\{\begin{array}{cc} \sigma, & \frac{\sigma\delta(\alpha-1)}{\alpha}\frac{n}{m^{\alpha-1}} \leq 1 \\
\frac{\sigma\delta(\alpha-1)}{\alpha}\frac{n}{m^{\alpha-1}}, & \frac{\sigma\delta(\alpha-1)}{\alpha}\frac{n}{m^{\alpha-1}} > 1 \end{array}\right., \label{capodanno3}\\
z &= \left\{\begin{array}{cc} 1, & \frac{\sigma\delta(\alpha-1)}{\alpha}\frac{n}{m^{\alpha-1}} \leq 1 \\
1, & \frac{\sigma\delta(\alpha-1)}{\alpha}\frac{n}{m^{\alpha-1}} > 1, \frac{\sigma\delta(\alpha-1)}{\alpha}\frac{m}{M} < 2 \\
\left\lfloor \frac{m}{2M} \right\rfloor, & \frac{\sigma\delta(\alpha-1)}{\alpha}\frac{n}{m^{\alpha-1}} > 1, \frac{\sigma\delta(\alpha-1)}{\alpha}\frac{m}{M} \geq 2 \end{array}\right., \label{capodanno4}
\end{align}
with $0 < \delta < 1$, and $ 0<\sigma <1$  positive constants determined in the following.  
Next,  we compute each term in (\ref{eq: general lower bound}) individually. In doing this, we consider to further regions 
 $\frac{\sigma\delta(\alpha-1)}{\alpha}\frac{n}{m^{\alpha-1}} \leq 1$ and $\frac{\sigma\delta(\alpha-1)}{\alpha}\frac{n}{m^{\alpha-1}} > 1$ (see Fig. \ref{fig: regimes_M_2}).
 
\begin{itemize}

\item If  $\frac{\sigma\delta(\alpha-1)}{\alpha}\frac{n}{m^{\alpha-1}} \leq 1$, using  \eqref{capodanno1} and Lemma \ref{lemma: H}, 
{following similar steps as in \eqref{eq: alpha<1 EIU 1}  and \eqref{eq: alpha<1 EIU 2}}, we have
\be
\label{eq: EIU M>1 12}
\frac{\alpha-1}{\alpha}\frac{n}{m^{\alpha-1}} + o\left( \frac{n}{m^{\alpha-1}} \right) \leq n\ell q_\ell \leq (\alpha-1)\frac{n}{m^{\alpha-1}} + o\left(\frac{n}{m^{\alpha-1}}\right),
\ee
from which using (\ref{eq: P1})  and \eqref{capodanno2}, we obtain
\begin{eqnarray}
\label{eq: P IU M>1 12}
%\PP(\Nsf \geq r  ) &\geq& \PP\left(\Nsf \geq \widetilde r\right) \notag\\
P_1(\ell,r)&=& 1- \exp\left(- \frac{\left( n\ell q_\ell - \frac{\delta(\alpha-1)}{\alpha}\frac{n}{m^{\alpha-1}}\right)^2}{2 n\ell q_\ell}\right) \notag\\
&\geq& 1- \exp\left(- \frac{\left(\frac{(\alpha-1)}{\alpha}\frac{n}{m^{\alpha-1}} - \frac{\delta(\alpha-1)}{\alpha}\frac{n}{m^{\alpha-1}} + o\left( \frac{n}{m^{\alpha-1}} \right)\right)^2}{2\left( (\alpha-1)\frac{n}{m^{\alpha-1}} + o\left(\frac{n}{m^{\alpha-1}}\right) \right)}\right) \notag\\
&=& 1- \exp\left(- \frac{\left( (1-\delta)\frac{(\alpha-1)}{\alpha}\frac{n}{m^{\alpha-1}} \right)^2}{2\left( (\alpha-1)\frac{n}{m^{\alpha-1}} \right)}\right) \notag\\
&\buildrel (a) \over \geq& 1 - \exp\left(- \frac{(1-\delta)^2(\alpha-1)}{2\alpha^2} \right),
\end{eqnarray}
where (a) follows from the fact that  $\Theta(1)=\frac{n}{m^{\alpha-1}} \geq \frac{m}{M} \geq 1$.
Furthermore, using   using (\ref{eq: P2}) and \eqref{capodanno3},
we have
\begin{eqnarray}
%\PP(\Zsf \geq \widetilde{z}) 
\label{eq: P boh M>1 12}
P_2(\ell, 1, \widetilde z)= 1 - \exp\left(-\frac{(1-\sigma)^2}{2}\right).
\end{eqnarray} 
Replacing   \eqref{capodanno1}-\eqref{capodanno3},  \eqref{eq: P IU M>1 12} and \eqref{eq: P boh M>1 12} in 
the second term of  (\ref{eq: general lower bound}), we obtain
\begin{eqnarray}
\label{eq: converse M>1 12}
&&R^{\rm lb}(n,m,M,\qv)  \notag\\
&& \geq  P_1(\ell,r) P_2(\ell, 1, \widetilde z) (1-M/{\ell}) \notag\\
%&& \geq \left( 1 - \exp\left(- \frac{(1-\delta)^2(\alpha-1)}{2\alpha^2} \right)\right)(1-M/{\ell}) \notag\\
&& \geq \left( 1 - \exp\left(- \frac{(1-\delta)^2(\alpha-1)}{2\alpha^2} \right)\right) \left(1 - \frac{M}{m} \right),
\end{eqnarray}
from which using (\ref{eq: achievable M>1 11}), we have
\begin{eqnarray}
\frac{R^{\rm ub}(n,m,M,\qv, \widetilde m)}{R^{\rm lb}(n,m,M,\qv) } &=& \frac{\frac{m}{M}-1 + o(1)}{\left( 1 - \exp\left(- \frac{(1-\delta)^2(\alpha-1)}{2\alpha^2} \right)\right)\left(1 - \frac{M}{m}\right)} \notag\\
&\leq& \frac{1}{\left( 1 - \exp\left(- \frac{(1-\delta)^2(\alpha-1)}{2\alpha^2} \right)\right)}\frac{m}{M} + o(1) \notag\\
&\buildrel (a) \over \leq& \frac{\frac{\sigma\delta(\alpha-1)}{\alpha}}{\left( 1 - \exp\left(- \frac{(1-\delta)^2(\alpha-1)}{2\alpha^2} \right)\right)} + o(1), 
\end{eqnarray}
where (a) is because that $\frac{m}{M} \leq \frac{n}{m^{\alpha-1}} \leq \frac{\alpha}{\sigma\delta(\alpha-1)}$.

\item If $\frac{\sigma\delta(\alpha-1)}{\alpha}\frac{n}{m^{\alpha-1}} > 1$, 
using (\ref{capodanno1})  and (\ref{capodanno2}), and following the same procedure as (\ref{eq: special 1}), %and (\ref{eq: special 2}), 
via Taylor expansion,  we obtain
\begin{eqnarray}
\label{eq: EZb M>1 12 1}
\ell\left(1-\left(1-\frac{1}{\ell}\right)^r\right) 
&=& m \left(1 - \left(1 - \frac{1}{m}\right)^{\frac{\delta(\alpha-1)}{\alpha}\frac{n}{m^{\alpha-1}}} \right) \notag\\
&\buildrel (a) \over =& \frac{\delta(\alpha-1)}{\alpha}\frac{n}{m^{\alpha-1}} + o(1), %\notag\\
\end{eqnarray}
where (a) is due to $\frac{\delta(\alpha-1)}{\alpha}\frac{n}{m^{\alpha-1}} = \Theta(1)$.
Hence, replacing (\ref{capodanno1}), (\ref{capodanno2}),  and (\ref{eq: EIU M>1 12}) in  (\ref{eq: P1}), we obtain: 
\begin{eqnarray}
P_1(\ell,r) &=& 1- \exp\left(- \frac{\left(n\ell q_\ell - \frac{\delta(\alpha-1)}{\alpha}\frac{n}{m^{\alpha-1}}\right)^2}{2n\ell q_\ell }\right) \notag\\
%&\geq& 1- \exp\left(- \frac{\left(\frac{(\alpha-1)}{\alpha}\frac{n}{m^{\alpha-1}} - \frac{\delta_{12}(\alpha-1)}{\alpha}\frac{n}{m^{\alpha-1}} \right)^2}{2\left( (\alpha-1)\frac{n}{m^{\alpha-1}} + o\left(\frac{n}{m^{\alpha-1}}\right) \right)}\right) \notag\\
&=& 1- \exp\left(- \frac{\left( (1-\delta)\frac{(\alpha-1)}{\alpha}\frac{n}{m^{\alpha-1}} \right)^2}{2\left( (\alpha-1)\frac{n}{m^{\alpha-1}} + o\left(\frac{n}{m^{\alpha-1}}\right) \right)}\right) \notag\\
%&\geq& 1 - \exp\left(-\frac{(1-\delta_{12})^2(\alpha-1)}{2\alpha^2}\frac{n}{m^{\alpha-1}} \right) \notag\\
&\buildrel (a) \over  \geq& 1 - \exp\left(-\frac{(1-\delta)^2(\alpha-1)}{2\alpha^2}\frac{\alpha}{\sigma\delta(\alpha-1)} \right) \notag\\
&=& 1 - \exp\left(- \frac{(1-\delta)^2}{2\alpha\sigma\delta} \right),
\label{vigiliaP1}
\end{eqnarray}
where (a) is because that $\frac{n}{m^{\alpha-1}} \geq \frac{\alpha}{\sigma\delta(\alpha-1)}$, while 
replacing (\ref{capodanno3}) and \eqref{eq: EZb M>1 12 1} in  (\ref{eq: P2}), we obtain: 
%
%Let $\widetilde{z} = \min\left\{\frac{\sigma_{12}\delta_{12}(\alpha-1)}{\alpha}\frac{n}{m^{\alpha-1}}, \frac{\delta_{12}(\alpha-1)}{\alpha}\frac{n}{m^{\alpha-1}}\right\} = \frac{\sigma_{12}\delta_{12}(\alpha-1)}{\alpha}\frac{n}{m^{\alpha-1}}$, then by using (\ref{eq: P2}), we have
\begin{eqnarray}
&&P_2(\ell,r,\widetilde z) \notag \\
%&& = 1- \exp\left(- \frac{\left(\ell\left(1-\left(1-\frac{1}{\ell}\right)^r\right) - \widetilde{z}\right)^2}{2\ell\left(1-\left(1-\frac{1}{\ell}\right)^r\right)}\right) \notag\\
%&& \geq 1- \exp\left(- \frac{\left(\frac{\delta_{12}(\alpha-1)}{\alpha}\frac{n}{m^{\alpha-1}} + o\left(\frac{n}{m^{\alpha-1}}\right) - \frac{\sigma_{12}\delta_{12}(\alpha-1)}{\alpha}\frac{n}{m^{\alpha-1}} \right)^2}{2\left( \frac{\delta_{12}(\alpha-1)}{\alpha}\frac{n}{m^{\alpha-1}} + o\left(\frac{n}{m^{\alpha-1}}\right) \right)}\right) \notag\\
&&  \geq 1 - \exp\left( \frac{\left((1-\sigma)\frac{\delta(\alpha-1)}{\alpha}\frac{n}{m^{\alpha-1}}  + o(\frac{n}{m^{\alpha-1}} )\right)^2}{2\left( \frac{\delta(\alpha-1)}{\alpha}\frac{n}{m^{\alpha-1}} + o\left(\frac{n}{m^{\alpha-1}}\right) \right)} \right). \notag\\
&& = 1- \exp\left(- \frac{(1-\sigma)^2\delta(\alpha-1)}{2\alpha} \frac{n}{m^{\alpha-1}}\right) \notag\\
&& \geq 1- \exp\left(- \frac{(1-\sigma)^2\delta(\alpha-1)}{2\alpha} \frac{\alpha}{\sigma\delta(\alpha-1)}\right) \notag\\
&& =1- \exp\left(- \frac{(1-\sigma)^2}{2\sigma} \right).
\label{vigiliaP2}
\end{eqnarray}
Replacing (\ref{capodanno1})-(\ref{capodanno4}), \eqref{vigiliaP1},  and \eqref{vigiliaP2}  in the first term of
 (\ref{eq: general lower bound}),  and noticing, from \eqref{capodanno3},  that, when $\frac{\sigma\delta(\alpha-1)}{\alpha}\frac{n}{m^{\alpha-1}} > 1$,    by definition,  $\widetilde z < r$,   we obtain
\begin{eqnarray}
\label{eq: converse 12 11}
&&R^{\rm lb}(n,m,M,\qv)  \notag\\
&& \geq  P_1(\ell,r) P_2(\ell,r,\widetilde z) \max_{z \in \{1, \cdots,  \lceil \widetilde{z}\rceil\}} (z-zM/{\lfloor \ell/z \rfloor}) \notag\\
&& \geq \left( 1 - \exp\left(- \frac{(1-\delta)^2}{2\alpha\sigma \delta} \right)\right)\left( 1- \exp\left(- \frac{(1-\sigma)^2}{2\sigma} \right) \right) \max_{z \in \{1, \cdots,  \lceil \widetilde{z}\rceil\}} z(1-M/{\lfloor \ell/z \rfloor}). \notag\\
\end{eqnarray}
Replacing    
(\ref{capodanno1}) and  (\ref{capodanno4}) in  (\ref{eq: converse 12 11}), we have
\begin{subequations}
\begin{empheq}[left={\vspace{-5cm}\displaystyle \max_{z \in \{1, \cdots,  \lceil \widetilde{z} \rceil \}} z  \left(1-\frac{M}{\lfloor \ell/z \rfloor} \right ) > }\empheqlbrace]{align}
&\left(1 - \frac{M}{m}\right),\,\, \,\,  \,\, \,\,\frac{\sigma\delta(\alpha-1)}{\alpha}\frac{m}{M} < 2 \label{rustica1} \\
&\frac{1}{2}\left(\frac{m}{2M} - 1\right) + o(1) , \,\, \,\, \,\, \,\, \,\, \,\,\frac{\sigma\delta(\alpha-1)}{\alpha}\frac{m}{M} \geq 2   \label{rustica2} 
 \end{empheq}
\label{rustica0}  
\end{subequations}
from which replacing  \eqref{rustica0}   in  (\ref{eq: converse 12 11}) and using  (\ref{eq: achievable M>1 11}) we obtain 
\begin{subequations}
\begin{empheq}[left={\vspace{-5cm}\frac{R^{\rm ub}(n,m,M,\qv, \widetilde m)}{R^{\rm lb}(n,m,M,\qv) }  \leq }\empheqlbrace]{align}
& \frac{2\alpha}{\sigma\delta(\alpha-1)\left( 1 - \exp\left(- \frac{(1-\delta)^2}{2\alpha\sigma\delta} \right)\right)\left( 1- \exp\left(- \frac{(1-\sigma)^2}{2\sigma} \right) \right)} + o(1), 
%\frac{\sigma\delta(\alpha-1)}{\alpha}\frac{m}{M} < 2 
\label{rompiC1} \\
& \frac{2}{\left( 1 - \exp\left(- \frac{(1-\delta)^2}{2\alpha\sigma\delta} \right)\right)\left( 1- \exp\left(- \frac{(1-\sigma)^2}{2\sigma} \right) \right) \left(\frac{1}{2} - \frac{\sigma\delta(\alpha-1)}{2\alpha}\right)} + o(1) , 
%\frac{\sigma\delta(\alpha-1)}{\alpha}\frac{m}{M} \geq 2  
 \label{rompiC2} 
 \end{empheq}
\label{rompiC0}  
\end{subequations}
where \eqref{rompiC1}  holds for  $\frac{\sigma\delta(\alpha-1)}{\alpha}\frac{m}{M} < 2 $,  while 
\eqref{rompiC2}  holds for  $\frac{\sigma\delta(\alpha-1)}{\alpha}\frac{m}{M} \geq 2 $. 
Note that  \eqref{rustica1} follows from lower bounding $z$ by  $\frac{m}{2M}-1$, while 
\eqref{rompiC0}  shows the order-optimality of the achievable expected rate for RLFU and RAP.

\end{itemize}

\section{Proof of Table \ref{table: table_1_2} in Theorem \ref{theorem: gamma > 1 achievable 3}}
\label{sec: proof of table II}

\begin{figure}[ht]
\centerline{\includegraphics[width=12cm]{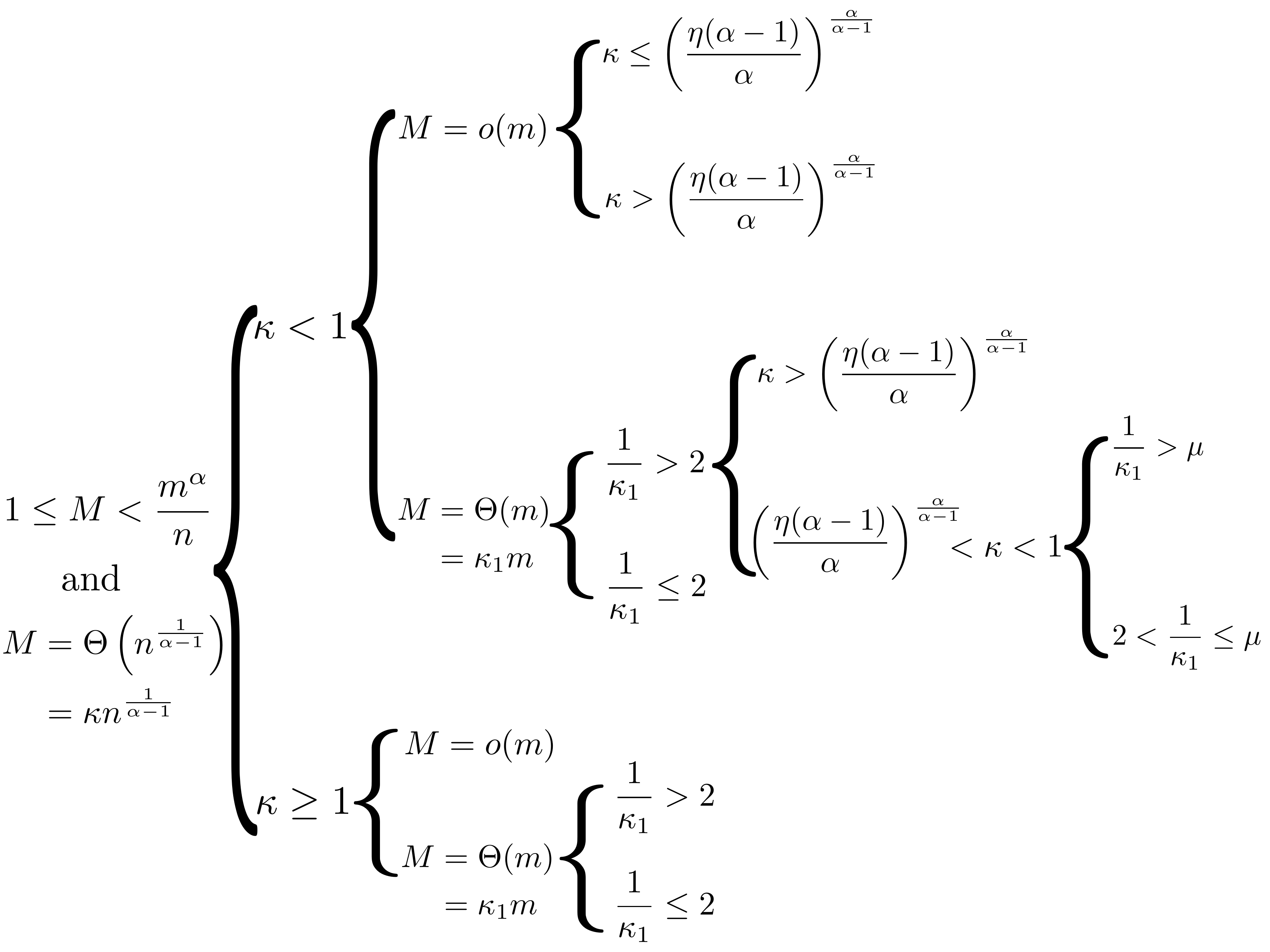}}
\caption{The sub-cases of the regimes of $M$ when $1 \leq M < \frac{m^{\alpha}}{n}$ and $M = \Theta(n^{\frac{1}{\alpha-1}})$, where $\kappa$, $\kappa_1$, $\eta$ are some constants which will be determined later.}
\label{fig: regimes_M_4}
\end{figure}

In this section, we prove the order-optimality of RLFU for the case when $n = o\left(m^{\alpha}\right)$ and the memory is such that 
 $1\leq M < \frac{m^\alpha}{n}$ and $M = \Theta(n^{\frac{1}{\alpha-1}}) = \kappa n^{\frac{1}{\alpha-1}} + o(n^{\frac{1}{\alpha-1}})< \frac{m^{\alpha}}{n}$ for some positive constant $\kappa$. %} 
In this case, we have two cases to consider, which are $\kappa \geq 1$ and $\kappa <1$ (see Fig. \ref{fig: regimes_M_4}). 
All the subregions of $M$ are illustrated in Fig. \ref{fig: regimes_M_4} and will be treated separately in the following proofs.  

In the following, since we focus on the asymptotic regime ($n \rightarrow \infty$),    in
$M = \kappa n^{\frac{1}{\alpha-1}} + o(n^{\frac{1}{\alpha-1}})$,  we  ignore   $o(n^{\frac{1}{\alpha-1}})$ and we write directly $M = \kappa n^{\frac{1}{\alpha-1}}$. 

\subsection{Region of  $\kappa < 1$} 
In this case,   we need to consider the scenarios $M=o(m)$ and $M=\Theta(m)$ separately (see Fig. \ref{fig: regimes_M_4})

\subsubsection {When $M = o(m)$} 
\label{When$M = o(m)$th6}
Within this region, we need to consider two subregions again,  $\kappa \leq \left(\frac{\eta (\alpha-1)}{\alpha}\right)^{\frac{\alpha}{\alpha-1}}$ and $\kappa > \left(\frac{\eta (\alpha-1)}{\alpha}\right)^{\frac{\alpha}{\alpha-1}}$ (see Fig. \ref{fig: regimes_M_4}), where $\eta =\sigma\delta$  with   $0 < \sigma <1$ and  $ 0 <\delta<1$ positive constants  determined later.

\begin{itemize}
\item If $\kappa \leq \left(\frac{\eta (\alpha-1)}{\alpha}\right)^{\frac{\alpha}{\alpha-1}}$ , letting $\widetilde m = M^{\frac{1}{\alpha}}n^{\frac{1}{\alpha}}= \kappa^{\frac{1}{\alpha}}n^{\frac{1}{\alpha-1}}$,  using (\ref{eq: general achievable 1}) and Lemma \ref{lemma: H}, we have
\begin{eqnarray}
\label{eq: achievable M>1 4}
R^{\rm ub}(n,m,M,\qv, \widetilde m)
\leq \frac{2M^{\frac{1}{\alpha}}n^{\frac{1}{\alpha}}}{M} + o\left(\frac{M^{\frac{1}{\alpha}}n^{\frac{1}{\alpha}}}{M} \right) = 2\kappa^{\frac{1}{\alpha}-1} - 1+o(1).
\end{eqnarray}

To compute the converse, similar as in previous sections, we use Theorem \ref{theorem: general lower bound}. 
The parameters required in Theorem \ref{theorem: general lower bound} are summarized in the following.
\begin{align}
\ell &= \widetilde m = M^{\frac{1}{\alpha}}n^{\frac{1}{\alpha}}= \kappa^{\frac{1}{\alpha}}n^{\frac{1}{\alpha-1}}, \label{santclaus1}\\
r &= \frac{\delta(\alpha-1)}{\alpha}\kappa ^{\frac{1}{\alpha}-1}, \label{santclaus2}\\
\widetilde z &=   \frac{\sigma\delta(\alpha-1)}{\alpha}\kappa ^{\frac{1}{\alpha}-1}, \label{santclaus3}\\
z &= \lfloor \widetilde z \rfloor, \label{santclaus4}
\end{align}
with $0 < \delta < 1$, and $ 0<\sigma <1$  positive constants determined in the following.  Note that by definition 
$ \widetilde z  < r$. Next,  we compute each term in (\ref{eq: general lower bound}) individually.
To this end, using  \eqref{santclaus1} and
\eqref{santclaus2},  we first find an expression  for 
$n\ell q_\ell$ and $\ell\left(1-\left(1-\frac{1}{\ell}\right)^r\right)$ in terms of  $\delta, \sigma,m,n$ and $\alpha$. 
Specifically, using   \eqref{santclaus1}  and Lemma \ref{lemma: H}, 
following similar steps as in \eqref{eq: IU M>1 1} and \eqref{eq: IU M>1 2}, we have
%%%%%%  {eq: alpha<1 EIU 1}  {eq: alpha<1 EIU 2} %%%%
% and 
%
%Let $\ell = \widetilde m = M^{\frac{1}{\alpha}}n^{\frac{1}{\alpha}} = \kappa^{\frac{1}{\alpha}}n^{\frac{1}{\alpha-1}}$, where the last equality is because $M = \kappa n^{\frac{1}{\alpha-1}}$, 
%by %using (\ref{eq: EIU}) and 
%Lemma \ref{lemma: H},  
%following the same procedure as (\ref{eq: IU M<1 3}), then we can write
%\begin{eqnarray}
%\label{eq: EIU M>1 4 1}
%n\ell q_\ell &=& n \cdot \widetilde{m} \cdot \frac{{ \left(\widetilde{m}\right)}^{-\alpha}}{H(\alpha, 1, m)} \notag\\
%&\geq& \frac{n\widetilde{m}^{1-\alpha}}{\frac{1}{1-\alpha}m^{1-\alpha} - \frac{1}{1-\alpha} + 1} \notag\\
%&=& \frac{n\left(\kappa^{\frac{1}{\alpha}}n^{\frac{1}{\alpha-1}}\right)^{1-\alpha}m^{\alpha-1}}{\frac{1}{1-\alpha} - \frac{1}{1-\alpha}m^{\alpha - 1} + m^{\alpha - 1}} \notag\\
%&=& \frac{\alpha-1}{\alpha}\kappa ^{\frac{1}{\alpha}-1} + o(1).
%\end{eqnarray}
%and
%\begin{eqnarray}
%\label{eq: EIU M>1 4 2}
%n\ell q_\ell &=& n \cdot \widetilde{m} \cdot \frac{{ \left(\widetilde{m}\right)}^{-\alpha}}{H(\alpha, 1, m)} \notag\\
%&\leq& \frac{n\widetilde{m}^{1-\alpha}}{\frac{1}{1-\alpha}(m+1)^{1-\alpha} - \frac{1}{1-\alpha}} \notag\\
%&=& \frac{\left(\kappa^{\frac{1}{\alpha}}n^{\frac{1}{\alpha-1}}\right)^{1-\alpha}nm^{\alpha-1}}{\frac{1}{\alpha-1} m^{\alpha - 1} - \frac{1}{\alpha-1} \left(\frac{m}{m+1}\right)^{\alpha-1}} \notag\\
%&=& (\alpha-1)\kappa ^{\frac{1}{\alpha}-1} + o(1).
%\end{eqnarray}
%
%Thus, we have
\begin{eqnarray}
\label{eq: EIU M>1 4 3}
\frac{\alpha-1}{\alpha}\kappa ^{\frac{1}{\alpha}-1} + o(1) \leq n\ell q_\ell \leq (\alpha-1)\kappa ^{\frac{1}{\alpha}-1} + o(1), 
\end{eqnarray}
while using \eqref{santclaus1} and \eqref{santclaus2},  via Taylor expansion,  we obtain
\begin{eqnarray}
\label{eq: EZb M>1 4}
\ell\left(1-\left(1-\frac{1}{\ell}\right)^r\right) &=& \widetilde m \left(1 - \left(1 - \frac{1}{\widetilde m}\right)^{\frac{\delta(\alpha-1)}{\alpha}\kappa ^{\frac{1}{\alpha}-1}} \right) \notag\\
&=& \frac{\delta(\alpha-1)}{\alpha}\kappa ^{\frac{1}{\alpha}-1} + o(1).
\end{eqnarray}  

%Let $\eta =\sigma_4\delta_4$, since $\kappa \leq \left(\frac{\eta (\alpha-1)}{\alpha}\right)^{\frac{\alpha}{\alpha-1}}$, then $\frac{\sigma_4\delta_4(\alpha-1)}{\alpha}\kappa ^{\frac{1}{\alpha}-1} \geq 1$ for some $0 < \sigma_4 <1$, which is an arbitrary constant in $(0,1)$. Let $r = \frac{\delta_4(\alpha-1)}{\alpha}\kappa ^{\frac{1}{\alpha}-1}$, where $0 < \delta_4 < 1$,
%then %by using (\ref{eq: EZb}), 
%we can write
%\begin{eqnarray}
%\label{eq: EZb M>1 4}
%\ell\left(1-\left(1-\frac{1}{\ell}\right)^r\right) &=& \widetilde m \left(1 - \left(1 - \frac{1}{\widetilde m}\right)^{\frac{\delta_4(\alpha-1)}{\alpha}\kappa ^{\frac{1}{\alpha}-1}} \right) \notag\\
%&=& \frac{\delta_4(\alpha-1)}{\alpha}\kappa ^{\frac{1}{\alpha}-1} + o(1).
%\end{eqnarray}  

Replacing  \eqref{santclaus1}-\eqref{santclaus3}, \eqref{eq: EIU M>1 4 3} and \eqref{eq: EZb M>1 4}  in 
 (\ref{eq: P1}) and  in (\ref{eq: P2}),  we obtain
\begin{eqnarray}
\label{eq: P IU M>1 4}
P_1(\ell, r) 
%&=& 1- \exp\left(- \frac{\left(n\ell q_\ell - \frac{\delta(\alpha-1)}{\alpha}\kappa ^{\frac{1}{\alpha}-1}\right)^2}{2n\ell q_\ell}\right) \notag\\
&\geq& 1- \exp\left(- \frac{\left(\frac{\alpha-1}{\alpha}\kappa ^{\frac{1}{\alpha}-1} + o\left(1\right) - \frac{\delta(\alpha-1)}{\alpha}\kappa ^{\frac{1}{\alpha}-1}\right)^2}{2\left((\alpha-1)\kappa ^{\frac{1}{\alpha}-1} + o\left(1\right)\right)}\right) \notag\\
&=& 1 - \exp\left(-\frac{(1-\delta)^2(\alpha-1)}{2\alpha^2}\kappa ^{\frac{1}{\alpha}-1}\right) + o(1) \notag\\
&\buildrel (a) \over \geq & 1 - \exp\left(-\frac{(1-\delta)^2(\alpha-1)}{2\alpha^2}\right),
\end{eqnarray}
and 
%Let  
%\begin{eqnarray}
%\widetilde{z} &=& \min\left\{\sigma_4\left(\frac{\delta_4(\alpha-1)}{\alpha}\kappa ^{\frac{1}{\alpha}-1} \right), \frac{\delta_4(\alpha-1)}{\alpha}\kappa ^{\frac{1}{\alpha}-1} \right\} \notag\\
%&=&  \frac{\sigma_4\delta_4(\alpha-1)}{\alpha}\kappa ^{\frac{1}{\alpha}-1},
%\end{eqnarray}
%where $0<\sigma_4 <1$ is a constant such that $\widetilde{z} \leq r$.
%Then by using (\ref{eq: P2}) and (\ref{eq: EZb M>1 4}) we have
\begin{eqnarray}
\label{eq: P Zb M>1 4}
P_2(\ell, r, \widetilde z) 
%&& = 1 - \exp\left(- \frac{\left((1-\sigma_4)\left(\frac{\delta_4(\alpha-1)}{\alpha}\kappa ^{\frac{1}{\alpha}-1} + o(1)\right) \right)^2}{2\left( \frac{\delta_4(\alpha-1)}{\alpha}\kappa ^{\frac{1}{\alpha}-1} + o(1) \right)} \right). \notag\\
\geq 1- \exp\left(-\frac{(1-\sigma)^2(\alpha-1)\delta}{2\alpha}\right),
% + o(1)
\end{eqnarray} 
where (a)  in  \eqref{eq: P IU M>1 4}  follows from ignoring  $o(1)$ and from the fact that $\kappa < 1$.
Finally,  replacing 
Eqs. \eqref{santclaus1}-\eqref{santclaus4},  \eqref{eq: P IU M>1 4} and  \eqref{eq: P Zb M>1 4}  in Theorem \ref{theorem: general lower bound}, using  footnote \ref{fnm:1}, 
we obtain:
\begin{eqnarray}
R^{\rm lb}(n,m,M,\qv)  &\geq&  P_1(\ell,r) P_2(\ell, r, \widetilde z)\max_{z \in \{1, \cdots, \lceil \widetilde{z} \rceil 
\}} (z-zM/{\lfloor \ell/z \rfloor}) \notag\\
%&& \geq \left(1 - \exp\left(-\frac{(1-\delta_4)^2(\alpha-1)}{2\alpha^2}\right) + o(1)\right) \notag\\
%&& \cdot \left(1- \exp\left(-\frac{(1-\sigma_4)^2(\alpha-1)\delta_4}{2\alpha}\right) + o(1)\right)  
& \geq &  \Xi  \max_{z \in \{1, \cdots,  \lceil \widetilde{z} \rceil\}}  \lfloor \widetilde{z} \rfloor  \left(1 - \frac{M}{\left\lfloor \frac{ \widetilde{m} }{ \lfloor \widetilde{z} \rfloor } 
\right\rfloor } \right)\notag\\
&&\buildrel (a) \over \geq  \Xi \cdot \widetilde{z} \left(1 - \frac{M}{\left\lfloor \frac{ \widetilde{m} }{\widetilde{z}} \right\rfloor } \right) \notag\\
&& \geq\Xi \cdot  \frac{\sigma\delta (\alpha-1)}{\alpha}\kappa^{\frac{1}{\alpha}-1}  \left(1 - \frac{M}{\left\lfloor\frac{\kappa^{\frac{1}{\alpha}}n^{\frac{1}{\alpha-1}}}{\frac{\sigma\delta(\alpha-1)}{\alpha}\kappa^{\frac{1-\alpha}{\alpha}}}\right\rfloor} \right) \notag\\
&& \geq  \Xi \cdot  \frac{\sigma\delta(\alpha-1)}{\alpha}\kappa^{\frac{1}{\alpha}-1} 
 \left(1 - \frac{M}{\frac{\alpha}{\sigma\delta(\alpha-1)}\kappa n^{\frac{1}{\alpha-1}} - 1} \right)
%\left(1 - \frac{M}{\frac{\kappa^{\frac{1}{\alpha}}n^{\frac{1}{\alpha-1}}}{\frac{\sigma\delta(\alpha-1)}{\alpha}\kappa^{\frac{1-\alpha}{\alpha}}} - 1} \right) 
\notag\\
%&& = \left(1 - \exp\left(-\frac{(1-\delta_4)^2(\alpha-1)}{2\alpha^2}\right)\right) \left(1- \exp\left(-\frac{(1-\sigma_4)^2(\alpha-1)\delta_4}{2\alpha}\right)\right) \frac{\sigma_4\delta_4(\alpha-1)}{\alpha}\kappa^{\frac{1}{\alpha}-1} \notag\\
%&& \left(1 - \frac{M}{\frac{\alpha}{\sigma_4\delta_4(\alpha-1)}\kappa n^{\frac{1}{\alpha-1}} - 1} \right) \notag \\
%\end{eqnarray}
%\begin{eqnarray}
%&& = \left(1 - \exp\left(-\frac{(1-\delta_4)^2(\alpha-1)}{2\alpha^2}\right)\right) \left(1- \exp\left(-\frac{(1-\sigma_4)^2(\alpha-1)\delta_4}{2\alpha}\right)\right) \frac{\sigma_4\delta_4(\alpha-1)}{\alpha}\kappa^{\frac{1}{\alpha}-1}  \notag\\
%&& \left(1 - \frac{1}{\frac{\alpha}{\sigma_4\delta_4(\alpha-1)}\frac{\kappa n^{\frac{1}{\alpha-1}}}{M} - \frac{1}{M}} \right) \notag\\
%&& = \left(1 - \exp\left(-\frac{(1-\delta_4)^2(\alpha-1)}{2\alpha^2}\right)\right) \left(1- \exp\left(-\frac{(1-\sigma_4)^2(\alpha-1)\delta_4}{2\alpha}\right)\right) \frac{\sigma_4\delta_4(\alpha-1)}{\alpha}\kappa^{\frac{1}{\alpha}-1}  \notag\\
%&& \left(1 - \frac{1}{\frac{\alpha}{\sigma_4\delta_4(\alpha-1)} - \frac{1}{M}} \right) \notag\\
&&\buildrel (b) \over =\Xi \cdot  \frac{\sigma\delta(\alpha-1)}{\alpha}\kappa^{\frac{1}{\alpha}-1} 
\left(1 - \frac{1}{\frac{\alpha}{\sigma\delta(\alpha-1)} - \frac{1}{M}} \right) \notag\\
%&& = \left(1 - \exp\left(-\frac{(1-\delta_4)^2(\alpha-1)}{2\alpha^2}\right)\right) \left(1- \exp\left(-\frac{(1-\sigma_4)^2(\alpha-1)\delta_4}{2\alpha}\right)\right) \frac{\sigma_4\delta_4(\alpha-1)}{\alpha}\kappa^{\frac{1}{\alpha}-1}  \notag\\
&&=\Xi \cdot  \frac{\sigma\delta(\alpha-1)}{\alpha}\kappa^{\frac{1}{\alpha}-1} 
\left(1 - \frac{1}{\frac{\alpha}{\sigma\delta(\alpha-1)}-o(1)} \right), \notag\\
\label{eq: converse M>1 4}
\end{eqnarray}
where in (a) we have defined  
$$
\Xi=   \left(1- \exp\left(-\frac{(1-\sigma)^2(\alpha-1)\delta}{2\alpha}\right) \right)  \left( 1 - \exp\left(-\frac{(1-\delta)^2(\alpha-1)}{2\alpha^2}\right) \right),
$$
while (b)  follows from the fact that $M = \kappa n^{\frac{1}{\alpha-1}}$.
Finally  using (\ref{eq: achievable M>1 4}) and (\ref{eq: converse M>1 4}), we obtain
\begin{eqnarray}
 \frac{R^{\rm ub}(n,m,M,\qv, \widetilde m)}{R^{\rm lb}(n,m,M,\qv) }\leq \frac{2}{ \Xi  \cdot \frac{\sigma\delta(\alpha-1)}{\alpha} \left(1 - \frac{1}{\frac{\alpha}{\sigma\delta(\alpha-1)}-o(1)} \right)}. \notag\\
\end{eqnarray}

\item If $\kappa > \left(\frac{\eta (\alpha-1)}{\alpha}\right)^{\frac{\alpha}{\alpha-1}}$, 
%which implies $\frac{\sigma_4\delta_4(\alpha-1)}{\alpha}\kappa ^{\frac{1}{\alpha}-1} < 1$ for some $0 < \sigma_4 <1$. 
letting $\widetilde m = M = \kappa n^{\frac{1}{\alpha-1}}$,\footnote{It can be easily seen that if we let $\widetilde m = M^{\frac{1}{\alpha}}n^{\frac{1}{\alpha}}$, then the achievable expected rate is also order-optimal. 
%We can determine $\widetilde m$ by comparing the two achievable expected rate. 
Here we only illustrate the case of $\widetilde m = M = \kappa n^{\frac{1}{\alpha-1}}$.}  using (\ref{eq: general achievable 1}) and Lemma \ref{lemma: H}, we have
\begin{eqnarray}
\label{eq: achievable M>1 5}
R^{\rm ub}(n,m,M,\qv, \widetilde m)
&\leq& %\left(\frac{{\widetilde m}} {M}-1\right)\left( 1 -\left(1-\frac{M}{\widetilde m} \right)^{ n \, G_{\widetilde m}} \right) +  (1-G_{\widetilde m})\,n \notag\\
%&&= 
 (1-G_{\widetilde m})\,n \notag\\
&\leq& \frac{n}{M^{\alpha-1}} + o\left(\frac{n}{M^{\alpha-1}}\right) = \kappa^{1-\alpha} + o(1).
\end{eqnarray}

To compute the converse, similar as before, we use the second term of  (\ref{eq: general lower bound})  in Theorem \ref{theorem: general lower bound},  where the required parameters are summarized in the following:
\begin{align}
\ell &= cM, \label{mamma1}\\
r &= \frac{\delta(\alpha-1)c^{1-\alpha}\kappa^{1-\alpha}}{\alpha}, \label{mamma2}\\
\widetilde z &=   \sigma,  \label{mamma3} %\notag\\
%z &= \widetilde z, \notag
\end{align}
with $1<c$,  $0 < \delta < 1$, and $ 0<\sigma <1$  positive constants determined in the following.  
 Next,  we compute each term in (\ref{eq: general lower bound}) individually. 
 To this end,  using   \eqref{mamma1}  and Lemma \ref{lemma: H}, 
 following the same steps as in \eqref{eq: EIU M>1 9 1} and \eqref{eq: EIU M>1 9 2}, and recalling that 
 $M = \kappa n^{\frac{1}{\alpha-1}}$ we have:
\be
\label{eq: EIU M>1 5}
\frac{(\alpha-1)c^{ 1-\alpha}\kappa^{1-\alpha}}{\alpha} + o(1) \leq n\ell q_\ell
\leq (\alpha-1)c^{1-\alpha}\kappa^{1-\alpha} + o(1),
\ee
from which using \eqref{mamma1}  and  (\ref{eq: P1})  we have: 
%Then, let $r = \frac{\delta_5(\alpha-1)c^{1-\alpha}\kappa^{1-\alpha}}{\alpha}$, where $0 < \delta_5 < 1$, then since 
%$\kappa > \left(\frac{\eta (\alpha-1)}{\alpha}\right)^{\frac{\alpha}{\alpha-1}}$
% and $c >  \left(\frac{\eta (\alpha-1)}{\alpha}\right)^{-
%\frac{\alpha}{\alpha-1}}$, then $r \leq 1$. By using (\ref{eq: P1}) and (\ref{eq: EIU M>1 5}), we obtain
\begin{eqnarray}
\label{eq: IU 10} 
P_1(\ell,r)
%&\buildrel (a) \over =& \PP\left(\Nsf \geq \widetilde r\right) \notag\\
%& = & 1- \exp\left(- \frac{\left(n\ell q_\ell - \frac{\delta(\alpha-1)c^{1-\alpha}\kappa^{1-\alpha}}{\alpha}\right)^2}{2n\ell q_\ell}\right) \notag\\
&\geq& 1 - \exp\left(- \frac{(1-\delta)^2(\alpha-1)c^{ 1-\alpha}}{2\alpha^2}\kappa^{1-\alpha}\right) \notag\\
&\geq& 1 - \exp\left(- \frac{(1-\delta)^2(\alpha-1)c^{1-\alpha}}{2\alpha^2}\right). 
\end{eqnarray}

Moreover, using  (\ref{eq: P2})  and  \eqref{mamma3},  we have 
%where (a) is because that $\Nsf$ is an integer such that $\PP(\Nsf \geq 1) = \PP(\Nsf \geq \widetilde r)$ for $\widetilde r<1$.
%Let $\widetilde{z} = \sigma_5 $, where $0 < \sigma_5 <1$, by %and (\ref{eq: EZb M>1 5}), 
%we have
\begin{eqnarray}
\label{eq: P Zb M>1 5}
P_2(\ell, 1, \widetilde z) %&=& \PP(\Zsf \geq \sigma_5 | r) \notag \\
%&=& 1- \exp\left(- \frac{\left(\ell\left(1-\left(1-\frac{1}{\ell}\right)^r\right) - \sigma_5\right)^2}{2\EE[\Zsf]}\right) \notag\\
&=& 1 - \exp\left( \frac{\left(1-\sigma \right)^2}{2} \right),
\end{eqnarray} 
from which replacing \eqref{mamma1}-\eqref{mamma3},  \eqref{eq: IU 10}   and \eqref{eq: P Zb M>1 5} in the second term 
of  (\ref{eq: general lower bound})  we obtain: 
\begin{eqnarray}
\label{eq: converse M>1 5}
R^{\rm lb}(n,m,M,\qv)  &\geq & P_1(\ell,r) P_2(\ell, 1, \widetilde z)(1-M/{\ell}) \notag\\
& \geq & \left(1 - \exp\left(- \frac{(1-\delta)^2(\alpha-1)c^{  1-\alpha}}{2\alpha^2}\right)\right) \left(1 - \exp\left( \frac{\left(1-\sigma \right)^2}{2} \right)\right)\left(1-\frac{1}{c}\right). \notag\\
\end{eqnarray}

Using (\ref{eq: achievable M>1 5}) and (\ref{eq: converse M>1 5}), we obtain
\begin{eqnarray}
\frac{R^{\rm ub}(n,m,M, \qv, \widetilde m)}{R^{\rm lb}(n,m,M,\qv) } &\leq& \frac{\kappa^{1-\alpha}}{\left(1 - \exp\left( \frac{\left(1-\sigma\right)^2}{2} \right)\right)\left(1 - \frac{1}{c} \right)\left(1 - \exp\left(- \frac{(1-\delta)^2(\alpha-1)c^{ 1-\alpha}}{2\alpha^2}\right)\right)} \notag\\
&\buildrel (a) \over  \leq   & \frac{\left(\frac{\alpha}{\sigma\delta(\alpha-1)}\right)^{\alpha}}{\left(1 - \exp\left( \frac{\left(1-\sigma \right)^2}{2} \right)\right)\left(1 - \frac{1}{c} \right)\left(1 - \exp\left(- \frac{(1-\delta)^2(\alpha-1)c^{  1-\alpha}}{2\alpha^2}\right)\right)},
\end{eqnarray}
{where (a) follows from the fact that  $\kappa > \left(\frac{\eta (\alpha-1)}{\alpha}\right)^{\frac{\alpha}{\alpha-1}}$ with $\eta=\sigma \delta$.}
\end{itemize}

\subsubsection {When $M=\Theta(m) = \kappa_1m + o(m) = \kappa n^{\frac{1}{\alpha-1}} + o(n^{\frac{1}{\alpha-1}})$,  with $0 < \kappa_1 \leq 1$ and $0 < \kappa \leq 1$} In this case, we distinguish between two cases: $\frac{m}{M} = \frac{1}{\kappa_1} \leq 2$ and $\frac{m}{M} = \frac{1}{\kappa_1} > 2$ (see Fig. \ref{fig: regimes_M_4}).

\begin{itemize}
\item If $\frac{m}{M} = \frac{1}{\kappa_1} > 2$, we need  to consider, two further cases: $\kappa \leq \left(\frac{\eta (\alpha-1)}{\alpha}\right)^{\frac{\alpha}{\alpha-1}}$ and $\kappa > \left(\frac{\eta (\alpha-1)}{\alpha}\right)^{\frac{\alpha}{\alpha-1}}$ (See Fig. \ref{fig: regimes_M_4}).
\begin{itemize}
\item Case $\kappa \leq \left(\frac{\eta (\alpha-1)}{\alpha}\right)^{\frac{\alpha}{\alpha-1}}$: In this case, letting $\widetilde m = M^{\frac{1}{\alpha}}n^{\frac{1}{\alpha}}= \kappa^{\frac{1}{\alpha}}n^{\frac{1}{\alpha-1}}$, and following exactly the same procedure adopted in Section \ref{When$M = o(m)$th6}  when $M=o(m)$ and $\kappa \leq \left(\frac{\eta (\alpha-1)}{\alpha}\right)^{\frac{\alpha}{\alpha-1}}$, we prove the order optimality of RLFU-GCC  with  $\widetilde m = M^{\frac{1}{\alpha}}n^{\frac{1}{\alpha}}= \kappa^{\frac{1}{\alpha}}n^{\frac{1}{\alpha-1}}$. 

\item Case $\kappa > \left(\frac{\eta (\alpha-1)}{\alpha}\right)^{\frac{\alpha}{\alpha-1}}$: For this scenario we  consider two further sub-cases: $\frac{1}{\kappa_1} > \mu$ and $\frac{1}{\kappa_1} \leq \mu$ (See Fig. \ref{fig: regimes_M_4}), where $\mu>2$ is an arbitrary positive constant.

\begin{enumerate}
\item When $\frac{1}{\kappa_1} > \mu$, letting  $\widetilde m = M$ or $\widetilde m = M^{\frac{1}{\alpha}}n^{\frac{1}{\alpha}}$ and  following exactly the same procedure adopted in Section \ref{When$M = o(m)$th6}  when $M=o(m)$ and $\kappa >\left(\frac{\eta (\alpha-1)}{\alpha}\right)^{\frac{\alpha}{\alpha-1}}$,
we prove the order optimality of RLFU-GCC  with $\widetilde m = M$ or $\widetilde m = M^{\frac{1}{\alpha}}n^{\frac{1}{\alpha}}$. Note that, in this case, differently from the case when  $M=o(m)$ 
the constant $c=\ell/M$ given in \eqref{mamma1} is constrained to take values in $1 < c < \mu$ in order to guarantee that $\ell < m$.

\item When $\frac{1}{\kappa_1} \leq  \mu$, letting  $\widetilde m = M$ or $\widetilde m = M^{\frac{1}{\alpha}}n^{\frac{1}{\alpha}}$ and  following exactly the same procedure adopted in Section \ref{When$M = o(m)$th6}  when $M=o(m)$ and $\kappa >\left(\frac{\eta (\alpha-1)}{\alpha}\right)^{\frac{\alpha}{\alpha-1}}$,
we prove the order optimality of  RLFU-GCC  with $\widetilde m = M$ or $\widetilde m = M^{\frac{1}{\alpha}}n^{\frac{1}{\alpha}}$.  
Note that in this scenario, in order to guarantee that $\ell < m$, recalling that $\frac{m}{M} = \frac{1}{\kappa_1} > 2$, the constant $c$ given in \eqref{mamma1} is constrained to take values in $1 < c < 2$.

Moreover, in this case ($\frac{1}{\kappa_1} \leq \mu$), we can prove that UP-GCC is also order-optimal.  
In the following, we derive the converse and the order-optimality of UP-GCC in this regime.  
Using (\ref{eq: general achievable 1}) and Lemma \ref{lemma: H}, we have
\begin{eqnarray}
\label{eq: achievable M>1 6}
R^{\rm ub}(n,m,M,\qv, \widetilde m) \leq \frac{m}{M}-1 + o(1) = \frac{1}{\kappa_1} - 1 + o(1).
\end{eqnarray}
To compute the converse, we use the second term in (\ref{eq: general lower bound}), where the required parameters are summarized in the following:
\begin{align}
\ell &= m, \label{peterpan1} \\
r &= \frac{\alpha-1}{\alpha}\delta\left(\frac{\kappa_1}{\kappa}\right)^{\alpha-1}, \label{peterpan2}\\
\widetilde z &=   \sigma, \label{peterpan3}
%z &= \widetilde z,
\end{align}
with $0 < \delta < 1$, and $ 0<\sigma <1$  positive constants determined in the following.  
Next,  we compute each term in (\ref{eq: general lower bound}) individually.  
To this end,  using   \eqref{peterpan1}, Lemma \ref{lemma: H},  and the fact that, by assumption,    $M= \kappa_1m = \kappa n^{\frac{1}{\alpha-1}}$, following the same steps as in \eqref{eq: alpha<1 EIU 1} and \eqref{eq: alpha<1 EIU 2}, %\eqref{eq: IU M<1 1} and \eqref{eq: IU M<1 2}, 
 we have:
\begin{eqnarray}
\label{eq: EIU M>1 6 3}
\frac{\alpha-1}{\alpha}\left(\frac{\kappa_1}{\kappa}\right)^{\alpha-1} + o(1) \leq n\ell q_\ell 
\leq (\alpha-1)\left(\frac{\kappa_1}{\kappa}\right)^{\alpha-1} + o(1),
\end{eqnarray}
from which  using   (\ref{eq: P1})  and  \eqref{peterpan2}, we obtain 
\begin{eqnarray}
\label{eq: P IU M>1 6}
%&& \PP\left(\Nsf \geq 1\right) \notag\\
%&& \buildrel (a) \over \geq \PP\left(\Nsf \geq \widetilde r \right) \notag\\
P_1(\ell,r)
%& = & 1- \exp\left(- \frac{\left(n\ell q_\ell - \frac{\alpha-1}{\alpha}\delta_6\left(\frac{\kappa_1}{\kappa}\right)^{\alpha-1}\right)^2}{2n\ell q_\ell}\right) \notag\\
&\geq& 1 - \exp\left(- \frac{(1-\delta)^2(\alpha-1)}{2\alpha^2}\left(\frac{\kappa_1}{\kappa}\right)^{\alpha-1}\right) \notag\\
&\buildrel (a) \over \geq& 1 - \exp\left(- \frac{(1-\delta)^2(\alpha-1)}{2\alpha^2}\left(\frac{1}{\mu}\right)^{\alpha-1}\right)
\end{eqnarray}

where %(a) is because that $\Nsf$ is an integer such that $\PP(\Nsf \geq 1) = \PP(\Nsf \geq \widetilde r)$ for $\widetilde r<1$ and $\PP(\Nsf \geq 1) \geq \PP(\Nsf \geq \widetilde r)$ for $\widetilde r \geq 1$ and 
(a) is because $\frac{\kappa_1}{\kappa} \geq \frac{1}{\mu}$, while  
using (\ref{eq: P2}),   and \eqref{peterpan3} 
%and (\ref{eq: EZb M>1 6}), 
we have
\begin{eqnarray}
\label{eq: P Zb M>1 6}
P_2(\ell, 1, \widetilde z) %&=& \PP(\Zsf \geq \sigma_6) \notag \\
%&\geq& 1- \exp\left(- \frac{\left(\ell\left(1-\left(1-\frac{1}{\ell}\right)^r\right) - \sigma_6\right)^2}{2\EE[\Zsf]}\right) \notag\\
&=& 1 - \exp\left( \frac{\left(1-\sigma \right)^2}{2} \right). 
\end{eqnarray} 
Replacing   \eqref{peterpan1}-\eqref{peterpan3},   \eqref{eq: P IU M>1 6}  and  \eqref{eq: P Zb M>1 6} in the second term 
of  \eqref{eq: general lower bound} given in   Theorem \ref{theorem: general lower bound}, we obtain
\begin{eqnarray}
\label{eq: converse M>1 6}
R^{\rm lb}(n,m,M,\qv)  & \geq&  P_1(\ell,r) P_2(\ell, r, \widetilde z)(1-M/{\ell}) \notag\\
& \buildrel (a) \over  \geq & \Xi \cdot \left(1-\frac{M}{m}\right) \notag\\
& \buildrel (b) \over  \geq &\Xi  \cdot \left(1-\frac{1}{2}\right),
\end{eqnarray}
where in (a) we have defined $\Xi$ as: 
$$\Xi= \left(1 - \exp\left(- \frac{(1-\delta)^2(\alpha-1)}{2\alpha^2}\left(\frac{1}{\mu}\right)^{\alpha-1}\right)\right) \left(1 - \exp\left( \frac{\left(1-\sigma \right)^2}{2} \right)\right),
$$
while in (b) we have used the fact that  $\frac{m}{M} = \frac{1}{\kappa_1} > 2$. 
From \eqref{eq: converse M>1 6} using (\ref{eq: achievable M>1 6}), we have 
\begin{eqnarray}
\frac{R^{\rm ub}(n,m,M,\qv, \widetilde m)}{R^{\rm lb}(n,m,M,\qv) } &\leq& \frac{\frac{1}{\kappa_1} + o(1).}{\Xi \cdot  \left(1-\frac{1}{2}\right)} \notag\\
&\buildrel (a) \over  \leq & \frac{2 \mu}{\Xi}, \notag\\
\end{eqnarray}
where in (a) we have used the fact that  $\frac{1}{\kappa_1} \leq  \mu$.

\end{enumerate}

\end{itemize}

\item {If  $\frac{m}{M} = \frac{1}{\kappa_1} \leq 2$ (see Fig. \ref{fig: regimes_M_4}) letting $\widetilde{m} = m$, 
and using (\ref{eq: general achievable 1}), we have
\begin{eqnarray}
\label{eq: achievable M>1 6 1}
R^{\rm ub}(n,m,M,\qv, \widetilde m) \leq \frac{m}{M}-1 + o(1) = \frac{1}{\kappa_1} - 1 + o(1).
\end{eqnarray}
To derive the converse, as before,  we use the second term of (\ref{eq: general lower bound})
given in Theorem \ref{theorem: general lower bound}  where the parameters $\ell, r, \widetilde z$ are given as in 
\eqref{peterpan1}-\eqref{peterpan3}. Following the same steps adopted in Eqs. \eqref{eq: EIU M>1 6 3}-\eqref{eq: converse M>1 6}, we obtain 
%
%%$\PP(\Zsf \geq 1)$ is given by (\ref{eq: P Zb M>1 6}) and 
%%by using (\ref{eq: P IU M>1 6}), 
%we obtain
%\begin{eqnarray}
%\label{eq: P IU M>1 6 1}
%&& P_1(\ell,r) \notag\\
%%&&\geq \PP\left(\Nsf \geq \widetilde r \right) \notag\\
%&& = 1- \exp\left(- \frac{\left(n\ell q_\ell - \frac{\alpha-1}{\alpha}\delta_6\left(\frac{\kappa_1}{\kappa}\right)^{\alpha-1}\right)^2}{2n\ell q_\ell}\right) \notag\\
%&&\geq 1 - \exp\left(- \frac{(1-\delta_6)^2(\alpha-1)}{2\alpha^2}\left(\frac{\kappa_1}{\kappa}\right)^{\alpha-1}\right) \notag\\
%&& \geq 1 - \exp\left(- \frac{(1-\delta_6)^2(\alpha-1)}{2\alpha^2}\left(\frac{1}{2}\right)^{\alpha-1}\right)
%\end{eqnarray}
%
%
%Let %$z=1$ and 
%Given all the required parameters $\ell, z_a, r, \widetilde z$, by using Theorem \ref{theorem: general lower bound}, (\ref{eq: P IU M>1 6 1}) and (\ref{eq: P Zb M>1 6}), we obtain
%\begin{eqnarray}
%\label{eq: converse M>1 6 1}
%&&R^{\rm lb}(n,m,M,\qv)  \notag\\
%&& \geq  P_1(\ell,r) P_2(\ell, 1, \widetilde z)(1-M/{\ell}) \notag\\
%&& \geq \left(1 - \exp\left(- \frac{(1-\delta_6)^2(\alpha-1)}{2\alpha^2}\left(\frac{1}{2}\right)^{\alpha-1}\right)\right) \notag\\
%&& \quad  \left(1 - \exp\left( \frac{\left(1-\sigma_6 \right)^2}{2} \right)\right) \left(1-\frac{M}{m}\right). \notag\\
%\end{eqnarray}
%Then, by using (\ref{eq: achievable M>1 6 1}) and (\ref{eq: converse M>1 6 1}), we have
\begin{eqnarray}
\frac{R^{\rm ub}(n,m,M,\qv, \widetilde m)}{R^{\rm lb}(n,m,M,\qv) } &\leq& \frac{\frac{1}{\kappa_1} + o(1).}{\Xi \cdot  \left(1-\frac{1}{2}\right)} \notag\\
&\buildrel (a) \over  \leq & \frac{2 }{\Xi}, \notag\\
\end{eqnarray}
where in (a) we have used the fact that  $\frac{1}{\kappa_1} \leq  2$.
}

%\begin{eqnarray}
%\frac{R^{\rm ub}(n,m,M,\widetilde \pv, \qv)}{R^{\rm lb}(n,m,M,\qv) } &\leq& \frac{\frac{m}{M} - 1 + o\left(1\right)}{\left(1 - \exp\left(- \frac{(1-\delta_6)^2(\alpha-1)}{2\alpha^2}\left(\frac{1}{2}\right)^{\alpha-1}\right)\right)\left(1 - \exp\left( \frac{\left(1-\sigma_6 \right)^2}{2} \right)\right) \left(1-\frac{M}{m}\right)} \notag\\
%&\leq& \frac{\frac{m}{M}\left(1-\frac{M}{m}\right)}{\left(1 - \exp\left(- \frac{(1-\delta_6)^2(\alpha-1)}{2\alpha^2}\left(\frac{1}{2}\right)^{\alpha-1}\right)\right)\left(1 - \exp\left( \frac{\left(1-\sigma_6 \right)^2}{2} \right)\right) \left(1-\frac{M}{m}\right)} \notag\\
%&\buildrel (a) \over \leq& \frac{2}{\left(1 - \exp\left(- \frac{(1-\delta_6)^2(\alpha-1)}{2\alpha^2}\left(\frac{1}{2}\right)^{\alpha-1}\right)\right)\left(1 - \exp\left( \frac{\left(1-\sigma_6 \right)^2}{2} \right)\right)},
%\end{eqnarray}
%which shows the order-optimality of the achievable expected rate, and where (a) is because $\frac{m}{M} \leq 2$.
%

\end{itemize}

\subsection{Region of  $\kappa \geq 1$}  
%(under the case $M = \Theta(n^{\frac{1}{\alpha-1}}) = \kappa n^{\frac{1}{\alpha-1}} + o(n^{\frac{1}{\alpha-1}})$ 
In this case we need to consider two further sub-cases: $M=o(m)$ and $M=\Theta(m)$ (see Fig. \ref{fig: regimes_M_4})).

\subsubsection{When $M = o(m)$} 
\label{When$M=o(m)kappa>1$}
Letting $\widetilde{m} = M = \kappa n^{\frac{1}{\alpha-1}} + o(n^{\frac{1}{\alpha-1}})$, by using (\ref{eq: general achievable 1}) and Lemma \ref{lemma: H}, 
%and the fact that  $G_{\widetilde m} = \sum_{f=1}^{\widetilde m}q_f$, 
we obtain
\begin{eqnarray}
\label{eq: achievable M>1 7}
R^{\rm ub}(n,m,M,\qv, \widetilde m)
%&\leq& \left(\frac{{\widetilde m}} {M}-1\right)\left( 1 -\left(1-\frac{M}{\widetilde m} \right)^{ n \, G_{\widetilde m}} \right) +  (1-G_{\widetilde m})\,n \notag\\
&\leq&  (1-G_{\widetilde m})\,n \notag\\
&\leq& \frac{n}{M^{\alpha-1}} + o\left(\frac{n}{M^{\alpha-1}}\right) \notag\\
&=& \kappa^{1-\alpha} + o(1).
\end{eqnarray}
To compute the converse, similar as before, we use the second term of  (\ref{eq: general lower bound}) in Theorem \ref{theorem: general lower bound}, where the required parameters are summarized in the following:
\begin{align}
\ell &= cM, \label{sola1}\\
r &= \frac{\delta(\alpha-1)c^{1-\alpha}\kappa^{1-\alpha}}{\alpha}, \label{sola2}\\
\widetilde z &=   \sigma, \label{sola3}
%z &= \widetilde z, \notag
\end{align}
with  $c>1$, $0 < \delta < 1$, and $ 0<\sigma <1$  positive constants determined in the following.  
Next,  we compute each term in (\ref{eq: general lower bound}) individually.  
To this end,  using   \eqref{sola1}, Lemma \ref{lemma: H},  and the fact that, by assumption,  
$M = \kappa n^{\frac{1}{\alpha-1}} + o(n^{\frac{1}{\alpha-1}})$,
 following the same steps as in \eqref{eq: EIU M>1 9 1} and \eqref{eq: EIU M>1 9 2},  we have:
%Let  $\ell = cM$, where $c>1$ is an arbitrary positive number larger than one, by using %(\ref{eq: EIU}) and 
%Lemma \ref{lemma: H}, following the the same steps as (\ref{eq: EIU M>1 4 1}) and (\ref{eq: EIU M>1 4 2}), then we can write
\be
\label{eq: EIU M>1 7}
\frac{(\alpha-1)c^{1-\alpha}\kappa^{1-\alpha}}{\alpha} + o(1) \leq n\ell q_\ell
\leq (\alpha-1)c^{1-\alpha}\kappa^{1-\alpha} + o(1),
\ee
%We let $r = 1$, then we obtain
%\begin{eqnarray}
%&& \ell\left(1-\left(1-\frac{1}{\ell}\right)^r\right) \notag\\ 
%&& = m \left(1 - \left(1 - \frac{1}{m}\right)^1 \right) \notag\\
%&& = 1.
%\end{eqnarray}  
from which 
%Then, let $r = \frac{\delta_7(\alpha-1)c^{1-\alpha}\kappa^{1-\alpha}}{\alpha}$, where $0 < \delta_7  < 1$, we observe that $r<1$. By 
using (\ref{eq: P1})  and  (\ref{sola2}), we obtain
\begin{eqnarray}
\label{eq: P IU M>1 7}
&& P_1(\ell,r) \notag\\
%&& \buildrel (a) \over = \PP\left(\Nsf \geq \widetilde r\right) \notag\\
%&& = 1- \exp\left(- \frac{\left(n\ell q_\ell - \frac{\delta_7(\alpha-1)c^{1-\alpha}\kappa^{1-\alpha}}{\alpha}\right)^2}{2n\ell q_\ell}\right) \notag\\
&&\geq 1 - \exp\left(- \frac{(1-\delta)^2(\alpha-1)c^{1-\alpha}}{2\alpha^2}\kappa^{1-\alpha}\right) \notag\\
&& \buildrel (a) \over \geq \frac{(1-\delta)^2(\alpha-1)c^{1-\alpha}}{2\alpha^2}\kappa^{1-\alpha} - \frac{1}{2}\left(\frac{(1-\delta)^2(\alpha-1)c^{1-\alpha}}{2\alpha^2}\kappa^{1-\alpha}\right)^2,
\end{eqnarray}
where %(a) is because that $\Nsf$ is an integer such that $\PP(\Nsf \geq 1) = \PP(\Nsf \geq \widetilde r)$ for $\widetilde r < 1$ and 
(a) follows from the fact that  $1-e^{-x} \geq x - \frac{x^2}{2}$. Furthermore,  using (\ref{eq: P2}) and (\ref{sola3}), we have
\begin{eqnarray}
\label{eq: P Zb M>1 7}
P_2(\ell, 1, \widetilde z) %&=& \PP(\Zsf \geq \sigma_7) \notag \\
%&\geq& 1- \exp\left(- \frac{\left(\ell\left(1-\left(1-\frac{1}{\ell}\right)^r\right) - \widetilde{z}\right)^2}{2\EE[\Zsf]}\right) \notag\\
&=& 1 - \exp\left(-\frac{\left(1-\sigma \right)^2}{2} \right),
\end{eqnarray} 
from which replacing   (\ref{sola1})-(\ref{sola3}),  \eqref{eq: P IU M>1 7} and  \eqref{eq: P Zb M>1 7} in  the second term of  
 (\ref{eq: general lower bound}), we obtain:
\begin{eqnarray}
\label{eq: converse M>1 7}
R^{\rm lb}(n,m,M,\qv)   &\geq&  P_1(\ell,r) P_2(\ell, 1, \widetilde z)(1-M/{\ell}) \notag\\
%&& \geq \left(\frac{\rho_7(1-\delta_7)^2(\alpha-1)c^{1-\alpha}}{2\alpha^2}\kappa^{1-\alpha} - \frac{1}{2}\left(\frac{\rho_7(1-\delta_7)^2(\alpha-1)c^{1-\alpha}}{2\alpha^2}\kappa^{1-\alpha}\right)^2\right) \notag\\
%&& \quad \left(1 - \exp\left( \frac{\left(1-\sigma_7 \right)^2}{2} \right)\right)\left(1-\frac{1}{c}\right) \notag\\
&\geq& \Xi \cdot \left(1-\frac{1}{c}\right), \notag\\
\end{eqnarray}
with 
$$
\Xi = \left(\frac{(1-\delta)^2(\alpha-1)c^{1-\alpha}}{2\alpha^2}\kappa^{1-\alpha} - \frac{1}{2}\left(\frac{(1-\delta)^2(\alpha-1)c^{1-\alpha}}{2\alpha^2}\right)^2\left(\kappa^{1-\alpha}\right)^2\right) \left(1 - \exp\left( \frac{\left(1-\sigma \right)^2}{2} \right)\right).
$$
Using (\ref{eq: achievable M>1 7}) and (\ref{eq: converse M>1 7}), we obtain
\begin{eqnarray}
 \frac{R^{\rm ub}(n,m,M, \qv, \widetilde m)}{R^{\rm lb}(n,m,M,\qv) } &\leq &\frac{\kappa^{1-\alpha}}{ \Xi \cdot \left(1-\frac{1}{c}\right)} \notag\\
%&& = \frac{\frac{1}{(1-o(1))}}{\left(\frac{\rho_7(1-\delta_7)^2(\alpha-1)c^{1-\alpha}}{2\alpha^2} - \frac{1}{2}\left(\frac{\rho_7(1-\delta_7)^2(\alpha-1)c^{1-\alpha}}{2\alpha^2}\right)^2\left(\kappa^{1-\alpha}\right)\right)\left(1 - \exp\left( \frac{\left(1-\sigma_7 \right)^2}{2} \right)\right)\left(1-\frac{1}{c}\right)} \notag\\
& \buildrel (a) \over \leq & \frac{1}{\left(\frac{(1-\delta)^2(\alpha-1)c^{1-\alpha}}{2\alpha^2} - \frac{1}{2}\left(\frac{(1-\delta)^2(\alpha-1)c^{1-\alpha}}{2\alpha^2}\right)^2\right)\left(1 - \exp\left( \frac{\left(1-\sigma \right)^2}{2} \right)\right)\left(1-\frac{1}{c}\right)}, \notag
\end{eqnarray}
where (a) is because $\kappa \geq 1$.

\subsubsection{When $M = \Theta(m) = \kappa_1m + o(m)$ with  $\kappa \geq 1$} In this case, we distinguish between two cases: $\frac{m}{M} = \frac{1}{\kappa_1} \leq 2$ and $\frac{m}{M} = \frac{1}{\kappa_1} > 2$, where $0 < \kappa_1 < 1$ (see Fig. \ref{fig: regimes_M_4}).

\begin{itemize}
\item If $\frac{m}{M} = \frac{1}{\kappa_1} > 2$, letting  $\widetilde m = M$ %and for the achievable rate and 
and  following exactly the same procedure adopted in Section \ref{When$M=o(m)kappa>1$}, 
we prove the order optimality of RLFU-GCC  with $\widetilde m = M$.  
Note that in this scenario,  differently from Section \ref{When$M=o(m)kappa>1$}, in order to guarantee that $\ell < m$, recalling that $\frac{m}{M} = \frac{1}{\kappa_1} > 2$, the constant $c$ given in \eqref{mamma1} is constrained to take values in $1 < c < 2$.

\item If $\frac{m}{M} = \frac{1}{\kappa_1} \leq 2$,  
leting $\widetilde{m} = M$, using (\ref{eq: general achievable 1}) and Lemma \ref{lemma: H}, we have
\begin{eqnarray}
\label{eq: achievable M>1 8}
R^{\rm ub}(n,m,M,\qv, \widetilde m) &\leq& (1-G_{\widetilde m})\,n \notag\\ 
&=&  n\sum_{f=\widetilde m+1}^{m}q_f \notag\\
&=& n \frac{H(\alpha,\widetilde m+1, m)}{H(\alpha,1, m)}\notag\\
&\leq& n\frac{\frac{1}{1-\alpha}m^{1-\alpha} - \frac{1}{1-\alpha}(\widetilde m+1)^{1-\alpha} + \frac{1}{(\widetilde m+1)^{\alpha}}}{\frac{1}{1-\alpha}(m+1)^{1-\alpha} - \frac{1}{1-\alpha}} \notag\\
&=& n (-m^{1-\alpha} + (\kappa_1 m + 1)^{1-\alpha}) + o(1) \notag\\
&\leq& \left(\kappa_1^{1-\alpha} - 1\right)nm^{1-\alpha} + o\left(1\right)  + o(1)\notag\\
&=& \left(\kappa_1^{1-\alpha} - 1\right)n\left(\frac{\kappa}{\kappa_1} n ^{\frac{1}{\alpha-1}}\right)^{1-\alpha}  + o(1)\notag\\
&=& \left(\kappa_1^{1-\alpha} - 1\right)\kappa^{1-\alpha} \left(\frac{1}{\kappa_1}\right)^{1-\alpha}  + o(1)\notag\\
&=& \kappa^{1-\alpha}\left(1 - \kappa_1^{\alpha-1}\right) + o(1).
\end{eqnarray}

To compute the converse, similar as before, we use the second term of  (\ref{eq: general lower bound}) in Theorem \ref{theorem: general lower bound}, where the required parameters are summarized in the following:
\begin{align}
\ell &= m, \label{marc1}\\
r &= \frac{\alpha-1}{\alpha}\delta\left(\frac{\kappa_1}{\kappa}\right)^{\alpha-1}, \label{marc2}\\
\widetilde z &=   \sigma, \label{marc3} %\notag\\
%z &= \widetilde z, \notag
\end{align}
with  $0 < \delta < 1$, and $ 0<\sigma <1$  positive constants determined in the following.   Recall here that, by assumption, 
$\kappa>1$ while $0 < \kappa_1 < 1$. Next,  we compute each term in (\ref{eq: general lower bound}) individually.  
To this end,  using   \eqref{marc1}, Lemma \ref{lemma: H},  and the fact that, by assumption,  
$M = \kappa n^{\frac{1}{\alpha-1}} =\kappa_1m$ (see Fig. \ref{fig: regimes_M_4}),
 following the same steps as in \eqref{eq: EIU M>1 9 1} and \eqref{eq: EIU M>1 9 2},  we have:
 \begin{eqnarray}
\label{eq: EIU M>1 8 3}\frac{\alpha-1}{\alpha}\left(\frac{\kappa_1}{\kappa}\right)^{\alpha-1} + o(1) \leq n\ell q_\ell \leq (\alpha-1)\left(\frac{\kappa_1}{\kappa}\right)^{\alpha-1} + o(1),
\end{eqnarray}
%Let $r = 1$, then by using (\ref{eq: EZb}), we obtain
%\begin{eqnarray}
%\label{eq: EZb M>1 8}
%\ell\left(1-\left(1-\frac{1}{\ell}\right)^r\right) = 1.
%\end{eqnarray}  
from which using  using (\ref{eq: P1} and  \eqref{marc2}, we obtain 
%Then, let $\widetilde r =\frac{\alpha-1}{\alpha}\delta_8\left(\frac{\kappa_1}{\kappa}\right)^{\alpha-1}$, where $0 < \delta_8 < 1$ is some positive constant determined later, we can observe that $r<1$. By using (\ref{eq: P1}) and (\ref{eq: EIU M>1 8 3}), we obtain
\begin{eqnarray}
\label{eq: P IU M>1 8}
&& P_1(\ell,r) \notag\\
%&& \buildrel (a) \over = \PP\left(\Nsf \geq \widetilde r\right) \notag\\
%&& = 1- \exp\left(- \frac{\left(n\ell q_\ell - \frac{\alpha-1}{\alpha}\delta_8\left(\frac{\kappa_1}{\kappa}\right)^{\alpha-1}\right)^2}{2n\ell q_\ell}\right) \notag\\
&&\geq 1 - \exp\left(- \frac{(1-\delta)^2(\alpha-1)}{2\alpha^2}\left(\frac{\kappa_1}{\kappa}\right)^{\alpha-1}\right) \notag\\
&&\buildrel (a) \over \geq \frac{(1-\delta)^2(\alpha-1)}{2\alpha^2}\left(\frac{\kappa_1}{\kappa}\right)^{\alpha-1} - \frac{1}{2}\left(\frac{(1-\delta)^2(\alpha-1)}{2\alpha^2}\left(\frac{\kappa_1}{\kappa}\right)^{\alpha-1}\right)^2,
\end{eqnarray}
where (a) follows from the fact that  $1-e^{-x} \geq x - \frac{x^2}{2}$.
%where (a) is because that $\Nsf$ is an integer such that $P_1(\ell,r) \geq \PP\left(\Nsf \geq \widetilde r\right)$ for $\widetilde r<1$.
Furthemore, using using (\ref{eq: P2}) and \eqref{marc3}, we have:
\begin{eqnarray}
\label{eq: P Zb M>1 8}
P_2(\ell, 1, \widetilde z) %&=& \PP(\Zsf \geq \sigma_8) \notag \\
%&\geq& 1- \exp\left(- \frac{\left(\ell\left(1-\left(1-\frac{1}{\ell}\right)^r\right) - \sigma_8\right)^2}{2\EE[\Zsf]}\right) \notag\\
&=& 1 - \exp\left( \frac{\left(1-\sigma\right)^2}{2} \right). 
\end{eqnarray} 
from which,  replacing \eqref{marc1}-\eqref{marc3}, \eqref{eq: P IU M>1 8}, and \eqref{eq: P Zb M>1 8} in the second term 
of (\ref{eq: general lower bound}),   we obtain
\begin{eqnarray}
\label{eq: converse M>1 8}
&&R^{\rm lb}(n,m,M,\qv)  \notag\\
&& \geq  P_1(\ell,r) P_2(\ell, 1, \widetilde z)(1-M/{\ell}) \notag\\
&& \geq  \Xi \cdot  \left(1-\frac{M}{m}\right),
\end{eqnarray}
with 
$$
\Xi= \left( \frac{(1-\delta)^2(\alpha-1)}{2\alpha^2}\left(\frac{\kappa_1}{\kappa}\right)^{\alpha-1} - \frac{1}{2}\left(\frac{(1-\delta)^2(\alpha-1)}{2\alpha^2}\left(\frac{\kappa_1}{\kappa}\right)^{\alpha-1}\right)^2\right) \left(1 - \exp\left( \frac{\left(1-\sigma \right)^2}{2} \right)\right)
$$
Using (\ref{eq: achievable M>1 8}) and (\ref{eq: converse M>1 8}), we obtain
\begin{eqnarray}
\label{eq: gap M>1 8}
 \frac{R^{\rm ub}(n,m,M, \qv, \widetilde m)}{R^{\rm lb}(n,m,M,\qv) } &\leq &
 \frac{\kappa^{1-\alpha}\left(1 - \kappa_1^{\alpha-1}\right)}{\Xi \left(1-\frac{M}{m}\right)} \notag\\
& \leq& \frac{\kappa^{1-\alpha}\left(1 - \left(\frac{M}{m}\right)^{\alpha-1}\right)}{\Xi \left(1-\frac{M}{m}\right)} \notag\\
%&& \leq \frac{\left(1 - \left(\frac{M}{m}\right)^{\alpha-1}\right)}{\left( \frac{(1-\delta_8)^2(\alpha-1)}{2\alpha^2}\kappa_1^{\alpha-1} - \frac{1}{2}\left(\frac{(1-\delta_8)^2(\alpha-1)}{2\alpha^2}\right)^2\kappa_1^{2(\alpha-1)}\kappa^{1-\alpha}\right) \left(1 - \exp\left( \frac{\left(1-\sigma_8 \right)^2}{2} \right)\right)\left(1-\frac{M}{m}\right)}
 \notag\\
& \leq &\frac{\kappa^{1-\alpha}}{\Xi} \frac{1 - \left(\frac{M}{m}\right)^{\alpha-1}}{1-\frac{M}{m}}. \notag\\
\end{eqnarray}
Next note that, using the fact that  $\kappa>1$ and  $0 < \kappa_1 < 1$,
\be
\frac{\kappa^{1-\alpha}}{\Xi}  \leq \frac{(1-\delta)^2(\alpha-1)}{2\alpha^2}\frac{1}{2^{\alpha-1}} - \frac{1}{2}\left(\frac{(1-\delta)^2(\alpha-1)}{2\alpha^2}\right)^2, 
\label{eq: gap M>1 8 2}
\ee
with  $\delta$ selected such that the right-hand side of \eqref{eq: gap M>1 8 2} is positive. 
%\be
%\frac{(1-\delta_8)^2(\alpha-1)}{2\alpha^2}\frac{1}{2^{\alpha-1}} - \frac{1}{2}\left(\frac{(1-\delta_8)^2(\alpha-1)}{2\alpha^2}\right)^2 > 0.
%\ee
Furthermore 
\begin{subequations}
\begin{empheq}[left={\vspace{-5cm}\displaystyle \frac{1 - \left(\frac{M}{m}\right)^{\alpha-1}}{1-\frac{M}{m}}  \leq }\empheqlbrace]{align}
&1,\,\, \,\,  \,\, \,\,\alpha \leq 2 \label{pazza1} \\
&\alpha - 1 , \,\, \,\, \,\, \,\, \,\, \,\,\alpha > 2    \label{pazza2} 
 \end{empheq}
\label{pazza0}  
\end{subequations}
% when $\alpha \leq 2$, we have
%\be
%\label{eq: gap M>1 8 2}
%\frac{1 - \left(\frac{M}{m}\right)^{\alpha-1}}{1-\frac{M}{m}} \leq 1.
%\ee
%When $\alpha > 2$, we have
%\be
%\label{eq: gap M>1 8 3}
%\frac{1 - \left(\frac{M}{m}\right)^{\alpha-1}}{1-\frac{M}{m}} \leq \alpha - 1.
%\ee
from which, using (\ref{eq: gap M>1 8}) and  (\ref{pazza0}), we obtain
\be
\frac{R^{\rm ub}(n,m,M, \qv, \widetilde m)}{R^{\rm lb}(n,m,M,\qv) } \leq \frac{\max\{1, \alpha-1\}}{(1-o(1))\left( \frac{(1-\delta)^2(\alpha-1)}{2\alpha^2}\frac{1}{2^{\alpha-1}} - \frac{1}{2}\left(\frac{(1-\delta)^2(\alpha-1)}{2\alpha^2}\right)^2\right)\left(1 - \exp\left( \frac{\left(1-\sigma \right)^2}{2} \right)\right)}. 
\notag
\ee 
\end{itemize}

\section{Proof of Corollary \ref{corollary: m constant}}
\label{sec: proof of corollary m constant}

To show Corollary \ref{corollary: m constant}, we follow the same procedure as the proof of Theorem \ref{theorem: gamma < 1}. By using Lemma \ref{lemma: achievable}, it is straightforward to see
\be
\label{eq: up m constant}
R^{\rm ub}(n,m,M, \qv,\widetilde m) = \min\left\{\frac{m}{M}-1, m\right\}.
\ee
To evaluate the converse in Theorem \ref{theorem: general lower bound},
we compute each term in  (\ref{eq: general lower bound})  individually using  $\ell$, $r$, $\widetilde z$, $z$,  as  summarized in the following.
\begin{align}
\ell &= m, \label{dentista1}\\
r &= \delta\frac{m^{1-\alpha}}{H(\alpha, 1, m)} n,\label{dentista2}\\
\widetilde{z} 
&=   \max\{(1-\epsilon) m, 1\}, 
\label{dentista3}
\end{align}
with  $0 < \delta < 1$  positive constants determined in the following and $0<\epsilon<\frac{1}{2}$  an arbitrarily small constant. Note that by assumption, due to the fact that $n \rightarrow \infty$ and $m$ is kept constant, we have $\widetilde z < r$.
Next,  we compute each term in (\ref{eq: general lower bound}) individually.  
To this end,  using   \eqref{sola1},  we have:
\begin{eqnarray}
\label{eq: alpha<1 EIU 1 c}
n\ell q_\ell% &=& n m \frac{m^{-\alpha}}{H(\alpha, 1, m)} \notag\\
&=& n \frac{m^{1-\alpha}}{H(\alpha, 1, m)},
\end{eqnarray}
from which, using (\ref{eq: P1}) and  using \eqref{sola2}, we obtain
\begin{eqnarray}
\label{eq: IU 5 c}
%P_1(\ell,r) 
P_1(\ell,r) %&=& 1- \exp\left(- \frac{(n\ell q_\ell - \delta_{16}\frac{m^{1-\alpha}}{H(\alpha, 1, m)} n)^2}{2n\ell q_\ell}\right) \notag\\
&=& 1- \exp\left(- \frac{(1- \delta)^2m^{1-\alpha}}{2 H(\alpha, 1, m)} n\right) \notag\\
%&=& 1- \exp\left(- \frac{((1-\delta)(1-\alpha)n + o(n))^2}{2(\delta(1-\alpha)n + o(n))}\right) \notag\\
&=& 1 - o(1).
\end{eqnarray}

%From \eqref{dentista1} Theorem \ref{theorem: general lower bound}, %in Section \ref{sec: The General Lower Bound of the Achievable Rate}, 
%we can see that it requires 
%$n\ell q_\ell$ and $\ell\left(1-\left(1-\frac{1}{\ell}\right)^r\right)$. 
%
%Let $\ell = m$, 
%we can write
%\begin{eqnarray}
%\label{eq: alpha<1 EIU 1 c}
%n\ell q_\ell &=& n m \frac{m^{-\alpha}}{H(\alpha, 1, m)} \notag\\
%&=& n \frac{m^{1-\alpha}}{H(\alpha, 1, m)}.
%\end{eqnarray}
%and
%\begin{eqnarray}
%\label{eq: alpha<1 EIU 2}
%n\ell q_\ell &=& n m \frac{m^{-\alpha}}{H(\alpha, 1, m)} \notag\\
%&\leq& \frac{nm^{1-\alpha}}{\frac{1}{1-\alpha}(m+1)^{1-\alpha} - \frac{1}{1-\alpha}} \notag\\
%&=& (1-\alpha)n + o(n).
%\end{eqnarray}
%Thus, by using (\ref{eq: alpha<1 EIU 1}) and (\ref{eq: alpha<1 EIU 2}), we have
%\be
%\label{eq: IU 5}
%n\ell q_\ell = (1-\alpha)n + o(n).
%\ee
%
Using  \eqref{sola1} and  \eqref{sola2},  we have %by using (\ref{eq: EZb}), we can write
\begin{eqnarray}
\ell\left(1-\left(1-\frac{1}{\ell}\right)^r\right) %&=& \ell\left(1-\left(1-\frac{1}{\ell}\right)^r\right) \notag\\
&=& m \left(1 - \left(1 - \frac{1}{m}\right)^{\delta\frac{m^{1-\alpha}}{H(\alpha, 1, m)} n} \right) \notag\\
&=& m \left(1 - o(1) \right),
\end{eqnarray} 
from which using  (\ref{eq: P2}) and \eqref{sola3}, we have
%\be
%\label{eq: gamma < 1 zb 1 c}
%\widetilde{z} =  \max\{(1-\epsilon) m, 1\}, %\min\left\{\sigma m \left(1 - \exp\left(-\delta(1-\alpha)\frac{n}{m}\right)\right), \delta(1-\alpha)n\right\} = \sigma m \left(1 - \exp\left(-\delta(1-\alpha)\frac{n}{m}\right)\right),
%\ee 
%where  
%then by, we have
\begin{eqnarray}
\label{eq: Zb 6 c}
P_2(\ell, r, \widetilde z) = 1- o(1).
\end{eqnarray} 
Replacing \eqref{sola1}-\eqref{sola3},  \eqref{eq: IU 5 c} and \eqref{eq: Zb 6 c} in  (\ref{eq: general lower bound}),
%Thus, by using Theorem \ref{theorem: general lower bound}, %and (\ref{eq: R lower ground}), 
we obtain
\begin{eqnarray}
\label{eq: R lower 1 c}
&&R^{\rm lb}(n,m,M,\qv) \notag\\
%&&\buildrel (a) \over \geq 
%P_1(\ell,r) P_2(\ell, r, \widetilde z)\max_{z \in \{1, \cdots, \widetilde{z}\}} (z-zM/{\lfloor \ell/z \rfloor}) \notag\\
%&& \geq (1-o(1)) (1-o(1)) \max_{z \in \{1, \cdots, \widetilde{z}\}} (z-zM/{\lfloor \ell/z \rfloor}) \notag\\
&& \geq (1-o(1))^2\max_{z \in \{1, \cdots,  \lceil\widetilde{z} \rceil \}} z(1-M/{\lfloor \ell/z \rfloor}) \notag\\
&& \geq (1-o(1))^2\max_{z \in \{1, \cdots,  \lceil \max\{(1-\epsilon) m, 1\} \rceil \}} z(1-M/{\lfloor \ell/z \rfloor}).
\end{eqnarray}

%where (a) is because $\widetilde{z} \leq r$. %In the following, we consider two cases, which are $n = \omega(m)$ and $n=O(m)$.
%Since $n = \omega(m)$, by using (\ref{eq: gamma < 1 zb 1}), we can compute $\widetilde{z}$ as
%\begin{eqnarray}
%%\label{eq: gamma < 1 zb 1}
%\widetilde{z} %&=& \min\left\{\sigma m \left(1 - \exp\left(- \frac{n}{m}\delta(1-\alpha)\right)\right), \delta(1-\alpha)n\right\} \notag\\
%\buildrel (a) \over = \sigma m \left(1 - o(1)\right),
%\end{eqnarray}
%where (a) is because that $\frac{n}{m} = \omega(1)$. Then, by using (\ref{eq: R lower 1}), we have
%\begin{eqnarray}
%\label{eq: converse alpha<1 1}
%R^{\rm lb}(n,m,M,\qv)  \geq (1-o(1))^2\max_{z \in \{1, \cdots, \sigma m \left(1 - o(1)\right)\}} (z-zM/{\lfloor \ell/z \rfloor})
%\end{eqnarray}

\subsection{When $M \leq \frac{1}{2}$}
In this case, letting $z = \max\{(1-\epsilon) m, 1\}$, and  using (\ref{eq: R lower 1 c}), we have
\be
\label{eq: up m constant 1}
R^{\rm lb}(n,m,M,\qv)  \geq (1-o(1))^2  \max\{(1-\epsilon) m, 1\} \left(1-M\right). 
\ee
Then, by using (\ref{eq: up m constant}) and (\ref{eq: up m constant 1}), we obtain
\begin{eqnarray}
\label{eq: gap alpha<1 2 c}
\frac{R^{\rm ub}(n,m,M, \qv, \widetilde m)}{R^{\rm lb}(n,m,M,\qv) } &\leq& \frac{m}{(1-o(1))^2  \max\{(1-\epsilon) m, 1\} \left(1-M\right)} \notag\\
&\leq& \frac{m}{\frac{1}{2} \max\{(1-\epsilon) m, 1\} } \notag\\
&\leq& \frac{2}{1-\epsilon}.
\end{eqnarray}

\subsection{When $\frac{1}{2} < M \leq 1+\epsilon$}

If $m<3$, letting $z=1$, by using (\ref{eq: R lower 1 c}),  we have
\be
\label{eq: lb m constant}
R^{\rm lb}(n,m,M,\qv)  \geq (1-o(1))^2 \left(1-\frac{M}{m}\right).
\ee
By using (\ref{eq: up m constant}) and (\ref{eq: lb m constant}), we obtain
\be
\frac{R^{\rm ub}(n,m,M, \qv, \widetilde m)}{R^{\rm lb}(n,m,M,\qv) } \leq \frac{\frac{m}{M}-1}{(1-o(1))^2 \left(1-\frac{M}{m}\right)} = \frac{m}{M} < 6.
\ee

If $m \geq 3$, letting $z = \lfloor \frac{m}{2}\rfloor$,  and  using (\ref{eq: R lower 1 c}), we have
\be
\label{eq: lb m constant 2}
R^{\rm lb}(n,m,M,\qv)  \geq (1-o(1))^2 \left\lfloor \frac{m}{2} \right\rfloor \left(1 - \frac{M}{2}\right).
\ee
By using (\ref{eq: up m constant}) and (\ref{eq: lb m constant 2}), we obtain
\begin{eqnarray}
\frac{R^{\rm ub}(n,m,M, \qv, \widetilde m)}{R^{\rm lb}(n,m,M,\qv) } &\leq&  \frac{m}{(1-o(1))^2 \left\lfloor \frac{m}{2} \right\rfloor \left(1 - \frac{M}{2}\right)} \notag\\
&\leq& \frac{m}{\left(\frac{m}{2}-1\right)\left(1-\frac{M}{2}\right)} \notag\\
&=& \frac{1}{\left(\frac{1}{2} - \frac{1}{m}\right)\left(1-\frac{M}{2}\right)} \notag\\
&\geq& \frac{1}{\left(\frac{1}{2} - \frac{1}{3}\right)\left(1-\frac{1 + \epsilon}{2}\right)} \notag\\ 
&=& \frac{12}{1-\epsilon}. 
\end{eqnarray}

\subsection{When $1+\epsilon < M \leq \frac{m}{6}$}

Letting $z = \left\lfloor \frac{m}{2M} \right\rfloor$, and  using (\ref{eq: R lower 1 c}), we obtain
\begin{eqnarray}
\label{eq: lb m constant 3}
R^{\rm lb}(n,m,M,\qv) % &&\geq& (1-o(1))^2\left(z - \frac{z}{\left\lfloor \frac{m}{z} \right\rfloor} M\right) \notag\\
%&=& (1-o(1))^2\left(\left\lfloor \frac{m}{2M} \right\rfloor - \frac{\left\lfloor \frac{m}{2M} \right\rfloor}{\left\lfloor \frac{m}{\left\lfloor \frac{m}{2M} \right\rfloor} \right\rfloor} M\right) \notag\\
&\geq&  (1-o(1))^2 \left(\frac{m}{2M} - 1\right)\left(1 - \frac{M}{\lfloor 2M \rfloor}\right) \notag\\
&\buildrel (a) \over\geq& (1-o(1))^2 \left(\frac{m}{2M} - 1\right)\left(1 - \frac{3}{4}\right) \notag\\
&=& (1-o(1))^2 \left(\frac{m}{2M} - 1\right)\frac{1}{4}, 
\end{eqnarray}
where (a) is because that when $M > 1+\epsilon$, $\frac{M}{\lfloor 2M \rfloor} \leq \frac{3}{4}$. 
Then, by using (\ref{eq: up m constant}) and (\ref{eq: lb m constant 3}), we have
\begin{eqnarray}
\frac{R^{\rm ub}(n,m,M, \qv, \widetilde m)}{R^{\rm lb}(n,m,M,\qv) } &\leq& \frac{\frac{m}{M}}{(1-o(1))^2 \left(\frac{m}{2M} - 1\right)\frac{1}{4}} \notag\\
&\leq& \frac{4}{\left(\frac{1}{2} - \frac{M}{m}\right)} \notag\\
&\leq& 12.
\end{eqnarray}

\subsection{When $M > \frac{m}{6}$} 

Letting $z = 1$, and using (\ref{eq: R lower 1 c}), we obtain
\be
\label{eq: lb m constant 4}
R^{\rm lb}(n,m,M,\qv)   \geq (1-o(1))^2\left(1-\frac{M}{m}\right).
\ee
Hence, using (\ref{eq: up m constant}) and (\ref{eq: lb m constant 4}), we have
\be
\label{eq: gap sl 2bis}
\frac{R^{\rm ub}(n,m,M,\qv,\widetilde{m})}{R^{\rm lb}(n,m,M,\qv) } \leq \frac{\frac{m}{M} - 1}{(1-o(1))^2\left(1-\frac{M}{m}\right)} \leq \frac{m}{M} \leq 6. 
\ee

\newpage

\bibliographystyle{IEEEbib}
\bibliography{references_GC,references_d2d}

\end{document}

%% file: macros.tex
\long\def\comment#1{}

\newfont{\bbb}{msbm10 scaled 700}

\newfont{\bb}{msbm10 scaled 1100}

\newcommand{\PP}{\mbox{\bb P}}
\newcommand{\RR}{\mbox{\bb R}}

\newcommand{\FF}{\mbox{\bb F}}

\newcommand{\EE}{\mbox{\bb E}}

\newcommand{\dv}{{\bf d}}

\newcommand{\fv}{{\bf f}}

\newcommand{\pv}{{\bf p}}
\newcommand{\qv}{{\bf q}}

\newcommand{\Cm}{{\bf C}}

\newcommand{\Qm}{{\bf Q}}

\newcommand{\Dc}{{\cal D}}

\newcommand{\Fc}{{\cal F}}

\newcommand{\Hc}{{\cal H}}
\newcommand{\Ic}{{\cal I}}
\newcommand{\Jc}{{\cal J}}

\newcommand{\Sc}{{\cal S}}

\newcommand{\Uc}{{\cal U}}

\newcommand{\Vc}{{\cal V}}
\newcommand{\Xc}{{\cal X}}

\newcommand{\fsf}{{\sf f}}

\newcommand{\Csf}{{\sf C}}

\newcommand{\Nsf}{{\sf N}}

\newcommand{\Qsf}{{\sf Q}}

\newcommand{\Xsf}{{\sf X}}
\newcommand{\Ysf}{{\sf Y}}
\newcommand{\Zsf}{{\sf Z}}

\renewcommand{\arg}{{\hbox{arg}}}

\newcommand{\eqdef}{\stackrel{\Delta}{=}}

\newcommand{\be}{\begin{equation}}
\newcommand{\ee}{\end{equation}}
\newcommand{\bea}{\begin{eqnarray}}
\newcommand{\eea}{\end{eqnarray}}

\def\fsf{ {\sf f}}

\def\Var{{\rm Var}}